\pdfoutput=1
\documentclass[usenatbib]{mn2e}
\usepackage{hyperref}
\usepackage{apjfonts}
\usepackage{amssymb}
\usepackage{amsmath}
\usepackage{ctable}
\usepackage{url}
\usepackage{breakurl}
\usepackage{fixltx2e} %% forces figure order to remain fixed (otherwise figure*'s can move out-of-order)

\newcommand{\be}{\begin{equation}}
\newcommand{\ee}{\end{equation}}

\newcommand{\msun}{M_{\sun}}

\newcommand{\wengenurl}{\burl{http://www.astrosim.net/code/doku.php}}
\newcommand{\gizmourl}{\burl{http://www.tapir.caltech.edu/~phopkins/Site/GIZMO.html}}
\newcommand{\Ddim}{\nu}
\newcommand{\vspacerpostplot}{\vspace{-0.4cm}}

\newcommand\plotonesize[2]
 {\centering \leavevmode \includegraphics[width={#2\columnwidth}]{#1}}
\newcommand{\plotsidesize}[2]
 {\centering \leavevmode \includegraphics[width={#2\textwidth}]{#1}}
\newcommand{\acknowledgments}{\begin{small}\section*{Acknowledgments}\end{small}}
\newcommand\altaffilmark[1]{$^{#1}$}
\newcommand\altaffiltext[1]{$^{#1}$}
\title[Conservative \&\ Consistent Meshfree Methods]{A New Class of Accurate, Mesh-Free Hydrodynamic Simulation Methods
\vspace{-0.5cm}}

\author[Hopkins et al.]{
\parbox[t]{\textwidth}{ 
Philip F. Hopkins\altaffilmark{1}\thanks{E-mail:phopkins@caltech.edu}
} 
\vspace*{6pt} \\
\altaffiltext{1}{TAPIR, Mailcode 350-17, California Institute of Technology, Pasadena, CA 91125, USA\vspace{-1.1cm}} \\
}

\date{Submitted to MNRAS, August, 2014\vspace{-0.6cm}}
\begin{document}
\maketitle
\label{firstpage}

\begin{abstract}

We present two new Lagrangian methods for hydrodynamics, in a systematic comparison with moving-mesh, SPH, and stationary (non-moving) grid methods. The new methods are designed to simultaneously capture advantages of both smoothed-particle hydrodynamics (SPH) and grid-based/adaptive mesh refinement (AMR) schemes. They are based on a kernel discretization of the volume coupled to a high-order matrix gradient estimator and a Riemann solver acting over the volume ``overlap.'' We implement and test a parallel, second-order version of the method with self-gravity \&\ cosmological integration, in the code {\small GIZMO}:$^{\ref{foot:gizmourl}}$ this maintains exact mass, energy and momentum conservation; exhibits superior angular momentum conservation compared to all other methods we study; does not require ``artificial diffusion'' terms; and allows the fluid elements to move with the flow so resolution is automatically adaptive. We consider a large suite of test problems, and find that on all problems the new methods appear competitive with moving-mesh schemes, with some advantages (particularly in angular momentum conservation), at the cost of enhanced noise. The new methods have many advantages vs.\ SPH: proper convergence, good capturing of fluid-mixing instabilities, dramatically reduced ``particle noise'' \&\ numerical viscosity, more accurate sub-sonic flow evolution, \&\ sharp shock-capturing. Advantages vs.\ non-moving meshes include: automatic adaptivity, dramatically reduced advection errors \&\ numerical overmixing, velocity-independent errors, accurate coupling to gravity, good angular momentum conservation and elimination of ``grid alignment'' effects. We can, for example, follow hundreds of orbits of gaseous disks, while AMR and SPH methods break down in a few orbits. However, fixed meshes minimize ``grid noise.'' These differences are important for a  range of astrophysical problems.

\end{abstract}

\begin{keywords}
methods: numerical --- hydrodynamics --- instabilities --- turbulence --- cosmology: theory
\vspace{-1.0cm}
\end{keywords}

\vspace{-1.1cm}
\section{Introduction: The Challenge of Existing Numerical Methods}
\label{sec:intro}

Numerical hydrodynamics is an essential tool of modern astrophysics, but poses many challenges. A variety of different numerical methods are used, but to date, most hydrodynamic simulations in astrophysics (with some interesting exceptions) are based on one of two popular methods: smoothed-particle hydrodynamics (SPH; \citealt{lucy:1977.sph,gingold.monaghan:1977.sph}), or stationary-grids, which can be either ``fixed mesh'' codes where a time-invariant mesh covers the domain \citep[e.g.][]{zeus:a}, or ``adaptive mesh refinement'' (AMR) where the meshes are static and stationary except when new sub-cells are created or destroyed within parent cells \citep{berger.colella:1989.amr}. 

These methods, as well as other more exotic schemes \citep[e.g.][]{xu:1997.unstructured.mesh.cosmo,zhang:1997.hierarchical.grid.code}, have advantages and disadvantages. Unfortunately, even on simple test problems involving ideal fluid dynamics, they often give conflicting results. This limits their predictive power: in many comparisons, it is unclear whether differences seen owe to physical, or to purely numerical effects (for an example, see e.g.\ the comparison of cosmological galaxy formation in the {\em Aquila} project; \citealt{scannapieco:2012.aquila.cosmo.sim.compare}). Unfortunately, both SPH and AMR have fundamental problems which make them inaccurate for certain problems -- because of this, the ``correct'' answer is often unknown in these comparisons.

In Table~\ref{tbl:methods}, we attempt a cursory summary of some of the methods being used in astrophysics today, making note of some of the strengths and weaknesses of each.\footnote{\label{foot:gizmourl}A public version of the {\small GIZMO} code (which couples the hydrodynamic algorithms described here to a heavily modified version of the parallelization and tree gravity solver of {\small GADGET-3}; \citealt{springel:gadget}) together with movies and additional figures, an extensive user's guide, and the files and instructions needed to run the test problems in this paper, are available at\\ \gizmourl} Below, we describe these in more detail.

\vspace{-0.5cm}
\subsection{Smoothed-Particle Hydrodynamics (SPH)}
\label{sec:intro:sph}

In SPH, quasi-Lagrangian mass elements are followed -- the conserved quantities are discretized into particles (like an N-body code), and a kernel function is used to ``smooth'' their volumetric distributions to determine equations of motion. SPH is numerically stable, Lagrangian (follows the fluid), provides continuous adaptive resolution, has truncation errors which are independent of the fluid velocity, couples trivially to N-body gravity schemes, exactly solves the particle continuity equation, and the equations of motion can be exactly derived from the particle Lagrangian \citep{springel:entropy} giving it excellent conservation properties.\footnote{The ``particle Lagrangian'' and ``particle continuity equation'' are the Lagrangian/continuity equation of a {\em discretized} particle field, where each particle occupies an infinitely small volume. Exactly solving the continuity equation of a continuous fluid, of course, requires infinite resolution. This often leads to SPH being described as correct in the ``molecular limit.'' But at finite resolution, the more relevant limit is actually the ``fluid limit.''} This has led to widespread application of SPH in many fields \citep[for reviews see][]{rosswog:2009.sph.review,springel:2010.sph.review,price:2012.sph.review}.

However, it is well-known that ``traditional'' SPH algorithms have a number of problems. They suppress certain fluid mixing instabilities \citep[e.g.\ Kelvin-Helmholtz instabilities;][]{morris:1996.sph.stability,dilts:1999.sph.stability,ritchie.thomas:2001.egy.wtd.sph,marri:2003.mod.sph.cosmo.sims,okamoto:2003.shear.sph.flows,agertz:2007.sph.grid.mixing}, corrupt sub-sonic (pressure-dominated) turbulence \citep{kitsionas:2009.grid.sph.compare.turbulence,price:2010.grid.sph.compare.turbulence,bauer:2011.sph.vs.arepo.shocks,sijacki:2011.gadget.arepo.hydro.tests}, produce orders-of-magnitude higher numerical viscosity in flows which should be inviscid (leading to artificial angular momentum transfer; \citealt{cullen:2010.inviscid.sph}), over-smooth shocks and discontinuities, introduce noise in smooth fields, and numerically converge very slowly. 

The sources of these errors are known, however, and heroic efforts have been made to reduce them in ``modern'' SPH. First, the SPH equations of motion are inherently inviscid, so require some artificial viscosity to capture shocks and make the method stable; this generally leads to excessive diffusion (eliminating on of SPH's main advantages). One improvement is to simply insert a Riemann solver between particles (so-called ``Godunov SPH''; see \citealt{inutsuka:2002.godunov.sph,cha:2003.godunov.sph}), but this is not stable on many problems; another improvement is to use higher-order switches (based on the velocity gradients and their time derivatives) for the diffusion terms \citep{cullen:2010.inviscid.sph,read:2012.sph.w.dissipation.switches}. 

Second, a significant part of SPH's suppression of fluid mixing comes from a ``surface-tension''-like error at contact discontinuities and free surfaces, which can be eliminated by kernel-smoothing all quantities (e.g.\ pressure), not just density, in the equations of motion \citep{ritchie.thomas:2001.egy.wtd.sph,saitoh:2012.dens.indep.sph,hopkins:lagrangian.pressure.sph}. Recently, it has also been realized that artificial diffusion terms should to be added for other quantities such as entropy \citep{price:2008.sph.contact.discontinuities,wadsley:2008.sph.mixing.cosmology}; and these further suppress errors at discontinuities by ``smearing'' them. 

Third, and perhaps most fundamental, SPH suffers from low-order errors, in particular the so-called ``E0'' zeroth-order error
\citep{morris:1996.sph.stability,dilts:1999.sph.stability,read:2010.sph.mixing.optimization}. It is straightforward to show that the discretized SPH equations are not {\em consistent} at any order, meaning they cannot correctly reproduce even a constant (zeroth-order) field, unless the particles obey {\em exactly} certain very specific geometric arrangements. This produces noise, often swamping real low-amplitude effects. Various ``corrected'' SPH methods have been proposed which eliminate some of these errors in the equation of motion \citep[e.g.][]{morris:1996.sph.stability,abel:2011.sph.pressure.gradient.est,garciasenz:2012.integral.sph}: however, thus far {\em all} such methods require numerically unstable violations of energy and momentum conservation, leading to exponentially growing errors in realistic problems \citep[see e.g.][]{price:2012.sph.review}. Adding terms to force them to be conservative re-instates the original problem by violating consistency at zeroth order, although it can still improve accuracy compared to other choices for the SPH equations of motion \citep{garciasenz:2012.integral.sph,rosswog:2014.sph.accuracy}. But in any case, these fixes also do not eliminate all the low-order inconsistencies. The only way to decrease all such errors is to increase the number of neighbors in the SPH kernel; using higher-order kernels with several hundred neighbors, instead of the ``traditional'' $\sim 32$ \citep{read:2010.sph.mixing.optimization,dehnen.aly:2012.sph.kernels}.\footnote{It is sometimes said that ``SPH does not converge,'' or that ``SPH is a second-order method'' (i.e.\ converges as $N^{-2}$ in a smooth 1D problem). Both of these are incorrect. SPH does converge at second order, but {\em only} in the limit where the number of neighbors inside the smoothing kernel goes to infinity ($N_{\rm NGB} \rightarrow \infty$), which eliminates the zeroth-order terms that do not converge away with increasing total particle number $N$ alone. However increasing $N_{\rm NGB}$ is both expensive and leads to a loss of resolution (and in most actual practice is not actually done correctly as $N$ increases). So the practical convergence rates of SPH are very slow \citep[see][]{zhu:2014.sph.convergence}.}

However, all of these improvements have costs. As such, it is unclear how ``modern'' SPH schemes compare with other methods.

\vspace{-0.5cm}
\subsection{Stationary-Grid Methods}
\label{sec:intro:grid}

In grid-based methods, the volume is discretized into points or cells, and the fluid equations are solved across these elements. These methods are well-developed, with decades of work in computational fluid dynamics. The most popular modern approach is embodied in finite-volume Godunov schemes,\footnote{Older, finite-difference methods simply discretized the relevant equations onto interpolation points in a lattice, but these methods often do not conserve quantities like mass, momentum, and energy, require artificial viscosity/diffusion/hyperviscosity terms (as in SPH), and can be numerically unstable under long integrations for sufficiently complicated problems. As such they have proven useful mostly for weak linear-regime flows where strong shocks are absent and growth of e.g.\ momentum errors will not corrupt the entire domain; here there can be significant advantages from the fact that such methods very easily generalize to higher-order.} which offer higher-order consistency,\footnote{Typically second-order, or third-order in the case of PPM methods. Some schemes claim much higher-order; however, it is almost always the case that this is true only for a sub-set of the scheme (e.g.\ a gradient estimator). In our convention, the order represents the convergence rate, which is limited by the lowest-order aspect of the method.} numerical stability and relatively low diffusivity, and conservation of mass, linear momentum, and energy. 

However, there are errors in these methods as well. At fixed resolution, grid codes have much larger advection errors compared to quasi-Lagrangian methods, when fluids (especially with sharp gradients) move across cells. These errors produce  artificial diffusion, and can manifest as unphysical forces. For example, rotating disks are ``torqued'' into alignment with the grid cardinal axes (``grid alignment''; see e.g.\ \citealt{hahn:2010.disk.gal.orientations.ramses}), shocks preferentially heat, propagate along, and ``break out of'' the grid axes (``carbuncle'' instabilities; \citealt{peery:carbuncle.discovery}), and contact discontinuities are ``smeared out'' upon advection. Related to this, angular momentum is not conserved: at realistic resolutions for many problems, gaseous orbits can be degraded within a couple orbital times (unless special coordinates are used, which is only possible if the problem geometry is known ahead of time). The errors in these methods are also velocity-dependent: unlike SPH, ``boosting'' the fluid (so it uniformly moves across the grid) increases diffusion across the problem \citep[see][]{wadsley:2008.sph.mixing.cosmology,tasker:2008.gas.turb.vs.gal.prop,springel:arepo} and suppresses fluid mixing instabilities \citep{springel:arepo}. Grid methods also require special fixes (e.g.\ energy-entropy switches; \citealt{ryu:1993.entropy.switch.cosmo.grid.hydro,bryan:1995.cosmo.ppm}) to deal with highly super-sonic flows. Free surfaces and steep ``edges'' (e.g.\ water flow in air, or sharp surfaces of planets/stars) require extremely high resolution to maintain. The inherent mis-match between particle-based N-body methods and cell-based hydro methods means that various errors appear when the hydrodynamics are coupled to gravity, making it difficult for such methods to handle simple situations like self-gravitating hydrostatic equilibrium \citep[see][]{muller:1995.grid.code.gravity.problems,leveque:1998.godunov.source.term.balance,zingale:2002.grid.hydro.eqm.issues}. Worse, these errors can introduce spurious instabilities \citep{truelove:1997.jeans.condition,truelove:1998.gmc.frag}.

In AMR methods, the fact that refinement boundaries are necessarily dis-continuous entails a significant loss of accuracy at the boundaries (the method becomes effectively lower-order).  This means convergence is slower. When coupled to gravity, various studies have shown these errors suppress low-amplitude gravitational instabilities (e.g.\ those that seed cosmological structure formation), and violate conservation in the long-range forces whenever cells are refined or de-refined \citep{oshea:sph.tests,heitmann:2008.cosmic.code.comparison,springel:arepo}. 

Again, significant effort has gone into attempts to reduce these sources of error. Higher-order WENO-type schemes for gradients can help reduce edge effects. Various authors have implemented partially-Lagrangian or ``abitrary Lagrange-Eulerian'' (ALE) schemes where partial distortion of the mesh is allowed, but then the meshes are re-mapped to regular meshes; or ``patch'' schemes in which sub-meshes are allowed to move, then mapped to larger ``parent meshes'' \citep[see e.g.][]{gnedin:1995.quasi.lagrangian.patch.scheme,pen:1998.moving.amr.patch.scheme,trac:2004.moving.frame.grid.code,murphy:2008.ale.code.sne}. However, these approaches usually require foreknowledge of the exact problem geometry to work well. And the re-mapping is a diffusive operation, so some of the errors above are actually enhanced.

\vspace{-0.5cm}
\subsection{Moving, Unstructured Meshes}
\label{sec:intro:movingmesh}

Recently, there has been a surge in interest in moving, unstructured mesh methods. These methods are well-known in engineering \citep[see e.g.][]{mavriplis:1997.unstructured.grid.review}, and there have been earlier applications in astrophysics \citep[e.g.][]{whitehurst:1995.moving.mesh.protocode,xu:1997.unstructured.mesh.cosmo}, but recently considerable effort has gone into development of more flexible examples \citep{springel:arepo,duffell:2011.TESS,gaburov:2012.public.moving.mesh.code}. These use a finite-volume Godunov method, but partition the volume into non-regular cells using e.g.\ a Voronoi tessellation, and allow the cells to move and deform continuously.

In many ways, moving meshes capture the advantages of both SPH and AMR codes: like SPH they can be Lagrangian and adapt resolution continuously, feature velocity-independent truncation errors, couple well to gravity, and avoid preferred directions, while also like AMR treat shocks, shear flows, and fluid instabilities with high accuracy and eliminate many sources of noise, low-order errors, and artificial diffusion terms. 

However, such methods are new, and still need to be tested to determine their advantages and disadvantages. It is by no means obvious that they are optimal for all problems, nor that they are the ``best compromise'' between Lagrangian (e.g.\ SPH) and Eulerian (e.g.\ grid) methods. And there are problems the method does not resolve. Angular momentum is still not formally conserved in moving meshes, and it is not obvious (if the cell shapes are sufficiently irregular) how much it improves on stationary-grid codes. ``Mesh-deformation'' and ``reconnection'' in which distortions to the mesh lead to highly irregular cell shapes, is inevitable in complicated flows. This can lead to errors which effectively reduce the accuracy and convergence of the method, and would eventually crash the code. This is dealt with by some ``mesh regularization,'' by which the cells are re-shaped or prevented from deforming (i.e.\ made ``stiff'' or resistant to deformations). But such regularization obviously risks re-introducing some of the errors of stationary-grid methods which the moving-mesh method tries to avoid (the limit of a sufficiently stiff grid is simply a fixed-grid code with a uniform drift). And dis-continuous cell refinement/de-refinement or ``re-connection'' is inevitable when the fluid motion is complicated, introducing some of the same errors as in AMR.

\vspace{-0.5cm}
\subsection{Structure of This Paper}
\label{sec:intro:gizmo}

Given the above, the intent of this paper is two-fold. 

First, we will introduce and develop two new methods for solving the equations of hydrodynamics which attempt to simultaneously capture advantages of both Lagrangian and Eulerian methods (\S~\ref{sec:methods}). The methods build on recent developments in the fluid dynamics community, especially \citet{lanson.vila:2008.meshfree.consistency,lanson.vila:2008.meshfree.convergence}, but have not generally been considered in astrophysics, except for recent efforts by \citet{gaburov:2011.meshless.dg.particle.method}. The methods move with the flow in a Lagrangian manner, adapt resolution continuously, eliminate velocity-dependent truncation errors, couple simply to N-body gravity methods, have no preferred directions, do not require artificial diffusion terms, capture shocks, shear flows, and fluid instabilities with high accuracy, and exhibit remarkably good angular momentum conservation. We will show how these methods can be implemented into {\small GIZMO}, a new (heavily modified) version of the flexible, parallel {\small GADGET-3} code.\footnote{Detailed attribution of different algorithms in the code, and descriptions of how routines originally written for {\small GADGET-3} have been modified, can be found in the public source code.}

Second, we will consider a systematic survey of a wide range of test problems (\S~\ref{sec:hydro.test}-\ref{sec:performance}), comparing both new methods, moving-mesh, modern stationary-grid, and both ``traditional'' and ``modern'' SPH methods. This is intended not just to validate our new methods, but also to test the existing major classes of numerical methods on a wide range of problems, to assess some of their relative strengths and weaknesses in different contexts.

%We briefly describe the methods (\S~\ref{sec:reference}), and then consider their performance in a series of tests, spanning equilibrium/steady-state configurations (\S~\ref{sec:tests:equilibrium}), strong shocks and highly non-linear evolution (\S~\ref{sec:tests:shocks}), a range of fluid mixing instabilities (\S~\ref{sec:tests:mixing}), and tests with strong self-gravity, cosmological integrations, and non-linear additional physics such as cooling, star formation, stellar explosions, and radiation-gas coupling (\S~\ref{sec:tests:gravity}). In \S~\ref{sec:performance}, we summarize the CPU performance (speed) of the different methods on the different classes of test problems.
%Finally, in \S~\ref{sec:discussion}, we summarize our conclusions and advantages/disadvantages of the new methods compared to SPH, AMR, and moving-mesh codes, and conclude with a discussion of possibilities for further improvements of the method. 

\begin{footnotesize}
\ctable[
  caption={{\normalsize Summary of Some Popular Numerical Hydrodynamics Methods}\label{tbl:methods}},center,star
  ]{lcccccccc}{
\tnote[ ]{A crude description of various numerical methods which are referenced throughout the text. Note that this list is necessarily incomplete, and specific sub-versions of many codes listed have been developed which do not match the exact descriptions listed. They are only meant to broadly categorize methods and outline certain basic properties.\\
{\bf (1)} Method Name: Methods are grouped into broad categories. For each we give more specific sub-categories, with a few examples of commonly-used codes this category is intended to describe.\\
{\bf (2)} Order: Order of consistency of the method, for smooth flows (zero means the method cannot reproduce a constant).  ``Corrected'' SPH is first-order in the pressure force equation, but zeroth-order otherwise. Those with 2-3 listed depend on whether PPM methods are used for reconstruction (they are not 3rd order in all respects). Note that all the high-order methods become 1st-order at discontinuities (this includes refinement boundaries in AMR).\\
{\bf (3)} Conservative: States whether the method manifestly conserves mass, energy, and linear momentum ($\checkmark$), or is only conservative up to integration accuracy ($\times$).\\
{\bf (4)} Angular Momentum: Describes the {\em local} angular momentum (AM) conservation properties, {\em when the AM vector is unknown or not fixed in the simulation}. In this regime, no method which is numerically stable exactly conserves local AM (even if {\em global} AM is conserved). Either the method has no AM conservation ($\times$), or conserves AM up to certain errors, such as the artificial viscosity and gradient errors in SPH. If the AM vector is known and fixed (e.g.\ for test masses around a single non-moving point-mass), it is always possible to construct a method (using cylindrical coordinates, explicitly advecting AM, etc.) which perfectly conserves it.\\
{\bf (5)} Numerical Dissipation: Source of numerical dissipation in e.g.\ shocks. Either this comes from an up-wind/Riemann solver type scheme (where diffusion comes primarily from the slope-limiting scheme; \citealt{toro:2009.unstructured.mesh.reconstruction}), or artificial viscosity/conductivity/hyperdiffusion terms.\\
{\bf (6)} Integration Stability: States whether the method has long-term integration stability (i.e.\ errors do not grow unstably).\\
{\bf (7)} Number of Neighbors: Typical number of neighbors between which hydrodynamic interactions must be computed. For meshless methods this is the number in the kernel. For mesh methods this can be either the number of faces (geometric) when a low-order method is used or a larger number representing the stencil for higher-order methods.\\
{\bf (8)} Known Difficulties: Short summary of some known problems/errors common to the method. An incomplete and non-representative list! These are described in actual detail in the text. ``Velocity-dependence'' (as well as comments about noise and lack of conservation) here refers to the property of the {\em errors}, not the converged solutions. Any well-behaved code is conservative (of mass/energy/momentum/angular momentum), Galilean-invariant, noise-free, and captures the correct level of fluid mixing instabilities in the fully-converged (infinite-resolution) limit.
}
}{
\hline\hline
\multicolumn{1}{c}{} &
\multicolumn{1}{c}{} & 
\multicolumn{1}{c}{Conservative?} & 
\multicolumn{1}{c}{Conserves} & 
\multicolumn{1}{c}{} & 
\multicolumn{1}{c}{Long-Time} & 
\multicolumn{1}{c}{Number} &
\multicolumn{1}{c}{} \\
\multicolumn{1}{c}{Method} &
\multicolumn{1}{c}{Consistency} & 
\multicolumn{1}{c}{(Mass/Energy} & 
\multicolumn{1}{c}{Angular} & 
\multicolumn{1}{c}{Numerical} & 
\multicolumn{1}{c}{Integration} & 
\multicolumn{1}{c}{of} &
\multicolumn{1}{c}{Known} \\
\multicolumn{1}{c}{Name} &
\multicolumn{1}{c}{/Order} & 
\multicolumn{1}{c}{/Momentum)} & 
\multicolumn{1}{c}{Momentum} & 
\multicolumn{1}{c}{Dissipation} & 
\multicolumn{1}{c}{Stability?} & 
\multicolumn{1}{c}{Neighbors} &
\multicolumn{1}{c}{Difficulties} \\
\hline\\
{\bf Smoothed-Particle Hydro.\ (SPH)} & & & & & & &  \\
\hline
``Traditional'' SPH & 0 & $\checkmark$ & up to AV & artificial  & $\checkmark$ & $\sim32$ & fluid mixing, noise,  \\
({\small GADGET}, {\small TSPH}) & \, & \, & \, & viscosity (AV) & \, & \, & E0 errors \\
\hline
``Modern'' SPH & 0 & $\checkmark$ & up to AV & AV+conduction  & $\checkmark$ & $\sim128-442$ & excess diffusion,  \\
({\small P-SPH}, {\small SPHS}, {\small PHANTOM}, {\small SPHGal}) & \, & \, & \, & +switches & \, & \, & E0 errors \\
\hline
``Corrected'' SPH & 0-1 & $\times$ & $\times$ & artificial  & $\times$ & $\sim32$ & errors grow  \\
({\small rpSPH}, {\small Integral-SPH}, {\small Morris96 SPH}, & \, & \, & \, & viscosity & \, & \, & non-linearly, \\
{\small Moving-Least-Squares SPH}) & \, & \, & \, & \, & \, & \, & ``self-acceleration''  \\
\hline
``Godunov'' SPH & 0 & $\checkmark$ & up to & Riemann  & $\checkmark$ & $\sim300$ & instability,  \\
({\small GSPH}, {\small GSPH-I02}, {\small Cha03 SPH}) & \, & \, & gradient & solver + & \, & \, & expense,  \\
\, & \, & \, & errors & slope-limiter\, & \, & \, & E0 errors remain \\
\hline\hline\\
{\bf Finite-Difference Methods} & & & & & & &  \\
\hline
Gridded/Lattice Finite Difference  & 2-3 & $\times$ & $\times$ & artificial  & $\times$ & $\sim8-128$ & instability,  \\
({\small ZEUS} [some versions], {\small Pencil} code) & \, & \, & \, & viscosity & \, & \, & lack of \\
Lagrangian Finite Difference & \, & \, & \, & \, & \, & $\sim60$ & conservation, \\
({\small PHURBAS}, {\small FPM}) & \, & \, & \, & \, & \, & \, & advection errors \\
\hline\hline\\
{\bf Finite-Volume Godunov Methods} & & & & & & &  \\
\hline
Static Grids  & 2-3 & $\checkmark$ & $\times$ & Riemann  & $\checkmark$ & $\sim8$ & over-mixing,  \\
({\small ATHENA}, {\small PLUTO}) & \, & \, & \, & solver + & \, & (geometric) & ang.\, mom.,  \\
\, & \, & \, & \, & slope-limiter & \, & $\sim8-125$ & velocity-dependent \\
\, & \, & \, & \, & \, & \, & (stencil) & errors (VDE) \\
\hline
Adaptive-Mesh Refinement (AMR)  & 2-3 & $\checkmark$ & $\times$ & Riemann  & $\checkmark$ & $\sim8-48$ & over-mixing,  \\
({\small ENZO}, {\small RAMSES}, {\small FLASH}) & (1) & \, & \, & solver + & \, & $\sim24-216$ & ang.\, mom., VDE, \\
\, & \, & \, & \, & slope-limiter & \, & \, & refinement criteria \\
\hline
Moving-Mesh Methods  & 2 & $\checkmark$ & $\times$ & Riemann  & $\checkmark$ & $\sim13-30$ & mesh deformation,  \\
({\small AREPO}, {\small TESS}, {\small FVMHD3D}) & \, & \, & \, & solver + & \, & \, & ang.\, mom.\, (?),  \\
\, & \, & \, & \, & slope-limiter & \, & \, & ``mesh noise'' \\
\hline\hline\\
{\bf New Methods In This Paper} & & & & & & &  \\
\hline
Meshless Finite-Mass  & 2 & $\checkmark$ & up to & Riemann  & $\checkmark$ & $\sim32$ & partition noise  \\
\&\ Meshless Finite-Volume & \, & \, & gradient & solver + & \, & \, & ? \\
({\small MFM}, {\small MFV}) & \, & \, & errors & slope-limiter & \, & \, & (TBD) \\
\hline\hline\\
}
\end{footnotesize}

\vspace{-0.5cm}
\section{A New Numerical Methodology for Hydrodynamics}
\label{sec:methods}

In the last two decades, there has been tremendous effort in the computer science, engineering, and fluid dynamics literature, directed towards the development of new mesh-free algorithms for hydrodynamics; but much of this has not been widely recognized in astrophysics \citep[see e.g.][]{hietel:2000.finite.volume.particle.method}. 
Various authors have pointed out how matrix and least-squares methods can be used to define consistent, higher-order gradient operators, and renormalization schemes can be used to eliminate the zeroth-order errors of methods like SPH \citep[see e.g.][]{dilts:1999.sph.stability,onate:1996.fpm,kuhnert:2003.finite.pointset.method,tiwari:2003.finite.pointset.method,liu:2005.finite.particle.method}. Most of this has propagated into astrophysics in the form of ``corrected'' SPH methods, which partially-implement such methods as ``fixes'' to certain operators \citep[e.g.][]{garciasenz:2012.integral.sph,rosswog:2014.sph.accuracy} or in finite point methods, which simply treat all points as finite difference-like interpolation points rather than assigning conserved quantities \citep{maron:2003.gradient.particle.mhd,maron:2012.phurbas.algorithm}. However these implementations often sacrifice conservation (of quantities like mass, momentum, and energy) and numerical stability. Meanwhile, other authors have realized that the uncertain and poorly-defined artificial diffusion operators can be eliminated by appropriate solution of a Riemann problem between particle faces; this has generally appeared in the form of so-called ``Godunov SPH'' \citep{cha:2003.godunov.sph,inutsuka:2002.godunov.sph,cha:2010.godunov.sph,murante:2011.godunov.sph}. However on its own this does not eliminate other low-order SPH errors, and those errors can de-stabilize the solutions. 

A particularly intriguing development was put forward by \citet{lanson.vila:2008.meshfree.consistency,lanson.vila:2008.meshfree.convergence}. These authors showed that the advances above could be synthesized into a new, meshfree finite-volume method which is both consistent and fully conservative. This is a fundamentally different method from any SPH ``variant'' above; it is much closer in spirit to moving-mesh methods. Critically, rather than just attaching individual fixes piece-wise to an existing method, they re-derived the discrete operators from a consistent mathematical basis. A first attempt to implement these methods in an astrophysical context was presented in \citet{gaburov:2011.meshless.dg.particle.method},\footnote{The source code from that study ({\small WPMHD}) is available at\\
\url{https://github.com/egaburov/wpmhd/tree/orig}} and the results there for both hydrodynamic and magneto-hydrodynamic test problems appeared extremely encouraging. We therefore explore and extend two closely-related versions of this method here.

\begin{figure*}
    \plotsidesize{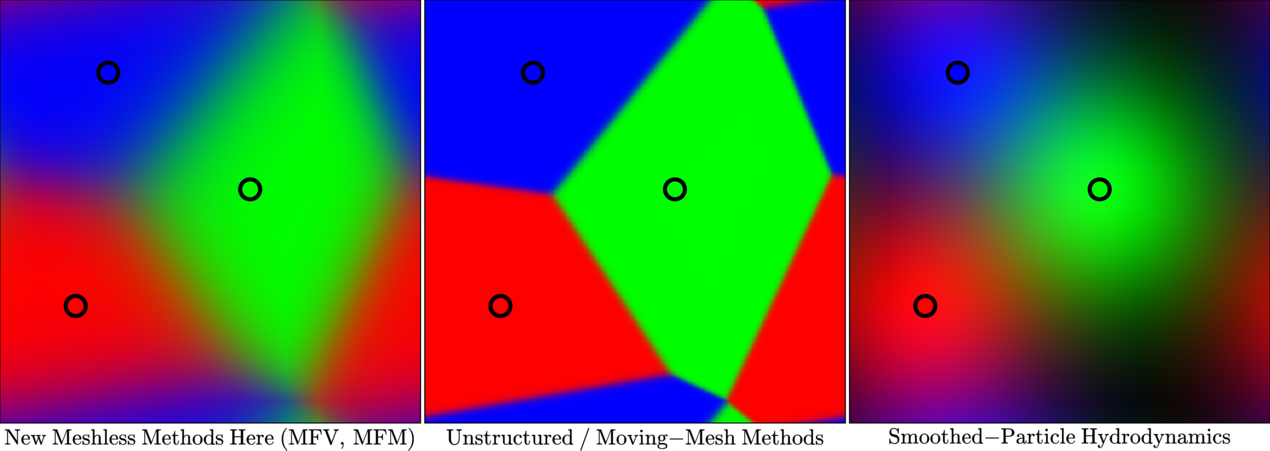}{0.98}
    \vspace{-0.1cm}
    \caption{Illustration of key conceptual differences between some of the methods here. For an irregularly distributed set of sampling/grid points or ``particles'' (black circles) with locations ${\bf x}_{i}$, we require a way to partition the volume to solve the equations of hydrodynamics between them. {\em Left:} The meshless finite-mass (MFM) and meshless finite-volume (MFV) methods here. The volume partition is given by the weighted kernel at each point (Eq.~\ref{eqn:wt.fn.kernel}); here the red/green/blue color channels represent the fraction of the volume at each point associated with the corresponding particle ($\psi_{i}({\bf x})$). Here we apply the same kernel function and typical kernel ``width'' as in the text. Note that this returns a Voronoi tessellation with the boundaries ``smoothed.'' Despite the {\em kernel} function being spherical, the {\em domains} associated with each particle are not, and the entire volume is represented. The fluid equations are then solved by integrating over the domain of each particle/cell.
    {\em Center:} The unstructured/moving-mesh partition. Now the boundaries are strict step functions at the faces given by the tessellation. Note that this is (exactly) the limit of our MFM/MFV method for an infinitely sharply-peaked kernel function; technically the moving-mesh method is a special case of the MFV method. The volume integrals are then reduced to surface integrals across the faces.
    {\em Right:} The SPH partition. In SPH the contribution to volume integrals behaves as the kernel, centered on each particle location; the whole volume is ``counted'' only when the kernel size is infinitely large compared to the inter-particle spacing (number of neighbors is infinite). The equations of motion are evaluated at the particle locations ${\bf x}_{i}$, using the weighted-average volumetric quantities from the volume partition.\vspacerpostplot
    \label{fig:method}}
\end{figure*}

\vspace{-0.5cm}
\subsection{Derivation of the Meshless Equations of Motion}
\label{sec:methods:deriv}

We begin with a derivation of the discretized equations governing the new numerical schemes. This will closely follow \citet{gaburov:2011.meshless.dg.particle.method}, and is aimed towards practical aspects of implementation. A fully rigorous mathematical formulation of the method, with proofs of various consistency, conservation, and convergence theorems, is presented in \citet{lanson.vila:2008.meshfree.consistency,lanson.vila:2008.meshfree.convergence,ivanova:2013.gaburov.appendix.meshless.derivations}. 

The homogeneous Euler equations for hydrodynamics are ultimately a set of conservation laws for mass, momentum, and energy, which form a system of hyperbolic partial differential equations in a frame moving with velocity ${\bf v}_{\rm frame}$ of the form 
\begin{align}
\label{eqn:conservation}
\frac{\partial {\bf U}}{\partial t} + \nabla\cdot({\bf F}-{\bf v}_{\rm frame}\otimes{\bf U}) &= 0
\end{align}
where $\nabla\cdot {\bf F}$ refers to the inner product between the gradient operator and tensor ${\bf F}$, $\otimes$ is the outer product, ${\bf U}$ is the ``state vector'' of conserved (in the absence of sources) variables, 
\begin{align}
{\bf U}  &= 
\left(
\begin{array}{c}
\rho \\
\rho\,{\bf v} \\
\rho\,e \\
\end{array}
\right)
=
\left(
\begin{array}{c}
\rho \\
\rho\,{\bf v} \\
\rho\,u + \frac{1}{2}\,\rho\,|{\bf v}|^{2} \\
\end{array}
\right)
=
\left(
\begin{array}{c}
\rho \\
\rho\,{v_{x}} \\
\rho\,{v_{y}} \\
\rho\,{v_{z}} \\
\rho\,u + \frac{1}{2}\,\rho\,|{\bf v}|^{2} \\
\end{array}
\right)
\end{align}
(where $\rho$ is mass density, $e$ is the total specific energy, $u$ the specific internal energy, and the last equality expands the compact form of ${\bf v}$ in 3 dimensions), 
and the tensor ${\bf F}$ is the flux of conserved variables
\begin{align}
{\bf F}  &= 
\left(
\begin{array}{c}
\rho\,{\bf v} \\
\rho\,{\bf v}\otimes{\bf v} + P\,\mathcal{I}\\
(\rho\,e + P)\,{\bf v} \\
\end{array}
\right)
\end{align}
where $P$ is the pressure, and $\mathcal{I}$ is the identity tensor.

As in the usual Galerkin-type method, to deal with non-linear and discontinuous flows we begin by determining the weak solution to the conservation equation. We multiply Eq.~\ref{eqn:conservation} by a test function $\phi$, integrate over the domain $\Omega$ (in space such that $d\Omega = d^{\Ddim}{\bf x}$, where $\Ddim$ is the number of spatial dimensions), and follow an integration by parts of the $\phi\,\nabla\cdot {\bf F}$ term to obtain:
\begin{align}
0 &= \int_{\Omega}\left(\frac{d{{\bf U}}}{dt}\,\phi - {\bf F}\cdot\nabla\phi\right)\,d\Omega + \int_{\partial \Omega} ({\bf F}\,\phi)\cdot \hat{{\bf n}}_{\partial \Omega}\,d\,\partial \Omega
\end{align}
where $d{f}/dt \equiv \partial f/\partial t + {\bf v}_{\rm frame}({\bf x},\,t)\cdot\nabla f$ is the co-moving derivative of any function $f$, and  $\hat{{\bf n}}_{\partial \Omega}$ is the normal vector to the surface $\partial \Omega$. 
The test function $\phi=\phi({\bf x},\,t)$ is taken to be an arbitrary (differentiable) Lagrangian function ($d\phi/dt = 0$). Assuming the fluxes and/or $\phi$ vanish at infinity, we can eliminate the boundary term and pull the time derivative outside of the integral \citep[see][]{luo:2008.dg.methods} to obtain 
\begin{align}
\label{eqn:dg.solution}
0 &= \frac{d}{dt}\int_{\Omega} {{\bf U}}({\bf x},\,t)\,{\phi}\ d^{\Ddim}{\bf x} - \int_{\Omega}{\bf F}({\bf U},\,{\bf x},\,t)\cdot \nabla\,\phi\ d^{\Ddim}{\bf x}
\end{align}

To discretize this integral, we must now choose how to discretize the domain volume onto a set of points/cells/particles $i$ with coordinates ${\bf x}_{i}$. If we chose to do so by partitioning the volume between the ${\bf x}_{i}$ with a Voronoi mesh, we would obtain the moving-mesh method of codes like {\small AREPO} with the more accurate gradient estimators implemented in \citet{mocz:2014.galerkin.arepo}. Here we consider a mesh-free alternative, following \citet{lanson.vila:2008.meshfree.consistency,lanson.vila:2008.meshfree.convergence,gaburov:2011.meshless.dg.particle.method}. Consider a differential volume $d^{\Ddim}{\bf x}$, at arbitrary coordinates ${\bf x}$; we can partition that differential volume fractionally among the nearest particles/cells\footnote{In this paper, we will use the terms ``particles'' and ``cells'' interchangeably when referring to our MFM and MFV methods, since each ``particle'' can just as well be thought of as a mesh-generating point which defines the volume domain (or ``cell'') whose mean fluid properties are represented by the particle/cell-carried quantities.} through the use of a weighting function $W$, i.e.\ associate a fraction $\psi_{i}({\bf x})$ of the volume $d^{\Ddim}{\bf x}$ with particle $i$ according to a function $W({\bf x}-{\bf x}_{i},\,h({\bf x}))$: 
\begin{align}
\label{eqn:wt.fn.kernel}
\psi_{i}({\bf x}) &\equiv \frac{1}{\omega({\bf x})}\,W({\bf x}-{\bf x}_{i},\,h({\bf x})) \\
\omega({\bf x}) &\equiv \sum_{j} W({\bf x}-{\bf x}_{j},\,h({\bf x}))
\end{align}
where $h({\bf x})$ is some ``kernel size'' that enters $W$. In other words, the weighting function determines how the volume at any point ${\bf x}$ should be partitioned among the volumes ``associated with'' the tracer points $i$. Note that $W$ can be, in principle, any arbitrary function; the term $\omega({\bf x})^{-1}$ normalizes the weights such that the total volume always sums correctly (i.e.\ the sum of fractional weights must always be unity at every point). That said, to ensure the second-order accuracy of the method, conservation of linear and angular momentum, and locality of the hydrodynamic operations, the function $W({\bf x}-{\bf x}_{i},\,h({\bf x}))$ must be continuous, have compact support (i.e.\ have $W= 0$ for sufficiently large $|{\bf x}-{\bf x}_{i}| \gg h({\bf x})$), and be symmetric (i.e.\ depend only on the absolute value of the coordinate differences $|x - x_{i}|$, $|y-y_{i}|$, etc.). Because of the normalization by $\omega({\bf x})$, the absolute normalization of $W$ is irrelevant; so without loss of generality we take it (for convenience) to be normalized such that $1=\int W({\bf x}-{\bf x}^{\prime},\,h({\bf x}))\,d^{\Ddim}{\bf x}^{\prime}$. 

An example of this is shown in Fig.~\ref{fig:method}, with (for comparison), the volume partitions used in moving-mesh and SPH methods. We construct a two-dimensional periodic box of side-length unity with three randomly placed particles, and use a cubic spline kernel for $W$ with kernel length $h$ set to the equivalent of what would contain $\approx 32$ neighbors in 3D. We confirm that the entire volume is indeed partitioned correctly, like a Voronoi tessellation with the ``edges'' between particles smoothed (avoiding discontinuities in the ``mesh deformation'' as particles move\footnote{Throughout, when we refer to ``mesh deformation,'' we refer to the fact that when particles move, the volume partition -- i.e.\ the map between position and association of a given volume element with different particles/cells -- changes. This occurs constantly in Lagrangian codes (SPH/MFM/MFV, and moving meshes), regardless of whether or not the partition is explicitly re-constructed each timestep or differentially ``advected.''}). In the limit where $W$ is sufficiently sharply-peaked, we can see from Eq.~\ref{eqn:wt.fn.kernel} that we should recover {\em exactly} a Voronoi tessellation, because $100\%$ of the weight ($\psi({\bf x})$) will be associated with the nearest particle. In fact, technically speaking, Voronoi-based moving-mesh methods are a special case of the method here, where the function $W$ is taken to the limit of a delta function and the volume quadrature is evaluated exactly.\footnote{In practice, the reconstruction step (\S~\ref{sec:methods:reconstruction}) differs slightly in most Voronoi-mesh schemes, because they reconstruct the primitive quantities at the centroid of the face, rather than at the point along the face intersecting the line between the two points sharing said face.}

We now insert this definition of the volume partition into Eq.~\ref{eqn:dg.solution}, and Taylor-expand all terms to second order accuracy in the kernel length $h({\bf x})$ (e.g.\ $f({\bf x}) = f_{i}({\bf x}_{i}) + h({\bf x}_{i})\,\nabla f({\bf x}={\bf x}_{i})\cdot({\bf x}-{\bf x}_{i})/h({\bf x}_{i}) + \mathcal{O}(h({\bf x}_{i})^{2})$; the algebra is somewhat tedious but straightforward). Note that $1=\sum_{i}\,\psi_{i}({\bf x})$, and since the kernel has compact support, $|{\bf x}-{\bf x}_{i}| \sim \mathcal{O}(h({\bf x}_{i}))$ where $W \ne 0$. If we apply this to the integral of an arbitrary function (and assume the kernel function is continuous, symmetric, and compact) we obtain 
\begin{align}
\int f({\bf x})\,d^{\Ddim}{\bf x} &= \sum_{i} \int f({\bf x})\,\psi_{i}({\bf x})\,d^{\Ddim}{\bf x} \\ 
& = \sum_{i} f_{i}({\bf x}_{i})\,\int \psi_{i}\,d^{\Ddim}{\bf x} + \mathcal{O}(h_{i}({\bf x}_{i})^{2}) \\
& \equiv \sum_{i}\,f_{i}\,V_{i} + \mathcal{O}(h_{i}^{2})
\end{align}
where $V_{i} = \int \psi_{i}({\bf x})\,d^{\Ddim}\,{\bf x}$ is the ``effective volume'' of particle $i$ (i.e.\ the integral of its volume partition over all of space). 
Applying the same to Eq.~\ref{eqn:dg.solution}, evaluating the spatial integral, and dropping the $\mathcal{O}(h^{2})$ terms, we obtain
\begin{align}
\label{eqn:dg.discrete}
0 &= \frac{d}{dt}\sum_{i} V_{i}\,{{\bf U}}_{i}\,{\phi}_{i} - \sum_{i}\,V_{i}\,{\bf F}_{i} \cdot (\nabla \phi)_{{\bf x}={\bf x}_{i}}\\ 
\nonumber &= \sum_{i}\left[ \phi_{i}\,\frac{d}{dt}(V_{i}\,{\bf U}_{i}) - V_{i}\,{\bf F}_{i} \cdot (\nabla \phi)_{{\bf x}={\bf x}_{i}} \right]
\end{align}
where ${\bf F}_{i} \cdot (\nabla \phi)_{{\bf x}={\bf x}_{i}}$ refers to the product of the tensor ${\bf F}$ with the gradient of $\phi$ evaluated at ${\bf x}_{i}$. 

To go further, and remain consistent, we require a second-order accurate discrete gradient estimator. Here, we can use locally-centered least-squares matrix gradient operators, which have been described in many previous numerical studies \citep{dilts:1999.sph.stability,onate:1996.fpm,kuhnert:2003.finite.pointset.method,maron:2003.gradient.particle.mhd,maron:2012.phurbas.algorithm,tiwari:2003.finite.pointset.method,liu:2005.finite.particle.method,luo:2008.compressible.flow.galerkin,lanson.vila:2008.meshfree.consistency,lanson.vila:2008.meshfree.convergence}. Essentially, for any arbitrary configuration of points, we can use the weighted moments to defined a least-squares best-fit to the Taylor expansion of any fluid quantity at a central point $i$, which amounts to a simple (small) matrix calculation; the matrix can trivially be designed to give an arbitrarily high-order consistent result, meaning this method will, by construction, {\em exactly} reproduce polynomial functions across the particles/cells up to the desired order, {\em independent} of their spatial configuration. The second-order accurate expression is: 
\begin{align}
\label{eqn:gradient} (\nabla f)_{i}^{\alpha} &= \sum_{j}\sum^{\beta=\Ddim}_{\beta=1}(f_{j} - f_{i})\,{\bf B}_{i}^{\alpha\beta}\,({\bf x}_{j}-{\bf x}_{i})^{\beta}\,\psi_{j}({\bf x}_{i}) + \mathcal{O}(h_{i}^{2})\\
\nonumber &\equiv \sum_{j} (f_{j}-f_{i})\,\tilde{\psi}_{j}^{\alpha}({\bf x}_{i}) \\ 
\nonumber \tilde{\psi}_{j}^{\alpha}({\bf x}_{i}) &\equiv \sum^{\beta=\Ddim}_{\beta=1} {\bf B}_{i}^{\alpha\beta}\,({\bf x}_{j}-{\bf x}_{i})^{\beta}\,\psi_{j}({\bf x}_{i})
\equiv {\bf B}_{i}^{\alpha\beta}\,({\bf x}_{j}-{\bf x}_{i})^{\beta}\,\psi_{j}({\bf x}_{i}) 
\end{align}
where the we assume an Einstein summation convention over the Greek indices $\alpha$ and $\beta$ representing the elements of the relevant vectors/matrices, and the matrix ${\bf B}_{i}$ is evaluated at each $i$ by taking the inverse of another matrix ${\bf E}_{i}$: 
\begin{align}
{\bf B}_{i} &\equiv {\bf E}_{i}^{-1} \\ 
\label{eqn:gradient.matrix} {\bf E}_{i}^{\alpha\beta} &\equiv \sum_{j}\,({\bf x}_{j}-{\bf x}_{i})^{\alpha}\,({\bf x}_{j}-{\bf x}_{i})^{\beta}\,\psi_{j}({\bf x}_{i})
\end{align}
Note that in Eqs.~\ref{eqn:gradient}-\ref{eqn:gradient.matrix}, we could replace the $\psi_{j}({\bf x}_{i})$ with any other function $\xi_{j}({\bf x}_{i})$, so long as that function $\xi$ is also continuous and compact. However, it is computationally convenient, and physically corresponds to a volume-weighting convention in determining the least-squares best-fit, to adopt $\xi_{j}({\bf x}_{i}) = \psi_{j}({\bf x}_{i})$, so we will follow this convention. It is straightforward to verify that when the $f_{j}$ follow a linear function in $N$-dimensions ($f_{j} = f_{i} + \nabla f_{\rm true}\cdot({\bf x}_{j}-{\bf x}_{i})$), this estimator {\em exactly} recovers the correct gradients (hence, the method is consistent up to second order).

Now, inserting this into Eq.~\ref{eqn:dg.discrete}, and noting that: 
\begin{align}
\sum_{i} V_{i}\,{\bf F}_{i}^{\alpha}\,(\nabla \phi)_{i}^{\alpha} &= 
\sum_{i}\sum_{j} V_{i}\,{\bf F}_{i}^{\alpha}\,(\phi_{j}-\phi_{i})\,\tilde{\psi}_{j}^{\alpha}({\bf x}_{i}) \\ 
\nonumber &= -\sum_{i}\,\phi_{i}\,\sum_{j}\,(V_{i}\,{\bf F}_{i}^{\alpha}\,\tilde{\psi}_{j}^{\alpha}({\bf x}_{i}) - 
V_{j}\,{\bf F}_{j}^{\alpha}\,\tilde{\psi}_{i}^{\alpha}({\bf x}_{j}))
\end{align}
we obtain
\begin{align}
\label{eqn:integral.form2}
0 = \sum_{i}\,\phi_{i}\,{\Bigl(} &\frac{d}{dt} (V_{i}\,{\bf U}_{i})  + \sum_{j}
[V_{i}\,{\bf F}_{i}^{\alpha}\,\tilde{\psi}_{j}^{\alpha}({\bf x}_{i}) - V_{j}\,{\bf F}_{j}^{\alpha}\,\tilde{\psi}_{i}^{\alpha}({\bf x}_{j})]
{\Bigr)}
\end{align}
This must hold for an arbitrary test function $\phi$; so therefore the expression inside the parenthesis must vanish, i.e.\ 
\begin{align}
\frac{d}{dt}(V_{i}\,{\bf U}_{i}) + \sum_{j}[V_{i}\,{\bf F}_{i}^{\alpha}\,\tilde{\psi}_{j}^{\alpha}({\bf x}_{i}) - V_{j}\,{\bf F}_{j}^{\alpha}\,\tilde{\psi}_{i}^{\alpha}({\bf x}_{j})] = 0
\end{align}

Now, rather than take the flux functions ${\bf F}$ directly at the particle location and time of $i$ or $j$, in which case the scheme would require some ad-hoc artificial dissipation terms (viscosity and conductivity) to be stable, we can replace the fluxes with the solution of an appropriate time-centered Riemann problem between the particles/cells $i$ and $j$, which automatically includes the dissipation terms. We define the flux as $\tilde{\bf F}_{ij}$; this replaces both ${\bf F}_{i}$ and ${\bf F}_{j}$ since the solution is necessarily the same for both $i$ and $j$ ``sides'' of the problem;\footnote{Note that this replacement of ${\bf F}_{i}$ and ${\bf F}_{j}$ can be directly derived, as well, by replacing the ${\bf F}$ in the integral Eq.~\ref{eqn:integral.form2} with a Taylor expansion in space and time, multiplying the terms inside by $1=\sum \psi_{i}$, centering the expansion about the symmetric quadrature point between $i$ and $j$ and centering it at the mid-point in time for a discretized time integral, and then evaluating the integrals to second order. For details, see \citet{lanson.vila:2008.meshfree.consistency}.} this gives 
\begin{align}
\frac{d}{dt}(V_{i}\,{\bf U}_{i}) + \sum_{j}\,\tilde{{\bf F}}_{ij}^{\alpha}\,[V_{i}\,\tilde{\psi}_{j}^{\alpha}({\bf x}_{i}) - V_{j}\,\tilde{\psi}_{i}^{\alpha}({\bf x}_{j})] = 0
\end{align}
Now, we can define the vector ${\bf A}_{ij} = |A|_{ij}\,\hat{A}_{ij}$ where ${\bf A}_{ij}^{\alpha} \equiv V_{i}\,\tilde{\psi}_{j}^{\alpha}({\bf x}_{i}) - V_{j}\,\tilde{\psi}_{i}^{\alpha}({\bf x}_{j})$, and the equations become: 
\begin{align}
\label{eqn:final}
\frac{d}{dt}(V_{i}\,{\bf U}_{i}) + \sum_{j}\,\tilde{{\bf F}}_{ij}\cdot {\bf A}_{ij} = 0
\end{align}

This should be immediately recognizable as the form of the Godunov-type finite-volume equations. The term $V_{i}\,{\bf U}_{i}$ is simply the particle-volume integrated value of the conserved quantity to be carried with particle $i$ (e.g.\ the total mass $m_{i}=V_{i}\,\rho_{i}$, momentum, or energy associated with the particle $i$); its time rate of change is given by the sum of the fluxes $\tilde{{\bf F}}_{ij}$ into/out of an ``effective face area'' ${\bf A}_{ij}$.

We note that our method is not, strictly, a traditional Godunov scheme as defined by some authors, since we do not actually calculate a geometric particle face and transform a volume integral into a surface integral in deriving Eq.~\ref{eqn:final}; rather, the ``effective face'' comes from solving the {\em actual volume integral}, over the partition defined by the weighting function, and this is simply the numerical quadrature rule that arises. But from this point onwards, it can be treated identically to Godunov-type schemes.

\vspace{-0.5cm}
\subsection{Conservation Properties}

It should be immediately clear from Eq.~\ref{eqn:final}, that since we ultimately calculate fluxes of conserved quantities directly between particles/cells, the conserved quantities themselves (total mass, linear momentum, and energy) will be conserved to machine accuracy {\em independent of the time-step, integration accuracy, and particle distribution}. Moreover, it is trivial to verify that ${\bf A}_{ij} = -{\bf A}_{ji}$, i.e.\ the fluxes are antisymmetric, so the flux ``from $i$ to $j$'' is always the negative of the flux ``from $j$ to $i$'' at the same time, and the discrete equations are manifestly conservative.

\vspace{-0.5cm}
\subsection{Pathological Particle/Cell Configurations}

We note that if very specific pathological particle configurations appear (e.g.\ if all particles in the kernel ``line up'' perfectly), our gradient estimator (the matrix ${\bf B}$ in Eq.~\ref{eqn:gradient}) becomes ill-conditioned. However it is straightforward to deal with this by expanding the neighbor search until cells are found in the perpendicular direction. Appendix~\ref{sec:condition.number} describes a simple, novel algorithm we use which resolves these (very rare) special cases.

\vspace{-0.5cm}
\subsection{Solving the Discrete Equations}
\label{sec:methods:reconstruction}

The approach to solving the discretized equations of the form in Eq.~\ref{eqn:final} is well-studied; we can use essentially the same schemes used in grid-based Godunov methods. Specifically, we will employ a second-order accurate (in space and time) MUSCL-Hancock type scheme \citep{vanleer:1984.slopelimiters,toro:1997.reimann.solver.book}, as used in state-of-the-art grid methods such as \citet{teyssier:2002.RAMSES,fromang:2006.ramses.amr.schemes,mignone:2007.pluto.code.methods,cunningham:2009.ct.mhd.code,springel:arepo}. This involves a slope-limited, linear reconstruction of face-centered quantities from each particle/cell, a first-order drift/predict step for evolution over half a timestep, and then the application of a Riemann solver to estimate the time-averaged inter-particle fluxes for the timestep. Details of the procedure are given in Appendix~\ref{sec:methods:fluxes}.

\vspace{-0.5cm}
\subsubsection{Gradient Estimation}

In order to perform particle drift operations and reconstruct quantities for the Riemann problem, we require gradients. But we have already defined an arbitrarily high-order method for obtaining gradients using the least-squares matrix method in Eq.~\ref{eqn:gradient}-\ref{eqn:gradient.matrix}. We will use the second-order accurate version of this to define the gradient of a quantity $(\nabla f)_{i}$ at position ${\bf x}_{i}$; recall that these are exact for linear gradients and always give the least-squares minimizing gradient in other situations. As noted by \citet{mocz:2014.galerkin.arepo}, this gradient definition has a number advantages over the usual finite-volume definition (based on cell-to-cell differences).

\vspace{-0.5cm}
\subsubsection{Slope-Limiting in Mesh-Free Methods}

However, as in all Riemann-problem based methods, some slope-limiting procedure is required to avoid numerical instabilities near discontinuities, where the reconstruction can ``overshoot'' or ``undershoot'' and create new extrema \citep[see e.g.][]{barth.jesperson:1989.upwind.schemes.for.unstructured.meshes}. Therefore, in the reconstruction step (only), the gradient $(\nabla f)_{i}$ above is replaced by an appropriately slope-limited gradient $(\nabla f)_{{\rm lim},\,i}$, using the limiting procedure in Appendix~\ref{sec:slopelimiters}. 

We have experimented with a number of standard slope-limiters like that in \citet{gaburov:2011.meshless.dg.particle.method} and find generally similar, stable behavior. However, as noted by \citet{mocz:2014.galerkin.arepo}, for unstructured point configurations in discontinuous Galerkin methods, there are some subtle improvements which can be obtained from more flexible slope-limiters. We find signifiant improvement (albeit no major changes in the results here) if we adopt the new, more flexible (and closer to total variation-diminishing) slope-limiting procedure described  in Appendix~\ref{sec:slopelimiters}.

\vspace{-0.5cm}
\subsubsection{Reconstruction: Projection to the Effective Face}

In Eq.~\ref{eqn:final}, only the projection of the flux onto $\hat{A}_{ij}$ is required; therefore the relevant flux $\tilde{\bf F}_{ij}\cdot \hat{A}_{ij}$ can be obtained by solving a one-dimensional, unsplit Riemann problem in the frame of the quadrature point between the two particles/cells. Because of the symmetry of the kernel, the relevant quadrature point at this order (the point where the volume partition between the two particles is equal) is the location along the line connecting the two which is an equal fraction of the kernel length $h$ from each particle, i.e.\ 
\begin{align}
{\bf x}_{ij} &\equiv {\bf x}_{i} + \frac{h_{i}}{h_{i}+h_{j}}\,\left( {\bf x}_{j} - {\bf x}_{i} \right)
\end{align}
This quadrature point moves with velocity ${\bf v}_{\rm frame}$ (at second order)
\begin{align}
\label{eqn:vframe}
{\bf v}_{{\rm frame},\,ij} = {\bf v}_{i} + ({\bf v}_{j}-{\bf v}_{i})\,{\Bigl[}\frac{({\bf x}_{ij}-{\bf x}_{i})\cdot ({\bf x}_{j}-{\bf x}_{i})} {|{\bf x}_{j} - {\bf x}_{i}|^{2}}{\Bigr]}
\end{align}

However, we note that we see very little difference in all the test problems here using this or the first-order quadrature point ${\bf x}_{ij} = ({\bf x}_{i} + {\bf x}_{j}) / 2$ (which can sometimes be more stable, albeit less accurate).\footnote{As shown in \citet{inutsuka:2002.godunov.sph}, a higher-order quadrature rule between particles using the ``traditional'' SPH volume partition implies a quadrature point which is offset from the midpoint at $\mathcal{O}(h^{2})$. It is straightforward to derive an analogous rule here, and we have experimented with this. However, we find no significant improvement in accuracy, presumably because the rest of the reconstruction we adopt is only second-order. Moreover, because \citet{inutsuka:2002.godunov.sph} derive this assuming there is always an exact linear gradient connecting the particles and extrapolate this to infinity beyond them, this can lead to serious numerical instabilities in the Riemann problem when there is some particle disorder.}. 

So we must reconstruct the left and right states of the Riemann problem at this location: for a second-order method, we only require a linear reconstruction in primitive variables, so we require gradients and reconstructions of the density $\rho$, pressure $P$ (and internal energy for a non-ideal equation of state), and velocity ${\bf v}$. For an interacting pair of particles $i$ and $j$, the linearly-reconstructed value of a quantity $f$ at a position ${\bf x}$, reconstructed from the particle $i$, is 
$f_{{\rm rec},\,i} = f_{i} + ({\bf x}-{\bf x}_{i})\cdot (\nabla f)_{i}$
and likewise for the reconstruction from particle $j$; these define the left and right states of the Riemann problem at the ``interface'' ${\bf x}_{ij}$. Details of the reconstruction and Riemann solver are given in Appendix~\ref{sec:methods:fluxes}.

\vspace{-0.5cm}
\subsubsection{The Riemann Solver}

Having obtained the left and right time-centered states, we then solve the unsplit Riemann problem to obtain the fluxes $\tilde{{\bf F}}_{ij}$. We have experimented with both an exact Riemann solver and the common approximate HLLC Riemann solver \citep{toro:1999.reimann.solvers.book}; we see no differences in any of the test problems here. So in this paper we adopt the more flexible HLLC solver with Roe-averaged wave-speed estimates as our ``default'' Riemann solver (see Appendix~\ref{sec:methods:fluxes}).

\vspace{-0.5cm}
\subsubsection{Time Integration}

The time integration scheme here closely follows that in \citet{springel:arepo}, and additional details are given in Appendix~\ref{sec:methods:timesteps}. 

We use the fluxes $\tilde{{\bf F}}_{ij}$ to obtain single-stage second-order accurate time integration as in \citet{colella:1990.upwind.methods.conservation.laws.multid,stone:2008.athena}. For a vector of conserved quantities ${\bf Q}_{i} = (V\,{\bf U})_{i}$, 
\begin{align}
\label{eqn:timestep.basic}
{\bf Q}_{i}^{(n+1)} &= {\bf Q}_{i}^{(n)} + \Delta t\, \left\langle \frac{d {\bf Q}_{i}}{d t} \right\rangle
\equiv {\bf Q}_{i}^{(n)} + \Delta t\,\frac{d {\bf Q}_{i}}{d t}^{(n+1/2)} \\ 
&= {\bf Q}_{i}^{(n)} - \Delta t\,\sum_{j}\,{\bf A}_{ij}\cdot \tilde{{\bf F}}_{ij}^{(n+1/2)}
\end{align}

We employ a local Courant-Fridrisch-Levy (CFL) timestep criterion; for consistency with previous work we define it as 
\begin{align}
\label{eqn:CFL} \Delta t_{{\rm CFL},\,i} &= 2\,C_{\rm CFL}\,\frac{h_{i}}{|v_{{\rm sig},\,i}|} \\ 
v_{{\rm sig},\,i} &= {\rm MAX}_{j}{\Bigl[} c_{s,\,i} + c_{s,\,j} - 
{\rm MIN}{\Bigl(}0,\,\frac{({\bf v}_{i}-{\bf v}_{j})\cdot ({\bf x}_{i}-{\bf x}_{j})}{|{\bf x}_{i}-{\bf x}_{j}|}{\Bigr)}
{\Bigr]}
\end{align}
where $h_{i}$ is the kernel length defined above, ${\rm MAX}_{j}$ refers to the maximum over all interacting neighbors $j$ of $i$,  and $|v_{\rm sig}|$ is the signal velocity \citep{whitehurst:1995.moving.mesh.protocode,monaghan:1997.sph.solvers.viscosities}.\footnote{Note that the normalization convention here is familiar in SPH, but different from most grid codes (in part because $h_{i}$ is not  exactly the same as the ``cell size''). For our standard choice of kernel, a choice $C_{\rm CFL}=0.2$ (our default in this paper) is equivalent to $C_{\rm CFL}=0.8$ in an AMR code with the convention $\Delta t_{\rm CFL} = C_{\rm CFL}\,\Delta x_{\rm cell} / (c_{s}+ |{\bf v}_{\rm gas}|)$.} We combine this with a limiter based on \citet{saitoh.makino:2009.timestep.limiter} to prevent neighboring particles/cells from having very different timesteps (see Appendix~\ref{sec:methods:timesteps}). 

We follow \citet{springel:arepo} to maintain manifest conservation conservation of mass, momentum, and energy even while using adaptive (individual) timesteps (as opposed to a single, global timestep, which imposes a severe cost penalty on high-dynamic range problems). This amounts to discretizing timesteps into a power-of-two hierarchy and always updating fluxes of conserved quantities across inter-particle faces synchronously. See Appendix~\ref{sec:methods:timesteps} for details. 

Because our method is Lagrangian, when the bulk velocity of the flow is super-sonic ($|{\bf v}| \gg c_{s}$), the signal velocity is typically still close to $c_{s}$. Contrast this to stationary grid methods, where $v_{\rm sig}$ must include the velocity of the flow across the grid \citep{ryu:1993.entropy.switch.cosmo.grid.hydro}. As a result, we can take much larger timesteps (factor $\sim1000$ in some test problems below) without loss of accuracy.

We also note that, like all conservative methods based on a Riemann solver, when flows are totally dominated by kinetic energy, small residual errors can appear in the thermal energy which are large compared to the correct thermal energy solution. This is a well-known problem, and there are various means to deal with it, but we adopt the combination of the ``dual energy'' formalism and energy-entropy switches described in Appendix~\ref{sec:methods:switches}. It is worth noting here, though, that the Lagrangian nature of our method minimizes this class of errors compared to stationary-grid codes.

\vspace{-0.5cm}
\subsection{Setting Particle Velocities: The Arbitrary-Lagrangian-Eulerian Nature of the Method}
\label{sec:methods:particle.velocities}

Note that so far, we have dealt primarily with the {\em fluid} velocity ${\bf v}={\bf v}({\bf x})$. We have not actually specified the velocity of the {\em particles} (e.g.\ the ${\bf v}_{i}$ which enters in determining the velocity of the frame in Eq.~\ref{eqn:vframe}). It is, for example, perfectly possible to solve the above equations, with the particle positions {\em fixed}; everything above is identical except the frame velocity is zero and the particles/cells are not moved between timesteps. This makes the method fully Eulerian; since the particle volumes depend only on their positions, they do not change in time, and we could choose an initial particle distribution to reproduce a stationary mesh method. On the other hand, we could set the velocities equal to the fluid velocity, in which case we obtain a Lagrangian method. Our derivation thus far describes a truly Arbitrary Lagrangian-Eulerian (ALE) method. 

In this paper, we will -- with a couple of noted exceptions shown for demonstration purposes -- set the particle velocities ${\bf v}_{i}$ to match the fluid velocities. Specifically, we choose velocities such that the ``particle momentum'' $m_{i}\,{\bf v}_{i}$ is equal to the fluid momentum integrated over the volume associated with the particle. This amounts numerically to treating the fluid and particle velocities at the particle positions as the same quantity. This is the Lagrangian mode of the method, which has a number of advantages. In pure Eulerian form, most of the advantages of the new methods here compared to stationary grid codes are lost.

That said, more complicated and flexible schemes are possible, and may be advantageous under some circumstances. For example, the particles could move with a ``smoothed'' fluid velocity, which would capture bulk flows but could reduce noise in complicated flows \citep[an idea which has been explored in both SPH and moving-mesh codes; see][]{imaeda:2002.sph.shear.flow.tests,duffell:2014.smooth.moving.mesh.motion}.

\vspace{-0.5cm}
\subsection{What Motion of the ``Face'' Means: The Difference Between MFV and MFM Assumptions}
\label{sec:methods:differences}

In our flux calculation, the projection of states to the ``face'' is well-defined. However, the distortions of the effective volume with time are more complex. When we solve the Riemann problem in Eq.~\ref{eqn:final}, we have to ask how the volumes assigned to one particle vs.\ the other are ``shifting'' during the timestep. 

One choice is to assume, that since we boosted to a frame moving with the velocity of the quadrature point assuming the time-variation in kernel lengths was second-order, the ``face'' is exactly stationary in this frame (${\bf v}_{\rm eff}^{\rm frame} = 0$). This is what we would obtain in e.g.\ a moving-mesh finite-volume method, where the face motion can be chosen (in principle) arbitrarily and the faces are locally simple, flat, ``planes'' of arbitrary extent in the directions perpendicular to the quadrature point. This was the choice made in \citet{gaburov:2011.meshless.dg.particle.method}, for example. We will consider this, and it defines what we call our ``Meshless Finite-Volume'' or MFV method. This is analogous to the finite-volume method: we solve the Riemann problem across a plane whose relative position to the mesh-generating points is (instantaneously) fixed.

However, when the fluid flow is complicated, there is relative particle motion which changes the domain, leading to higher-order corrections. Moreover, assuming ${\bf v}_{\rm eff}^{\rm frame} = 0$ does not necessarily capture the true {up-wind} motion of the face. Since we derived this method with the assumption that the particles/cells move with the fluid, we could instead assume that the Lagrangian volume is distorting with the mean (time-centered and face-area averaged) motion of the volume partition, such that the mass on either ``side'' of the state is conserved. In practice, this amounts to an identical procedure as in our MFV case, but in the Riemann problem itself, we assume the face has a residual motion ${\bf v}_{\rm eff}^{\rm frame} = S_{\ast}$, where $S_{\ast}$ is the usual ``star state'' velocity (the speed of the contact wave in the Riemann problem), on either side of which mass is conserved. Note that this does not require that we modify our boost/de-boost procedure, since the frame we solve the problem in is ultimately arbitrary so long as we correct the quantities appropriately. This assumption defines what we call our ``Meshless Finite-Mass'' (MFM) method, because it has the practical effect of eliminating mass fluxes between particles. This choice is analogous to the finite-element method: we are solving the Riemann problem across a complicated Lagrangian boundary distorting with the relative fluid flow. We stress that this is {\em only} a valid choice if the particles are moved with the fluid velocity; otherwise the MFM choice has a zeroth-order error (obvious in the case where particles are not moving but the fluid is). 

A couple of points are important to note. First, for a smooth flow (with only linear gradients), it is straightforward to show that the MFM and MFV reduce to each other (they become {\em exactly} identical). So the difference is only at second-order, which is the order of accuracy in our method in any case. Second, it is true that, in situations with complicated flows, because we cannot perfectly follow the distortion of Lagrangian faces, the assumption made in the MFM method for the motion of the face in the Riemann problem will not {\em exactly} match the ``real'' motion of the face calculated by directly time-differencing the positions estimated for it across two timesteps. However, the error made is second-order (in smooth flows); and moreover, this is true for MFV and most moving-mesh finite-volume methods as well.

So both methods have different finite numerical errors. We will systematically compare both, to study how these affect certain problems. Not surprisingly, we will show that the differences are maximized at discontinuities, where the methods become lower-order, and the two should not be identical.

\vspace{-0.5cm}
\subsection{Kernel Sizes and Particle ``Volumes''}
\label{sec:methods:smoothing}

In our method, the kernel length $h$ does not play any ``inherent'' role in the dynamics and we are free to define it as we like; however it does closely relate to the ``effective volume'' of a particle. This suggests setting it so that some (relatively small) number of neighbors is enclosed by the compact kernel function centered at each particle; this also makes the resolution intrinsically adaptive. It is also implicit in our derivation and required for second-order accuracy that $h({\bf x})$ vary smoothly across the flow. Therefore, like in most modern SPH schemes, we do not set the particle-centered $h_{i} = h({\bf x}_{i})$ to enclose some actual discrete ``number of neighbors,'' (which is discontinuous) but rather follow \citet{hopkins:lagrangian.pressure.sph} and \citet{gaburov:2011.meshless.dg.particle.method} and constrain it by the smoothed particle number density $n_{i} \equiv n({\bf x}_{i}) = \omega({\bf x}_{i})$ \citep[see e.g.][]{springel:entropy,monaghan:2002.sph.turbulence}. In $\Ddim$ dimensions, this is 
\begin{align}
\label{eqn:nngb}
N_{\rm NGB} = C_{\Ddim}\,n_{i}\,h_{i}^{\Ddim} = C_{\Ddim}\,h_{i}^{\Ddim}\,\sum_{j}\,W({\bf x}_{j}-{\bf x}_{i},\,h_{i})
\end{align}
where $C_{\Ddim}=1$, $\pi$, $4\pi/3$ for $\Ddim=1,\,2,\,3$, and $N_{\rm NGB}$ is a constant we set, which is the ``effective'' neighbor number (close to, but not necessarily equal to, the discrete number of neighbors inside $h_{i}$). Just as in \citet{hopkins:lagrangian.pressure.sph} this is solved iteratively (but the iteration imposes negligible cost).\footnote{We use an iteration scheme originally based on that in \citet{springel:entropy}, which uses the continuity equation to guess a corrected $h_{i}$ each timestep, then uses the simultaneously computed derivatives of the particle number density to converge rapidly; and we have further optimized the scheme following \citet{cullen:2010.inviscid.sph} and our own SPH experiments in \citet{hopkins:lagrangian.pressure.sph,hopkins:2013.fire}. This means that once a solution for $h_{i}$ is obtained on the first timestep, usually $<1\%$ of active particles require multiple iteration (beyond a first-pass) in future timesteps and almost none require second iterations, so the CPU cost compared to a single-sweep is negligible (and the gains in accuracy are very large).} Note that, unlike a ``constant mass in kernel'' approach as in \citet{springel:entropy}, this choice is independent of the local fluid properties (depends only on particle positions) and eliminates any discontinuities in the kernel length.\footnote{We should also note that because the kernel length is ultimately arbitrary, so long as it is continuous (as far as the formal consistency, conservation, and accuracy properties of our method are concerned), the particle number density estimator in Eq.~\ref{eqn:nngb} does not actually have to reproduce the ``true'' particle number density, just a continuous and finite approximation.}

With this definition of $h$, we can also now calculate the effective volume of a particle, $V_{i} = \int \psi_{i}({\bf x})\,d^{\Ddim}{\bf x}$. Inserting the above, and keeping terms up to second order accuracy, we find 
\begin{align}
\label{eqn:volumes}
V_{i} &= \int \psi_{i}({\bf x})\,d^{\Ddim}{\bf x} = \omega({\bf x}_{i})^{-1}\left(1 + \mathcal{O}(h^{2})\right)
\end{align}
In fact, this expression is {exact} if $h$ is locally constant (does not vary over the kernel domain). For $h$ being a general function of position, it is not possible to analytically solve for the exact $V_{i}$; however this expression is second-order accurate so long as the variation of $h$ across the kernel obeys $|(\nabla h)_{i}| \lesssim 1$, which our definition maintains (except at discontinuities, where this drops to first order-accurate). We stress that the method is still a ``partition of unity'' method \citep[see][]{lanson.vila:2008.meshfree.consistency,lanson.vila:2008.meshfree.convergence}, since our equations of motion were derived from an exact and conservative analytic/continuum expression for the volume partition; this is distinct from SPH where even in the continuum limit, volume is not conserved. However, the quadrature rule we use on-the-fly to estimate the volume integral associated with a given particle is only accurate in our scheme to the same order as the integration accuracy. This does not, therefore, reduce the order of the method; however, we will show that it does lead to noise in some fields, compared to methods with an exact discretized volume partition (e.g.\ meshes). If desired, an arbitrarily high-order numerical quadrature rule could be applied to evaluate Eq.~\ref{eqn:volumes}; this would be more expensive but reduce noise.

We should stress that while some authors have advocated using the continuity equation ($d h/dt = \Ddim^{-1}\,h\,\nabla \cdot {\bf v}$) to evolve the kernel lengths (and this is done in e.g.\ {\small GASOLINE}), it is not a stable or accurate choice for this method. As noted by \citet{gaburov:2011.meshless.dg.particle.method}, the results of such an exercise depend on the discretization of the divergence operator in a way that is not necessarily consistent, and more worryingly, this will inevitably produce dis-continuous kernel lengths in sufficiently complex flows, reducing the accuracy and consistency of the method.

\vspace{-0.5cm}
\subsection{Higher-Order Versions of the Scheme}
\label{sec:methods:high.order}

It is straightforward to extend most elements of this method to higher-order. The moving, weighted-least squares gradient estimators can be trivially extended to arbitrarily high order if higher-order gradients are desired; it simply increases the size of the matrix which must be calculated between neighbors \citep[see][]{bilotta:2011.mls.sph.higher.order.renorm}. As discussed in \citet{gaburov:2011.meshless.dg.particle.method}, Appendix A, this makes it straightforward to perform the reconstruction for the Riemann at equivalent higher-order, for example they explicitly show the equations needed to make this a piecewise parabolic method (PPM). From the literature on finite-volume Godunov methods, there are also well-defined schemes for higher-order time-integration accuracy, which can be implemented in this code in an identical manner to stationary-grid codes. However, if we wish to make the method completely third-order (or higher) at all levels, we also need to re-derive an appropriate quadrature rule; that can trivially be done numerically via Gaussian quadrature \citep[see e.g.][]{maron:2012.phurbas.algorithm}, however it is computationally expensive, so an analytic quadrature expression would be more desirable. Finally, this quadrature rule, if used, should also be used to re-discretize the equation of motion (i.e.\ correct the ``effective face'' terms in \S~\ref{sec:methods:deriv}), to complete the method.

\vspace{-0.5cm}
\subsection{Gravity \&\ Cosmology}
\label{sec:methods:gravity:brief}

In astrophysics, gravity is almost always an important force. Indeed, as stressed by \citet{springel:arepo}, there is essentially no point in solving the hydrodynamic equations more accurately in most astrophysical problems if gravity is treated less accurately, since the errors from gravity will quickly overwhelm the real solution. A major advantage, therefore, of the new methods proposed here is that they, like SPH, couple naturally, efficiently, and accurately to $N$-body based gravitational solvers. 

Details of the implementation of self-gravity and cosmological integrations are reviewed in Appendix~\ref{sec:methods:gravity}. Briefly, the $N$-body solver used in our code follows {\small GADGET-3}, but with several important changes. Like {\small GADGET-3}, we have the option of using a hybrid Tree or Tree-Particle Mesh (TreePM) scheme; these schemes are computationally efficient, allow automatic and continuous adaptivity of the gravitational resolution in collapsing or expanding structure, and can be computed very accurately. Following \citet{springel:arepo}, the gravity is coupled to the hydrodynamics via operator splitting (see Eq.~\ref{eqn:grav1}-\ref{eqn:grav2}); if mass fluxes are present, appropriate terms are added to the energy equation to represent the gravitational work (\S~\ref{sec:methods:gravity:hydrocoupling}). This makes the coupling spatially and temporally second-order accurate in forces (exact for a linear gravitational force law) and third-order accurate in the gravitational potential.

An advantage of our particle-based method is that it removes many of the ambiguities in coupling gravity to finite-volume systems. Essentially all $N$-body solvers implicitly neglect mass fluxes in calculating the forces; our Lagrangian methods either completely eliminate or radically reduce these fluxes, eliminating or reducing second-order errors in the forces.

To treat the self-gravity of the gas, we must account for the full gas mass distribution at second order. For any configuration other than a uniform grid, this cannot be accomplished using a constant gravitational softening or a particle-mesh grid. However, in Appendix~\ref{sec:methods:gravity}, we show that the gravitational force obtained by integrating the exact mass distribution ``associated with'' a given particle/cell ($d m_{i} = d^{\Ddim}{\bf x}\,\rho({\bf x})\,\omega^{-1}({\bf x})\,{W({\bf x}-{\bf x}_{i},\,h({\bf x}))}$) can be represented at second order with the following gravitational force law: 
\begin{align}
m_{i}\,\frac{d {\bf v}_{i}}{d t}{\Bigr|}_{\rm grav} &= - \nabla_{i} E_{\rm grav} \\
\nonumber &= -\sum_{j}\,\frac{G\,m_{i}\,m_{j}}{2}\,\left(\frac{\partial \phi(r,\,h_{i})}{\partial r}{\Bigr|}_{r_{ij}} + \frac{\partial \phi(r,\,h_{j})}{\partial r}{\Bigr|}_{r_{ij}} \right)\,\frac{{\bf r}_{ij}}{r_{ij}} \\ 
\nonumber &\ \ \ \ -\sum_{j}\,\frac{G}{2}\left( \zeta_{i}\,\frac{\partial W(r,\,h_{i})}{\partial r}{\Bigr|}_{r_{ij}}
+\zeta_{j}\,\frac{\partial W(r,\,h_{j})}{\partial r}{\Bigr|}_{r_{ij}} \right)\,\frac{{\bf r}_{ij}}{r_{ij}} \\
\zeta_{a} &\equiv m_{a}\,\frac{h_{a}}{n_{a}\,\Ddim}\,\frac{1}{\Omega_{a}}\,\sum_{b}\,m_{b}\,\frac{\partial{\phi(r_{ab},\,h)}}{\partial h}{\Bigr|}_{h=h_{a}}\\
\Omega_{a} &\equiv 1 + \frac{h_{a}}{n_{a}\,\Ddim}\,\frac{\partial n_{i}}{\partial h_{i}}\\
\nonumber &= 1 - \frac{h_{a}}{n_{a}\,\Ddim}\,\sum_{b}\left(\frac{r_{ab}}{h_{a}}\,\frac{\partial W(r,\,h_{a})}{\partial r}{\Bigr|}_{r_{ab}} + \frac{\Ddim}{h_{a}}\,W(r_{ab},\,h_{a})\right)
\end{align}
where ${\bf r}_{ij} = {\bf x}_{i} - {\bf x}_{j}$, $\phi$ is defined by $\Phi_{i} \equiv G\,m_{i}\,\phi_{i}$ where $\Phi_{i}$ is the 
the gravitational potential given by integrating Poisson's equation for a mass distribution $d m({\bf x}) = m_{i}\,W({\bf x}-{\bf x}_{i})\,d^{\nu}{\bf x}$, and the $\zeta$ terms account for the temporal and spatial derivatives of the kernel lengths $h$. This is derived following \citet{price:2007.lagrangian.adaptive.softening}, although the final equations differ owing to different definitions of the kernel length and volume partition. We emphasize that these equations manifestly conserve momentum and energy, and are exact to all orders if $\phi$ exactly represents the mass distribution. We also note that a similar volume partition rule, and corresponding second-order accurate adaptive gravitational softenings, can be applied to other volume-filling fluids (e.g.\ dark matter) in the code.

Essentially, this makes the gravitational softening adaptive, in a way that represents the hydrodynamic volume partition. In other words, the resolution of gravity and hydrodynamics is always equal and the two actually use the {\em same}, consistent set of assumptions about the mass distribution. This also avoids the ambiguities and associated errors of many mesh-based codes, which must usually adopt some ad-hoc assumptions to treat Cartesian cells or complicated Voronoi cells as simplified ``spheres'' with some ``effective'' radius.\footnote{Strictly speaking, at small separations, this mis-match in mesh-based methods leads to low-order errors in the sense that the gravitational forces calculated from the cell deviate from the true force associated with that cell geometry. If the field is well-resolved so that the gravitational forces require the collective effect of many cells, the errors are diminished rapidly, but for self-gravitating regions near the resolution limit, the errors can be significant (and increase as cells become less spherical).}

\vspace{-0.5cm}
\section{Smoothed-Particle Hydrodynamics}
\label{sec:sph}

\subsection{Implementing SPH as an Alternative Hydro Solver} 

Having implemented our new methods, we note that, with a few straightforward modifications, we can also run our code as an SPH code. The details of ``SPH mode'' in {\small GIZMO} are given in Appendix~\ref{sec:sph.methods} (note these are distinct in several ways from the ``default'' {\small GADGET-3} SPH algorithms). In ``SPH mode,'' much of the code is identical including: gravity, time-integration and timestep limitation, and definition of the kernel lengths. The essential changes are in (1) the computation of volumetric quantities (e.g.\ density, pressure), and (2) the computation of ``fluxes'' (there is no mass flux in SPH and no Riemann problem, but we can use the same apparatus for time integration, treating the results from the standard SPH equation of motion as momentum and energy fluxes). We implement both a ``traditional'' and ``modern'' SPH, as described below.

\vspace{-0.5cm}
\subsection{Differences Between our New Methods and SPH} 

Although it should be obvious from \S~\ref{sec:methods} and Fig.~\ref{fig:method}, we wish to clarify here that our MFM and MFV methods are {\em not} any form of SPH. Formally, they are arbitrary Lagrangian-Eulerian finite-volume Godunov methods. The derivation of the equations of motion, their final functional form, and even their dimensional scaling (with quantities like the kernel length and particle separations) are {\em qualitatively different} between our new methods and SPH. The gradient estimators are also different -- note that kernel gradients never appear in our equation of motion, while they are fundamental in SPH. ``Particles'' in the MFM/MFV methods are really just moving cells -- they represent a finite volume, with a well-defined volume partition (and our equations come from explicit volume integrals). As noted in Fig.~\ref{fig:method}, the SPH volume partition is not well-defined at all, and ``particles'' in that method are represent true point-particles (their ``volume'' is collapsed to a delta function in deriving quadrature rules), hence the common statement that SPH represents the ``molecular limit.'' As an important consequence of this, our MFM/MFV methods are second-order consistent, while SPH is not even zeroth-order consistent. Although so-called ``Godunov SPH'' introduces a Riemann problem between particles to eliminate artificial viscosity terms, it does not change any of these other aspects of the method, so other than the existence of a ``kernel function'' (which has a fundamentally different meaning) and ``Riemann solver,'' has nothing in common with our MFM/MFV methods. These differences will manifest in our test-problem comparisons below. In fact, the our MFM/MFV methods are most closely related to Voronoi-based moving-mesh methods (which are formally a special case of the MFV method).

\vspace{-0.5cm}
\section{Test Problems}
\label{sec:hydro.test}

In this section, we compare results from the different methods we have discussed in a number of pure hydrodynamic test problems. We will frequently compare both of our new proposed methods, both ``traditional'' and ``modern'' SPH, as well as moving mesh and stationary grid codes. 

\vspace{-0.5cm}
\subsection{Reference Methods for Test Problems}
\label{sec:reference}

In the tests below, we will generally consider six distinct numerical methods for treating the hydrodynamics (sometimes with individual variations). These and other methods are roughly summarized in Table~\ref{tbl:methods}. They include:

\begin{itemize}

\item{{\bf Meshless Finite-Volume (MFV)}: This refers to the meshless finite-volume formulation which we present in \S~\ref{sec:methods}. This is one of the two new methods used here, specifically the quasi-Lagrangian formulation which includes inter-particle mass fluxes. We use the implementation in {\small GIZMO} for all runs, with the details of the scheme following those outlined above. But we have confirmed that it gives very similar results on the same tests to the simplified, earlier implementation in \citet{gaburov:2011.meshless.dg.particle.method}.}

\item{{\bf Meshless Finite-Mass (MFM)}: This refers to the other of the two new methods developed here in \S~\ref{sec:methods}. Specifically, this is the Lagrangian formulation which conserves particle masses. As in the MFV case, we use the implementation in {\small GIZMO} for all runs. As such, up to the details of the frame in which the Riemann problem is solved (as discussed in \S~\ref{sec:methods:differences}), the code structure is exactly identical.}

\item{{\bf ``Traditional'' SPH (TSPH)}: This refers to ``traditional'' or ``old-fashioned'' SPH formulations (see \S~\ref{sec:intro} and \S~\ref{sec:sph.methods}). This is essentially the version of SPH in the most-commonly used versions of codes like {\small GADGET} \citep{springel:gadget}, {\small TREE-SPH} \citep{hernquistkatz:treesph}, and others. By default, it uses a Lagrangian, fully-conservative equation of motion \citep{springel:entropy}, a cubic spline smoothing kernel with $\sim 32$ neighbors, a standard (constant) \citet{morris:1997.sph.viscosity.switch} artificial viscosity with a \citet{balsara:1989.art.visc.switch} switch for shear flows, no artificial conductivity or other artificial diffusion terms, and the standard (error-prone) SPH gradient estimators. To make our methods comparison as exact as possible, we use the TSPH version implemented in {\small GIZMO} (\S~\ref{sec:sph.methods:tsph}), so that the code architecture is identical up to the details of the hydro solver. As such the results are not exactly identical to other TSPH-type codes; however, we have re-run several comparisons with {\small GADGET} and find the differences are very small.}

\item{{\bf ``Modern'' SPH (PSPH)}: This refers to ``modern'' SPH formulations (\S~\ref{sec:intro} and \S~\ref{sec:sph.methods:psph}). This is essentially the version of SPH in the {\small P-SPH} code used for the FIRE project simulations \citep{hopkins:lagrangian.pressure.sph,hopkins:2013.fire}, and the adapted version of {\small P-SPH} in {\small SPHGal} \citep{hu:2014.psph.galaxy.tests}. But it also gives very similar results to the SPH formulations in {\small SPHS} \citep{read:2012.sph.w.dissipation.switches}, {\small PHANTOM} \citep{price:2010.grid.sph.compare.turbulence}, and \citet{rosswog:2014.sph.accuracy}, and modern (post-2008) versions of {\small GASOLINE}. As above, to make our comparison as fair as possible, we use the version of PSPH implemented in {\small GIZMO}.}

\item{{\bf Moving-Mesh Method}: This refers to unstructured, moving-mesh finite-volume Godunov schemes (\S~\ref{sec:intro:movingmesh}). These are the schemes in {\small AREPO} \citep{springel:arepo}, {\small TESS} \citep{duffell:2011.TESS}, and {\small FVMHD3D} \citep{gaburov:2012.public.moving.mesh.code}\footnote{A public version of FVMHD3D is available at \\ \url{https://github.com/egaburov/fvmhd3d}}. While we have made partial progress in implementing moving meshes into {\small GIZMO}, this remains incomplete at present. Therefore we will use {\small AREPO} as our default comparison code for this method, instead. This is convenient, since both {\small GIZMO} and {\small AREPO} share a common evolutionary history; much of the underlying code architecture (for example, the parallelization, timestep scheme, and gravity solver) is similar (both share a similar amount of code with {\small GADGET-3}). Most of the {\small AREPO} results shown here are exactly those in the code methods paper \citep{springel:arepo}.}

\item{{\bf Stationary Grids}: This refers to non-moving grid codes (\S~\ref{sec:intro:grid}). There are many codes which adopt such a method, for example {\small ATHENA} \citep{stone:2008.athena} and {\small PLUTO} \citep{mignone:2007.pluto.code.methods} in the ``fixed grid'' sub-class and {\small ENZO} \citep{oshea:2004.enzo.introduction}, {\small RAMSES} \citep{teyssier:2002.RAMSES}, {\small ART} \citep{kravtsov:1997.ART} and {\small FLASH} in the AMR sub-class. The code {\small ATHENA} represents the state of the art and is often considered a ``gold standard'' for comparison studies; we therefore will use it as our example in most cases (unless otherwise specified). However, we have re-run a subset of our tests with {\small ENZO}, {\small RAMSES}, and {\small FLASH}, and confirm that we obtain very similar results. We stress that on almost every test here, AMR does not improve the results relative to fixed-grid codes {\em at fixed resolution} (and in fact it can introduce more noise and diffusion in several problems) -- it only improves things if we allow refinement to much larger cell number, in which case the same result would be obtained by simply increasing the fixed-grid resolution. This is because most of our test problems (with a couple exceptions) would require refinement everywhere. Unfortunately, none of these codes shares a common architecture in detail with {\small GIZMO} or {\small AREPO}; so we do our best to control for this by running these codes wherever possible with the same choice of Riemann solver, slope limiter, order of the reconstruction method, Courant factor and other timestep criteria, and of course resolution. Where possible, we have also compared runs with {\small AREPO} in ``fixed grid'' mode (the mesh is Cartesian and frozen in time); as shown in \citet{springel:arepo}, this gives very similar results to {\small ATHENA} on the problems studied.}

\end{itemize}

\vspace{-0.5cm}
\subsubsection{Comments on Other, Alternative Numerical Methods}
\label{sec:reference:alternatives}

Before going on, we briefly comment on a couple of the other methods discussed in Table~\ref{tbl:methods}, to which we have compared a limited sub-set of problems, but will not show a systematic comparison. 

First, we note that the SPH mode of our code can be modified to match different ``corrected'' SPH methods, and we have explicitly run several test problems using the {\small rpSPH} method \citep{abel:2011.sph.pressure.gradient.est}, the \citet{morris:1996.sph.stability} method, the {\small MLS-SPH}, \citep{dilts:1999.sph.stability}, and exact-integral SPH method \citep{garciasenz:2012.integral.sph,rosswog:2014.sph.accuracy}.\footnote{It is worth noting that there are many classes of ``corrected'' SPH, many of which can be summarized within the context of the ``reproducing kernel particle method'' as generally developed in \citet{liu:1995.reproducing.kernel.particle.methods}.} However, as noted in \S~\ref{sec:intro}, all of these methods sacrifice conservation of energy and linear momentum, and numerical integration stability. As shown in \citet{hopkins:lagrangian.pressure.sph} (see also \citealt{abel:2011.sph.pressure.gradient.est}), on many test problems (for example, a low-resolution Sedov blastwave), this leads to catastrophic errors that grow exponentially and totally dominate the solution. Most will crash on more complicated test problems (the Keplerian disk, interacting blastwave, Noh, Zeldovich, and Santa Barbara cluster). And even on the problems where they run stably, these methods do not eliminate all the low-order SPH errors (just the zeroth-order pressure gradient error in the equation of motion); therefore we do not see much performance improvement compared to PSPH (which is stable and conservative). We find similar results if we use the method IAD0 method in \citealt{garciasenz:2012.integral.sph,rosswog:2014.sph.accuracy} to restore conservation at the cost of first and zeroth-order consistency (though in sufficiently smooth flows, this does improve things over TSPH noticeably).  

Next, we can also modify the SPH mode of our code to behave as a ``Godunov SPH'' code. We have done so using both the ``standard'' implementation by \citet{cha:2003.godunov.sph}, and using the improved version from \citet{inutsuka:2002.godunov.sph,cha:2010.godunov.sph,murante:2011.godunov.sph} which uses a higher-order quadrature rule for the equation of motion. However, we note that while this eliminates the need for artificial viscosity and conductivity terms in SPH, it does not inherently resolve any of the other errors in the method (e.g.\ low-order errors and zeroth-order inconsistency, the surface tension error, etc.). Because of this, we find, as did \citet{cha:2010.godunov.sph}, that it does not yield any noticeable improvement for fluid mixing instabilities (e.g.\ the Kelvin-Helmholtz test); what {\em does} improve things in their tests is going to many more neighbors ($N_{\rm NGB}\gtrsim300$) and using a higher-order quadrature rule, but this is already implicit in PSPH. And accuracy on some tests is noticeably poor compared to PSPH (e.g.\ the ``blob'' test; see Fig.~12 in \citealt{murante:2011.godunov.sph}). And while Godunov SPH has certain formal stability properties, it is very difficult in practice to maintain non-linear stability in the Riemann solver given the low-order SPH errors which survive, unless huge neighbor numbers are used (this has been a significant barrier to implementation). This instability appears in mild form in large post-shock oscillations in shocktubes in the papers above, but for some of the complicated test problems here (e.g.\ the Zeldovich and Noh problems), the errors crash the method.

Finally, we have also compared several test problems to finite-difference codes. We have modified a version of our code to run as a Lagrangian finite-difference code, in which case the volumetric quantities (density, etc.) rather than the conserved quantities are evolved explicitly with our matrix gradient estimator; this makes it more similar to the method in {\small PHURBAS} \citep{maron:2012.phurbas.algorithm} or {\small FPM}.\footnote{By FPM, we refer to the ``Finite Point Method'' in e.g.\ \citet{onate:1996.fpm}; this includes methods referred to as ``finite pointset methods'' and ``Lagrangian finite point methods'' in e.g.\ \citet{kuhnert:2003.finite.pointset.method,tiwari:2003.finite.pointset.method}, as well as the ``finite particle method'' of \citet{liu:2005.finite.particle.method}.} We have also compared to fixed-grid Cartesian finite difference methods; specifically using the {\small ZEUS} code \citep{zeus:a}, but this is similar in many tests to other methods like those in the {\small PENCIL} code \citep{brandenburg:2002.PENCIL.code.ppr}. These methods perform quite well on some tests where the flow is smooth and density fluctuations are small, such as the (non-boosted) Gresho vortex test, or the sub-sonic Kelvin-Helmholtz test. However, as with ``corrected'' SPH, we find that the sacrifice of conservation in these methods can lead to serious non-linear errors in many of the other test problems we consider here. The methods fail on all the cosmological tests and strongly self-gravitating tests, as well as the Noh problem \citep[see][]{liska:2003.eulerian.code.test.ppr} and the interacting blastwave problem. Even where the methods run well, we do not see any major improvement compared to the Godunov method in {\small ATHENA} (as noted by the code authors themselves in e.g.\ \citealt{maron:2012.phurbas.algorithm}, \citealt{stone:2008.athena}). For a more thorough discussion of the conservation properties and stability of these methods, we refer to \citet{clarke:2010.zeus.reliability}.

Given these results, we conclude that these methods, in their present form, are not optimal for the kinds of problems we are interested in here. Although they may have advantages for specific sub-categories of problems (for example, sub-sonic, smooth flows where higher-order methods such as those in the {\small PENCIL} code can be easily implemented and stabilized), given their difficulties running the full set of test problems in our suite, will not consider them further.

\vspace{-0.5cm}
\subsection{Smooth Equilibrium Tests}
\label{sec:tests:equilibrium}

First we consider tests which should reflect equilibrium or steady-state configurations. Some of these turn out to be the most demanding tests of certain methods! 

\begin{figure}
    \plotonesize{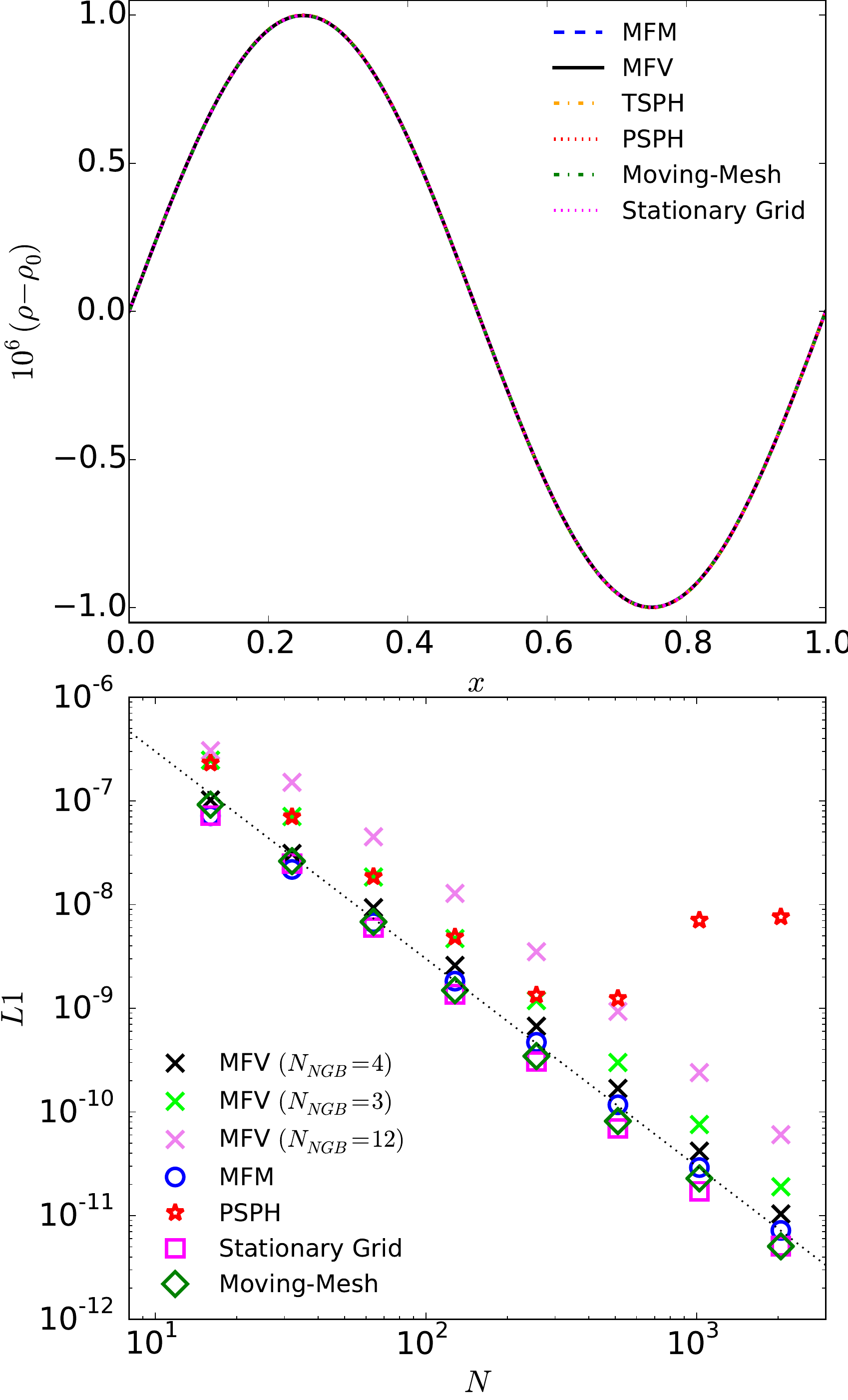}{0.98}
    \vspace{-0.25cm}
    \caption{Linear traveling soundwave test problem (\S~\ref{sec:soundwave}). {\em Top:} Soundwave evolved one period; we solve the problem with each of the methods shown, corresponding to our new Lagrangian meshless finite-mass method (MFM), new meshless finite-volume method (MFV), ``traditional'' SPH (TSPH), ``modern'' pressure-SPH (PSPH), a moving mesh ({\small AREPO}), and a fixed grid ({\small ATHENA}). All the codes give indistinuishable results from the analytic solution. {\em Bottom:} L1 error norm as a function of particle number $N$ (the L2 norm is nearly identical). Dotted line shows the ideal second-order ($L1\propto N^{-2}$) scaling. Both new methods (MFM \&\ MFV) are second-order, and have similar convergence to grid and moving-mesh methods on this problem. In all cases, the neighbor number $N_{\rm NGB}$ is fixed while $N$ is varied; we plot results for three choices of $N_{\rm NGB}$ for the MFV method. Convergence is independent of $N_{\rm NGB}$, but the normalization depends systematically on $N_{\rm NGB}$ (at $N_{\rm NGB}<4$ in 1D, there are too few neighbors and noise increases; at $N_{\rm NGB}>4$, the effective resolution systematically decreases). In {\small SPH} ({\small TSPH} or {\small PSPH}), L1 is very sensitive to the ``particle order'' of the initial ICs. Here, L1 decreases until the zeroth-order (E0) error from imperfect initial order dominates (fractional errors $\sim 10^{-3}-10^{-2}$), then the error actually increases.
    \vspacerpostplot 
    \label{fig:soundwave}}
\end{figure}

\vspace{-0.5cm}
\subsubsection{Linear Traveling Soundwave: Convergence Testing}
\label{sec:soundwave}

We begin by considering a simple linear one-dimensional soundwave.\footnote{See \burl{http://www.astro.princeton.edu/~jstone/Athena/tests/linear-waves/linear-waves.html}} This is problem is analytically trivial; however, since virtually all schemes are first-order for discontinuities such as shocks, smooth linear problems with known analytic solutions are the only way to measure and quantitatively test the accuracy and formal convergence rate of numerical algorithms. Following \citet{stone:2008.athena}, we initialize a periodic domain of unit length, with a polytropic $\gamma=5/3$ gas with unit mean density and sound speed (so pressure $P=3/5$). We then add to this a traveling soundwave with small amplitude $\delta \rho/\rho = 10^{-6}$ (to reduce non-linear effects) with unit wavelength. After the wave has propagated one wavelength, it should have returned exactly to its initial condition. 

Fig.~\ref{fig:soundwave} shows the results for each code after one period. Unsurprisingly, all the methods are able to accurately follow the soundwave. After one wave propagation period, we define the L1 error norm as 
\begin{align}
\label{eqn:L1}
{\rm L1} = \frac{1}{N}\,\sum_{i}\,|\rho_{i} - \rho(x_{i})|
\end{align}
where $N$ is the number of particles, $\rho_{i}$ is the numerical solution for cell $i$, and $\rho(x_{i})$ is the analytic solution (identical to the initial conditions for this problem). Fig~\ref{fig:soundwave} shows the error norm as a function of the particle number: for both the MFM and MFV methods, the results show second-order convergence (as expected for a smooth problem and a second-order accurate method). The MFM shows slightly smaller errors but the difference is not large. Note that the number of neighbors in the kernel is kept {\em fixed} as $N$ is increased: convergence does not require higher-$N$. For all kernel-based methods, we use $N_{\rm NGB}=4$ neighbors in one dimension unless otherwise specified, for this and all other 1D tests. 

For the MFM and MFV methods the rate of convergence (power-law slope) is insensitive to the choice of neighbor number. We show this explicitly by comparing the L1 norm for $N_{\rm NGB}=12$ and $N_{\rm NGB}=3$ for the MFV method (the MFM result is similar). At fixed $N$, the L1 norm becomes slightly larger for $N_{\rm NGB}\lesssim 3$, because there are not enough particles in the stencil (so there are slightly larger ``start up'' errors in the density \&\ velocity fields). At much larger $N_{\rm NGB} \gtrsim 5-6$, the L1 norm increases again with $N_{\rm NGB}$ simply because the effective resolution is lower: in Fig.~\ref{fig:soundwave}, the $N_{\rm NGB}=12$ case shows an L1 norm very similar to that with $N_{\rm NGB}=4$ and $N\rightarrow N/3$, exactly as expected for a kernel $3$ times larger than is ``needed.'' The optimal choice in 1D, $N_{\rm NGB}\approx 4$, is expected (this corresponds to $\approx 1$ neighbor within the Gaussian-like ``core'' of the kernel on each side of the searching particle). Note that we see identical behavior in the L2 error norm (${\rm L2} \equiv  \langle (\rho_{i}-\rho(x_{i}))^2 \rangle^{1/2}$). We have evolved the wave to $\sim 1000$ periods in the MFM and MFV methods, and see no visible diffusion at the resolution plotted (as expected). 

In SPH methods, it is more difficult to define the L1 norm for this problem, because it depends sensitively on the start-up conditions. Per \citet{springel:2010.sph.review}, if the initial particle order is imperfect, the E0 error totally dominates the L1 density norm (although the velocity norm can continue to show convergence). Here, we iteratively relax the initial grid and refine the smoothing lengths until ``good particle order'' is ensured (the absolute deviation from perfect equality in $\sum \partial W({\bf x}_{i}-{\bf x}_{j},\,h_{i})/\partial |{\bf x}_{i}-{\bf x}_{j}| = 0$ and Eq.~\ref{eqn:nngb} is $<10^{-15}$, with the correct initial densities at the particle locations). This eliminates the ``start up'' density field errors. We therefore do see some convergence at low resolutions; however, once the fractional errors become comparable to the E0 pressure gradient error introduced by the particle relative motion (part in $\sim 10^{-3}-10^{-2}$), convergence ceases, and in fact the errors actually grow at higher resolution. The behavior is qualitatively identical in TSPH and PSPH, for fixed $N_{\rm NGB}$. 

Repeating this test in 2D and 3D gives similar results for all codes.

\begin{figure*}
    \plotsidesize{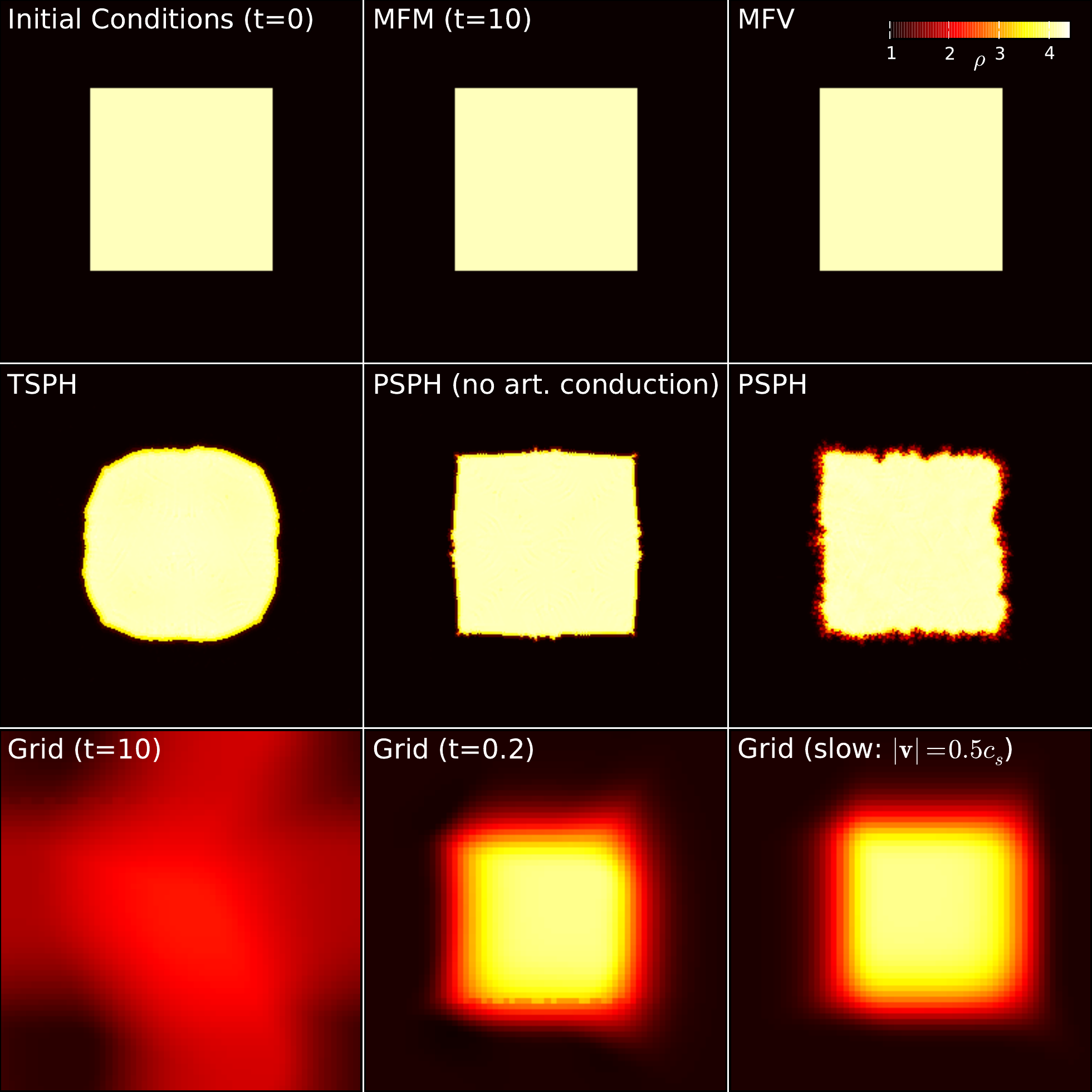}{0.98}
    \vspace{-0.15cm}
    \caption{Hydrostatic square advection test (\S~\ref{sec:square}). {\em Top Left:} Initial condition (yellow shows density $\rho=4$, black $\rho=1$): a high-density square in hydrostatic equilibrium, with all fluid moving at constant velocity ($|{\bf v}|\sim150$). The periodic box ($0<x<1$, $0<y<1$) is shown. {\em Top Center:} MFM solution at $t=10$; this reproduces the correct solution (identical to the IC) to machine precision. {\em Top Right:} MFV solution. This is also exact to machine precision. Moving-meshes should do the same. {\em Middle Left:} TSPH: Advection is handled well, but the known ``surface tension'' error forces the square gradually into a circle. {\em Middle Center:} PSPH, using the ``traditional'' SPH artificial viscosity and no artificial conductivity: this removes the surface tension but particle asymmetry around the contact discontinuity still produces spurious forces. {\em Middle Right:} ``standard'' PSPH: artificial conductivity produces excessive (and noisy) diffusion around the discontinuity. {\em Bottom Left:} Stationary grid (here {\small ATHENA}): advection errors completely destroy the square, despite forcing $\sim1000$ times smaller timesteps in this case. {\em Bottom Center:} Stationary grid result at time $t=0.2$, showing the magnitude of distortion after the square moves a few times its size. {\em Bottom Right:} Stationary grid result for a slower sub-sonic ($|{\bf v}|=0.5$) advection at $t=10$ (the square has traveled much less distance).\vspacerpostplot
    \label{fig:square}}
\end{figure*}

\vspace{-0.5cm}
\subsubsection{The Square Test: Advection \&\ Surface Tension Errors}
\label{sec:square}

We next consider the ``square'' test common for recent SPH studies \citep{cha:2010.godunov.sph,hes:2010.tesselation.sph,saitoh:2012.dens.indep.sph,hopkins:lagrangian.pressure.sph}. We initialize a two-dimensional fluid in a periodic box of length $L=1$ and uniform pressure $P=2.5$, $\gamma=1.4$, and density $\rho=4$ within a central square of side-length $L=1/2$ and $\rho=1$ outside. The entire fluid is given a uniform and abitrary large initial velocity $v_{x} = 142.3$, $v_{y} = -31.4$. We use $64^{2}$ particles. Fig.~\ref{fig:square} shows the initial condition and the resulting system evolved to a time $t = 10$, centered on the central square position at that time. The square should be in hydrostatic equilibrium, with the final state exactly identical to the initial state.

The MFM and MFV methods perform essentially perfectly here: in fact, it is straightforward to show that they solve this particular test problem exactly (to machine accuracy). The same is true of moving mesh codes, provided that the moving mesh also uses a gradient estimator which is exact for linear gradients and advects cells with the bulk fluid velocity.

It is well known (see the references above) that ``traditional'' SPH (all SPH methods where the density is kernel-smoothed but entropy or internal energy is particle-carried) have an error term which behaves as a physical surface tension: a repulsive force appears on either side of the contact discontinuity, opening the gap between the central square and outer medium which then deforms the square to minimize the surface area of the contact discontinuity (eventually becoming a circle). This is the same as the error which generates the well-known ``pressure blips'' in shocktube tests. We see exactly this behavior here. Perhaps most disturbing, the error {\em does not converge away} (it is zeroth-order). The pressure-entropy case minimizes this error \citep[see][]{hopkins:lagrangian.pressure.sph,hu:2014.psph.galaxy.tests}; however, there is still a ``rounding'' of the corners and substantial noise around the edges of the square. This owes to two factors: (1) the zeroth-order consistency (E0) error in SPH means that even when every particle has an exactly identical pressure, there are still net forces on the particles, especially when there is an asymmetry in the particle distribution as occurs near the contact discontinuity; (2) the artificial conduction terms in the modern SPH diffuse the contact discontinuity even when there is perfect stability. 

If the square is not moving, this problem is trivial for grid codes. However, if the square has any motion relative to the grid (and not perfectly aligned), then large advection errors appear. In {\small ATHENA}, each time the square moves its own length, it is both diffused and distorted (the magnitude of the distortion comparable to that in SPH ``per crossing''). Here we have used the second-order integration method to match the other codes; if we use a higher order PPM method we see some improvement but the qualitative behavior is the same. If we use a first-order grid method the square cannot be reliably advected even a single unit length. It is also worth noting that in the grid code, the timestep criterion should include the relative gas-grid motion: thus these errors appear despite the fact that the timesteps in the grid code are a factor of $\sim1000$ smaller than in all the other methods. And we stress that AMR methods cannot help here, without overall increasing the resolution (in which case they will still be less accurate than an MFM or MFV run at the same resolution), since the diffusion is uniform around the boundaries -- in fact running this test with {\small RAMSES} or {\small ENZO}, we actually see {\em more} diffusion if we refine at the contact discontinuity, because (as is well-known) the AMR scheme effectively becomes lower-order along refinement boundaries.

Note that in the 1D analog of this problem (advecting a constant-pressure, constant-velocity 1D contact discontinuity), MFM, MFV, and moving-mesh methods perform similarly well, and SPH has no problems (a pressure `blip' is present, but the surface tension-like instability only appears in higher dimensions). But it is well-known that non-moving grid codes will still excessively diffuse the discontinuity \citep[even though the motion is necessarily grid-aligned; see][]{springel:arepo}. In the 3D analog (advecting a cube), the results and differences between codes are essentially identical to the 2D test here (SPH deforms it into a sphere, fixed-grid codes diffuse along all edges, moving-mesh, MFM, and MFV codes are machine-accurate). 

\begin{figure}
    \plotonesize{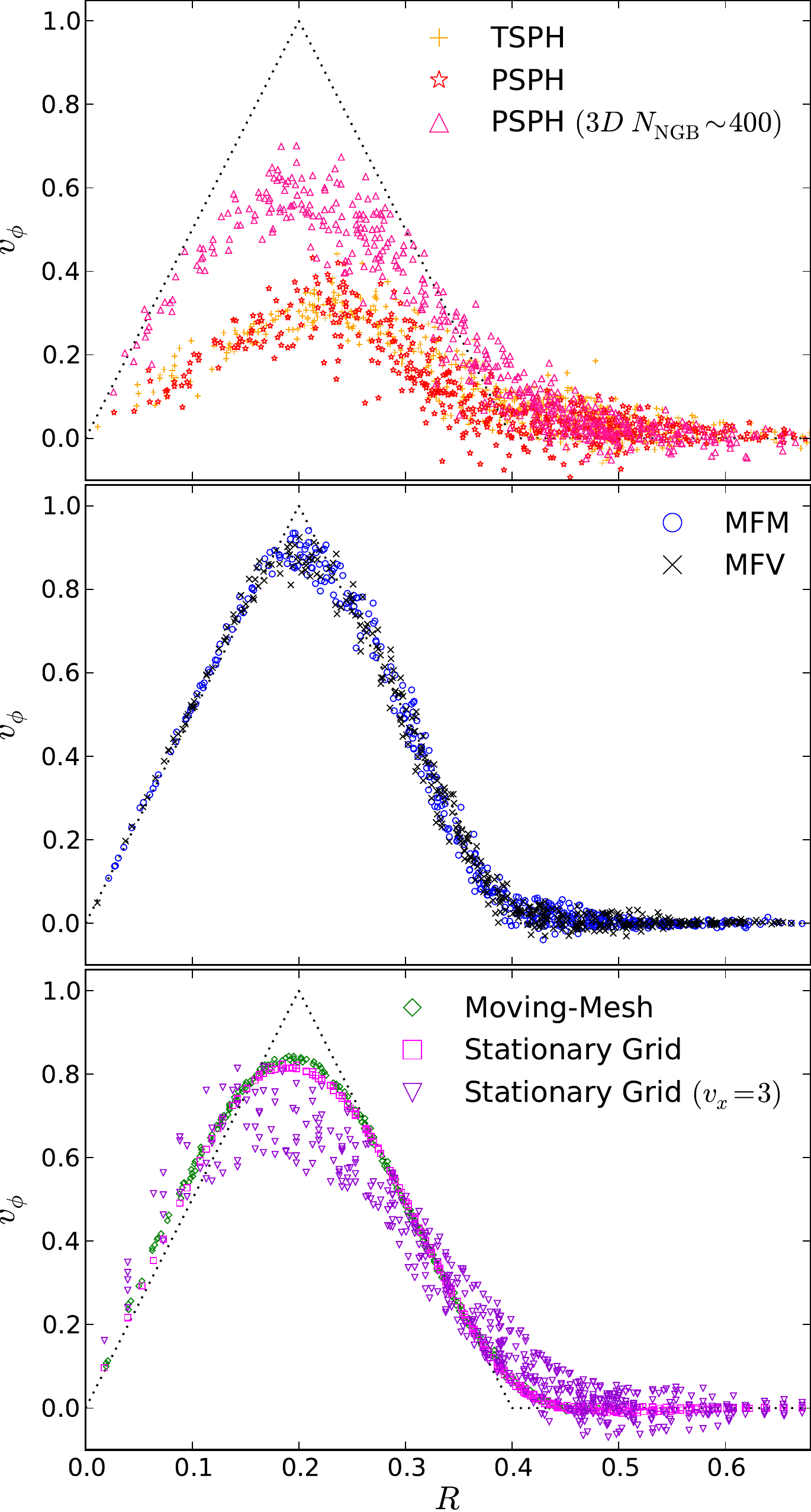}{0.98}
    \vspace{-0.25cm}
    \caption{The Gresho vortex (\S~\ref{sec:gresho}). The code should preserve a steady-state hydrodynamic vortex following the analytic solution (dotted line); we plot the azimuthal velocity versus radius for each code method at $t=3$, or $\sim1$ complete vortex orbit, at $40^{2}$ resolution (each point is one particle/cell; for clarity we plot a random subset of all cells). {\em Top:} SPH methods. This is known to be a very challenging test for SPH, and even the most sophisticated SPH methods generate large noise (from ``partition noise'' and the E0 error) and steadily degrade the vortex. Increasing the kernel neighbor number helps, but convergence is slow: we compare a test run with a higher-order Wendland kernel and the 3D equivalent of $400$ neighbor particles (vs.\ standard $32$). {\em Middle:} MFM and MFV methods. The two are very similar. Some (much smaller) noise is generated but the peak is preserved. {\em Bottom:} Moving-mesh and fixed-grid methods. These give very similar results when the vortex has zero mean velocity; both give much less noise (because the volume partition is exact, and mesh-deformation is reduced), though they dissipate the peak slightly more than MFM/MFV. We compare, however, the results if the vortex is moving (add a uniform velocity $v_{x}=3$); here advection errors lead to much larger noise in the fixed-grid solution (while the moving-mesh, MFM, MFV, and SPH results are invariant to such boosts).\vspacerpostplot  
    \label{fig:gresho}}
\end{figure}

\begin{figure}
    \plotonesize{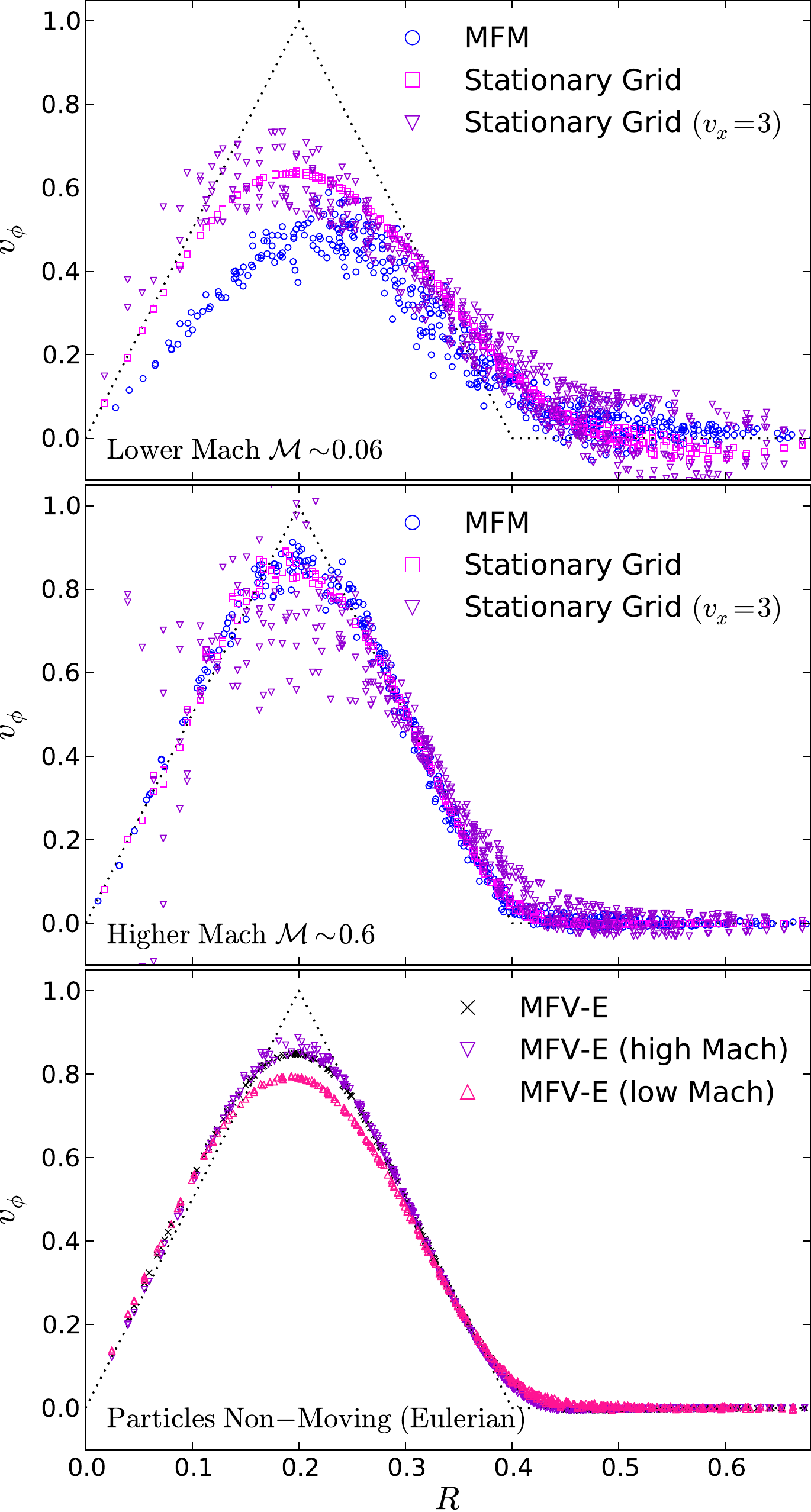}{0.98}
    \vspace{-0.25cm}
    \caption{Gresho vortex as Fig.~\ref{fig:gresho}, but varying the Mach number of the vortex. We compare the MFM and fixed-grid methods at $t=3$ (with and without a moving vortex, for the fixed-grid results). MFV is very similar to MFM in all cases so is not shown. {\em Top:} A highly sub-sonic (rms Mach number $\sim 0.06$) vortex (background pressure increased by $P_{0}=50$). The accuracy of all methods degrades, but the effect is more severe in the MFM/MFV methods. {\em Middle:} A trans-sonic (rms Mach $\sim 0.6$; $P_{0}=-5$) vortex. All solutions improve, except the noise in grid methods (especially for a moving vortex) gets larger. In all cases, both SPH methods (even with $N_{\rm NGB}=400$) perform significantly worse than any other method. {\em Bottom:} Same tests, run with our MFV method but with the particle positions {\em fixed}, so the fluid is purely advected between particles (the method becomes an Eulerian stationary-finite volume method). Here even at very low Mach number, the noise is totally eliminated; this confirms that the code behaves like a stationary-grid code in the limit of zero particle motion, and the noise we see comes from the second-order noise in the volume partition and mesh deformation. This change introduces the same errors in advection and velocity-dependence of errors as fixed-grid methods, however (the examples here have no bulk fluid flow).\vspacerpostplot
    \label{fig:gresho.P0}}
\end{figure}

\vspace{-0.5cm}
\subsubsection{The Gresho Vortex: Sub-Sonic Turbulence \&\ Angular Momentum}
\label{sec:gresho}

We next consider the triangular vortex of \citet{gresho:1990.vortex}. A two-dimensional gas with uniform $\rho=1$ is initialized in a periodic domain $0<x<1$, $0<y<1$, with zero radial velocity, pressure 
\begin{align}
\label{eqn:gresho.pressure} P(R) &= 
\begin{cases}
	{\displaystyle 5 + 12.5\,R^{2}\ \ \ \ \ \hfill { (0 \le R < 0.2)}} \\
	{\displaystyle 9 + 12.5\,R^{2} - 20\,R + 4\,\ln{(5\,R)}\ \ \ \hfill { (0.2 \le R < 0.4)}} \\
	{\displaystyle 3 + 4\,\ln{2}\ \ \ \ \ \hfill { (R \ge 0.4)}} 
\end{cases}
\end{align}
and azimuthal velocity 
\begin{align}
v_{\phi}(R) &=
\begin{cases}
	{\displaystyle 5\,R\ \ \ \ \ \hfill { (0 \le R < 0.2)}} \\
	{\displaystyle 2-5\,R\ \ \ \ \ \hfill { (0.2 \le R < 0.4)}} \\
	{\displaystyle 0\ \ \ \ \ \hfill { (R \ge 0.4)}} 
\end{cases}
\end{align}
where $R^{2}=x^{2}+y^{2}$. This represents a steady-state equilibrium vortex. We initialize the vortex with $64^{2}$ elements. Fig.~\ref{fig:gresho} shows the results, evolved to time $t=3$, or about $2.4$ orbits of the vortex ``peak'' ($1.2$ orbits of the outermost vortex edge). There is no ``1D vortex'' analogue of this problem; but we discuss the 3D analogue (the ``vortex tube'') below.\footnote{We have also considered the isentropic vortex from \citet{yee:1999.isentropic.vortex.test}, which involves a similar setup but with a smooth (albeit non-linear) pressure and density variation balancing the vortex rotation. Our qualitative conclusions are the same as for the Gresho vortex, for all the methods we compare.}

It is well-known that SPH has serious difficulties with this test (in fact, in most SPH tests in the literature, the vortex is not evolved beyond $t=1$). The shear motion of particles leads to a constant implicit ``mesh deformation'' and re-calculation of the effective particle ``volume.'' Because this ``volume'' is not conserved in SPH, but conserved quantities (particle masses and energies) are locally carried, this leads to a sort of ``volume partition noise'' (henceforth, simply ``partition noise'') in the volumetric fields (i.e.\ pressure), hence ultimately in the velocity field. The velocity noise is damped by artificial viscosity, diffusing the vortex. We confirm this: with both TSPH and PSPH, the results are very noisy, and the damping of the peak velocity is severe, as is the velocity diffusion out to larger radii (beyond the original vortex). Improved artificial viscosity switches do not do much to change this. Various authors have pointed out that increasing the kernel neighbor number does help here; \citet{read:2012.sph.w.dissipation.switches} (see Fig.~5-6 there) and \citet{dehnen.aly:2012.sph.kernels} (Fig.~9-10 therein) advocate going to $N_{\rm NGB}>400$ neighbors in 3D. We have in fact repeated this test using the Wendland $C^{6}$ kernel with $N_{\rm NGB}=500$ or triangular kernel with $N_{\rm NGB}=442$ (to do so we repeated this with the 3D analogue of the test, which also helps to reduce noise). This does help, but not very much; as shown by both groups, the $L1$ norm decreases only as $\sim N_{\rm NGB}^{-0.5}$. In fact, the performance with both SPH methods even with $N_{\rm NGB}\sim 500$ is still significantly worse than any other method we consider. And the computational expense involved is large: depending on the kernel, the ``effective resolution'' goes down as something like $N_{\rm NGB}^{-1/2}$, so the CPU cost of the SPH computation for equivalent resolution scales something like $\sim N_{\rm NGB}^{3/2}$ \citep{zhu:2014.sph.convergence} -- i.e.\ this improvement entails a factor $>50$ CPU cost in the SPH loops over the standard $\sim 32$ neighbors! And both these authors verify that, because this problem is significantly affected by the E0 error in SPH, convergence with total particle number, even at high $N_{\rm NGB}$, is slow ($\sim N_{1D}^{-0.6}$).

The MFM and MFV methods show a tremendous improvement relative to SPH, despite using the simple cubic spline kernel with fixed $N_{\rm NGB} = 32$ in 3D ($N_{\rm NGB}=16$ in 2D, following \citealt{gaburov:2011.meshless.dg.particle.method}). The solution is less noisy than the SPH equivalent with $N_{\rm NGB} \gtrsim 500$, and the vortex has decayed much less rapidly. However there is still significant noise. This comes from a combination of the ``partition noise'' above (a much milder form, compared to SPH, still exists in these methods, because our discrete volume partition estimator is only accurate to second-order; see \S~\ref{sec:methods:smoothing}), as well as the usual ``grid noise'' associated with mesh motion/deformation (which can be significant here because, as discussed in \S~\ref{sec:methods:differences}, the implicit deformation of the ``effective faces'' can be more complicated than simple uniform motion of a flat geometric face; see \citealt{springel:2011.voronoi.tesselation.review,mcnally:2012.kh.test.comparison,munoz:2014.disk.planet.interaction.sims} for discussion of this noise in moving-mesh codes). We also find (not surprisingly) that the degree of vortex decay is very sensitive to our choice of slope-limiter: using a more conservative limit on monotonicity (see \S~\ref{sec:slopelimiters}) leads to a smoother solution but much stronger damping of the vortex peak. 

By comparison, the results from {\small ATHENA} show almost no noise, because the exact and time-independent volume partition means there is no ``partition noise'' or ``mesh deformation.'' There is more decay at this time compared to the MFM and MFV methods, but we find at later times the vortex is better preserved. However, while fixed-grid methods do very well in the basic version of this test, at least two simple modifications of the problem dramatically reduce their accuracy (while having no effect on the other methods we consider). The first is if we consider a 3D version of the problem, where the vortex is initialized as a cylinder with the same dependence of $v$ and $P$ on $R$ and infinite (periodic) in $z$, then rotate the problem geometry so it is not exactly aligned with the Cartesian grid axes. This creates significant errors which quickly damp the angular momentum until the vortex is realigned with the local grid (then, this realignment will slightly offset the vortices at different heights in the cylinder, which will interfere with each other via numerical viscosity until the structure is dissipated). We will consider such errors in the next test problem (\S~\ref{sec:keplerian}). The second modification is to simply set the problem in bulk motion. \citet{springel:arepo} consider this case in more detail, and show that the errors at fixed resolution then grow rapidly with the bulk motion: for a bulk motion with $v_{\rm bulk}\gtrsim v_{\rm vortex}$ (where $v_{\rm vortex}=1$ is the peak vortex velocity here), the noise in the fixed-grid solution becomes very large. We verify this here -- for modest bulk flow velocities relative to the grid (any velocity comparable to the vortex rotation velocity itself), the noise ``blows up,'' becoming worse than our MFM and MFV results (though still superior to the SPH results, until we reach vortex velocities $\gtrsim 30-50$).\footnote{We have verified that what matters for these errors in stationary-grid codes is the ratio of bulk velocity to vortex velocity, not the sounds speed or pressure. The errors which scale with the sound speed (discussed below) are almost entirely separable.}

Of course, in stationary grid codes this noise owing to mis-alignment or bulk-motion of vortices can be reduced by increasing the resolution, and will eventually converge away. However this means that at fixed resolution, their accuracy can be severely reduced, or equivalently that their ``effective resolution'' will be much lower for certain problems. By comparison, all the other methods we consider are manifestly invariant to both rotations of the vortex and bulk motions. So, for a mis-aligned vortex with bulk motion of $\sim v_{\rm vortex}$, for example, we require a resolution of $\sim 256^{2}$ to achieve similar accuracy to a $64^{2}$ simulation with the MFM or MFV methods. And since the whole volume is affected, AMR does not improve things.

This is a serious concern for realistic simulations with stationary grids, where the vortex position and motion cannot be exactly known ``ahead of time.'' Consider, for example, simulations of super-sonic turbulence. If we desire to resolve a modest Reynolds number of $\sim100$, then since the super-sonic cascade $|{\bf v}^{2}(\lambda)|^{1/2} \propto \lambda^{1/2}$ (where $\lambda$ is a parameter reflecting spatial scale; see \citealt{federrath:2010.obs.vs.sim.turb.compare}), we expect the smallest ``resolved'' eddies to be randomly advecting through the box with bulk motions set by the largest eddies, a factor $100^{3/8}\sim 6$ larger than their internal eddy velocities. If we ``boost'' the Gresho problem by this multiplier, we find we require a resolution $\sim 32^{2}-64^{2}$ across the eddy for its structure to survive to $t=3$, and $\sim 256^{2}$ for it to have comparable accuracy to a non-boosted $32^{2}$ simulation: so the smallest eddy we wish to resolve should actually be $\sim 32-256$ (depending on the desired accuracy) times larger along each axis than the grid scale! Quite similar criteria have been obtained in other studies with grid codes,  when the turbulence is driven by self-gravity and/or magnetic fields \citep{federrath:2011.resolution.criteria.dynamos}. This is more demanding than what is usually estimated based on examining the shape of the turbulent power spectrum, by a factor of a few (which should actually not be surprising, since here we are not just requiring the second moments be reasonable, which can be accomplished via noise, but that the eddy structure is reasonable). Because the errors grow with boost velocity, the resolution requirement grows super-linearly with the desired Reynolds number in stationary-grid simulations. 

The best compromise in this particular test problem appears to come from moving mesh methods. These give similarly accurate and smooth results to the second-order stationary grid methods with no bulk velocity, but are invariant to bulk motions of the vortex and to rotations. The advantage over the MFV and MFM methods here is the exact volume partition (which eliminates the ``partition noise'' described above), combined with the fact that the faces are simple, flat geometric objects (which decreases the ``mesh noise'' as these are deformed, although non-trivial motions such as mesh rotation still introduce grid-scale noise). 

All of these results, however, are sensitive to the Mach number of the vortex. Note that, mathematically, we can add any constant $P_{0}$ to the pressure everywhere in Eq.~\ref{eqn:gresho.pressure} and the dynamics should be identical. Numerically, however, none of the methods is invariant with respect to this choice. In {\em all} cases, lowering (raising) the background pressure ($P_{0}<0$ or $P_{0}>0$) leads to better (worse) conservation of the vortex; this is because small integration errors in the pressure gradients will launch spurious velocities that have magnitudes which scale with the sound speed. The minimum physical pressure for this problem, $P_{0}=-5$, corresponds to a vortex with Mach number $\mathcal{M}(R=0.2)\approx 1.1$ at the vortex ``peak'' (rms $\langle\mathcal{M}^{2}\rangle^{1/2}\approx 0.6$ over the vortex). The standard choice of \citet{gresho:1990.vortex} above ($P_{0}=0$) corresponds to $\mathcal{M}(R=0.2)\approx 1/3$ ($\langle\mathcal{M}^{2}\rangle^{1/2}\approx 0.2$). We also consider a much higher $P_{0}=50$, or $\mathcal{M}(R=0.2)\approx 0.1$ ($\langle\mathcal{M}^{2}\rangle^{1/2}\approx 0.06$). 

In all cases, the qualitative differences between the methods are similar. With $P_{0}=-5$, there is some improvement across all methods, but the comparison between methods is similar (although some surprising noise of unclear origin appears in the {\small ATHENA} solution even without a velocity boost). However, the meshless methods (SPH, MFM, and MFV) are much more sensitive to large $P_{0}$ than the stationary grid methods. It appears that the ``partition errors'' from the implicit mesh deformation grow super-linearly with sound speed. As a result, in SPH the vortex is completely ``wiped out'' by $t=3$ for $\langle\mathcal{M}^{2}\rangle^{1/2}\lesssim 0.2-0.3$; for MFM and MFV methods we see some vortex survive to $t=3$ down to $\langle\mathcal{M}^{2}\rangle^{1/2}\sim 0.03-0.05$; but with stationary grids we can reach $\langle\mathcal{M}^{2}\rangle^{1/2}\sim 0.001$ (perhaps even lower with a higher-order PPM method). As above, we can always improve this with increasing resolution, but for SPH the convergence is very slow (sub-linear), and even for the MFM and MFV the convergence is closer to linear than second-order (also seen for {\small AREPO}; see \citealt{springel:arepo}, Fig.~29). This is a serious concern for simulations of sub-sonic turbulence. The limitations of SPH in this regime are well-known \citep[see e.g.][]{price:2011.sph.turb.response,bauer:2011.sph.vs.arepo.shocks,vogelsberger:2011.arepo.vs.gadget.cosmo}; we confirm those results here. But while our MFM and MFV methods offer a tremendous improvement relative to SPH, and can converge, this test suggests they lose accuracy rapidly relative to stationary grid codes once the Mach numbers fall to $\sim 0.01$ (numerical noise starts to swamp ``real'' turbulent effects at reasonable resolution of the smaller eddies in a realistic simulation). For highly sub-sonic problems, then, the lack of a partition error suggests stationary grid codes offer a significant advantage. It remains to be seen how moving-mesh codes compare in this limit, since there still is a non-zero grid noise and mesh deformation (which depends on the ``mesh regularization'' procedure), but the volume partition is exact. 

To confirm that the major differences here are related to the volume partition and ``mesh deformation'' errors, we re-run our MFV simulations with each Mach number in Fig.~\ref{fig:gresho.P0}, but with a fixed-particle (Eulerian) form of the code. Recall (\S~\ref{sec:methods:particle.velocities}), in the MFV form (with particle-particle mass fluxes), our method allows arbitrary particle velocities in principle -- we do not have to set them to follow the fluid velocity in a Lagrangian manner (although this is our default choice). So here we re-run the method with the particle positions fixed; the fluid is entirely advected between particles, then, and the code is effectively a meshless, Eulerian stationary-grid method. At all Mach numbers, this totally eliminates the noise. Because particle ``volumes'' are conserved (since their relative positions do not change), there is no partition/deformation error, and the code looks very similar to the stationary-grid {\small ATHENA} results. This confirms that indeed, it is these errors introducing noise, and that this method is, in principle, flexible and just as capable of capturing many advantages of fixed-grid codes if they would be desired for certain problems. However, we emphasize that this change means the method is no longer boost-invariant: if the fluid is in bulk motion, it must be advected constantly, and we see similar (in fact slightly larger, because the particle positions are not uniform along the velocity direction) noise as in the stationary-grid method. Likewise all the other advection and diffusion errors of fixed-grid codes pertain to this method as well.

\begin{figure*}
    \plotsidesize{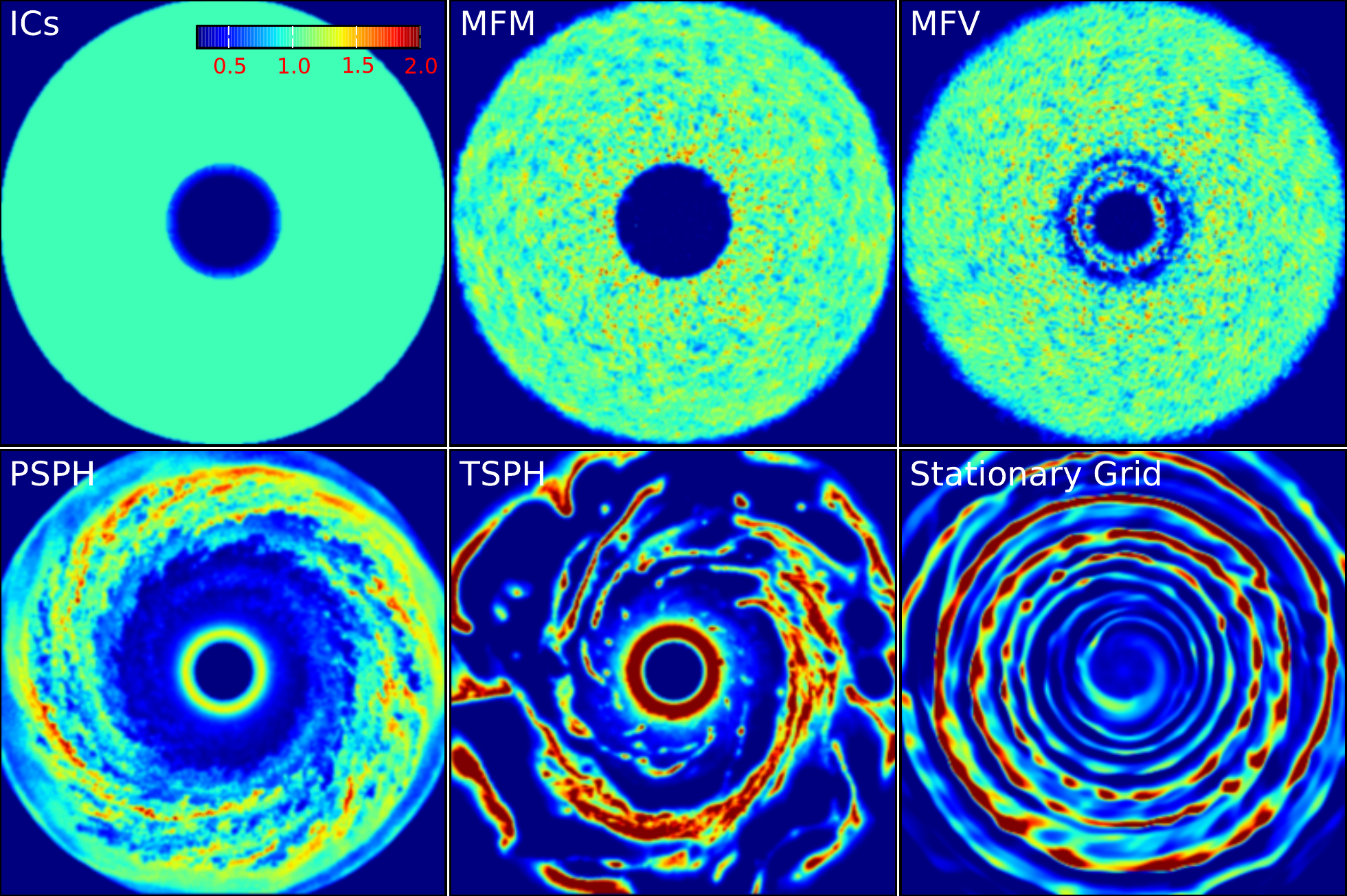}{0.98}
    \caption{Keplerian disk problem (\S~\ref{sec:keplerian}). {\em Top Left:} Initial Conditions. The gas is initialized with constant surface density from $r=0.5-2.0$, on circular orbits, with vanishing pressure, subject to an analytic Keplerian potential (without self-gravity; orbit time is $=2\pi\,r^{3/2}$), with effective $256^{3}$ resolution (the plotted domain extends from $-2<x<2$, $-2<y<2$). This should remain in equilibrium indefinitely, but numerical viscosity and advection errors steadily degrade the disk and transport angular momentum. We show the surface density evolved in each method to $t=120$ ($\sim 20$ orbits at $r=1$). {\em Top Middle:} MFM: the disk preservation is excellent (there is a small amount of noise in the density field, as in the Gresho test, but this does not degrade the orbits). We can continue to evolve the system for $\gg100$ orbits before the disk degrades. {\em Top Right:} MFV: mass fluxes lead to slightly less noise in the disk density, but a small amount of angular momentum transfer which begins to degrade the inner disk at $\gtrsim 30-50$ orbits. {\em Bottom Left:} PSPH: Using a high-order artificial viscosity switch, shear viscosities are sufficiently suppressed to allow good evolution to $\sim 5-8$ orbits, but the degradation is significant. {\em Bottom Center:} TSPH: Using ``traditional'' SPH artificial viscosity with a \citet{balsara:1989.art.visc.switch} switch leads to far too much shear viscosity, and the disk undergoes the viscous instability and disrupts within $\sim 2$ orbits. {\em Bottom Right:} Stationary (Cartesian) meshes: numerical viscosity is low but advection errors of circular orbits through a Cartesian mesh are significant and disrupt the disk in $\sim 1$ orbit.\vspacerpostplot
    \label{fig:keplerian}}
\end{figure*}

\begin{figure}
    \plotonesize{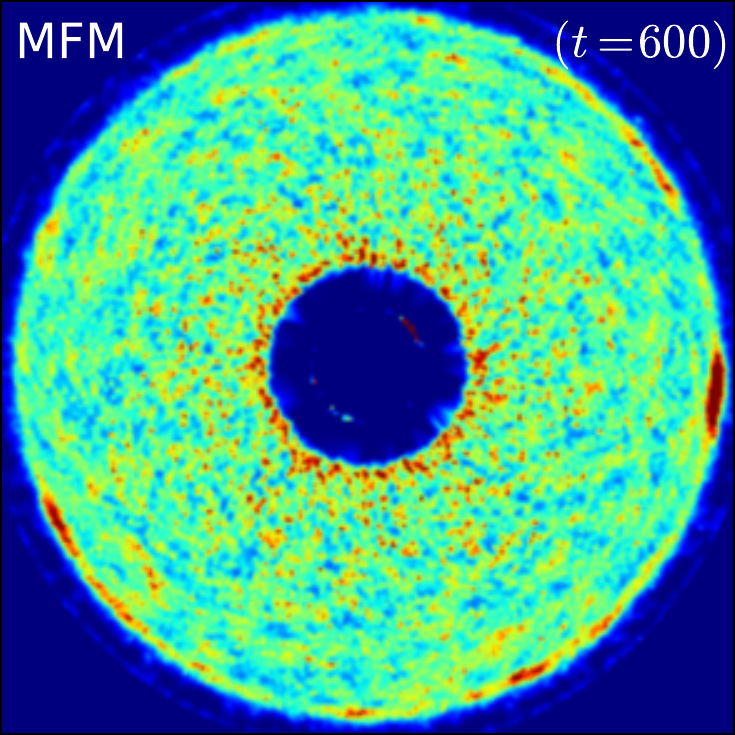}{0.98}
    \caption{Keplerian disk as Fig.~\ref{fig:keplerian}, at time $t=600$ (not a typo)! The inner ($r\sim0.5$) disk has executed $ > 250$ orbits, at this time, without decaying or disrupting.\vspacerpostplot
    \label{fig:keplerian.late}}
\end{figure}

\vspace{-0.5cm}
\subsubsection{Keplerian Disks: Angular Momentum Conservation, ``Grid Alignment,'' \&\ Stability of Cold Orbits}
\label{sec:keplerian}

\begin{figure}
    \plotonesize{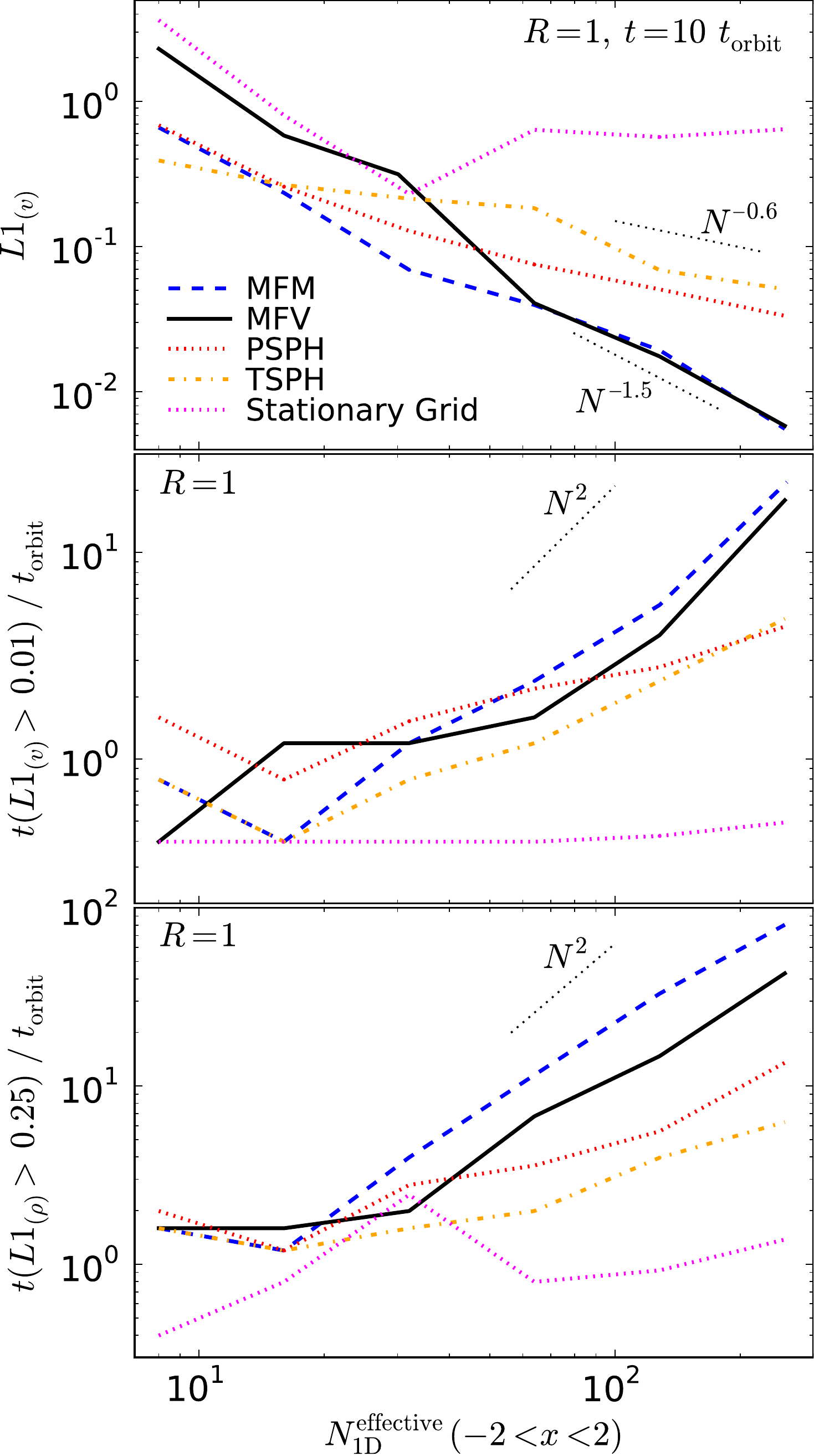}{0.92}
    \vspace{-0.2cm}
    \caption{Convergence tests of the Keplerian disk problem in Fig.~\ref{fig:keplerian}.  
    {\em Top:} $L1$ norm of the velocity error ($L1_{(v)}$) at radius $R\approx 1$ and time equal to $10$ orbital times, as a function of resolution (number of elements on a side if the initial conditions formed a uniform grid from $-2<x<2$, $-2<y<2$). MFM and MFV methods converge as $\sim N^{-1.5}$. TSPH and PSPH converge more slowly, as $\sim N^{-0.5}$. The stationary grid result does not show convergence except at the lowest resolutions. 
    {\em Middle:} Time (relative to the orbital time) at which the $L1_{(v)}$ velocity norm at radius $R=1$ first exceeds a threshold $=0.01$, as a function of resolution. This grows as $\sim N^{1.8}$ in MFM and MFV methods; more slowly $\sim N^{0.5}$ in TSPH/PSPH; and $\sim N^{0.1}$ in the stationary grid result.
    {\em Bottom:} Time at which the $L1_{(\rho)}$ norm of the gas density exceeds $0.25$, as a function of resolution. This grows as $\sim N^{1.8}$ in MFM/MFV; $\sim N^{0.5-0.6}$ for SPH; $\sim N^{0.2-0.4}$ for stationary grids. The higher threshold is chosen because small velocity errors in this problem lead to large density changes.\vspacerpostplot\label{fig:keplerian.convergence}}
\end{figure}

We now consider a cold Keplerian disk test problem. This is a critical problem for understanding the ability of codes to conserve angular momentum and follow gas in gravitational orbits for any length of time. Disks, in particular, are ubiquitous in astrophysics, and unlike the vortex problem above are dominated by gravitational acceleration rather than pressure forces (with the rotation velocity often super or trans-sonic); this focuses on that regime.

The problem is a simple variant of the well-studied Keplerian ring/disk test \citep{maddison:1996.sph.ring.instabilities,imaeda:2002.sph.shear.flow.tests,monaghan:2006.sph.shear.flows,cartwright:2009.poisson.noise.sph.shear,cullen:2010.inviscid.sph,hu:2014.psph.galaxy.tests}. We initialize a two-dimensional $\gamma=5/3$ disk with surface density 
\begin{align}
%\label{eqn:keplerian.setup} \Sigma &= \Sigma_{0} + \Sigma_{1}\times
%\begin{cases}
%	{\displaystyle (r / r_{\rm in})^{3}}\ \ \ \hfill { (r < r_{\rm in})} \\
%	{\displaystyle 1}\ \ \ \hfill { (r_{\rm in}\le r \le r_{\rm out})} \\
%	{\displaystyle \left[1 + (r - r_{\rm out})/\Delta r \right]^{-3}}\ \ \ \hfill { (r > r_{\rm out} )} \\
%\end{cases}
\label{eqn:keplerian.setup} \Sigma &= 0.01 + 
\begin{cases}
	{\displaystyle (r / 0.5)^{3}}\ \ \ \hfill { (r < 0.5)} \\
	{\displaystyle 1}\ \ \ \hfill { (0.5 \le r \le 2)} \\
	{\displaystyle \left[1 + (r - 2)/0.1 \right]^{-3}}\ \ \ \hfill { (r > 2 )} \\
\end{cases}
\end{align}
%where $r = |{\bf r}|$ is the distance from the origin, and $\Sigma_{1}=1$, $\Sigma_{0}=0.01\ll \Sigma_{1}$, $r_{\rm in} = 0.5$, $r_{\rm out}=2$, $\Delta r=0.1$. 
where $r = |{\bf r}|$ is the distance from the origin. The gas has vanishingly small, constant pressure $P=10^{-6}$, and is subject to the softened external Keplerian potential $\Phi = -(r^{2} + \epsilon^{2})^{-1/2}$ (acceleration $\ddot{{\bf r}} = - {\bf r}\,(r^{2} + \epsilon^{2})^{-3/2}$). It is initialized on stable circular orbits ($V_{c} = |{\bf r}|\,(r^{2} + \epsilon^{2})^{-3/4}$), with no self-gravity. The disk should maintain this initial configuration indefinitely.\footnote{We have also considered the Gaussian ring version of this test problem from \citet{cullen:2010.inviscid.sph}, where $\Sigma\propto \exp{[8\,(r-1)^{2}]}$ is peaked in a narrow, Gaussian ring, with $c_{s}=0.01$ everywhere. The qualitative behavior of every method is identical on this version of the test.}

We use an ``effective'' resolution of $256^{2}$ (i.e.\ for the particle methods the initial particles are evenly spaced such that the square domain from $-2<x<2$, $-2<y<2$ has $256^{2}$ particles; then particles are removed inside $r<0.5$ and outside $r>2$ to match the density profile; for the grid methods this is the same as a $320^{2}$ grid in 2D across the $-2.5<x<2.5$ domain, or $320^{3}$ in 3D).\footnote{Note that in particle-based schemes which easily handle vacuum boundary conditions, we could just initialize a constant-surface density ring on circular orbits in a pure Keplerian potential, and a simple ``edge'' at some inner and outer radius (say, $r=0.5$ and $r=2$). However, most mesh-based schemes require non-vacuum boundaries and smoothed ``edges,'' and so we introduce the small minimum $\Sigma\ge 0.01$ and softened edges of the disk, together with a softened potential to prevent numerical divergences. However, we find qualitatively identical results for this test for any small values of these quantities.}

Note that for the non-mesh (particle-based) codes, the fully three dimensional version of this problem is {manifestly} identical to a 2D problem where we initialize the gas in the $x-y$ plane with $\rho=1$. In other words the code and solution are invariant to rotations of the disk in any direction. However for any structured-grid code (fixed grid codes like {\small ATHENA} but also AMR codes, there is a difference if the disk is not moving exactly in the same plane as the grid cells (i.e. if we rotate the disk out of the $x-y$ plane so it is not perfectly grid-aligned). So we show the result for both cases.

Here, our MFM and MFV methods perform exceptionally well. Noise arises in the particle density and pressure distribution, as in the Gresho problem, but it has very weak effects on the dynamics. Total angular momentum and local orbits are well-conserved at the $\sim10$ orbits we have followed.\footnote{As noted by \citet{gaburov:2011.meshless.dg.particle.method}, although the exact angular momentum conservation properties of the MFM and MFV methods are unclear in the general case, they do exactly conserve angular momentum if the gas distribution is first-order (i.e.\ there are no second-order terms in the expansion of gas properties). Higher-order terms are generated by the noise here, and by the numerical viscosity of the method, but these do not grow rapidly. In practice, we find that the errors for SPH and fixed-grid codes are dominated by a combination of numerical viscosity and advection errors -- not the formal angular momentum conservation of the code. The MFM and MFV methods do the best job of {\em simultaneously} minimizing these errors, hence their good behavior in this test.} Of all the methods we study, these appear to exhibit the lowest numerical viscosity in this specific problem (not necessarily in all problems!). 

The MFV method generates some very small angular momentum errors, because there is a non-zero mass flux between particles; whenever this carries momentum flux not aligned with the line connecting the particle centers-of-mass, there can be weak violations. These begin to affect the disk evolution at $\sim 30-50$ orbits; hence we see in Fig.~\ref{fig:keplerian} that the very inner edge of the disk has experienced some angular momentum evolution (at any time, the errors will always be largest at the innermost radii, since the gas has executed more orbits, and the number of enclosed cells is smaller). The MFM method has no mass flux, hence identically zero advection errors in angular momentum; the only way it can dissipate angular momentum is via numerical viscosity. The combination of the Riemann solver and accurate gradient estimator make this very low (it primarily comes from the slope-limiter). Hence the angular momentum evolution is nearly perfect. In Fig.~\ref{fig:keplerian.late}, we show the evolution to time $t=600$, or $\sim 160$ orbits of the inner disk, and see the angular momentum conservation is still excellent! In fact, we have integrated as far as $\sim 1000$ orbits, and found that the angular momentum conservation in our MFM method is nearly as good as that for collisionless test particles.

On the other hand, we see a rapid and catastrophic breakup of the disk within $\sim 2$ orbits in our TSPH test. This is a well-known result \citep{maddison:1996.sph.ring.instabilities,imaeda:2002.sph.shear.flow.tests,cullen:2010.inviscid.sph}, and occurs because of the physical viscous instability \citep{lyubarskij:1994.viscous.instability}, except that the disk is supposed to be inviscid! The problem is the ``standard'' SPH artificial viscosity produces far too much shear viscosity.\footnote{Specifically, our TSPH example uses the ``standard'' \citet{gingold.monaghan:1983.artificial.viscosity} artificial viscosity with a \citet{balsara:1989.art.visc.switch} switch. This attempts to suppress numerical shear viscosities, but only does so by a modest amount, and is very noisy because it is based on the ``standard'' SPH gradient estimator that has large zeroth and first-order errors (especially in shear flows). \citet{cullen:2010.inviscid.sph} study several variations of SPH in this problem (there a Keplerian ring; see their Fig.~8), and confirm that the both the ``standard'' SPH artificial viscosity and the time-dependent viscosity method of \citet{morris:1997.sph.viscosity.switch}, with or without the \citet{balsara:1989.art.visc.switch} switch (e.g. methods in {\small PHANTOM}, {\small GADGET-2}, and many other codes), undergo catastrophic fragmentation within $\lesssim2-3$ orbits.}

Our PSPH method uses an improved artificial viscosity switch proposed by \citet{cullen:2010.inviscid.sph}; this uses a least-squares matrix-base gradient estimator (similar to our MFM and MFV methods), which is zeroth-order accurate. This dramatically improves the results, allowing semi-stable evolution to $\sim 5-10$ orbits; however, we still see the viscous instability appear. The artificial viscosities are still excessively large in shear flows, and the method still has zeroth and first-order errors in the hydrodynamic forces together with first-order errors in the velocity gradient estimator.\footnote{\citet{cullen:2010.inviscid.sph}, in their similar test problem (Fig.~8 therein), find that their method works well to $\sim 5$ orbits, which we confirm, but we should note several differences between the test problem there and here. They use an effectively higher resolution and a carefully chosen initial particle distribution following \citet{cartwright:2009.poisson.noise.sph.shear} which minimizes the artificial viscosity noise, both of which delay breakup. They also set the minimum artificial viscosity in their method to zero, which gives good results on this test but we find leads to significant particle disorder and potentially catastrophic particle-interpenetration (where particles ``move through'' each other) in poorly-resolved shocks (very common in real problems). We find that the numerical parameters required for stable evolution in all other test problems shown here lead to somewhat faster breakup than the ``ideal'' parameters for this test problem alone.}

Moreover, as noted in \citet{hu:2014.psph.galaxy.tests}, all SPH artificial viscosity methods also produce excessively high numerical viscosities and disk breakup if the disk has modest internal turbulence (enough to set a scale height $h/R\gtrsim0.1$), because then the artificial viscosity is ``triggered'' in the turbulent compressions, but cannot be ``removed'' instantly.\footnote{The standard prescription for ``damping'' artificial viscosity in PSPH, in a supersonic disk, operates more slowly than the local dynamical time, hence the viscous instability can grow.} Once any artificial viscosity appears, the viscous instability grows rapidly. \citet{hu:2014.psph.galaxy.tests} suppress this with an additional, stronger switch that leads to instantaneous post-shock viscosity decay. We have experimented with this, and find it helps here but does not eliminate the viscous instability, and it leads to significantly larger particle noise in all the shock problems we consider below. 

Of course, we can evolve this problem perfectly with SPH if we simply disable artificial viscosity entirely, but then the method is disastrously unstable in other problems!

\begin{figure}
 \begin{tabular}{c}
  \includegraphics[width=0.98\columnwidth]{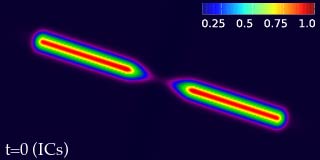} \\
  \includegraphics[width=0.98\columnwidth]{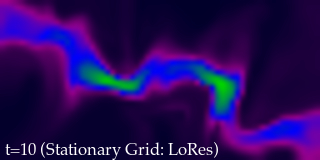} \\
  \includegraphics[width=0.98\columnwidth]{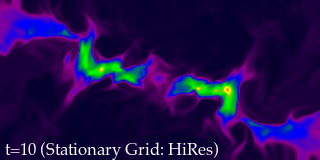}
 \end{tabular}
    \caption{Keplerian disk as Fig.~\ref{fig:keplerian}, but in 3D, with the thin disk rotated out of the $x-y$ plane by an angle $=\pi/20$. We show the gas density in a slice through the $x-z$ ($y=0$) plane (we show $-2.3<x<2.3$, $-1.15<z<1.15$). The Lagrangian particle-based methods (TSPH, PSPH, MFM, MFV) are invariant to such rotations, so we focus on the stationary-grid case. {\em Top:} The disk has constant height $h=0.1$ and is in equilibrium; it should be preserved at all times. {\em Middle:} Stationary-grid result at time $t=10$ ($1.6$ orbits at $r=1$), at lower resolution ($64\times64\times32$). {\em Bottom:} Same, at higher resolution ($256\times256\times128$). There is a strong grid-alignment effect (see \S~\ref{sec:keplerian}) whereby the disk is forced into alignment with the grid axes. This leads to more rapid angular momentum loss than in the exactly-aligned case in Fig.~\ref{fig:keplerian}. It also produces an unphysical warp which becomes a ``break'' or ``tear'' in the disk as the alignment occurs first locally (i.e.\ the disk aligns at different heights) then globally.\vspacerpostplot
    \label{fig:tilted.disk}}
\end{figure}

In grid methods, the numerical viscosity is much lower. However, as shown in \S~\ref{sec:square}, advection errors in non-moving grids are serious. We find (as have many others before) that these very quickly diffuse the disk, spreading the mass around and seriously distorting the shape of the disk before completely destroying its structure within $\sim 2$ orbits. The inner parts lose angular momentum until they form a hot, hydrostatic center, and the outer parts are flung out. If we use a first-order solver, the central parts diffuse rapidly outwards; if we use a second or third-order solver (shown here), some regions move in and some move out, leading to ``rings'' forming which then get broken into clumps. In either case, total and local angular momentum are poorly conserved even over $\sim1$ orbit (significantly more poorly than any other method we consider, including TSPH). This is well-known, and can be improved by going to higher resolution and higher-order methods, but even then the improvement is comparatively slow and the same qualitative effects occur. The problem is that the motion requires constant advection with the grid faces almost never aligned with the flow, in a circular trajectory which is not accurately approximated by second-order methods. Since the errors affect the whole disk volume simultaneously, going to AMR methods does not help \citep[see e.g.][]{de-val-borro:2006.disk.planet.interaction.comparison}.

These issues are even more severe if we rotate the disk relative to the axes (i.e.\ embed it in three dimensions, but tilt it out of the $x-y$ plane), as shown in Fig.~\ref{fig:tilted.disk}. The MFM, MFV, and SPH methods reproduce themselves to machine accuracy independent of such tilt. But in the structured grid codes, the advection errors above are compounded (by another mis-aligned axis). Moreover, the grid-alignment effect leads to an effective ``numerical torque'' which forces the orbits to align with the nearest coordinate axis (eventually pushing the disk back into the $x-y$ plane); this generates an unphysical large-scale warp in the disk on just $\sim1$ orbital timescale. Such grid alignment effects are well-known \citep[e.g.][]{davis:1984.rotating.upwind.eulerian.scheme}. For example, \citep{hahn:2010.disk.gal.orientations.ramses} study cosmological simulations of galaxies in AMR and find that the galaxy spin axes are strongly aligned with the grid axes by low redshift, even at ``effective'' resolutions of $\sim 128^{2}-512^{2}$ in the disk plane (particle numbers in the disk up to $5 \times 10^{5}$); \citet{byerly:2014.hybrid.cartesian.scheme.for.ang.mom} demonstrate similar grid alignment and disk destruction effects in AMR simulations of stellar evolution and binary orbits up to AMR resolutions of $\sim 1024^{3}$. A variety of coordinate ``patch'' schemes or hybrid advection schemes have been designed to reduce these errors, but these all rely on some prior knowledge of the computational geometry. 

For the problem here, of course, we would obtain the most accurate results by using a pre-defined cylindrical coordinate system translated and rotated to center on and align with the disk. We have explicitly confirmed this: running the exact same 2D setup in {\small ATHENA}, but with cylindrical coordinates, we can evolve the disk to $>1000$ orbits even at $32^{2}$ resolution, with the fraction deviations in $\Sigma(r)$ from the ICs remaining below a part in $10^{4}$ (independent of the boundary conditions, or use of a second or third-order method). But while useful for some idealized problems, we specifically wish to study the more general case, since there are a huge range of problems (e.g.\ turbulence, galaxy formation and evolution, stellar evolution with convection and rotation, binary mergers, accretion from eccentric or gravitationally unstable disks, asymmetric SNe explosions) where the flow geometry cannot be completely determined ahead of time, or adaptive meshes must be used, so rotation cannot be perfectly grid-aligned.

Using moving meshes helps reduce the angular momentum errors from advection in grid codes. We have run $>200$ iterations of this test problem using the public version of {\small FVMHD3D}, systematically varying choices like the mesh regularization scheme, mesh ``drifting'' (whether to use a strictly-Lagrangian drift, or locally smoothed velocity, or regularized drift), initial mesh geometry, and boundary conditions. In both {\small FVMHD3D} and more limited tests with {\small AREPO}, we find that running in the ``simplest'' initial configuration (an initial Cartesian mesh with outflow boundary conditions, with the default mesh regularization scheme used for all other test problems shown here), the disk goes unstable and the angular momentum evolution tends to be corrupted within a few orbits (similar to the fixed-grid cases). Unfortunately, some significant errors in angular momentum evolution are difficult to avoid in moving-mesh codes, as has been discussed extensively in e.g.\ \citet{duffell:2012.disco.method.protoplanetary.disk,ivanova:2013.gaburov.appendix.meshless.derivations,mocz:2014.galerkin.arepo,munoz:2014.disk.planet.interaction.sims}. In a shearing disk, if the cells adapt in a truly Lagrangian manner, then they are inevitably deformed into a highly sheared/irregular shape \citealt{munoz:2014.disk.planet.interaction.sims}. This leads to other errors. As soon as they become non-spherical (or more accurately fail to be radially symmetric about their own cell center of mass), then mass advection between cells necessarily leads to additional angular momentum errors (indeed, the angular momentum of an irregular cell cannot be defined exactly but only to the same order of integration accuracy as the local velocity gradient estimator). This is akin to the errors in our MFV method. More importantly, if some regularization procedure is used to keep the cell shapes ``regular'' (as is necessary in {\em any} moving-mesh code used for this problem), then the regularization means the cells cannot move entirely with the fluid and the gas must be advected over the cells. This re-introduces some of the same (more serious) errors we saw with stationary grid methods (specifically, see \citealt{ivanova:2013.gaburov.appendix.meshless.derivations}, Eq.~53). This means that the results for moving meshes are quite sensitive to choices like the mesh ``stiffness,'' regularization procedure, and in particular the choice of boundary conditions for the mesh-generating points (since the rigid Voronoi volume partition can lead to a ``mesh tension'' effect, whereby regularization-induced distortions in the central regions propagate outwards ``through'' the mesh; \citealt{springel:arepo}). So there are ways to improve the situation on this problem -- for this reason, we do not show a single ``standard'' moving-mesh result, because significantly different results are obtained if we make just small changes to the mesh-regularization procedure in each code. However, like with AMR codes, the most effective methods for eliminating angular momentum errors in moving meshes generally depend on knowing the problem geometry ahead of time. For example, \citet{duffell:2012.disco.method.protoplanetary.disk} design a moving grid which is a series of cylindrical shells free to rotate independently about a shared axis (the {\small DISCO} code); \citet{munoz:2013.viscous.flows.arepo} use a carefully-chosen initial grid configuration with a specially-designed boundary condition designed to prevent inward-propagation of ``mesh deformation''; these help considerably, but must be fine-tuned to the exact disk configuration.

Fig.~\ref{fig:keplerian.convergence} quantitatively compares the errors on this problem as a function of resolution and method. We define the $L1$ norm as the mean absolute fractional deviation from the expected value in either velocity $L1_{(v)} \equiv \langle |v_{\phi,\,i} - v_{K}[R_{i}]| / v_{K}[R_{i}] \rangle$ (where $R_{i}$ is the radial position of particle $i$, $v_{\phi,\,i}$ is its azimuthal velocity, and $v_{K}[R_{i}]$ is the Keplerian velocity at that radius) or density $L1_{(\rho)} \equiv \langle |\rho_{i} - 1| \rangle = \langle |\Sigma_{i} - 1| \rangle$. Because the orbits degrade on a timescale relative to the orbital time, it is useful to focus on a narrow radial annulus, here chosen to be $0.8 < R < 1.2$ ($R\sim1$). For either $L1$ norm, initially small errors tend to grow exponentially in all methods: if their growth timescale is related to the effective numerical viscosity and/or diffusion introduced by various methods, it should decrease with resolution. We therefore compare the $L1_{(v)}$ norm in this annulus ($R\approx1$), at a fixed time (here $t = 10\,t_{\rm orbit}(R=1) = 20\,\pi$), for different ``effective'' resolution defined as above (number of elements across a side from $-2<x<2$, for a stationary grid; so our standard case is $N_{1D}^{\rm eff} = 256$). For MFM and MFV methods, the errors decrease as $L1_{(v)}\propto N^{-1.5}$ (with, as expected, somewhat smaller errors systematically in the MFM case). This is somewhat slower than the ideal rate ($N^{-2}$), but we should recall that we are well into the non-linear regime of the problem; that there are discontinuities (shocks), at which all methods are lower-order, introduced by even very small velocity perturbations because the disk is cold; and convergence as $N^{-2}$ is not always gauranteed when external forces (here, gravity) dominate. Similar convergence rates are seen for the Gresho vortex test with these methods as well as moving-mesh and stationary-grid methods \citep[see][]{springel:arepo,gaburov:2011.meshless.dg.particle.method}. In both problems, for the MFM/MFV methods, the volume ``partition errors'' tend to dominate, so this is not surprising. TSPH and PSPH do show convergence in Fig.~\ref{fig:keplerian.convergence}, but much slower, as $\sim N^{-0.6}$; again a similar scaling is seen in the Gresho problem \citep[][Fig.~10]{dehnen.aly:2012.sph.kernels}. At much higher resolution, this should saturate at constant values because of the zeroth-order errors (unless we further increase the kernel size), but we estimate the resolution would need to be $\gtrsim 2048^{2}$ before we reach this threshold. More strikingly, stationary grids show no real convergence here; except some small improvement compared to the lowest resolution. This is somewhat ``noisy,'' however; if we average over many radial annulli and times we see a weak convergence trend $\sim N^{-(0.2-0.4)}$. This is quite different from the Gresho test, which converges as $\sim N^{-1.4}$ in stationary grids. Clearly the external forcing and large advection errors associated with the highly super-sonic, non-uniform flow are critical. 

A related test is to ask how long we can evolve the disk before the $L1$ norm exceeds some tolerance. We show this, for $L1$ measured at $R\sim1$ as above, for both thresholds $L1_{(v)}=0.01$ and $L1_{(\rho)}=0.25$. Note that, at similar times, $L1_{(\rho)}\gg L1_{(v)}$, because the problem setup is such that small velocity errors can lead to large density changes over many orbits. But otherwise the behavior with respect to both norms scales similarly. For MFM/MFV methods, we find the maximum time scales as $\sim N^{1.8}$, with MFM systematically reaching factor $\sim 2$ longer times (reaching $\sim 100$ orbits, or $t > 600$, for these thresholds at $R=1$, consistent with Fig.~\ref{fig:keplerian.late}). Note that some of the orbital differences we see in Fig.~\ref{fig:keplerian} with the MFV/MFM methods are not captured here, because they depend on smaller but more systematic differences. For TSPH/PSPH, we find the time increases as $\sim N^{0.5-0.6}$, with factor $\sim 2-3$ larger times for PSPH vs.\ TSPH. And for stationary grids we see the time grow slowly, as $\sim N^{0.1}$ in $L1_{(v)}$ and $\sim N^{0.2-0.4}$ in $L1_{(\rho)}$. \footnote{Consider this in terms of the CPU cost required to evolve the (2D) problem to some time $t$ with a given accuracy; this scales as $\sim N^{2}\,t/\Delta t$ where $\Delta t$ is the timestep, which for a problem dominated by external forcing should scale as $\Delta t \propto (\Delta x/|{\bf a}|)^{1/2} \propto N^{-1/2}$ (\S~\ref{sec:methods:timesteps}). This gives a scaling $\sim N^{2.4}$ for MFM/MFV, $\sim N^{6}$ for SPH, $\sim N^{8.5}$ for Cartesian grids.}

Finally, we note that again there is no 1D analogue of this problem, but if we were to repeat our experiments for a 3D analogue (a cylindrically symmetric rotating tube) we would reach all the same conclusions. The purely geometrical disk thickness is not important; ``thickness'' matters only in the sense of the relative importance of pressure support versus angular momentum. In the limit of a ``thicker disk'' meaning a more pressure-dominated disk, the problem becomes progressively more hydrostatic and therefore ``easier'' for all methods considered here.

\vspace{-0.5cm}
\subsection{Shock and Non-Linear Jump Tests}
\label{sec:tests:shocks}

We now consider several tests which probe the opposite regime: strong shocks.

\begin{figure*}
    \plotsidesize{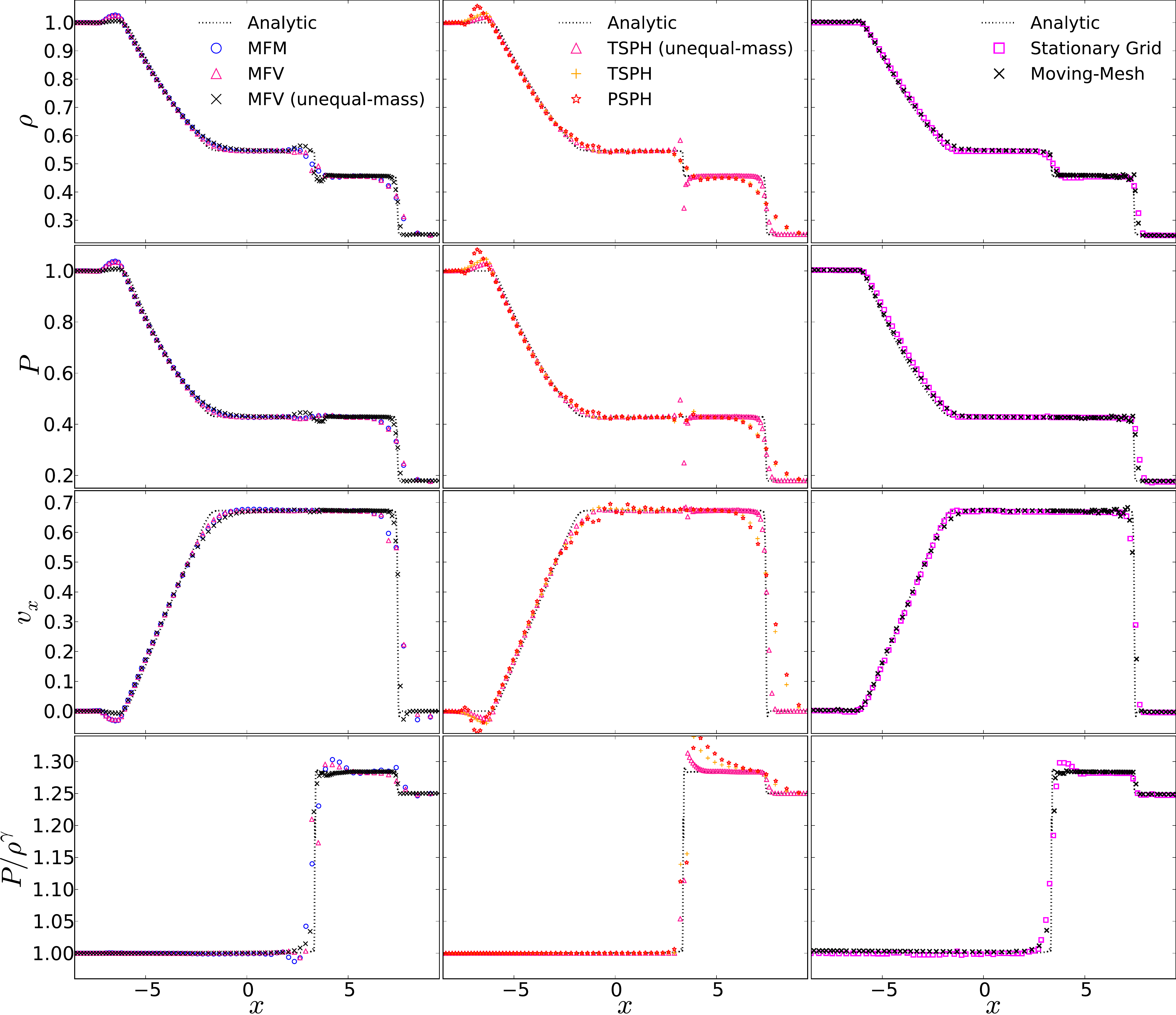}{0.98}
    \vspace{-0.25cm}
    \caption{One-dimensional Sod shock tube \S~\ref{sec:shocktube}. From top to bottom, we plot density $\rho$, pressure $P$, velocity $v_{x}$, and entropy $P/\rho^{\gamma}$. We show the analytic solution (dotted) for each, compared to different methods (all with $100$ resolution elements): all perform reasonably well with subtle differences. {\em Left:} MFM and MFV methods: both are very similar, with a small ``bump'' at the rarefaction and contact discontinuity due to the slope-limiter which rapidly converges away at higher resolution. ``Default'' cases shown assume equal-particle masses; in the ``unequal-mass'' case the ICs feature a factor $=4$ jump in particle mass at the contact discontinuity (hence sharper resolution in the low-density shock). {\em Middle:} SPH methods: The ``bumps'' are larger, especially using PSPH, shocks are more smoothed, there is some velocity ``ringing'' in the rarefaction, and there is the known ``pressure blip'' around the contact discontinuity which does not converge away. {\em Right:} Moving-mesh ({\small AREPO}) and fixed-mesh ({\small ATHENA}) methods: these have the sharpest shock-capturing; but still feature weak ``bumps'' ({\small ATHENA}) or post-shock oscillations ({\small AREPO}).\vspacerpostplot 
    \label{fig:shocktube}}
\end{figure*}

\vspace{-0.5cm}
\subsubsection{Sod Shock Tube: A Basic Riemann Problem}
\label{sec:shocktube}

We begin by considering one of the many simple Riemann problems used in standard code tests. We simulate a one-dimensional Sod shock tube with a left state ($x < 0$) described by $P_{1}=1$, $\rho_{1}=1$, $v_{1}=0$ and right state ($x\ge 0$) with $P_{2}=0.1795$, $\rho_{2}=0.25$, $v_{2}=0$, and $\gamma=1.4$. These parameters are used in many code tests \citep{hernquistkatz:treesph,rasio:1991.stellar.merger.hydro,wadsley:2004.GASOLINE,springel:gadget,springel:arepo}. We intentionally consider a ``low'' resolution test, in which we place an initial $100$ particles in the range $-10<x<10$ (spacing $\Delta x \approx 0.01$, $0.2$, respectively). We plot results at $t=5.0$. 

All calculations here capture the shock and jump conditions reasonably, but there are differences. For all the non-mesh methods, it makes a difference whether we initialize the problem with equal-mass particles, or with the initial discontinuity corresponding to a particle mass jump (in which case the particle masses change discontinuously by a factor $\sim4$ at the contact discontinuity). At the front at $x\sim-6$, all methods produce a small 'bump' in the density and corresponding dip in $v_{x}$; this is minimized in the grid codes and the unequal-mass particle MFV model. The `bump' is amplified by the PSPH method. In PSPH there is also added noise where the pressure becomes flat ($x\sim0$).

At the contact discontinuity ($x\sim3$), MFM and MFV methods with equal-mass particles behave well; the large particle-mass discontinuity in the unequal-mass case requires (because particle {\em volumes} are kernel-determined and vary smoothly) a `blip' in the density, which then appears in the pressure. In SPH, however, a comparable blip appears even with equal-mass particles, and it is much more severe with unequal-mass particles; most importantly, the `blip' converges away in the new methods (a modest-resolution MFV run with just $500$ particles is indistinguishable from the dotted line shown), while the SPH blip never converges away (it gets narrower but higher-amplitude at higher resolution). Fixed-grid and SPH methods also produce an ``entropy overshoot'' on the right side of the contact discontinuity; this is particularly strong in the equal-mass SPH examples. We should note that on this problem, non-conservative SPH methods (see Table~\ref{tbl:methods}) produce disastrous errors (easily order-of-magnitude deviations from the real solution, often with comparable particle-to-particle scatter) behind the shock front.

At the shock ($x\sim8$), the methods are similar when the resolution is similar. The ``equal particle mass'' case feature a broader shock, but only because the mass choice means the spatial resolution is lower in this region: the number of particles needed to capture the shock is actually very similar. For the MFV and MFM methods, this is $\sim3-4$ particles ($<1$ kernel), only slightly larger than the $\sim2-3$ grid cells needed in a second-order grid method; for SPH it is $\sim7-8$ particles ($\sim2$ kernels). As noted in \citet{springel:arepo}, the moving mesh exhibits significant post-shock velocity oscillations, despite the slope limiter employed (we see the same with no slope limiter in the MFV and MFM methods, so suspect it is sensitive to the slope-limiting procedure).

\begin{figure}
    \plotonesize{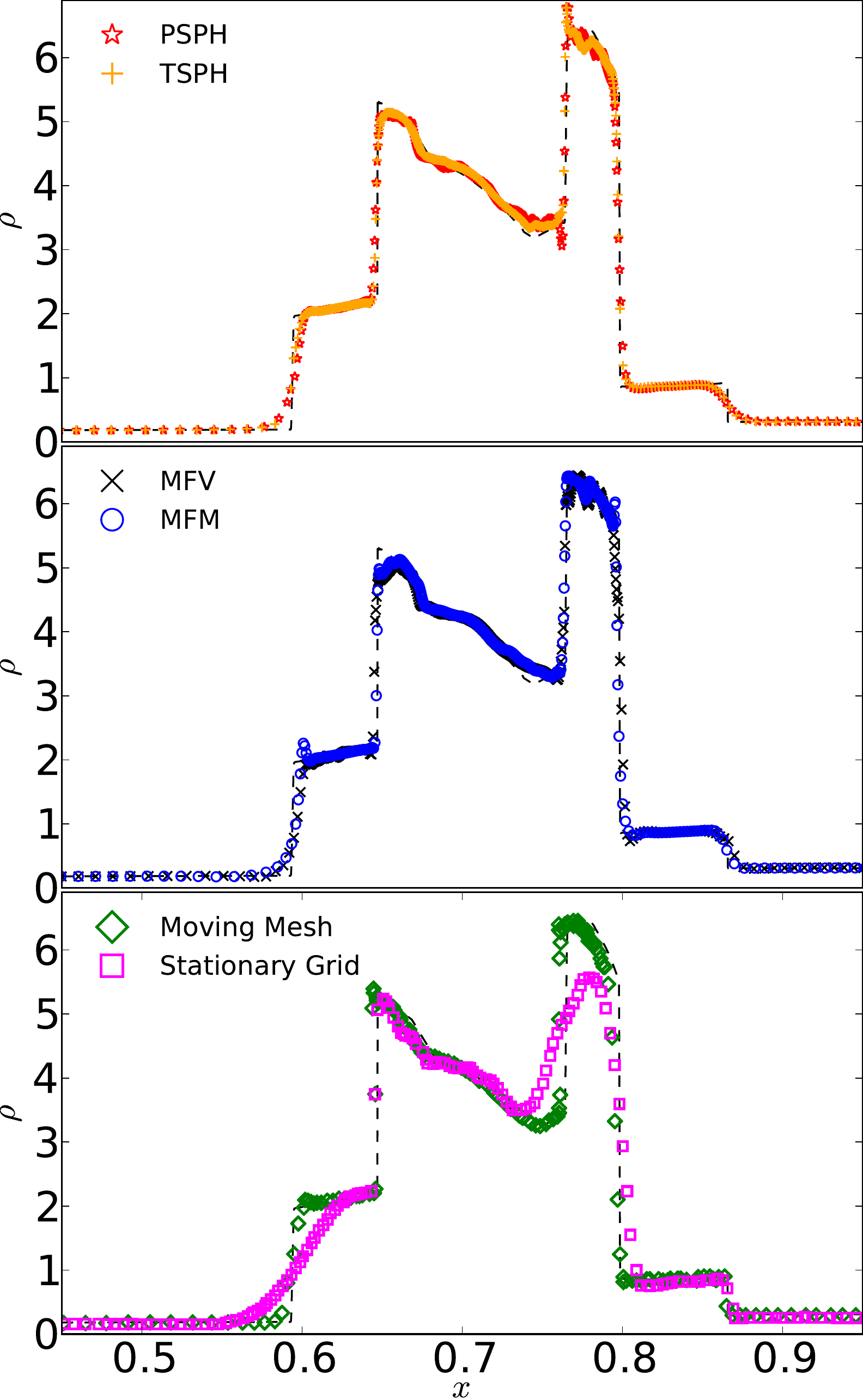}{0.98}
    \vspace{-0.25cm}
    \caption{One-dimensional interacting blastwave test \S~\ref{sec:interact}. We compare all methods (computed with $400$ resolution elements from $0<x<1$) to a reference solution computed using {\small ATHENA} with $10^{5}$ cells, a third-order PPM solver, an exact Riemann solver, and Courant factor $=0.1$. {\em Top:} SPH methods: these do well here. Contact discontinuities at $x\sim 0.6$ and $x\sim 0.85$ are noticeably smoothed and there is a ``pressure blip'' at $x\sim 0.75$, but the jumps are all captured. {\em Middle:} MFM \&\ MFV methods: These also do well. The discontinuities are slightly less smoothed than SPH, but the density ``dip'' at $x\sim 0.75$ is not quite as well-traced, and there is some smoothing of the jump at $x\sim 0.65$. {\em Bottom:} Moving mesh \&\ stationary mesh methods. Moving meshes do well, with the sharpest jumps and no ``wiggles'' in density at $x\sim 0.75-0.8$, but are slightly offset in the shock position. Stationary grids are noticeably less accurate than the other methods, severely smoothing the jump at $x\sim 0.6$ and the density peak from $x\sim 0.75-0.8$.\vspacerpostplot
    \label{fig:interact}}
\end{figure}

Note that for all codes, we obtain essentially identical results if we solve this problem in 2D or 3D (i.e. as a true ``tube'') with the fluid having constant properties along the $y$ and $z$ directions (and periodic boundaries). In fixed-grid codes, it is well-known that if we rotate the tube so it is not exactly aligned with a coordinate axis, the correct solution is still recovered but shock jumps and contact discontinuities are diffused across $\sim 2$ times as many cells in the direction of motion. The particle and moving-mesh methods are invariant to rotations of the tube.

\vspace{-0.5cm}
\subsubsection{Interacting Blastwaves: Complicated Jumps \&\ Extreme Riemann Problems in 1D}
\label{sec:interact}

Another related one-dimensional test problem is the interaction of two strong blast waves, from \citet{woodward:1984.interacting.blastwave.problem}. We initialize gas with density $\rho=1$, $\gamma=1.4$, $v=0$ in the domain $0<x<1$, with pressure $P=1000$ for $x<0.1$, $P=100$ for $x>0.9$, and $P=0.01$ elsewhere. The boundary conditions are reflective at $x=0$ and $x=1$. This features multiple interactions of strong shocks and rarefactions, and provides a powerful test of Riemann solvers (many crash on this test, and essentially all codes require some hierarchy of solvers as we have implemented). We use $400$ particles initially evenly spaced and equal-mass.

This is a problem where SPH does very well, actually. As in \S~\ref{sec:shocktube}, the shocks are smeared over more particles compared to other methods, and a small density ``blip'' appears near $x\sim0.75$, but the structure of the density peaks is captured well even at low resolution. Moving mesh codes perform extremely well, with sharper shock resolution (and no ``blip''), especially around the narrow peak at $x\sim 0.65$, and they also capture the full under-density around $x\sim0.75$.\footnote{One puzzling result is that, even at high-resolution, {\small AREPO} shows a slight offset in the position of the density jump at $x\sim0.8$; in contrast, MFV, MFM, SPH, fixed-grid ({\small ATHENA}), and a different moving-mesh simulation (using {\small TESS}) agree on the shock position. We suspect this has to do with either: a too-aggressive application of the entropy-energy switch (see \S~\ref{sec:methods:switches}) in {\small AREPO} (the switch does not trigger in our default runs with MFM/MFV, but if we modify to make the switch less conservative, we can reproduce the shock offset), or too-aggressive allowance in the code for adaptive timestepping (the pre-shock gas can have long timesteps, which lead to small offsets in time when they ``become active,'' hence an offset in shock position).}

At this resolution, both the MFM and MFV methods give similar results. The MFM method broadens the discontinuity at $x\sim0.6$ by slightly more and similarly smooths the leading edge of the discontinuity at $x\sim 0.85$. The major difference between these methods and moving-meshes is that MFM/MFV do not capture the full density dip without going to higher resolution (perhaps surprising given SPH's success, but this is where the fixed-grid method also has difficulty). But we confirm that at high resolution, the MFM and MFV methods converge to the same solution in good agreement with {\small AREPO}. 

The largest errors at fixed resolution come from the fixed grid code. As noted in \citet{springel:arepo}, both the discontinuity at $x\sim 0.6$ and the density peak/pair of discontinuities around $x\sim 0.75$ are severely smoothed, the jump at $x\sim0.8$ is more broadened than in any other method, and the density ``dip'' is captured but actually over-estimated. This stems largely from contact discontinuities being advected through the grid. 

As in \S~\ref{sec:shocktube}, we obtain identical results solving this problem as a 2D or 3D ``tube'', except that if the tube is not exactly aligned with the grid, non-moving grid methods will diffuse it even more severely.

\begin{figure}
    \plotonesize{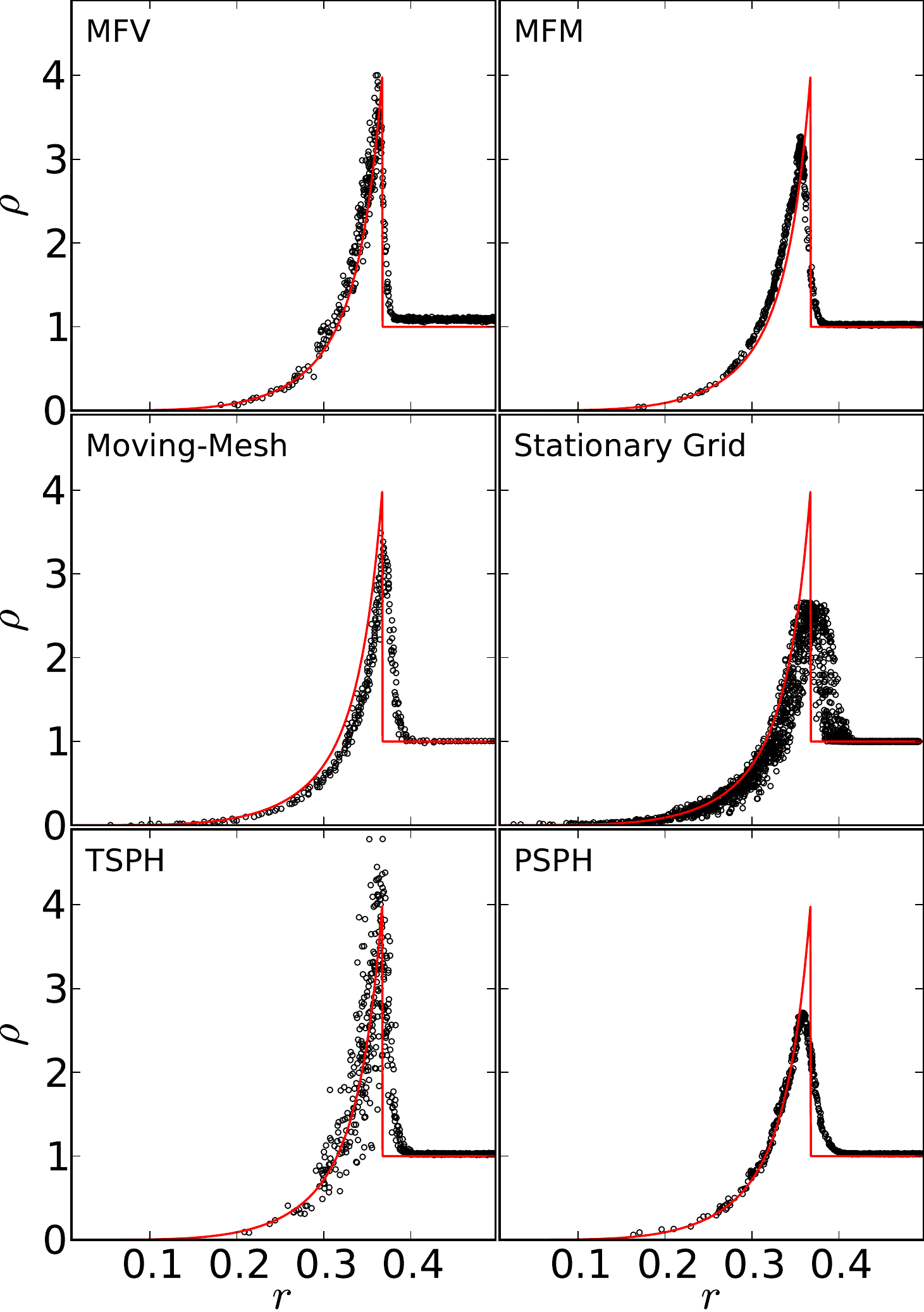}{0.98}
    %\plotonesize{sedov2.pdf}{0.98}
    \vspace{-0.3cm}
    \caption{Three-dimensional Sedov-Taylor blastwave (\S~\ref{sec:sedov}). We plot the radial density profile at time $t=0.06$; each point is one particle/cell (for clarity we plot only a random subset of cells) at $64^{3}$ resolution; red line is the analytic solution. {\em Top:} MFV \&\ MFM solutions: the MFV shows excellent capturing of the shock jump, but is slightly noisier than MFM. {\em Middle:} Moving-mesh ({\small AREPO}) and stationary-mesh ({\small ATHENA}) solutions: the moving-mesh solution lies ``in between'' our MFM and MFV solutions (there is a slight offset in shock position, which may result from the particular timestep scheme); the stationary-mesh solution is substantially more noisy and diffuses the shock (suppresses the jump) significantly. {\em Bottom:} SPH solutions: TSPH captures the jump, but is much noisier than any other method (and speads the jump over more particles). PSPH suppresses this noise via artificial conductivity, but this suppresses the jump amplitude and diffuses the leading-edge of the shock.\vspacerpostplot
    \label{fig:sedov}}
\end{figure}

\begin{figure}
    \plotonesize{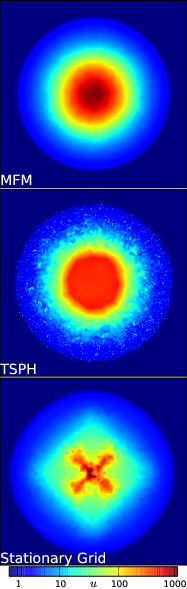}{0.69}
    \vspace{-0.3cm}
    \caption{Sedov blastwave from Fig.~\ref{fig:sedov}; here we plot the gas internal energy $u$ (log-scaled) in a 2D slice ($-0.45<x,\,y<0.45$) through $z=0$, at $t=0.06$. {\em Top:} MFM: The solution is smooth and shows good spherical symmetry. {\em Middle:} TSPH: The solution is spherical on average, but the severe noise is again visible. {\em Bottom:} Stationary-grid: Grid effects on the symmetry are clearly visible (the cross/diamond shapes).\vspacerpostplot
    \label{fig:sedov.im}}
\end{figure}

\begin{figure}
    \plotonesize{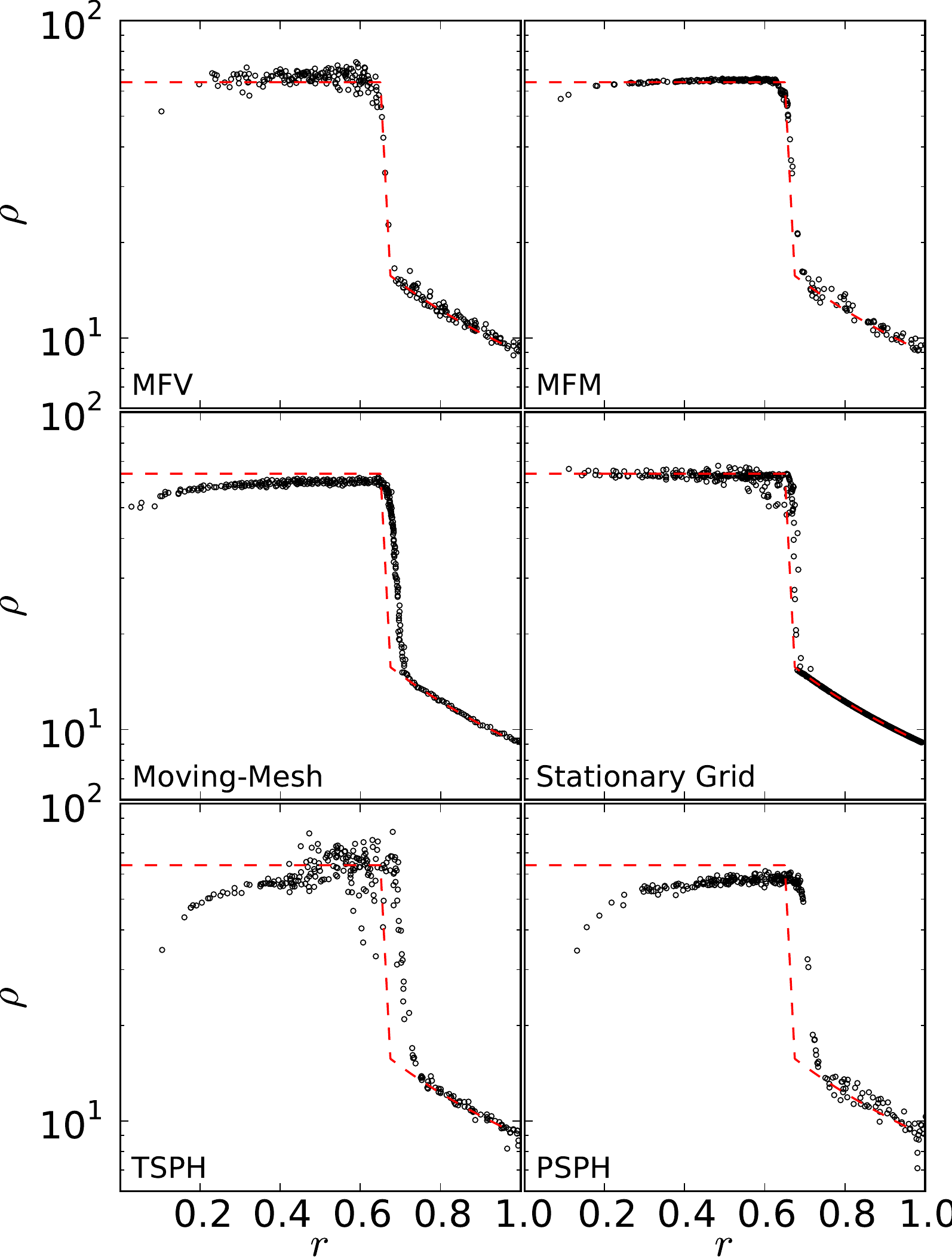}{0.98}
    \vspace{-0.3cm}
    \caption{Three-dimensional Noh implostion problem (\S~\ref{sec:noh}). We plot the radial density profile at time $t=2$; each point is one particle/cell (for clarity we plot only a random subset of cells) at $50^{3}$ resolution; red line is the analytic solution. {\em Top:} MFV \&\ MFM solutions: the MFV shows excellent capturing of the shock jump, but is noisier than MFM. {\em Middle:} Moving-mesh ({\small AREPO}) and stationary-mesh ({\small ATHENA}) solutions: the moving-mesh solution lies ``in between'' our MFM and MFV solutions in noise level, but the offset in shock position corresponds to a systematic under-estimate of the density jump, and the wall-heating is slightly more severe. The stationary-mesh solution gets the jump right (and is the only example without wall-heating), but with serious noise and asymmetry related to the carbuncle instability (see below). {\em Bottom:} SPH solutions: TSPH captures the jump but exhibits severe noise, shock-spreading, and wall-heating errors. PSPH suppresses the noise, but at the expense of more diffusion and enhanced wall-heating.\vspacerpostplot
    \label{fig:noh}}
\end{figure}

\begin{figure}
    \plotonesize{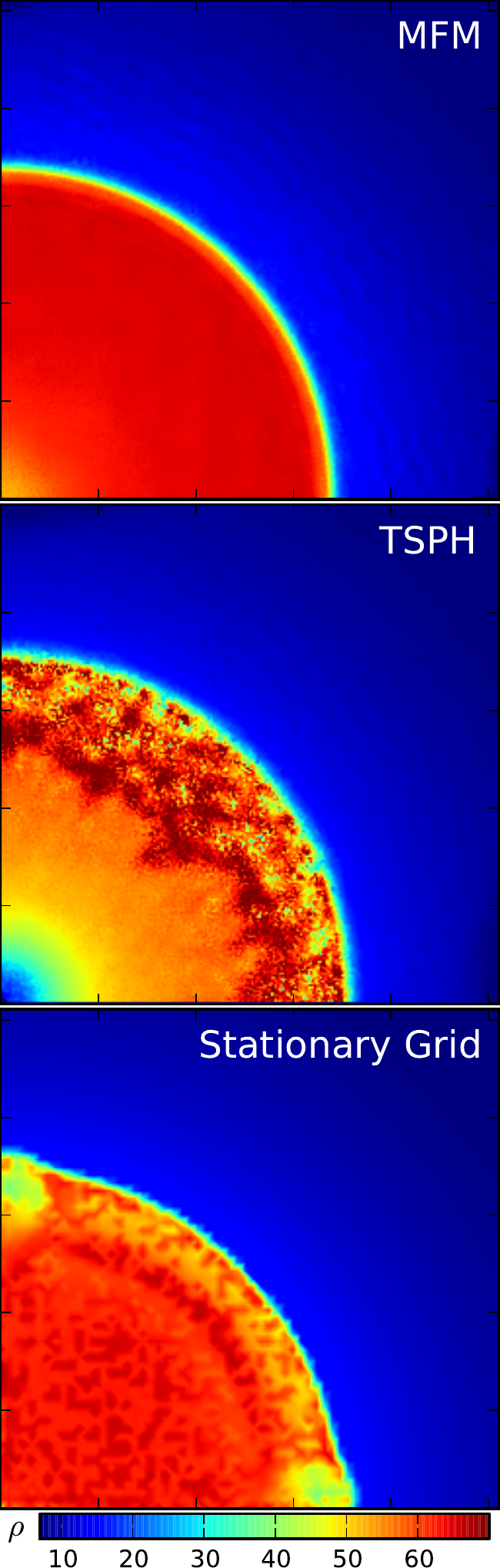}{0.69}
    \vspace{-0.3cm}
    \caption{Noh implosion test from Fig.~\ref{fig:noh}; we plot an image of the gas density, in a 2D slice (one quadrant: $0<x<1$, $0<y<1$, $z=0$), at $t=2$. {\em Top:} MFM: As in the Sedov test, the solution is smooth and shows good spherical symmetry. {\em Middle:} TSPH: The solution is spherical on average, but severe noise is again visible (there should be no internal structure here). {\em Bottom:} Stationary-grid: The carbuncle instability leads to the ``hot spots'' where the shock is propagating along the coordinate axes.\vspacerpostplot 
    \label{fig:noh.im}}
\end{figure}

\vspace{-0.5cm}
\subsubsection{Sedov Blastwaves: Conservation, Integration Stability, \&\ Symmetry}
\label{sec:sedov}

Here we consider a Sedov-Taylor blastwave, a point explosion with large Mach number. This provides a powerful test of the accuracy of code conservation, as well as of how well codes capture shock jumps and preserves symmetry in three dimensions. When adaptive (non-constant) timesteps are used (as they are in our code) this is also an important test of the integration stability of the method \citep[see][who show how various simple integration schemes become unstable]{saitoh.makino:2009.timestep.limiter}.

We initialize a large domain with $\rho=1$, $P=10^{-6}$ (small enough to be irrelevant), and $\gamma=5/3$, with $64^{3}$ particles in the domain affected by the blastwave; we inject an energy $E=1$ into the central particle. We compare results at $t=0.06$. A strong, spherically symmetric shock (of initially extremely high Mach number) should have formed, with a density jump of a factor $(\gamma+1)/(\gamma-1) = 4$. 

%Here we consider an extreme Sedov-Taylor blastwave, with very large mach number, designed to be a powerful test of conservation. A box of side-length $6\,$kpc is initially filled with $128^{3}$ equal-mass particles at constant density $n=0.5\,{\rm cm^{-3}}$ and temperature $10\,$K; $6.78\times10^{46}\,$J of energy is added to the central $64$ particles in a top-hat distribution. This triggers a blastwave with initial Mach number $\sim1000$, which we compare at $20\,$Myr, where the shock front should be at $r\approx1.19\,$kpc. In this test and below, unless otherwise specified, we assume a $\gamma=5/3$ gas equation of state.

As expected, at fixed particle/cell number, fixed-grid methods smooth the shock jump significantly compared to Lagrangian methods (which by definition end up with more resolution in the shock). Conversely the deep interior structure of the blastwave (where densities are low and temperatures high) is better-resolved in fixed-grid methods; it depends on the problem of interest whether this is an advantage or disadvantage. However all grid codes (AMR or fixed) also suffer from variations of the carbuncle instability, in which shocks preferentially propagate along the grid axes; we see that this has a significant effect on the blast geometry, giving it an ``eight pointed'' morphology along the grid axes which only decreases in time because diffusion tends to isotropize the blastwave. 

The MFM, MFV and moving mesh methods perform similarly well here. In all cases the jump is better captured (less ``smeared''), giving a maximum density $\sim3.5$ (compared to the perfect case $=4$) instead of $\sim2.7$. All maintain excellent spherical symmetry in the shock front. Although a carbuncle instability still exists for moving mesh codes, it is substantially suppressed here. The meshless methods (MFM, MFV, SPH) simply have no such instability because there is no preferred axis.

SPH methods generally do ok on this problem, except that the shock is spread out further (see \S~\ref{sec:shocktube}) and they give noisy solutions in the post-shock behavior unless some additional diffusion is added.\footnote{The noise arises from the E0 error when particles move through the shock.} PSPH substantially enhances this noise, in fact, without additional diffusion. Adding artificial conductivity dramatically reduces the noise in all implementations, but at the cost of suppressing the shock jump and creating an unphysical ``leading'' temperature jump (diffusing the entropy jump {\em ahead of} the shock).

\begin{figure}
    \plotonesize{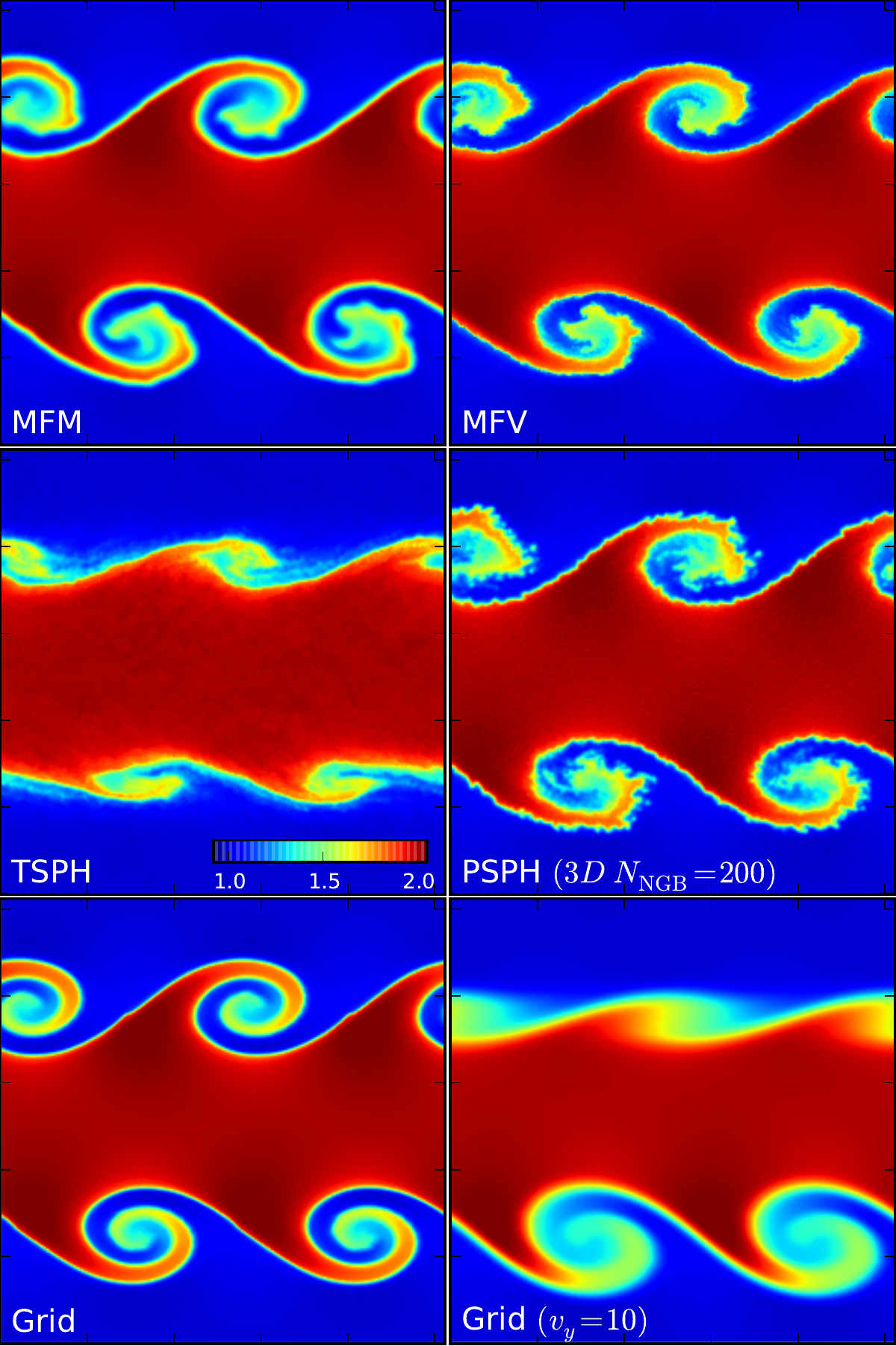}{1.0}
    \caption{Kelvin-Helmholtz instability (\S~\ref{sec:kh}). We compare the result of a 2D, $256^{2}$ KH test problem at $t=2.1$, where the rolls should be going non-linear. {\em Top:} In the MFM \&\ MFV methods, the rolls are well-captured (with just the standard, small neighbor number, a 3D equivalent of $N_{\rm NGB}=32$). There are small differences in the secondary structures developing, discussed below. {\em Middle:} SPH: In TSPH, a combination of surface tension and E0 errors suppress KH roll formation. In PSPH, the noise is large enough that eliminating the surface tension alone does not help; we must also go to very large neighbor number to see rolls. Even then, the small-scale structure is corrupted by E0 errors. {\em Bottom:} Fixed-grid (PPM). Symmetry is well-preserved, while diffusion suppresses small-scale (grid-seeded) modes, at the expense of structure inside the whorls. If we boost the fixed-grid run by a uniform $v_{y}=10$ ({\em right}), diffusion increases (at resolution $<128^{2}$, this ``wipes out'' the instability), and symmetry is broken.\vspacerpostplot
    \label{fig:kh.early}}
\end{figure}

\begin{figure}
    %\plotonesize{kh_nonlinear.pdf}{0.95}
    \plotonesize{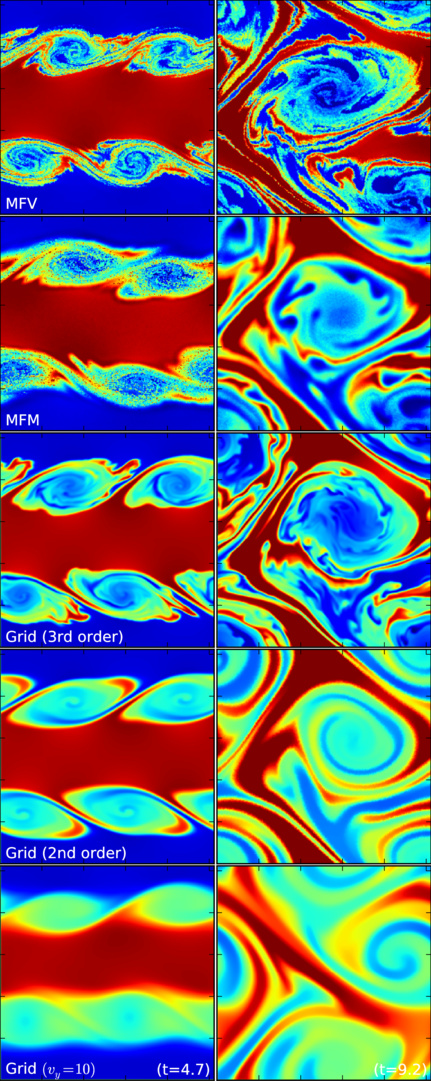}{0.95}
    \caption{Non-linear evolution of the KH instability in Fig.~\ref{fig:kh.early}, at $t=4.7$ and $t=9.2$. In MFV ({\em top}) \&\ MFM ({\em second from top}) calculations, the sub-structure of the rolls is well-preserved; so they continue to ``roll up'' until they overlap, leading to the entire box going non-linear. The sub-structure of the non-linear rolls is especially well-preserved in the MFV calculation (remember this is only $256^{2}$!). In stationary grid codes, the rolls diffuse into one another. This is minimized if we use a high-order (PPM) scheme ({\em middle}); nearly all sub-structure is lost with a typical, second-order grid method ({\em second-from bottom}); and even more severe diffusion appears if we apply a boost ({\em bottom}). Much higher resolution is required in grid codes to reduce this diffusion and see the same roll sub-structure at late times.\vspacerpostplot
    \label{fig:kh.nonlinear}}
\end{figure}

\begin{figure}
    \plotonesize{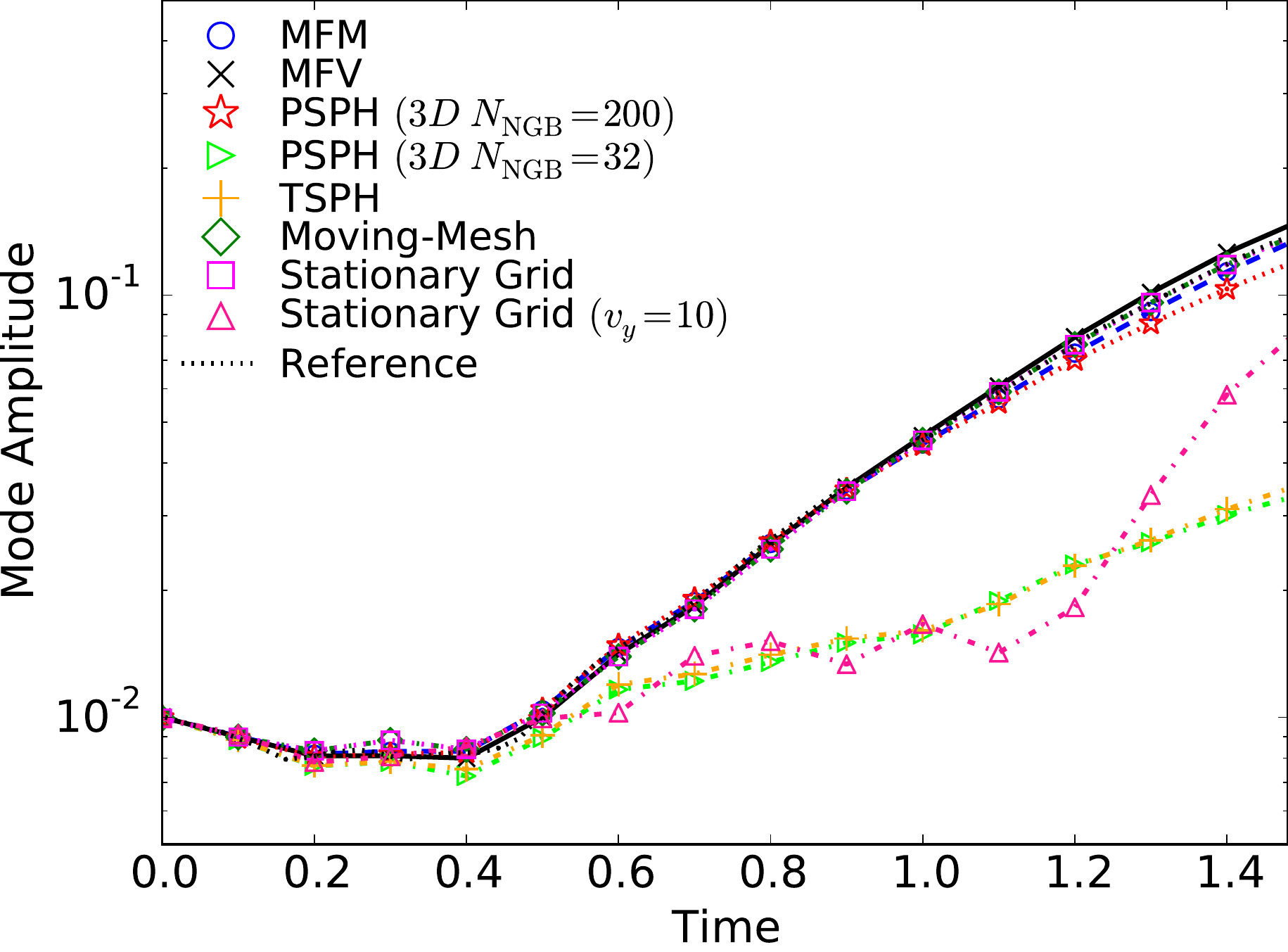}{0.98}
    \vspace{-0.3cm}
    \caption{KH mode amplitude as a function of time for the $256^{2}$ runs in Fig.~\ref{fig:kh.early}. We compare a reference solution at $4096^{2}$, from \citet{mcnally:2012.kh.test.comparison}, which is well-converged when linear evolution still dominates ($t\lesssim1.5$). MFM, MFV, and stationary grid results (with no velocity boost) are essentially identical at these times, and well-converged. PSPH with high neighbor number is similar, though begins to depart at $t\gtrsim 1.1$. Mode growth is strongly suppressed in TSPH or PSPH with modest neighbor number (comparable to a grid calculation at $\sim 32^{2}$ resolution). Grid results (here 3rd-order PPM, with corner-transport-upwind integration) with a velocity boost converge more slowly (errors are similar to a first-order or $\sim 50^{2}$ ``un-boosted'' grid calculation). 
    \vspacerpostplot
    \label{fig:kh.numbers}}
\end{figure}

A fairly extensive comparison of $\sim10$ different SPH variations for this problem is shown in \citet{hopkins:lagrangian.pressure.sph} (Figs.\ 1-3 therein). As shown there, using a ``consistent'' (``corrected'') but non-conservative SPH method almost immediately leads to large numerical errors dominating the real solution (and runaway growth of the momentum errors). Similar catastrophic errors appear if one uses adaptive timesteps but removes the timestep limiter from \citet{saitoh.makino:2009.timestep.limiter,durier:2012.timestep.limiter}. Using an SPH method which does not explicitly include correction terms for the spatial gradients of the smoothing length (as in SPHS and many other non-Lagrangian SPH codes) simply leads to the shock being in the wrong place, even if the code conserves energy. 

If we solve this problem in 2D the differences between methods are qualitatively identical, but slightly reduced in magnitude. A 1D analogue is essentially a Riemann problem (see \S~\ref{sec:shocktube}).

\begin{figure}
    \plotonesize{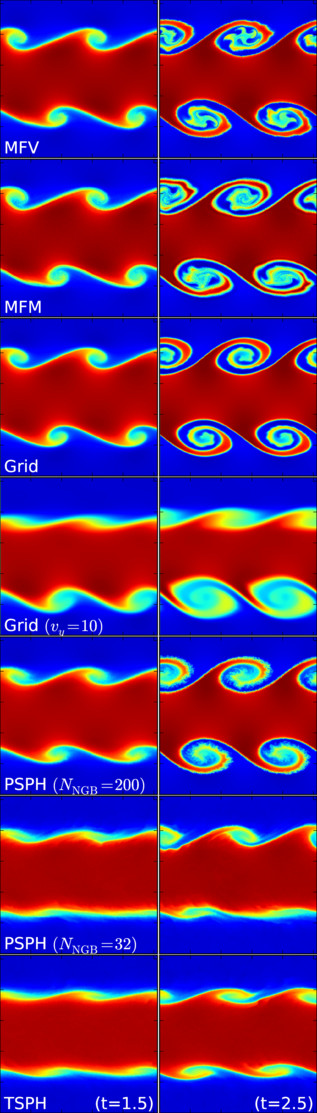}{0.73}
    \vspace{-0.1cm}
    \caption{Evolution of a 3D ($256^{2}$x$16$) version of the KH instability from Figs.~\ref{fig:kh.early}-\ref{fig:kh.nonlinear} at earlier times $t=1.5$ and $t=2.5$. The 3D instability is captured as well as to the 2D instability. Note that PSPH with low $N_{\rm NGB}$ (shown explicitly) still fails here. Also note that the early-time (linear and early non-linear) growth is nearly identical in MFV, MFM, and stationary-grid calculations (though the grid result degrades when ``boosted''); only later into the non-linear evolution do we see the differences from Figs.~\ref{fig:kh.early}-\ref{fig:kh.nonlinear}.\vspacerpostplot
    \label{fig:kh.3d.evol}}
\end{figure}

\begin{figure}
    \plotonesize{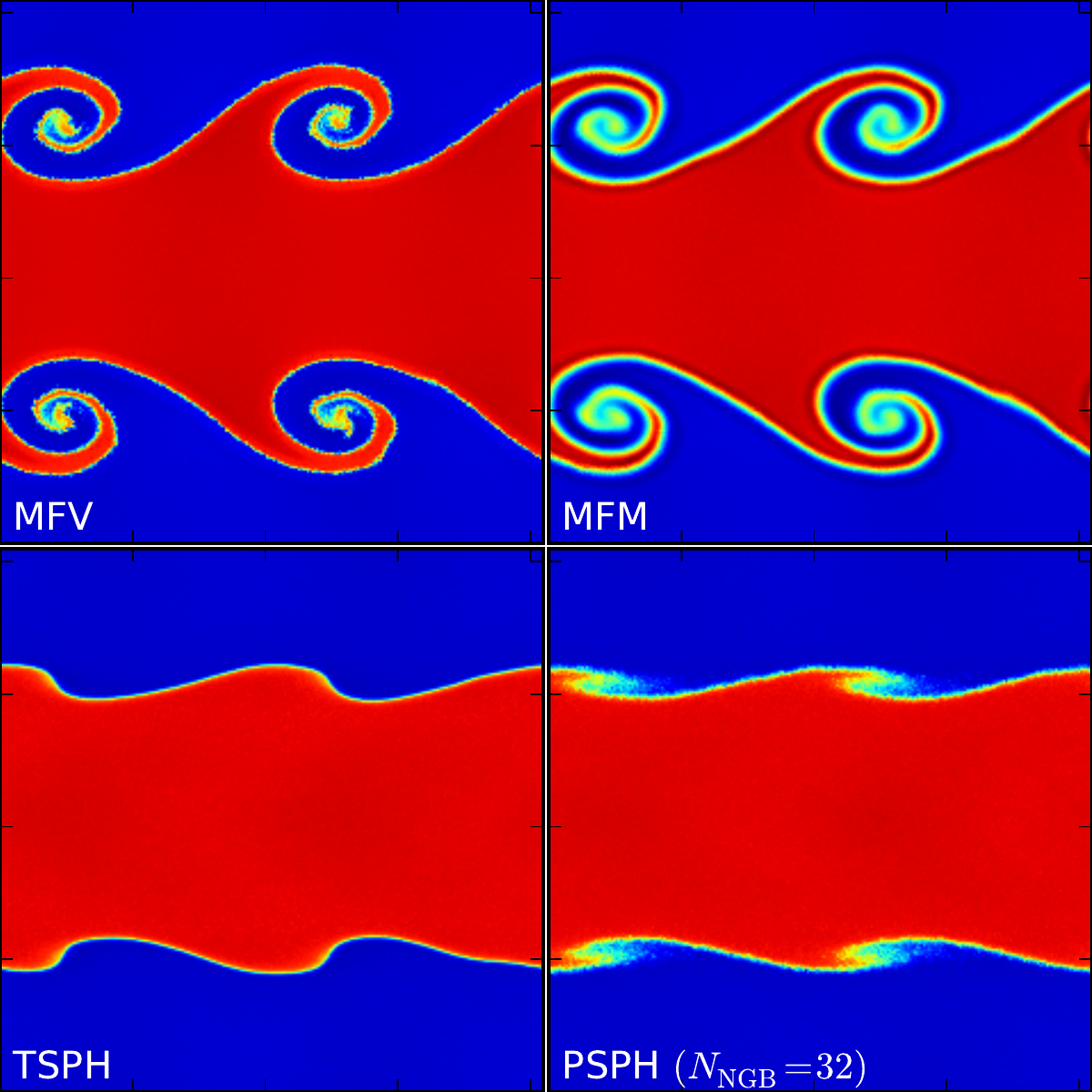}{1.0}
    \caption{Alternative 3D ($256^{2}$x$16$) KH instability test from the Wengen suite, where the ICs include a perfectly sharp contact discontinuity (as well as different shear \&\ seed modes from the previous test), at time $t=3.75\approx1.1\,\tau_{\rm KH}$. {\em Top:} MFV \&\ MFM results: the sharp discontinuity does not suppress mode growth (unlike in SPH and stationary-grid methods). Here the ICs are symmetric, and we see excellent preservation of symmetry even in the non-linear parts of the rolls. As before the MFM method smears the fluid phase boundaries slightly; the MFV method preserves a sharp contrast. {\em Bottom:} SPH results with the same neighbor number ($N_{\rm NGB}=32$); both TSPH and PSPH fail to capture the instability in this case.\vspacerpostplot
    \label{fig:kh.wengen}}
\end{figure}

\begin{figure}
    \plotonesize{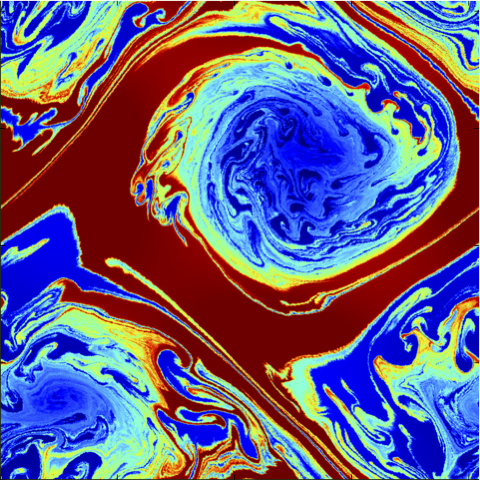}{1.0}
    \caption{2D KH instability at high-resolution ($1024^{2}$) with the MFV method, at time $t=10$.\vspacerpostplot  
    \label{fig:kh.hires}}
\end{figure}

\vspace{-0.5cm}
\subsubsection{The Noh (Spherical Collapse) Test: Extreme Shock Jumps to Break Your Solver}
\label{sec:noh}

Next consider the \citet{noh:1987.implosion.problem} test. This is a challenging test: many codes cannot run it without crashing, or run it but see catastrophic examples of the carbuncle instability. \citet{liska:2003.eulerian.code.test.ppr} noted only four of eight schemes they studied could actually run the problem. An arbitrarily large domain\footnote{For the particle codes, we simply use a huge domain so that we do not have to worry about boundary conditions. For the grid codes complicated explicit setting of inflow boundary conditions is possible and has been done here, but at fixed time it is identical to the result with a sufficiently large domain.} is initialized with $\rho = 1$, $\gamma=5/3$, vanishing pressure, and a radial inflow (directed towards the origin from all points) with $|{\bf v}| = 1$ ($v_{r} = -1$). The analytic solution involves an infinitely strong shock with constant internal density moving outwards at speed $=1/3$, with a density jump of $4^{3}=64$ at the shock in 3D. 

We focus on the 3D case since it is considered the most difficult. All our ``default'' setups run on the problem, but we confirm that several approximate Riemann solvers can fail at the shock (requiring a hierarchy of solvers). We also confirm the well-known result that in particle-based codes, an initial lattice is a pathological configuration (especially for this problem), leading to singular particle distributions (similar problems arise if initializes the moving mesh from a regular lattice); we therefore use a glass for our ICs. The density profile is shown quantitatively in Fig.~\ref{fig:noh}, and the spatial structure of the shock in Fig.~\ref{fig:noh.im}.

The MFM and MFV methods give similar results here. The shock position is recovered accurately, and the shock is appropriately spherical and smooth (there is no carbuncle instability or preferential shock propagation direction). The jump is recovered very well even at this low resolution. Both have some post-shock noise in $\rho$ owing to post-shock oscillations, but this is much weaker in the MFM result. The pre-shock $\rho$ field also has noise which is geometrically induced (since the initial particle/mesh distribution is a glass, as opposed to a perfectly spherically symmetric lattice). Both feature some (weak) suppression of the density near the origin owing to wall-heating \citep[as do many other codes, see][]{liska:2003.eulerian.code.test.ppr,stone:2008.athena}. 

In the moving-mesh method, the noise level lies between our MFM and MFV methods. However some details appear slightly less accurate than our MFV or MFM calculation. The jump with moving-meshes is slightly under-estimated; this does eventually converge to the correct jump but requires somewhat higher resolution. As we saw with the Sedov test, the shock position is slightly offset (leading the analytic solution); we suspect this owes more to the timestepping scheme than the numerical method. And the wall-heating is noticeably more severe than in the MFM or MFV methods. 

In the fixed-grid code, the carbuncle instability is particularly prominent -- this actually seeds most of the noise around the jump.\footnote{Note that we have run this with the ``standard'' version of {\small ATHENA}, which is very similar to {\small AREPO} in ``fixed grid'' mode, and gives similar results at fixed resolution to AMR codes like {\small RAMSES} (which we have also compared), {\small FLASH}, and {\small PLUTO}. As noted in \citet{stone:2008.athena}, this can be cured with the addition of problem-specific additional dissipation in the correct places (and the pre-packaged {\small ATHENA} Noh test problem uses this approach). However we wish to compare the more general behavior in their ``default'' mode for all codes here.} The instability is evident as the ``hot spots'' along the Cartesian grid axes, which at the time shown have begun to propagate faster than the rest of the shock. In {\small ATHENA} there is very little wall-heating, though this is not generally true of grid codes. 

As in the Sedov test, traditional SPH dramatically enhances the noise compared to all other methods. It has no carbuncle instability but seeds considerable spurious shock structure. It also has the most severe wall-heating. The noise is reduced by adding artificial conductivity and a larger kernel in PSPH, but still exceeds most other methods, and this makes the wall-heating more severe still. Both TSPH and PSPH spread the shock well ahead of the analytic solution: this weakens the shock jump, and it requires significantly higher resolution to capture the correct jump condition.

Finally, if we consider the 2D version of this problem, as in \S~\ref{sec:sedov}, the qualitative results are identical, but the shock jump is weaker ($4^{2}=16$ in density) and easier to capture, so the quantitative differences between methods are reduced, and all methods converge to the exact solution more rapidly. The 1D analogue (collapse along a line) is a much less interesting test because many of the challenges (pathological grid setups in particle methods, the carbuncle instability, the large density jump, preservation of symmetry in the face of grid noise) are eliminated.

\vspace{-0.5cm}
\subsection{Fluid Mixing Tests}
\label{sec:tests:mixing}

The next set of tests focuses on various fluid instabilities which are ubiquitous in astrophysics and many other areas of fluid dynamics, especially any regimes where turbulence and/or mixing are important. Considerable attention has been paid in the literature to difficulties of SPH methods in dealing with these instabilities \citep[see e.g.][]{morris:1996.sph.stability,dilts:1999.sph.stability,ritchie.thomas:2001.egy.wtd.sph,marri:2003.mod.sph.cosmo.sims,okamoto:2003.shear.sph.flows,agertz:2007.sph.grid.mixing,kitsionas:2009.grid.sph.compare.turbulence,price:2010.grid.sph.compare.turbulence,bauer:2011.sph.vs.arepo.shocks,sijacki:2011.gadget.arepo.hydro.tests}. And in response many improvements have been made to SPH, which allow it to better handle such instabilities \citep[see][]{monaghan:1997.sph.drag.viscosities,ritchie.thomas:2001.egy.wtd.sph,price:2008.sph.contact.discontinuities,wadsley:2008.sph.mixing.cosmology,read:2010.sph.mixing.optimization,read:2012.sph.w.dissipation.switches,abel:2011.sph.pressure.gradient.est,garciasenz:2012.integral.sph,saitoh:2012.dens.indep.sph,hopkins:lagrangian.pressure.sph,valdarnini:2012.sph.conductivity.doesnt.do.much}. However, as pointed out in \citet{springel:arepo}, comparatively little attention has been paid to difficulties faced by stationary-grid codes in this regime. As shown therein (see Figs.~33 \&\ 36 there), the fact that such codes have velocity-dependent truncation errors means that simply assigning the whole fluid a bulk velocity comparable to, say, the shear velocities (for a Kelvin-Helmholtz problem) or ``sinking'' velocity (for a Rayleigh-Taylor problem) will substantially change the errors and can even wipe out the instabilities entirely at low resolution. We therefore consider these in more detail below. 

\vspace{-0.5cm}
\subsubsection{Kelvin-Helmholtz Instabilities}
\label{sec:kh}

We will consider the Kelvin-Helmholtz (KH) instability in detail, since this has been the focus of most such tests of SPH and grid codes. 

First, we consider a two-dimensional setup from \citet{mcnally:2012.kh.test.comparison}. This is a KH initial condition with a non-zero thickness surface layer, and seeded mode, designed to behave identically in the linear regime in all well-behaved methods (as opposed to some setups, which depend on numerical error to seed the KH instability initially). The initial density and $x$ velocity depend on the $y$ direction as 
\begin{align}
\label{eqn:kh.setup} \rho(y) &= 
\begin{cases}
	{\displaystyle \rho_{2} - \Delta\rho\,\exp{[(y-0.25)/\Delta y]}\ \ \ \hfill { (0\le y < 0.25)}} \\
	{\displaystyle \rho_{1} + \Delta\rho\,\exp{[(0.25-y)/\Delta y]}\ \ \ \hfill { (0.25\le y \le 0.5)}} \\
	{\displaystyle \rho_{1} + \Delta\rho\,\exp{[(y-0.75)/\Delta y]}\ \ \ \hfill { (0.5\le y \le 0.75)}} \\
	{\displaystyle \rho_{2} - \Delta\rho\,\exp{[(0.75-y)/\Delta y]}\ \ \ \hfill { (0.75< y \le 1)}} \\
\end{cases}\\
v_{x}(y) &= 
\begin{cases}
	{\displaystyle -0.5 + 0.5\,\exp{[(y-0.25)/\Delta y]}\ \ \ \hfill { (0\le y < 0.25)}} \\
	{\displaystyle 0.5 - 0.5\,\exp{[(0.25-y)/\Delta y]}\ \ \ \hfill { (0.25\le y \le 0.5)}} \\
	{\displaystyle 0.5 - 0.5\,\exp{[(y-0.75)/\Delta y]}\ \ \ \hfill { (0.5\le y \le 0.75)}} \\
	{\displaystyle -0.5 + 0.5\,\exp{[(0.75-y)/\Delta y]}\ \ \ \hfill { (0.75< y \le 1)}} \\
\end{cases}
\end{align}
with $\rho_{2}=2$, $\rho_{1}=1$, $\Delta\rho = 0.5\,(\rho_{2}-\rho_{1})$, $\Delta y = 0.025$, and constant pressure $P=5/2$ with $\gamma=5/3$ throughout a periodic domain of size $0<x<1$, $0<y<1$. The system is seeded with an initial $y$ velocity mode: 
\begin{align}
v_{y}(x) &= \delta v_{y}^{0}\,\sin{(4\pi\,x)}
\end{align}
with $\delta v_{y}^{0} = 0.01$. The exponential terms above are designed to be the smoothing layer described above, so that the initial mode is well-defined; but essentially, this is a constant-pressure fluid with a density contrast of a factor $=2$ between two layers, with a relative shear velocity $=1$. The linear KH growth timescale is usually defined as 
\begin{align}
\tau_{\rm KH} &\equiv \frac{\lambda\,(\rho_{1}+\rho_{2})}{(\rho_{1}\,\rho_{2})^{1/2}\,|v_{x,\,1}-v_{x,\,2}|}
\end{align}
where $\lambda$ is the mode wavelength (here $=1/2$); so $\tau_{\rm KH}=2^{-1/2}\approx 0.71$. 

Fig.~\ref{fig:kh.early} shows the results at $t=2.1$ for a $256^{2}$ run. In the non-SPH methods, the mode behaves as expected. The linear growth phase is almost perfectly identical between the MFM, MFV, moving-mesh, and fixed-grid codes (we have compared quantitatively with the linear-growth curves in \citealt{mcnally:2012.kh.test.comparison}, and find all these methods behave similarly; see also Fig.~\ref{fig:kh.3d.evol}). The instability grows at the shear layer and the peaks of each fluid phase penetrate further, until the non-linear shear leads them to roll up into the well-known KH ``whorls.'' In the non-linear phase, we see differences begin to appear. This is further emphasized in Fig.~\ref{fig:kh.nonlinear}, where we compare later times. In Fig~\ref{fig:kh.numbers}, we quantitatively compare the amplitude of the $y$-velocity perturbation in the early (linear) phase, where we define the amplitude following \citet{mcnally:2012.kh.test.comparison} (their Eq.~6-13), and compare to the converged reference solution therein at $4096^{2}$ resolution.

In the MFM and MFV methods, the whorl height and linear growth is nearly identical to the stationary-grid results. However, unless the initial conditions in the particle codes are a perfect lattice (symmetrized exactly about the mode center and perturbation sinusoid), which is a pathological configuration, there is some seed asymmetry which we see amplified in these late times. We see in the non-linear phase, additional small-scale modes begin to grow (as they should). Here we can also begin to see that the MFV method, by allowing mass fluxes, can more sharply capture complicated contact discontinuities. In the late non-linear phases, it is truly remarkable how much fine-structure is captured by the MFV runs, given the relatively low resolution used. In these stages, we see the expected behavior: the rolls continue to grow until they overlap, at which point the box becomes non-linear and the two fluid layers ``kink'' leading to the merger of the rolls into bigger and more complex structures. This is consistent (and shows good convergence with) the behavior at higher resolution; Fig.~\ref{fig:kh.hires} shows the state of the box at $t=10$ in an MFV run at high resolution ($1024^{2}$), showing the same character and the exceptional degree of resolved sub-structure and small-scale modes. Very similar results are obtained with moving-mesh methods (see \citealt{springel:2011.voronoi.tesselation.review}, Fig.~8). 

Since the particle volume is continuous by definition, and initial particle masses are constant, the MFM method necessarily smooths the density field over $\sim1$ kernel length. This leads to less-detailed small-scale structure in the MFM method, and in the non-linear phase to enhanced diffusion. However the behavior on large scales is similar -- i.e. the MFM solution, even late into the non-linear phase, resembles a ``smoothed'' MFV solution, rather than departing from it. This is important since it demonstrates the second-order advection errors in the MFM method do not corrupt fluid mixing instabilities even in late-time, non-linear stages, where the true (physical) Lagrangian volumes of a fluid parcel would be distorted into arbitrarily complex shapes.

On the other hand, the symmetry of the ICs is manifest more obviously in the stationary-grid codes.\footnote{To ensure a fair comparison, we actually construct the ICs for the meshless methods first, then bilinearly interpolate them back to the grid for {\small ATHENA} (using scripts graciously provided by R.\ O'Leary, private communication). This is important because it ensures similar seed modes at the grid-scale in both codes. Otherwise, if {perfectly} symmetric, periodic ICs are constructed, one can obtain perfectly-canceling terms which artificially suppress non-linear mode growth in this problem.} However, the stationary grid methods are more diffusive: if we use a second-order method (the same order as our meshless methods), we see the internal structure of the whorls diffuse away after about one roll, and at all times there is a relatively large ``fuzzy'' layer in their boundaries. Especially at late times, this completely changes the character of the solution. We have to go to $\sim 2048^{2}$ resolution to see the same level of sub-structure as our meshless methods. Going to higher-order (here, {\small ATHENA} in third-order PPM mode) helps considerably, and allows much more accurate retention of the sub-structure; the diffusion level here is comparable to our MFM method. 

As noted by \citet{springel:arepo}, at any order in the stationary grid methods, if we ``boost'' the problem by adding a uniform velocity to all the gas (which has no effect on the Lagrangian methods), the diffusion and symmetry-breaking errors increase substantially, even in the early (linear) phase, where the errors are comparable to those from a lower-order method or a much lower-resolution simulation. Figs.~\ref{fig:kh.early}-\ref{fig:kh.numbers} show this explicitly. The additional diffusion is especially obvious in the non-linear (late-time) solutions. The diffusion is closely related to what we saw in the ``square'' test (\S~\ref{sec:square}): the ``rolling'' is the result of the contact discontinuity being stretched and distorted, and advected across cells in an increasingly irregular (non-grid aligned) fashion. Hence the diffusion grows as time passes and the rolls become more complicated. This also produces dramatic (unphysical) symmetry-breaking.\footnote{The excess diffusion and symmetry-breaking that appears when the stationary grid is boosted is similar at both 2nd and 3rd-order (PPM). However, the degree of symmetry-breaking is strongly sensitive to the integration scheme. A corner-transport-upwind (CTU) scheme substantially reduces (though does not eliminate) these errors, compared to more commonly-used van Leer integrators \citep[see][]{gardiner:2008.ctu.methods.paper}.} On the other hand, in Lagrangian, mesh-free methods, the arbitrary angles the rolls necessarily form as they ``roll up'' do not present any problems for advection of contact discontinuities.

SPH methods, as expected, have difficulty capturing the KH instability. It is well-known that TSPH suppresses this instability, owing to a combination of the surface tension error and E0 force errors swamping the low-amplitude mode. PSPH eliminates the surface tension term, but the E0 error cannot be eliminated in a conservative SPH scheme, only reduced by going to much higher neighbor number. So if we use a TSPH {\rm or} PSPH method with the same $N_{\rm NGB}$ as used for the MFV and MFM kernels, or as used in traditional SPH work, then we find in Fig.~\ref{fig:rt} that the mode simply does not grow (the E0 errors are still too large). {\em Only} if we use a higher-order kernel with more neighbors does the mode begins to grow appropriately: for this IC, we require a 3D-equivalent neighbor number $\gtrsim128$. However, we see that even in this case, the small-scale modes appear corrupted, with a ``shredded'' morphology. This is because the small-scale modes are corrupted in PSPH by the addition of the artificial conductivity term. Better-looking results can be obtained by using PSPH without conductivity, as in \citet{hopkins:lagrangian.pressure.sph} (Fig.~6 there); however, this comes at the cost of serious noise in problems with shocks/pressure discontinuities (much worse than the noise in TSPH, which we have already shown is worse than any other method we show here).

In Fig.~\ref{fig:kh.3d.evol}, we consider a three-dimensional version of this instability: to construct this we simply extend the ICs with constant properties in the $z$ direction, to a $256$x$256$x$16$ periodic box. Here we see essentially identical qualitative behavior, as expected. We explicitly show the earlier stages of the runs, to demonstrate again that the linear mode growth is identical in MFM, MFV, and stationary-grid methods (when the fluid has no net velocity). The transition to 3D causes no problems for either MFM or MFV methods (if anything, the extra dimension means the condition numbers of the gradient matrices tend to be slightly better-behaved, so the errors are slightly smaller). The stationary-grid results are also essentially identical. If the fluid is boosted in the stationary-grid method, we see the linear-phase mode growth is artificially suppressed (the whorls have not reached the same height at this resolution), diffusion is increased (especially at later times), and the symmetry is broken (the ``upper'' set of rolls now differ in amplitude from the ``lower'' set). PSPH is able to do reasonably well with large neighbor numbers; although the linear-phase growth is slightly slower than the converged solution from MFM/MFV/non-boosted grid methods, it is close, and the late-time solution looks reasonable. However, once again, with small neighbor number, both TSPH and PSPH fail to form rolls properly.

Finally, for the sake of completeness, we compare a different KH IC in Fig.~\ref{fig:kh.wengen}. Specifically, we consider the 3D KH test from the Wengen multiphase test suite\footnote{Available at \wengenurl} and described in \citet{agertz:2007.sph.grid.mixing,read:2010.sph.mixing.optimization}. Briefly, in a periodic box with size $256,\,256,\,16\,{\rm kpc}$ in the $x,\,y,\,z$ directions (centered on $0,\,0,\,0$), respectively, $\approx10^{6}$ equal-mass particles are initialized in a cubic lattice, with density, temperature, and $x$-velocity $=\rho_{1},\,T_{1},\,v_{1}$ for $|y|<64$ and $=\rho_{2}\,T_{2},\,v_{2}$ for $|y|>64$, with $\rho_{2}=0.5\,\rho_{1}$, $T_{2}=2.0\,T_{1}$, $v_{2}=-v_{1}=40\,{\rm km\,s^{-1}}$. The values for $T_{1}$ are chosen so the sound speed $c_{s,\,2}\approx 8\,|v_{2}|$; the system has constant initial pressure. To trigger instabilities, a sinusoidal velocity perturbation is applied to $v_{y}$ near the boundary, with amplitude $\delta v_{y} = 4\,{\rm km\,s^{-1}}$ and wavelength $\lambda=128\,{\rm kpc}$. 

As expected from the previous tests, both MFM and MFV methods capture the instability with high accuracy. One benefit of this version of the KH test is that the ICs are designed to have much better symmetry and less ``start-up noise'' for particle-based codes (while the \citealt{mcnally:2012.kh.test.comparison} IC is optimized for grid codes), and as a result we directly see that the symmetry in the MFV and MFM simulations is well-preserved, and the small-scale modes are (by design) slower to evolve (i.e.\ the loss of symmetry and appearance of small-scale ``grid noise'' in the previous simulation is not a result of the code, but of the ICs). Another useful aspect of this IC is that, unlike the previous IC, it has a true density discontinuity, across a single particle separation. We see that this is smoothed to $\sim1$ softening in the MFM method (the green ``edge''; still much less than in a stationary-grid code), and preserved nearly perfect in the MFV code, despite the rolls having executed multiple ``wraps.'' 

This discontinuity makes the problem even more challenging for SPH and stationary-grid methods, and we see that essentially no KH growth occurs in SPH without going to very large neighbor number.\footnote{In \citet{hopkins:lagrangian.pressure.sph} Figs.~8-9, we showed that PSPH was able to capture at least some ``whorl'' structure using a very similar IC to the Wengen IC, still using a simple cubic spline with $\sim 32$ neighbors, if the initial seed mode amplitude was larger ($\sim 10\%$, as opposed to $\sim 1\%$). This is because the E0 errors were then smaller than the seed mode.} As discussed in \citet{springel:2011.voronoi.tesselation.review} (see their Figs.~7-8), this is also more challenging for stationary-grid codes because of their difficulty advecting the contact discontinuity. This leads to an incorrect mode growth rate -- similar to the discrepancy seen our in Fig.~\ref{fig:kh.numbers} for a ``boosted'' grid solution, but still present even with zero ``boost.'' It also leads to much more pronounced grid noise than any other methods, because the non-smoothed contact discontinuity cannot be represented as soon as it becomes mis-aligned with the grid, and this artificially seeds secondary modes.

\begin{figure*}
    \plotsidesize{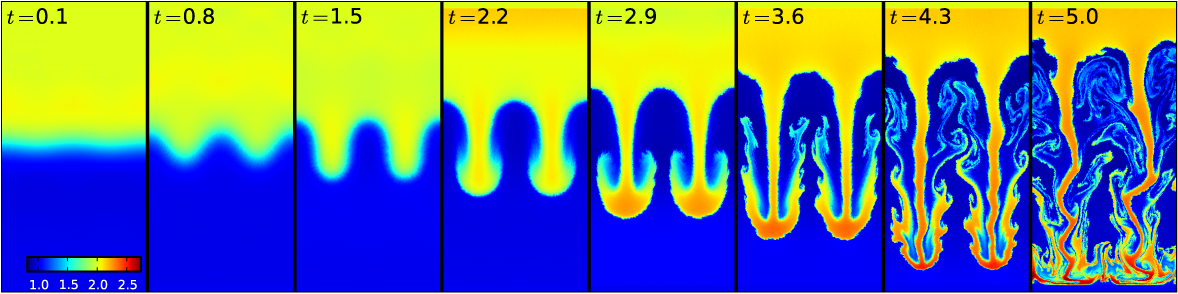}{1.0}
    \vspace{-0.5cm}
    \caption{Rayleigh-Taylor instability test (\S~\ref{sec:rt}). We plot density, from $0.8-2.8$ (black-red), in a two-dimensional simulation. Panels show the evolution of the RT instability using the MFV method at high resolution ($512$x$1024$), at different times. The linear growth of the instability is nearly identical in MFV, MFM, moving-mesh, and fixed-grid runs; in all cases it grows and secondary KH instabilities appear along the rising/sinking streams. Note the fine resolution of contact discontinuities and mixing. This run uses our standard number of particle neighbors: for both MFM and MFV runs, the instability develops regardless of the number of neighbors used (we have tested from $\sim 8-64$ in 2D). The breaking of symmetry in the non-linear phase is expected from the problem setup.\vspacerpostplot
    \label{fig:rt.evol}}
\end{figure*}

\begin{figure}
    %\plotonesize{rt.png}{0.75}
    \plotonesize{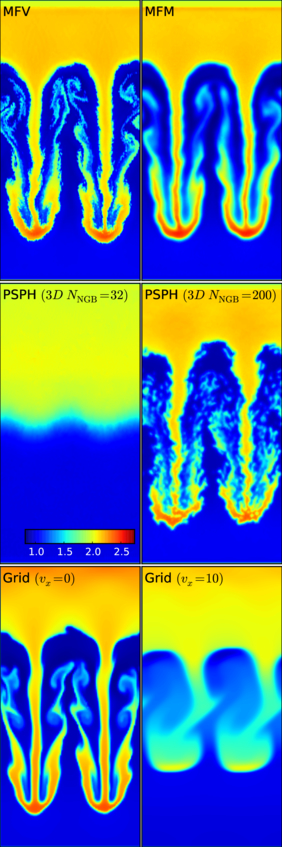}{0.75}
    \vspace{-0.2cm}
    \caption{RT instability as Fig.~\ref{fig:rt.evol} with different methods, at medium-resolution ($128$x$256$) \&\ time $t=4$. {\em Top Left:} MFV. Secondary instabilities are sharply resolved (as Fig.~\ref{fig:rt.evol}), even at lower resolution. {\em Top Right:} MFM. The evolution is similar to MFV, but contact discontinuities are not as sharply resolved. {\em Middle Left:} PSPH, with same neighbor number as MFM/MFV runs. No instability develops, despite other improvements over TSPH, because E0 errors swamp the mode growth. {\em Middle Right:} PSPH, with a higher-order kernel and increased neighbor number; this reduces E0 errors allowing the mode to grow. However non-linear evolution is corrupted by noise in the conduction scheme. {\em Bottom Left:} Stationary-grid ({\small ATHENA}) run, when the fluid has no bulk velocity relative to the grid; this gives sharply-defined features and excellent symmetry. {\em Bottom Right:} Stationary-grid, with a bulk velocity $v_{x}=10$. Velocity-dependent advection errors (most severe in Eulerian methods) substantially affect the symmetry and accuracy of the solution.\vspacerpostplot
    \label{fig:rt}}
\end{figure}

\vspace{-0.5cm}
\subsubsection{Rayleigh-Taylor Instabilities}
\label{sec:rt}

We now consider the Rayleigh-Taylor (RT) instability, with initial conditions from \citet{abel:2011.sph.pressure.gradient.est}. In a two-dimensional domain with $0<x<1/2$ (periodic boundaries) and $0<y<1$ (reflecting boundary with particles at initial $y<0.1$ or $y>0.9$ held fixed for the non-grid methods), we take $\gamma=1.4$ and initialize a density profile $\rho(y)=\rho_{1}+(\rho_{2}-\rho_{1})/(1+\exp{[-(y-0.5)/\Delta]})$ where $\rho_{1}=1$ and $\rho_{2}=2$ are the density ``below'' and ``above'' the contact discontinuity and $\Delta=0.025$ is its width; initial entropies are assigned so the pressure gradient is in hydrostatic equilibrium with a uniform gravitational acceleration $g=-1/2$ in the $y$ direction (at the interface, $P=\rho_{2}/\gamma=10/7$ so $c_{s}=1$). An initial $y$-velocity perturbation $v_{y} = \delta v_{y}\,(1+\cos{(8\pi\,(x+1/4))})\,(1+\cos{(5\pi\,(y-1/2))})$ with $\delta v_{y}=0.025$ is applied in the range $0.3<y<0.7$.

In Fig.~\ref{fig:rt.evol} we show the evolution of the instability in a high-resolution ($512$x$1024$) run with the MFV method. As expected, the initial velocity grows and buoyancy drives the lighter fluid to rise, driving bulk motions. Secondary KH instabilities form on the shear surface between the rising/sinking fluids. The linear growth of the instability is nearly identical in MFV, MFM, {\small ATHENA}, and {\small AREPO} runs; however the non-linear dynamics start to differ. For example, in the particle methods, the vertical symmetry is eventually broken, albeit weakly. This is discussed at length in \citet{springel:arepo}, but is completely expected here, because the initial particle distribution is not perfectly mirror-symmetric with the seed mode; for any seed asymmetry, growth of the non-linear KH modes making it less symmetric is the physically correct solution. The only way to force exact symmetry in these methods is to use a very specific and usually pathological initial particle distribution.\footnote{We do see here and in the KH tests that the MFV and MFM methods appear to preserve symmetry longer in time and more accurately than moving-meshes (compare Figs.~35-36 in \citealt{springel:arepo}). For the MFV and MFM methods, it is easily verified that the numerical equations are manifestly symmetry-preserving (provided the problem setup and initial particle distribution are symmetric). The growth of asymmetry in symmetric ICs stems purely from roundoff errors. In moving-mesh codes, however, the fact that mesh boundaries are ``sharp'' means that when cells are sufficiently deformed, they must eventually dis-continuously change their connectivity in a manner that does not necessarily preserve symmetry. This leads to a sort of ``mesh tension'' or ``mesh bending'' instability discussed in \citet{springel:arepo}.}

Fig.~\ref{fig:rt} compares the non-linear RT evolution across different methods, with the same initial conditions at medium resolution ($128$x$256$). The MFV and moving-mesh methods capture the most small-scale structure: this is because they are both Lagrangian and can follow contact discontinuities very sharply. The large-scale evolution of the MFM run is very similar to MFV; the growth of the RT mode is identical, but the structure of the secondary instabilities and boundaries is noticeably less sharp. As in the KH test, this is because the method enforces constant particle masses; so a contact discontinuity must necessarily be smoothed over at least one kernel kernel length (while in the MFV method it could be captured, in principle, across two particles). The result is similar if we apply a ``post-processing'' density kernel convolution to the MFV result. However both converge to the same result at high resolution.

We see the same problems with SPH as in the KH test: at low neighbor number, E0 errors and surface tension (in TSPH) suppress the growth of the instability entirely, and even in PSPH we require a 3D-equivalent $N_{\rm NGB}\gtrsim 128$ to see good linear growth. As in the KH problem, conductivity in PSPH helps the mode initially grow but corrupts the non-linear structure of small-scale KH modes (here the problem looks better without conductivity as shown in Fig.~11 of \citealt{hopkins:lagrangian.pressure.sph}, but as noted above this leads to excessive noise in other problems). 

If the fluid is not moving with respect to the grid, a stationary-grid method performs excellently on this problem. We note that the growth rate and even non-linear height of the light fluid is almost identical between MFV, MFM, {\small AREPO}, and {\small ATHENA} runs. However, the stationary-grid {\small ATHENA} run captures both fine detail in the secondary instabilities while maintaining perfect symmetry (here, the problem is set up so the grid is exactly symmetric about the perturbation; otherwise this would not hold). However, as soon as we set the fluid in motion with respect to the grid, advection errors become significant at this resolution. We show the results if we ``boost'' the entire system by a horizontal velocity $v_{x}=10$. Physically, this should leave the solution unchanged; and in all the Lagrangian methods it has no effect. But for stationary grids, it substantially slows down the mode growth rate (hence the RT plumes have not reached the correct locations), breaks the symmetry systematically (giving the fluid a ``drift'' which depends on the vertical location; this is a more serious error than random symmetry breaking because it implies a systematic shear velocity generated by the grid across the whole domain), and severely diffuses the fluid (wiping out the secondary structures). As in the KH test, because the whole volume is affected, an AMR scheme does not reduce this advection error.

As in the KH test, we note there is no 1D analogue of this test, but we see the essentially identical qualitative results whether we use 2D or 3D setups.

\begin{figure*}
    \plotsidesize{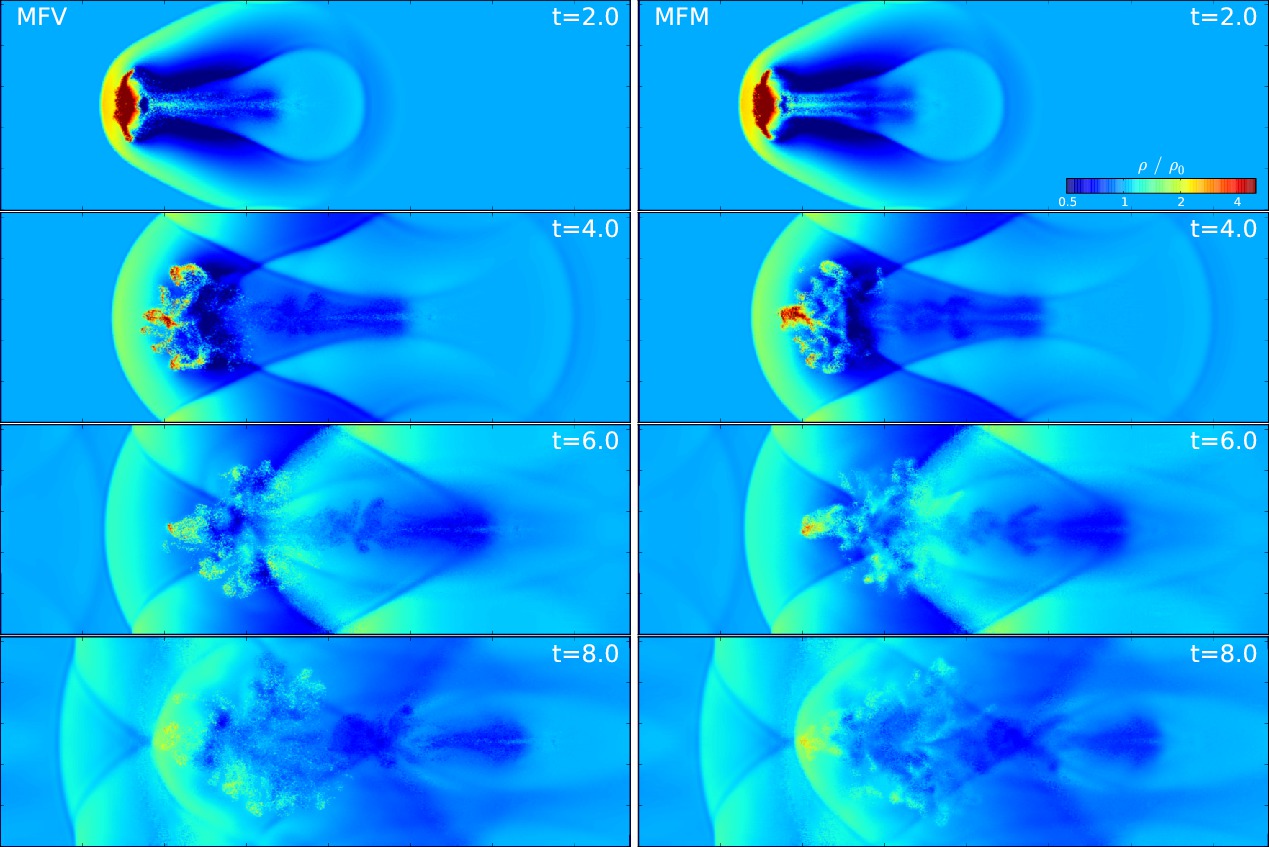}{1.0}
    \caption{The ``blob'' test (\S~\ref{sec:blob}). We plot density (in units of the initial ambient density of the background $\rho_{0}$) in a 2D slice ($0<y<6000$, $0<x<2000$, $z=0$) through the blob center, for the MFV and MFM runs at different code times (labeled). An initially dense, cold spherical cloud in pressure equilibrium is hit by a wind tunnel (moving left-to-right). The wind-cloud collision generates a bow shock and rapidly disrupts the cloud via RT and KH instabilities at the interface. We see good agreement between MFM/MFV methods; the cloud is rapidly ``shredded'' in both, and shocks are sharply-captured.\vspacerpostplot
    \label{fig:blob}}
\end{figure*}

\begin{figure}
    \plotonesize{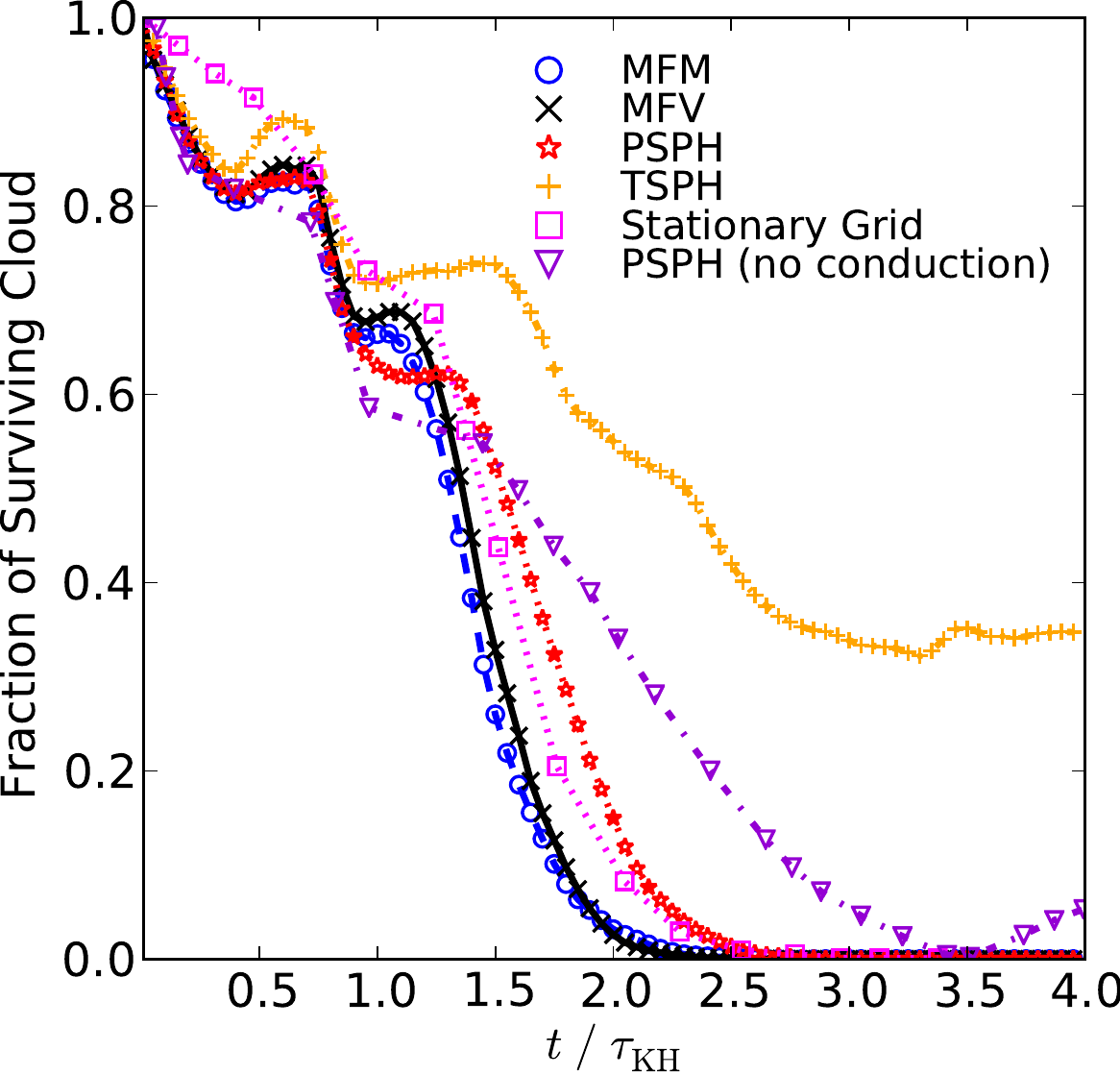}{0.9}
    \vspace{-0.2cm}
    \caption{Quantitative decay of the ``blob'' in Fig.~\ref{fig:blob}. We plot the total mass of cold, dense gas (normalized to the initial cloud mass) at each simulation time (normalized to the KH timescale $\tau_{\rm KH}=2$). Here, the grid result is from {\small ENZO}, an AMR code, but {\small ATHENA} agrees well, as do the MFM and MFV methods. In TSPH, surface tension effects and the suppression of mixing instabilities prevent the destruction of the cloud. In PSPH, most of these effects are eliminated so the cloud is much more well-mixed. However, without artificial conductivity, a ``tail'' of particles remain low-entropy and dense because there is no mechanism for generation of mixing entropy.\vspacerpostplot
    \label{fig:blob.decay}}.
\end{figure}

\vspace{-0.5cm}
\subsubsection{The ``Blob'' Test: KH \&\ RT Instabilities in a Supersonic, Astrophysical Situation}
\label{sec:blob}

Next we consider the ``blob'' test, which is designed to synthesize the fluid mixing instabilities above (as well as ram-pressure stripping) in a more ``realistic'' example of astrophysical interest representative of a multi-phase medium. The initial conditions come from the Wengen test suite and are described in \citet{agertz:2007.sph.grid.mixing}: we initialize a spherical cloud of uniform density in pressure equilibrium with an ambient medium, in a wind-tunnel with period boundaries. The imposed wind has Mach number $\mathcal{M}=2.7$ (relative to the ``ambient'' gas) with the cloud having a density $=10$ times larger than the ambient medium. The domain is a periodic rectangle with dimensions $x,\,y,\,z=2000,\,2000,\,6000\,$kpc (the absolute units are not important), with the cloud centered on $0,\,0,\,-2000\,$kpc; $9.6\times10^{6}$ particles/cells are initialized in a lattice (with equal-masses in the particle-based methods).

Fig.~\ref{fig:blob} shows the cloud morphology versus time. The wind-cloud collision generates a bow shock and begins to disrupt the cloud via KH and RT instabilities at the interface; within a few cloud-crossing timescales the dense material is well-mixed (the cloud is destroyed). Various additional shock fronts appear because of the periodic boundary conditions leading to the bow shock interacting with itself. The qualitative behavior is similar in our MFM and MFV results (see also \citealt{gaburov:2011.meshless.dg.particle.method}, Fig.~7-8, who find the same with their implementation of an MFV-like scheme), and in grid-based codes including moving meshes \citep[][Figs.~4-5]{sijacki:2011.gadget.arepo.hydro.tests}, fixed Cartesian grids, and AMR schemes (see \citealt{agertz:2007.sph.grid.mixing}, Figs.~4-10). Note in particular the good agreement between MFV and MFM results for the small-scale structure of the shredded cloud and the sharp capturing of the shock fronts. 

Quantitatively, Fig.~\ref{fig:blob.decay} follows \citet{agertz:2007.sph.grid.mixing} and measures the degree of mixing. At each time we measure the total mass in gas with $\rho>0.64\,\rho_{c}$ and $T<0.9\,T_{a}$ (where $\rho_{c}$ and $T_{a}$ are the initial cloud density and ambient temperature). We compare our results here with a compilation from other methods in \citet{agertz:2007.sph.grid.mixing}. For a stationary-grid result, we use the published result from {\small ENZO} (an AMR code), run with an effective resolution about equal to our runs here. The MFM, MFV, and stationary-grid results agree quite well. The cloud is ``completely mixed'' by this definition within a couple of KH timescales (note that there is essentially no ``residual'' beyond $t\sim 2.5$ at this resolution). The ``bumps'' at early times are real, and owe to the choice of boundary conditions (the repeated bow-shock self-interactions each time it crosses); we suspect they are suppressed in {\small ENZO} owing to a different implementation of the boundaries in that code. 

However, in ``traditional'' SPH the cloud is compressed to a ``pancake'' but surface tension prevents mixing and a sizeable fraction survives disruption for long timescales; tens of percents of the cold, dense mass survives. This is remedied in PSPH \citep{saitoh:2012.dens.indep.sph,hopkins:lagrangian.pressure.sph}. However, it is worth noting that if we neglect artificial conductivity, PSPH allows mixing in density, but entropy is still a particle-carried quantity which does not mix as easily as it should (see e.g.\ \citealt{wadsley:2008.sph.mixing.cosmology}); so the early-time behavior agrees well with the MFM, MFV, and grid methods, but there is a long ``tail'' of material which is not disrupted even at much later times ($\sim 1-10\%$ of the cloud). This is eliminated by adding an artificial conductivity or thermal diffusion term; however, there is some ambiguity (just as with artificial viscosity) regarding the ``best'' choice of switches for controlling the diffusion (hence controlling exactly how fast the cloud is mixed). Of course, we could tune parameters until the PSPH result agreed exactly with the other codes here, but given the complicated, non-linear nature of these switches, it is by no means clear that this would be appropriate for any other problem.

In 1D there are no KH or RT instabilities so the blob is not destroyed, this simply becomes a pair of Riemann problems easily solved by all methods. In 2D we see the same qualitative behavior in all cases.

\begin{figure}
    \plotonesize{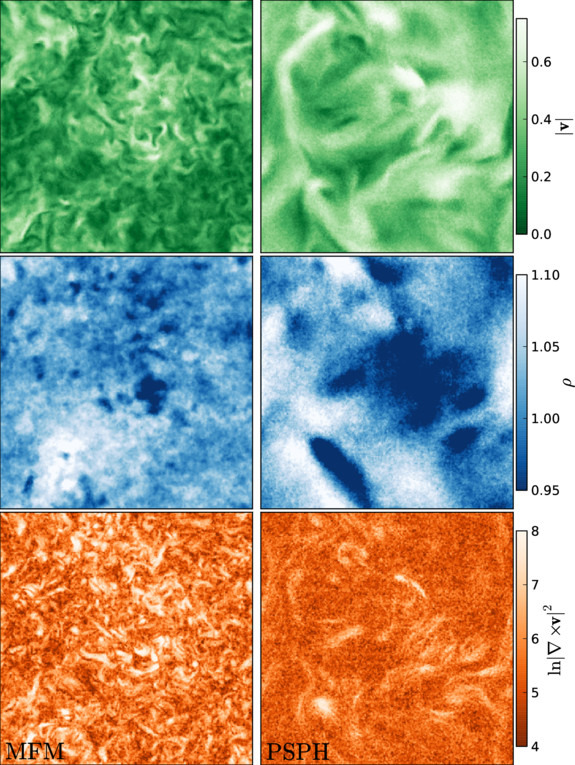}{1.07}
    %\plotonesize{turb_image.jpeg}{0.6}
    \vspace{-0.5cm}
    \caption{Driven sub-sonic turbulence (\S~\ref{sec:turbulence}), with our MFM ({\em left}) and PSPH ({\em right}) methods. We show the velocity ({\em top}), density ({\em middle}), and enstrophy ({\em bottom}), in 2D slices through the center of the 3D box. The resolution is $256^{3}$, and the time is chosen so the turbulence has reached quasi-steady-state with rms Mach number $\mathcal{M}\sim0.3$. The MFV and MFM results are nearly identical, and closely resemble stationary-grid and moving-mesh results (compare Fig.~4 in \citealt{bauer:2011.sph.vs.arepo.shocks}); note in particular the fine structure in vorticity which is captured. SPH tends to smear out some of the small-scale structure.\vspacerpostplot
    \label{fig:turb.image}}
\end{figure}

\begin{figure}
    \plotonesize{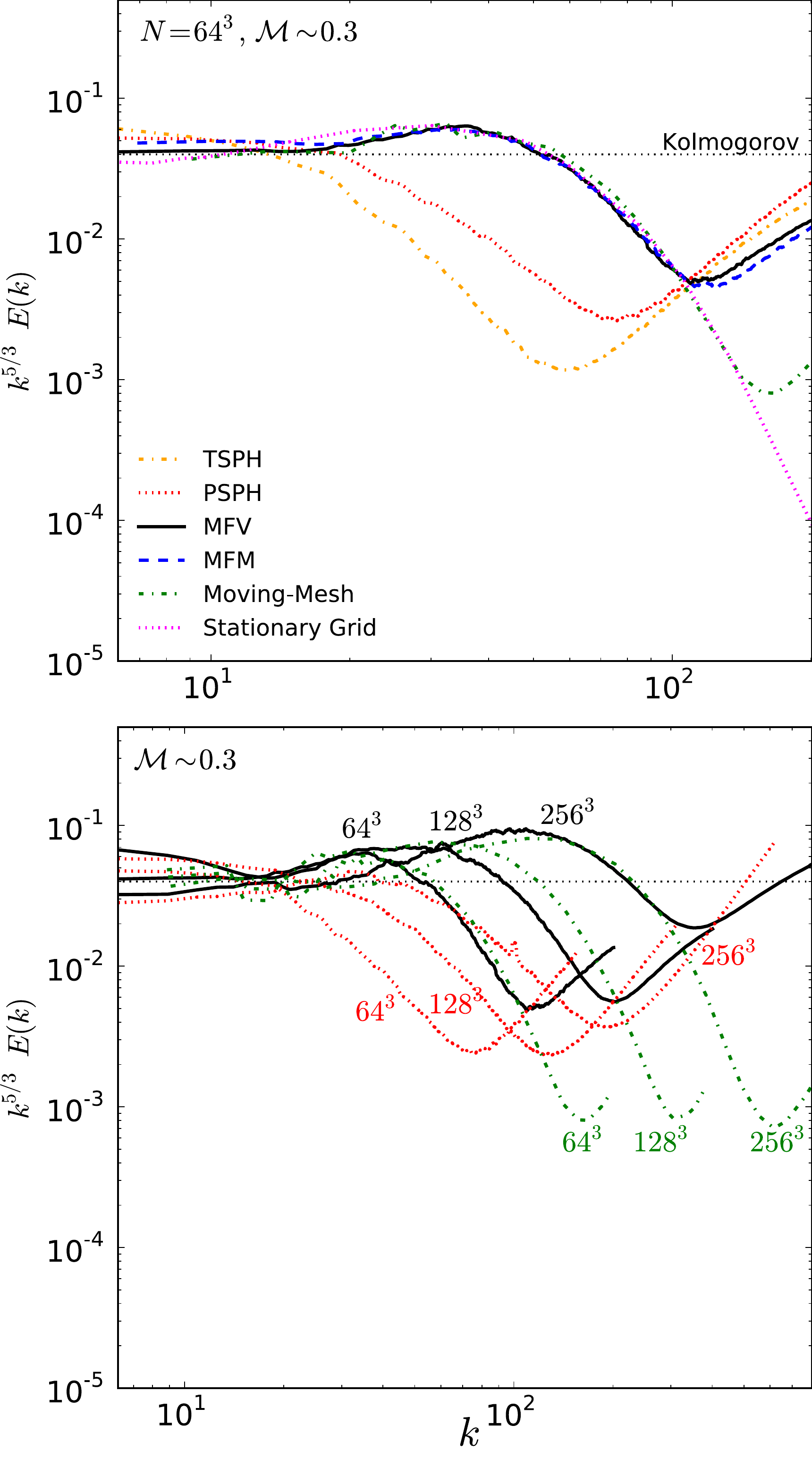}{0.9}
    \vspace{-0.3cm}
    \caption{Compensated velocity power spectra for the driven, sub-sonic turbulence in Fig.~\ref{fig:turb.image}. Dotted line shows the Kolmogorov $E(k)\propto k^{-5/3}$ law. {\em Top:} Different methods at low ($64^{3}$) resolution. MFM, MFV, moving-mesh, and stationary-grid methods are essentially identical down to the grid-scale, even including the bottleneck regime. The differences very close to the grid scale are not physically meaningful. TSPH fails to capture much power at all, on scales between the grid and driving scale. PSPH fares better, but still suppresses power in the velocity and vorticity fields on intermediate scales, compared to other methods (owing to noise in the gradient estimators). {\em Bottom:} MFV, moving-mesh, and PSPH spectra vs.\ resolution. MFV and moving-mesh (also MFM and stationary-grid) methods remain identical at higher resolution; these methods show good convergence. The dynamic range of the power captured in PSPH does increase with resolution, but more slowly.\vspacerpostplot
    \label{fig:turb.power}}
\end{figure}

\begin{figure}
    \plotonesize{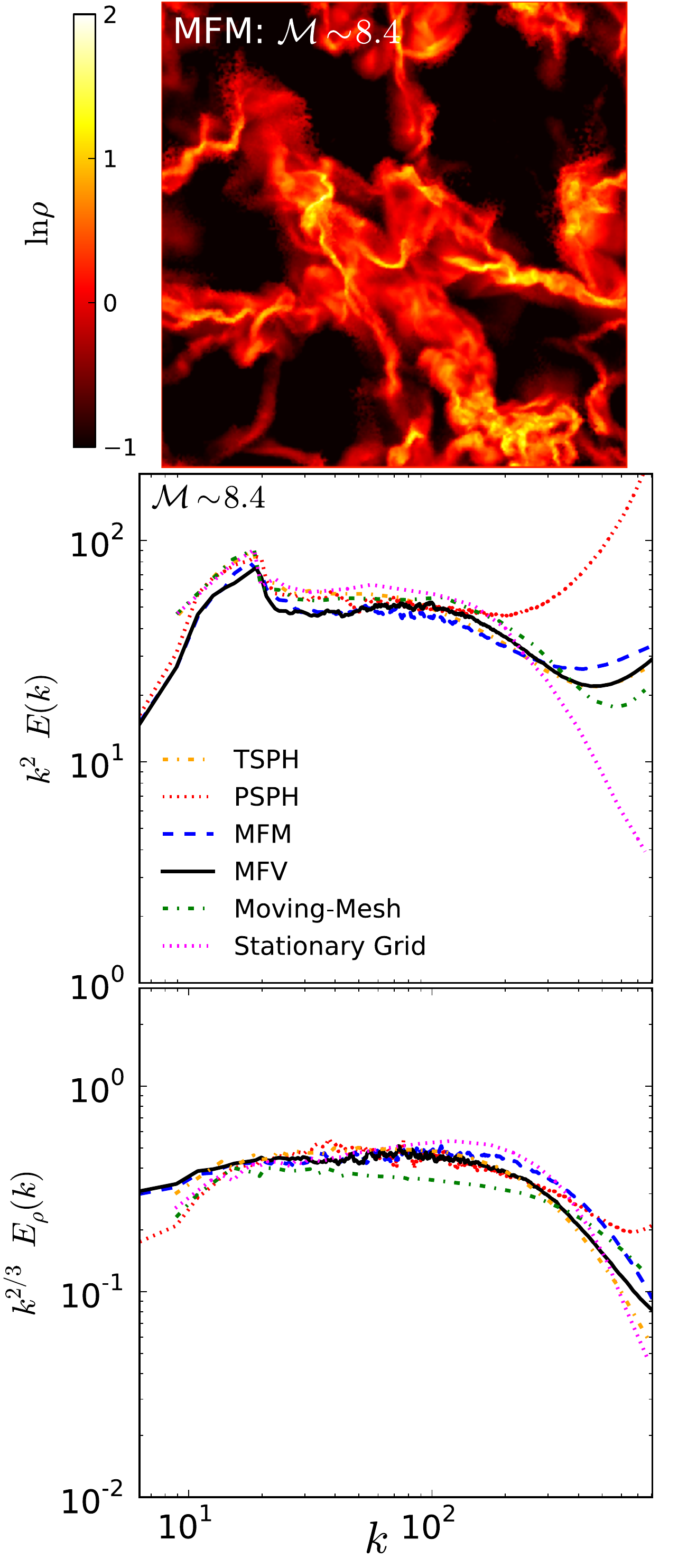}{0.8}
    \vspace{-0.3cm}
    \caption{As Figs.~\ref{fig:turb.image}-\ref{fig:turb.power}, but for super-sonic turbulence (Mach $\mathcal{M}\sim8.4$). We show the logarithmic density field ({\em top}), velocity power spectrum ({\em middle}), and linear density power spectrum ({\em bottom}), across methods at a resolution of $256^{3}$. Here, the differences between methods are smaller.\vspacerpostplot
    \label{fig:turb.supersonic}}
\end{figure}

\vspace{-0.5cm}
\subsubsection{Driven Turbulence: Sub-Sonic \&\ Super-Sonic Limits}
\label{sec:turbulence}

We next consider tests of driven, isothermal turbulence in a periodic box, in both the super-sonic and sub-sonic limits. This tests the numerical accuracy, convergence, shock-capturing, stability, and effective resolution of different methods, in a context directly relevant for almost all astrophysical problems. 

The turbulent driving routines are implemented here in an identical manner to \citet{bauer:2011.sph.vs.arepo.shocks}. Briefly, a periodic box of unit length $L=1$, density $\rho=1$, sound speed $c_{s}=1$, and isothermal equation of state $\gamma=1$ is stirred via the usual method in e.g.\ \citet{schmidt:2008.turb.structure.fns,federrath:2008.density.pdf.vs.forcingtype,price:2010.grid.sph.compare.turbulence}, where a small range of modes corresponding to wavelengths between $1/2-1$ times the box size are driven in Fourier space as an Ornstein-Uhlenbeck process, with the compressive part of the acceleration projected out via a Helmholtz decomposition in Fourier space so that the driving is purely incompressible/solenoidal (most appropriate for sub-sonic turbulence).  We use identical parameters to \citet{bauer:2011.sph.vs.arepo.shocks}, Table~4 for the driving, and consider two cases: a sub-sonic case where the driving is set such that the box maintains a quasi-steady-state rms Mach number $\mathcal{M}\sim 0.3$ and a super-sonic case with rms $\mathcal{M}\sim 8.4$. 

First, we consider the sub-sonic case, since \citet{bauer:2011.sph.vs.arepo.shocks} and others have noted this is more challenging for methods like SPH. Fig.~\ref{fig:turb.image} shows an image of the turbulent velocity, density, and vorticity/enstrophy fields, after the turbulence has reached a steady state ($t\gtrsim 5$); the image is based on a tri-linear interpolation of the particle field values from the nearest neighbors to a slice at the midplane of the $z$-axis. This can be compared to the similar Fig.~4 in \citet{bauer:2011.sph.vs.arepo.shocks}, which compares moving-mesh ({\small AREPO}), fixed-grid, and {\small GADGET-2} SPH results for the same setup. Fig.~\ref{fig:turb.power} compares the different methods quantitatively; we measure the velocity power spectra (following exactly the power-spectrum definition in \citealt{bauer:2011.sph.vs.arepo.shocks} for all methods) and show them as a function of methodology and resolution.

The results here from our MFV and MFM methods are very similar to the moving-mesh and stationary grid methods (both visually and quantitatively).\footnote{The larger apparent differences seen in the sub-sonic turbulence in the KH problem (\S~\ref{sec:kh}) clearly relate to advection of the contact discontinuities and strong density gradients, not the maintenance of vorticity. The noise seen in the Gresho problem (\S~\ref{sec:gresho}) in the particle-based methods as compared to the moving-mesh methods does not seem to be a problem here; it is small compared to the net circulation of the vortices, and furthermore they do not typically survive as long as the test problem there is run.}  In particular, we note the striking amount of small-scale structure which can be seen in the vorticity and velocity fields (Fig.~\ref{fig:turb.image}), and similarity in the predictions for the power spectrum (Fig.~\ref{fig:turb.power}). MFV and MFM methods give essentially indistinguishable results here, because the density gradients and associated mass fluxes in the MFV method are very weak. For the same reason, \citet{bauer:2011.sph.vs.arepo.shocks} found very little difference between moving-mesh and stationary-grid methods. In the power spectrum, all of these methods exhibit a similar ``bottleneck'' (the well-known feature whereby the deficit of physical viscosity leads to some excess power on scales just above the dissipation range) with a dropoff in power on the smallest scales. The only differences appear on very small scales near the Nyquist frequency (twice the inter-particle spacing, i.e.\ on scales smaller than the ``bottleneck''); the particle and moving-mesh methods show some upturn of power here, but as pointed out by \citet{price:2011.sph.turb.response}, this is dependent on how one defines the power spectrum and interpolates values for the Fourier transform (if we, for example, interpolate the particle-valued velocities onto a regular lattice, then perform the FFT, the feature goes away and the MFM and MFV methods look like the stationary grid result down to the Nyquist frequency). In any case, this all occurs below the scale where the cascade is no longer captured, so is not physically meaningful. All of these methods also show similar convergence; Fig.~\ref{fig:turb.power} shows this explicitly for MFV and moving-mesh methods but the results are again identical for MFM and stationary grids. Increasing the resolution directly translates to a larger dynamic range in the cascade; if we retain scales where the numerical result remains within a factor of $2$ of the ``expected'' power for a Kolmogorov cascade, then for a 3D ($N^{3}$-size) simulation, the power can be followed down to $k\sim 2\pi\,N/5$ (or, equivalently, the methods can meaningfully define some vorticity for structures as small as 5 elements -- either particles or cells -- across), or an ``effective'' Reynolds number of ${\rm Re}\sim (L_{\rm box}/L_{\rm diss})^{4/3}\sim 0.1\,N^{4/3}$. 

As expected, SPH performs less well here. Small-scale structure is lost in both TSPH and PSPH, owing to low-order gradient errors introducing noise and artificial viscosity not perfectly vanishing. In TSPH, the artificial viscosity is high everywhere, so there is almost no inertial range, and convergence is very slow. An extensive study of TSPH on this problem is presented in \citet{bauer:2011.sph.vs.arepo.shocks}; our conclusions are consistent with theirs. As shown by \citet{price:2011.sph.turb.response} and \citet{hopkins:lagrangian.pressure.sph}, an artificial viscosity switch improves the performance of SPH greatly, and even allows us to see some convergence, but at all resolutions we study the cascade in PSPH is still truncated compared to the non-SPH methods (by a factor $\sim 4$).

In Fig.~\ref{fig:turb.supersonic}, we repeat these experiments but now with highly super-sonic turbulence (rms $\mathcal{M}\sim8.4$). Consistent with many previous studies, we find smaller differences between SPH and all other methods \citep[see e.g.][]{kitsionas:2009.grid.sph.compare.turbulence,price:2010.grid.sph.compare.turbulence,bauer:2011.sph.vs.arepo.shocks}. The dynamic range of the velocity and density power is similar to the sub-sonic case, though an inertial range is less well-defined. Since the power on small scales in super-sonic turbulence is dominated by shocks and discontinuities \citep{burgers1973turbulence}, the essential property of the methods is that they can stably capture strong shocks and advection; in general, as long as the methods are conservative and numerically stable (true of all the methods here, although not of many finite-difference type methods), they do reasonably well in this limit. 

We have also repeated our experiments with compressively-driven turbulence, and find similar systematic differences between methods (with overall properties consistent with those well-known from previous studies; \citealt{schmidt:2009.isothermal.turb,federrath:2010.obs.vs.sim.turb.compare}); stationary-grid methods there perform slightly poorer since larger density gradients must be advected. And we have repeated our experiments in 2D, with the turbulence driven on small scales, to verify that we indeed recover the expected inverse cascade; our conclusions regarding the relative performance of different methods are identical, with all methods recovering a slightly larger inertial range. The 1D analogue (Burgers turbulence) is essentially just the randomly-driven version of the interacting blastwave problem.

\vspace{-0.5cm}
\subsection{Tests with Self-Gravity}
\label{sec:tests:gravity}

Now we consider several tests involving self-gravity and hydrodynamic forces on gas. Recall, the N-body gravity algorithm here is essentially identical to that in {\small GADGET} and {\small AREPO}, modulo well-tested improvements and optimizations, and this has been tested in a huge variety of situations \citep[see e.g.][and references therein]{springel:gadget,springel:millenium,hayward:arepo.gadget.mergers,vogelsberger:2011.arepo.vs.gadget.cosmo,scannapieco:2012.aquila.cosmo.sim.compare,kim:2013.AGORA}. We have confirmed these by re-running tests like the collisionless (dark matter) Zeldovich pancake, collisionless spherical collapse and virialization, and cosmological dark matter halo evolution using the public {\small AGORA} project initial conditions \citep[see][for details]{kim:2013.AGORA}. For our purposes here, therefore, it is not interesting to test the gravity solver in and of itself. However, it is important to test the coupling of hydrodynamics to self-gravity. This is both because complicated and interesting regimes can arise, quite distinct from those in any of the pure hydrodynamic test problems above, and because there are many different choices for how to solve the coupled hydro-gravity equations, some of which can corrupt the hydrodynamics (via e.g.\ noise from gravity, poor total energy conservation, etc.). It is also important to test that our implementation of a cosmological integration scheme appropriately handles the hydrodynamic quantities.

\begin{figure}
    \plotonesize{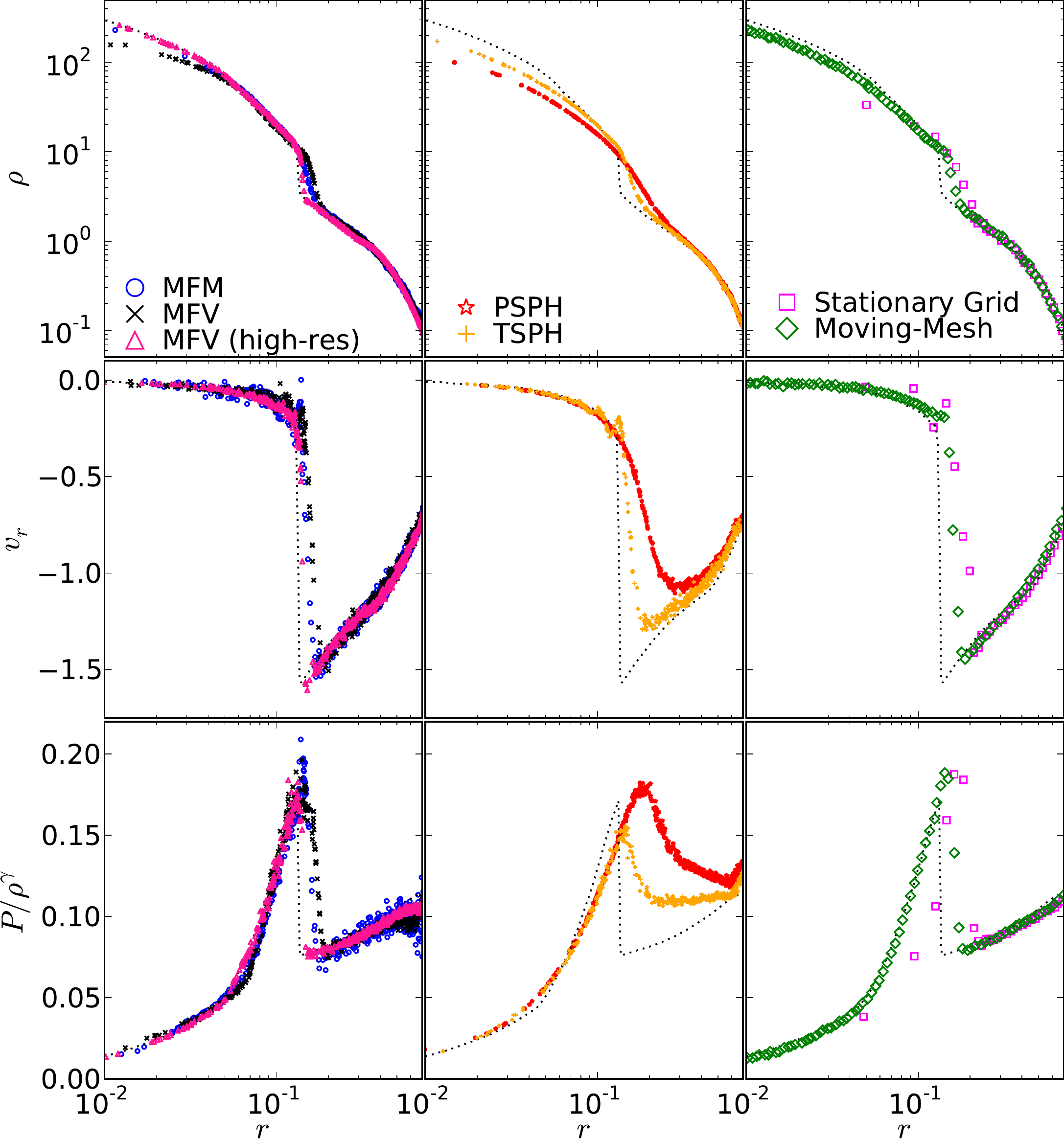}{1.0}
    \vspace{-0.6cm}
    \caption{Evrard test (\S~\ref{sec:evrard}); the collapse of a spherical, self-gravitating polytrope. We show the radial profile of density ({\em top}), velocity ({\em middle}), and entropy ({\em bottom}), in low-resolution ($30^{3}$) runs (all particles are shown). The collapse converts potential energy to kinetic, which sets up a strong shock with a virialized internal structure. At high resolution, the methods all converge to the exact solution (dotted); we demonstrate this with a high-res ($128^{3}$) MFV run. At low resolution MFM, MFV, and moving-mesh results are similar, with the former two more noisy (and exhibiting some post-shock ringing), but all leading the high-resolution shock location. Stationary grids poorly-resolve the shock interior, suppressing the internal entropy and density, because of lack of adaptive resolution. SPH smooths the shock front much more noticeably, especially P-SPH (because of the larger kernel size and added conduction terms).\vspacerpostplot
    \label{fig:evrard}}
\end{figure}

\vspace{-0.5cm}
\subsubsection{The Evrard (Spherical Collapse) Test: Gravity-Hydrodynamic Coupling \&\ Energy Conservation}
\label{sec:evrard}

We begin with the simple but very relevant test problem from \citet{evrard:1988.gas.collapse.problem}, which is commonly used to test SPH codes \citep{hernquistkatz:treesph,dave:1997.ptreesph,springel:cluster.subhalos,wadsley:2004.GASOLINE}, but until recently had not generally been used for grid methods. On an arbitrarily large (open) domain, we initialize a three-dimensional sphere of gas with adiabatic index $\gamma=5/3$, mass $M=1$, radius $R=1$, and initial density profile $\rho(r) = M/(2\pi\,R^{2}\,r) = 1/(2\pi\,r)$ for $r<R$ and $\rho=0$ outside the sphere. The gas is initially at rest and has thermal energy per unit mass $u=0.05$ (much less than the gravitational binding energy). When the simulation begins, the gas free-falls towards $r=0$ under self-gravity, until a strong shock occurs and the inner regions ``bounce'' back, sending the shock outwards through the infalling outer regions of the sphere. Eventually, the shock propagates across the whole sphere and the system settles into a hydrostatic virial equilibrium. The test is useful because it is typical of gravitational collapse of structures, and because it involves the conversion of gravitational energy to kinetic energy then to thermal energy; so it is quite sensitive to the total energy conservation of the code (particularly challenging for coupled gravity-hydro methods with adaptive timestepping, as we use here). 

Following \citet{springel:arepo}, we show in Fig.~\ref{fig:evrard} the radial profiles of density, velocity, and entropy at time $t=0.8$ (after the strong shock has formed but before the whole system is virialized), using a fixed number $\approx 30^{3}$ resolution elements for the initial sphere in all methods. There is no analytic solution here, but we use as a reference the result of a one-dimensional high-resolution, high-order (PPM) calculation in spherical coordinates; at sufficiently high resolution our MFM and MFV results are indistinguishable from this so it should be close to an exact solution. 

In every method, at limited resolution, the shock front is smoothed and leads the exact shock front slightly, but this is expected. All the methods capture the key qualitative features of the problem, but with important differences. 

The MFM, MFV, and moving-mesh results are similar. MFM appears to give a slightly more accurate shock location, and as a result more accurate post-shock density profile (the others are slightly depressed because the shock is moving ``too fast''). Both MFM and MFV methods exhibit some post-shock ``ringing,'' which owes to our particular choice of slope-limiter. Moving meshes give the least-noisy result, but slightly larger shock position offset. All capture the full entropy jump, to the same width as the density and velocity jumps. All converge similarly rapidly to the exact solution. For example, we show an MFV run with $128^{3}$ resolution, which is now almost indistinguishable from the exact solution (the same is true with MFM; for the same with moving meshes, see \citealt{springel:arepo}, Fig.~41). 

SPH captures the key behaviors, but with much more severe smoothing of the shock. In particular the entropy jump is flattened and spread over nearly $\sim1\,$dex in radius. Because of the artificial conduction terms and larger kernel size in PSPH, the smoothing effect is even more severe. In particular the artificial conduction leads to an entropy jump which is not only more smoothed, but actually leads the real shock position by a couple of smoothing lengths. 

Fixed-grids produce the least-accurate result in the shocked (interior) region.\footnote{In this section, because {\small ATHENA} does not have a flexible self-gravity solver which can be fairly compared to the other methods we use, we will use as our reference ``fixed grid'' solutions the published results from {\small AREPO} using a fixed, Cartesian grid (i.e.\ not allowing its mesh to move or deform with the fluid). As shown in \citet{springel:arepo} these are very similar to those from {\small ATHENA} and other grid codes on problems where they can overlap, so do not expect the subtle code differences to be as important as the basic aspects of the method.} This is mostly because at fixed resolution of the ICs, the ``effective'' resolution in the center of the collapsing region is much worse than the other methods (since the method is non-adaptive). But as we have shown, spherical inflow/outflow across a Cartesian grid also produces significant noise and advection errors aligned with the grid axes. As expected from our tests above, the solution quality with fixed-grids will further degrade if we set the sphere in motion across the grid. In fact, comparing an AMR result where the maximum refinement is limited so that the cell number not exceed the particle number of the Lagrangian methods by more than a factor of $\sim2$, the result is not improved (see e.g.\ Fig.~12 in \citealt{bryan:2014.enzo}, for an example with {\small ENZO}). 

We note that a 1D or 2D analogue of this problem is straightforward to construct, and produces the same qualitative behavior in all methods.

\begin{figure*}
    \plotsidesize{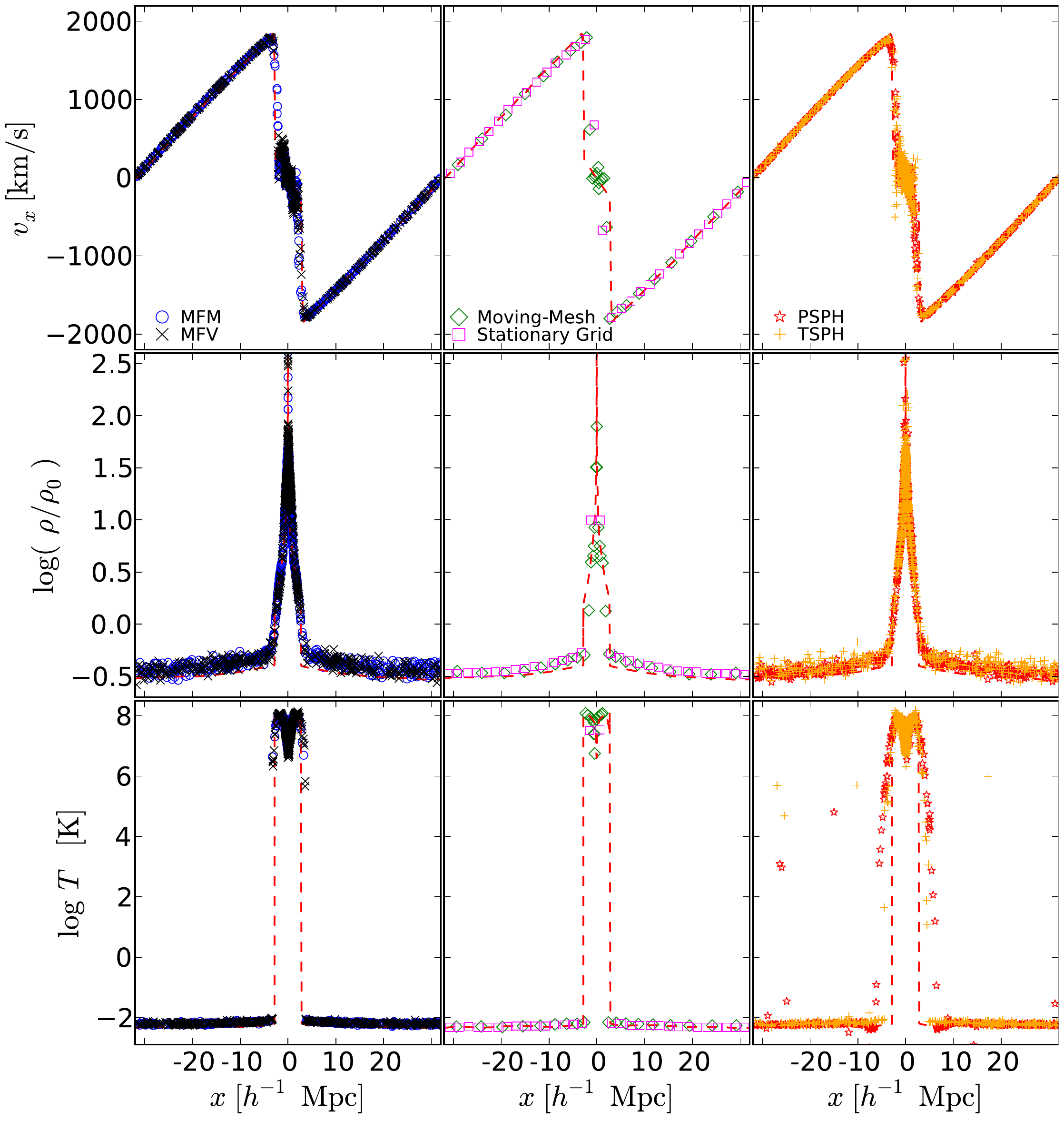}{0.9}
    \vspace{-0.3cm}
    %\plotsidesize{zeldovich_clean.pdf}{0.9}
    \caption{The Zeldovich pancake (\S~\ref{sec:zeldovich}). A density perturbation is initialized along the $x$-axis (in 3D space) at high redshift in an expanding, baryonic Einstein-de Sitter universe; it grows until collapsing into a caustic and shocking at redshift $z_{c}=1$. We plot $x$-velocity ({\em top}), density ({\em middle}), and temperature ({\em bottom}) at $z=0$, as a function of $x$ position, at $32^{3}$ resolution (the appearance of more elements in the non-mesh results is only because we use a glass IC, instead of a lattice, so the $x$-coordinates of particles spaced in other dimensions do not exactly overlap). Dashed red line shows a much higher-resolution (8192) 1D PPM calculation, which should be close to exact. All methods capture the key dynamics. Non-moving meshes under-resolve the shock interior at fixed element number (true even in AMR, in 3D, because the method does not allow anisotropic cells). SPH captures the shock and adiabatic evolution with no special treatment, but smooths the shock significantly and allows some particle-interpenetration (seen in $v_{x}$) due to imperfect application of artificial viscosity. MFM/MFV methods are similar to each other: there is noise in the low-density $\rho$-field, from small inhomogeneities in the glass ICs which are amplified cosmologically; but the interior shock structure, and steep shock jump, are well-captured. Moving-meshes are similar; less noisy but also less well-resolved in the shock center (vs.\ MFM/MFV) because of the mesh regularization procedure (see text).\vspacerpostplot
    \label{fig:zeldovich}}
\end{figure*}

\vspace{-0.5cm}
\subsubsection{The Zeldovich Pancake: Cosmological Integration, Anisotropic Geometries, \&\ Entropy Conservation}
\label{sec:zeldovich}

A standard test for cosmological integration is the ``Zeldovich pancake'': the evolution of a single Fourier mode density perturbation in an Einstein-de Sitter space. This is useful both as a ``single mode'' of large-scale structure formation in cosmology and for testing a code's ability to deal with cosmological integrations, small-amplitude perturbations, extremely high Mach-number flows and shocks, and highly anisotropic cell/particle arrangements. Following \citet{zeldovich:1970.pancakes}: assume initial (unperturbed) fluid elements have uniform density, represent Lagrangian patches, and have position $q$ along the $x$-axis at redshift $z\rightarrow \infty$ as well as an un-perturbed temperature $T_{i}$ at some arbitrarily large initial simulation redshift $z_{i}$, and $\gamma=5/3$. The perturbed comoving position $x$, density, peculiar velocity (also in the $x$-direction), and temperature are then 
\begin{align}
x(q,\,z) &= q - \frac{1+z_{c}}{1+z}\,\frac{\sin{(k\,q)}}{k} \\
\rho(q,\,z) &= \frac{\rho_{0}}{1 - \frac{1+z_{c}}{1+z}\,\cos{(k\,q)}} \\
v_{\rm pec}(x,\,z) &= -H_{0}\,\frac{1+z_{c}}{\sqrt{1+z}}\,\frac{\sin{(k\,q)}}{k} \\ 
T(x,\,z) &= T_{i}\,{\Bigl[} {\Bigl(}\frac{1+z}{1+z_{i}}{\Bigr)\,\frac{\rho(x,\,z)}{\rho_{0}}}^{3} {\Bigr]}^{2/3}
\end{align}
with $k=2\pi/\lambda$ the wavenumber of the perturbation (wavelength $\lambda$), $\rho_{0}$ the background (critical) density,  $H_{0}$ the Hubble constant (today), $z_{c}$ the redshift of ``caustic formation'' (i.e.\ non-linear collapse). This is the exact solution to the linearized perturbation equations. Following \citet{bryan:1995.cosmo.ppm,trac:2004.moving.frame.grid.code}, we set $\lambda=64\,h^{-1}\,{\rm Mpc}$ and $z_{c}=1$, and start the simulations at an initial redshift $z_{i}=100$ (in the linear regime) with $T_{i}=100$\,K (pressure forces are negligible outside the collapse region). We initialize this in a 3D periodic box of side-length $=\lambda$ (the density and temperature are uniform in the directions perpendicular to the $x$-axis, and the perpendicular components of the peculiar velocity are zero). This is done because the 3D version of the problem is most challenging, for reasons discussed below. For the particle-based methods, we initialize the particles in a glass rather than a lattice, since this is the ``standard'' for cosmological simulations; however, this seeds some small noise in the initial density fields.

Fig.~\ref{fig:zeldovich} shows the density, peculiar $x$-velocity, and temperature at redshift $z=0$, as a function of $x$ position, where we use a low-resolution initial condition of just $32^{3}$ particles in the domain (the results are similar, but with decreasing noise and sharper shock capturing, at $64^{3}$ and $128^{3}$). In early phases, $z \gg z_{c}$ (when pressure forces are negligible), the system simply traces the linear solution given above: this is captured well by all methods. The interesting dynamics occur after the caustic formation at $z_{c}$: the caustic collapses and forms a strong shock (factor $\sim 10^{10}$ temperature jump!), which propagates outwards, with a central temperature cavity that has (formally) divergent density at $x=0$ as the external pressure/temperature vanishes ($T_{i}\rightarrow0$). The un-shocked flow follows the extension of the linear solution. 

As we saw before, stationary-grid and moving-mesh methods show the least noise in the un-shocked flow. However, because of its non-Lagrangian nature, the stationary grid has the poorest resolution inside the shock, and so (at this resolution) it misses all the internal structure in the shocked region (the difference between the central divergence and outward-moving shock, for example), and suppresses the density jump by factors of $\sim 100$ relative to the particle-based methods.\footnote{In AMR methods, the outward jumps can be better captured with more refinement, of course, but it requires an effective refinement level of $\sim512^{3}-1024^{3}$ (five level-hierarchies or $2^{5}$ refinement in each dimension, increasing the total cell number and CPU cost by a factor of $\sim 1000$ in the 3D version of this problem) to achieve the same accuracy as the moving-mesh result (see e.g.\ Fig.~13 in \citealt{bryan:2014.enzo}).} The moving mesh does not suffer from this problem so captures some of the structure and obtains a factor $\sim 10$ higher density jump, but this is still over-smoothed by a factor of $\sim 10$ relative to the MFM, MFV, and SPH methods.\footnote{In fact, the moving-mesh and stationary-grid results here are actually 2D, at $32^{2}$ resolution, since that is what was provided by \citet{springel:arepo}. Since the stationary-grid is not AMR, the results should be identical in the $32^{3}$ case, except more expensive. For the moving-mesh case, if one forces the aspect ratios of cells to be regular and the same in both directions perpendicular to the $x$-axis, it should again be identical (just more expensive) in 3D, but as discussed below the mesh-deformation problems are more challenging in higher-dimensional versions of the problem. So this comparison may over-estimate the accuracy of moving-meshes on this problem.}

SPH methods do reasonably well on this problem, avoid the need for an entropy/energy switch, and capture the density peak. As expected, however, the shock jump is spread over multiple smoothing lengths, here about twice the ``true'' width of the shocked region. There is also more noise, especially in the un-shocked density and temperature fields: initial noise in the density field in this problem is (correctly) amplified as if it were the seeds of cosmological structure. Finally, in TSPH, notice that the velocity solution exhibits some points near $x\sim \pm5\,h^{-1}\,$Mpc which over/under-shoot the correct solution. This is a failure of the artificial viscosity switch (here, the constant, ``standard'' artificial viscosity of SPH) -- the artificial viscosity (even when ``always on'') is ``too weak'' to prevent particle interpenetration at these extremely super-sonic Mach numbers (particles ``punch through'' the shock). In PSPH, the higher-order artificial viscosity switches actually trip a stronger artificial viscosity term when a strong shock is detected, which eliminates this behavior. 

The MFM and MFV methods perform very well, with substantially reduced noise (especially in temperature) relative to the SPH solution. Note that if we use a regular lattice to initialize this problem instead of a glass, the noise is almost completely eliminated (as in the moving-mesh and fixed-mesh codes); however, the particle anisotropy in the shock is more severe (discussed below). In both MFM and MFV methods, the shock temperature jump is captured as well as in the moving-mesh code, with its internal structure and the density peak very well-resolved compared to both the moving-mesh and stationary-mesh methods. 

Two elements are key for good behavior on this problem. The first is some entropy-energy switch or explicit thermal energy evolution (see \S~\ref{sec:methods:switches}). {\em Whenever} a conservative Riemann method is used for the hydrodynamics on a problem like this (where the flows are extremely super-sonic, Mach number $\sim 10^{5}$), very small errors (part in $\sim 10^{10}$) in the momentum solution must (given energy conservation) appear in the temperature solution, which can lead to large deviations from the exact solution (although, by definition, these errors appear when the temperature is so low it has no effect on the dynamics, so this does not actually corrupt any other parts of the numerical solution). In stationary-mesh codes, the choice of entropy-energy switch totally controls the accuracy of the solution in the un-shocked regions. We find by systematic experimentation that the MFM and MFV methods are much less sensitive to this source of error compared to moving meshes and especially stationary-mesh codes (because the mass advection ``across cells'' is zero or reduced); however they are not free of it. Still, this reduced sensitivity allows us to use a much more conservative switch compared to even the choice used for this problem in {\small AREPO} (as described in \S~\ref{sec:methods:switches}). 

Second, the code must be able to deal with an extremely anisotropic geometry: the fluid is compressed enormously (factor $\sim 1000$) along the $x$ axis but not the other two axes. In stationary-meshes (including AMR), since the cells are always ``regular'' (usually cubical), this leads to a practical loss of resolution -- obviously non-AMR methods lose resolution when the fluid is compressed, but AMR methods which would try and ``refine'' near $x\sim 0$ in this problem (i.e.\ around/within the shock) are forced to refine in the $y$ and $z$ directions simultaneously. So to capture a factor $\sim 10$ compression in the $x$-direction, a factor $\sim 10^{3}$ more cells are required (filling in the ``plane'')! Practically, this means that these methods always, at fixed CPU cost, under-resolve these compressions in 3D. In a moving-mesh, as the compression becomes more anisotropic, the cell becomes more irregular (less cubical or spherical) in shape, which leads to larger and larger errors in the hydrodynamics and gravity (which assumes a regular cell); this will eventually destroy the solution or crash the code if some ``mesh regularization'' is not used to enforce more-regular cells (making the mesh ``stiff''; this is done in {\small AREPO}). But the more mesh regularity is enforced, the more it acts like an AMR code and suffers from loss of resolution (and advection errors) -- this is why the density peak is still suppressed by a factor of $\sim10$ in {\small AREPO} compared to the particle-based methods. In particle-based methods, there is a different problem: as the geometry is more compressed in $x$, the local particle distribution becomes highly anisotropic. In SPH, that increases the zeroth-order errors in the method (hence the larger noise). In the MFM and MFV methods, these errors are eliminated by the matrix-based gradient approach; however, if the particle distribution becomes sufficiently anisotropic, the gradient matrix becomes ill-conditioned. This is especially severe if we begin from a perfect particle lattice, in which case we can end up with the pathological particle distribution where all $N_{\rm NGB}$ neighbors lie exactly alone a line in the $x$-direction! To handle this, the adaptive checks described in \S~\ref{sec:condition.number} are necessary (or else the code will crash); for a glass IC, we find that the code adapts well and ends up finding well-conditioned matrices inside the shock region at $\sim 1.5-2$ times the ``default'' neighbor number; for the lattice IC, the initial caustic formation is the one case where the code has difficulty finding a well-conditioned matrix and resorts to the method in \S~\ref{sec:condition.number}. This, however, produces very small differences in the final solution. Note that all these problems are artificially masked (and we can make all methods appear much more accurate) if one studies a 1D version of the test problem.

\begin{figure}
    \plotonesize{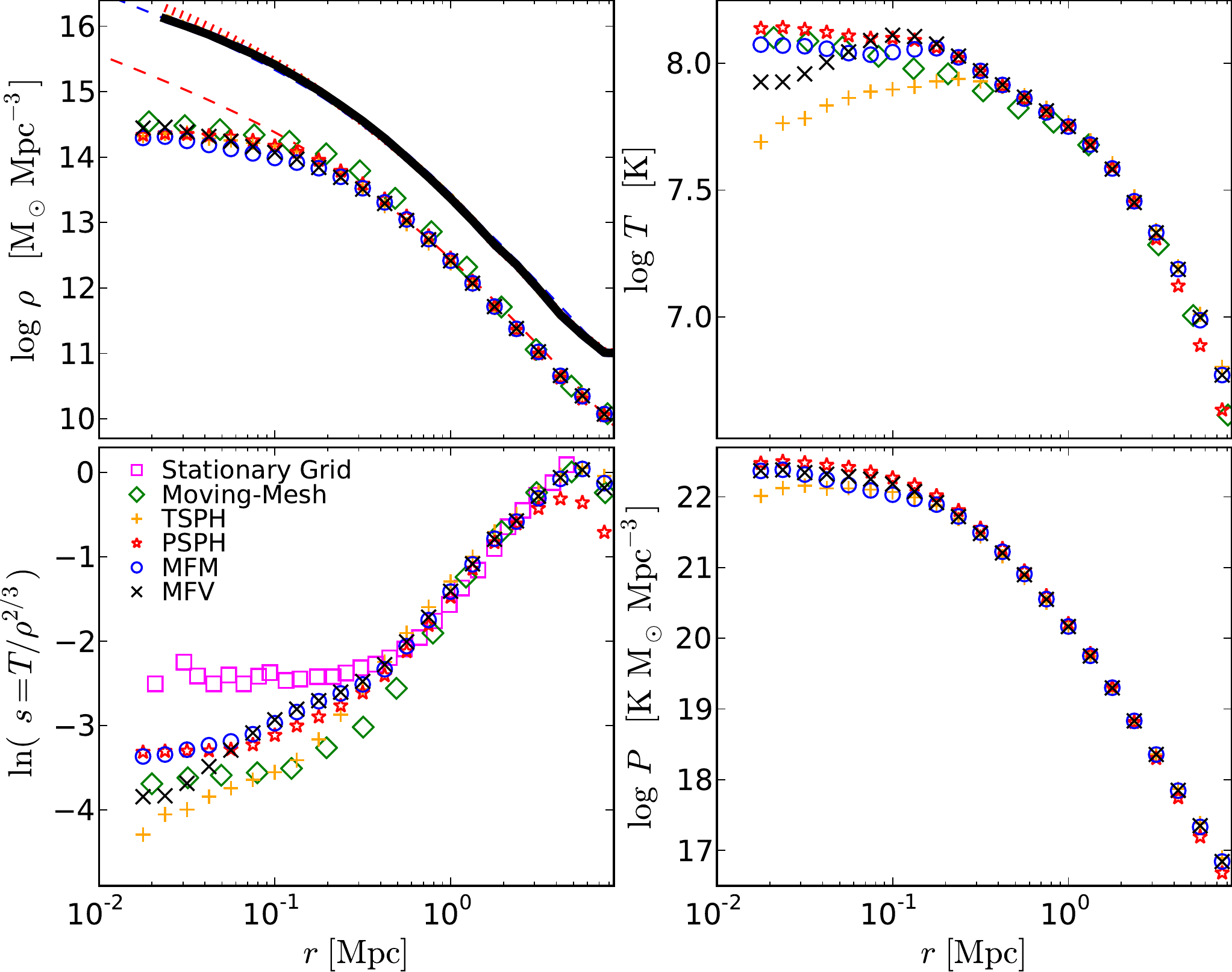}{0.98}
    \vspace{-0.3cm}
    \caption{The Santa Barbara cluster (\S~\ref{sec:sb.cluster}); a ``zoom-in'' simulation of the cosmological formation and collapse of a massive cluster, with collisionless dark matter and non-radiative gas. We plot radially-averaged profiles at $z=0$. {\em Top Left:} Gas (points) \&\ dark matter (thick lines) density. Dashed lines compare the best-fit NFW profile (blue) and it rescaled by the Universal baryon fraction (red). {\em Top Right:} Temperature. {\em Bottom Left:} Entropy. {\em Bottom Right:} Pressure. All methods agree well on the dark-matter structure, and reasonably well on the gas-pressure profile (determined by hydrostatic equilibrium vs.\ gravity). The important differences are in central entropy/temperature. Stationary grids (here, from the AMR code {\small RAMSES}) produce high-entropy ``cores.'' TSPH predicts a nearly power-law entropy decline. Moving-mesh, MFM, MFV, and PSPH methods all produce intermediate cases: some ``core'' but at a much weaker level than grid codes (closer to TSPH). TSPH appears lowest due to its suppression of fluid mixing, grids highest due to their tendency to over-mix.\vspacerpostplot
    \label{fig:sbcluster}}
\end{figure}

\begin{figure}
    \plotonesize{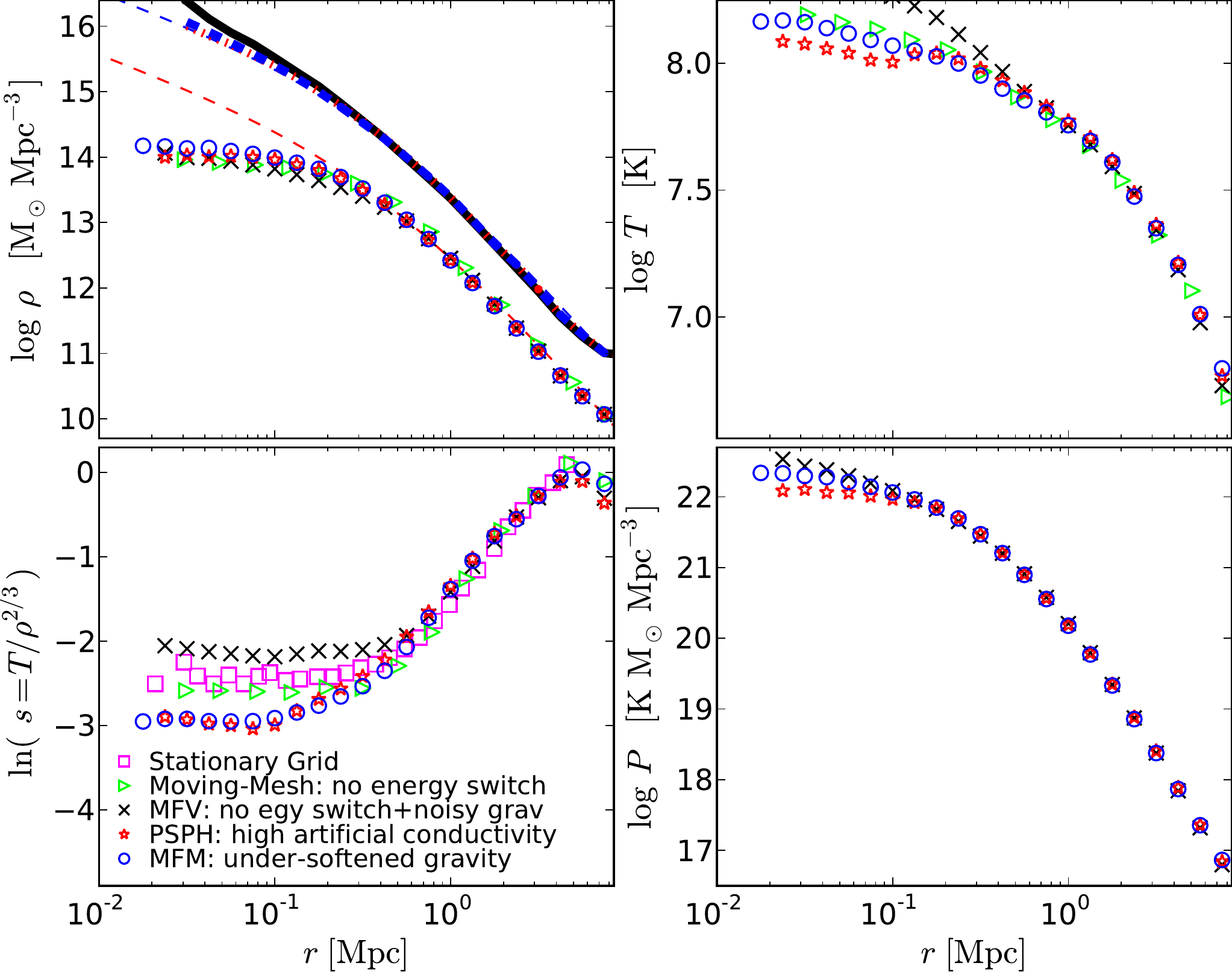}{0.98}
    \vspace{-0.3cm}
    \caption{Santa Barbara cluster as Fig.~\ref{fig:sbcluster}, but comparing the runs in our large parameter survey of $\sim 250$ test runs which most closely reproduce the grid-code results with the moving-mesh, MFM, MFV, and PSPH methods. Turning off the energy-entropy switch (i.e.\ allowing spurious heating from Riemann solver errors) in moving meshes, using too small a gravitational force softening in the MFM/MFV Riemann methods (so shot noise in the gravitational potential is translated into small shocks and entropy production), or forcing large artificial conductivity values in PSPH, all enable us to roughly reproduce the grid-code results. However all produce clearly unphysical artifacts in this and other test problems. We conclude that the high entropy cores in AMR codes are almost certainly over-estimated.\vspacerpostplot
    \label{fig:sbcluster.hientropy}}
\end{figure}

\vspace{-0.5cm}
\subsubsection{The Santa Barbara Cluster: Cosmological Hydrostatic Equilibrium, Inflow, \&\ Entropy Noise}
\label{sec:sb.cluster}

We next consider the ``Santa Barbara Cluster'' from the comparison project in \citet{frenk:1999.sb.cluster}. This is a standard reference test problem for which many codes have been compared. It is a ``zoom-in'' simulation in which a low-resolution cosmological background contains a higher-resolution Lagrangian region which will collapse to form an object of interest (and the region around it) by $z=0$; here chosen so the object represents a rich galaxy cluster in an Einstein-de Sitter Universe. The details of the cluster ICs are described there; briefly, a periodic box of side-length $64\,h^{-1}\,{\rm Mpc}$ is initialized at redshift $z=49$ ($a=1/(1+z)=0.02$),  in a flat Universe with dark matter density $\Omega_{\rm DM}=0.9$, baryonic $\Omega_{\rm b}=0.1$, Hubble constant $H_{0}=100\,h\,{\rm km\,s^{-1}\,Mpc^{-1}}$ with $h=0.5$, and negligible initial gas temperature $T=100\,$K. The gas is non-radiative (ideal) with $\gamma=5/3$. 

In Fig.~\ref{fig:sbcluster}, we show the (spherically mass-weighted average) radial profile of dark matter density, and gas density, temperature, pressure, and entropy, at $z=0$ (centered on the center-of-mass of the gas in the most massive system). The dark matter density we compare to an NFW profile with virial radius $R_{\rm vir}=2.734\,h^{-1}\,{\rm Mpc}$, and concentration $c=7.5$, which provides a reasonably good fit to all our simulations. Note that all the methods here, and indeed the other methods in the literature and even the original survey of methods in \citet{frenk:1999.sb.cluster}, agree fairly well on the dark matter profile and, in turn, the gas pressure profile (because the pressure gradient must balance gravity, which is primarily set by the dark matter profile). The methods also all agree well on the gas density/temperature/entropy profiles outside the cluster center ($\gtrsim 0.2\,h^{-1}\,{\rm Mpc}$).

The largest differences between methods reflect what \citet{frenk:1999.sb.cluster} originally identified as the main differences between SPH and grid methods: namely, that stationary grid methods tended to predict systematically higher central entropy ``cores'' as compared to SPH. The difference is discussed at length in \citet[][\S~9.3 therein]{springel:arepo}; briefly, SPH conserves particle entropy accurately (unlike grid methods), but suppresses fluid mixing, hence mixing entropy when averaging over finite scales. Grid codes, on the other hand, over-mix and diffuse entropy, and are subject to spurious ``grid heating'' (noise in the gravitational field from collisionless particles producing weak shocks which heat the gas). The difference persists even in modern, high-resolution comparisons: note that the state-of-the-art AMR result here from {\small RAMSES} \citep[see also][]{power:2013.sphs.entropy.cores} is very similar to other AMR codes like {\small ENZO} and the original \citet{frenk:1999.sb.cluster} fixed-grid results, and the TSPH result here is very similar to the \citet{frenk:1999.sb.cluster} SPH results and {\small GADGET-2}.

Interestingly, the moving-mesh, MFM, MFV, and PSPH results lie generally between the TSPH and stationary-grid result, but somewhat closer to TSPH. The {largest} central entropy among these methods is predicted by PSPH, actually, but we have shown that this method tends to over-diffuse entropy compared to MFM, MFV, and moving meshes. The MFM and MFV predictions agree well; interestingly, the moving-mesh result from {\small AREPO} is slightly closer to the TSPH result in entropy, but to MFM and MFV in temperature. 

To investigate this further, we have re-run an extensive suite of simulations of the cluster IC: $>50$ high-resolution ($128^{3}$) runs and $>200$ low-resolution ($64^{3}$) runs, in which we have systematically varied numerical aspects of the method like the choice of Riemann solver, slope-limiter, order of the reconstruction, gravitational softening (relative to the inter-particle separation), Courant factor/timestep criteria, energy/entropy switches, gravitational force accuracy, and (in SPH) artificial viscosity and conductivity parameters. The result of this extensive survey strongly supports the conclusions from \citet{springel:arepo}. Fig.~\ref{fig:sbcluster.hientropy} illustrates this with a few representative simulations: we show that we can reproduce the stationary-grid results {\em if} we artificially enhance the numerical diffusion and/or gravitational ``noise'' in each method. For example, \citet{springel:arepo} show that if they disable the energy/entropy switch used to suppress artificial heating of adiabatic flows with high bulk Mach number, they obtain a result very similar to the stationary grid; however this numerical method is clearly wrong, since it gives a seriously incorrect solution for the (analytically known) temperature of the IGM in the early Universe and produces too much entropy on other tests (e.g.\ the Zeldovich pancake). 

Similarly, in both our MFM and MFV methods, if we use a very strong slope limiter or a lower-order method (greatly increasing the numerical diffusion in other test problems), we can reproduce the stationary-grid result (with similar errors in the adiabatic phase). Alternatively, we can under-soften gravity for the dark matter -- i.e.\ reduce the gravitational softening for the dark matter to a value smaller than the inter-particle separation in the cluster center at this resolution -- in which case the noise seeded by individual particle motions is greatly enhanced. This leads to ``jostling'' of the particles, which in the Riemann solution produces entropy, and again leads to a result similar to stationary grids. In this case, even though the softening is clearly poorly-chosen, since the early Universe is more smooth, the early (adiabatic) phase of expansion is still captured correctly. However the late-time dark matter profile at small radii is corrupted by N-body transfer of energy from dark matter to gas particles; again indicating this solution is clearly incorrect. In TSPH, these sources of noise are suppressed by the particle-based entropy conservation (i.e.\ we might get a more correct answer in these limits, but for the wrong reasons). But in PSPH, the addition of an artificial conductivity term means the same noise sources lead to similar effects. Alternatively, in PSPH, we can simply choose to enhance the numerical entropy diffusion by making the artificial conductivity coefficient much larger: we show the results increasing this by a factor of $\sim5$, which leads to reasonable agreement with the stationary-grid results. However, this leads to seriously excessive diffusion in nearly every other test problem we consider (where PSPH was already one of the most diffusive methods). In short, essentially every run with parameters that give good results on the other test problems leads to an answer similar to those in Fig.~\ref{fig:sbcluster} -- i.e.\ much lower central entropies compared to stationary-grid codes; while {every} parameter choice we consider which gives good agreement with the stationary-grid codes on this test problem leads to a serious problem in some other test.

We therefore echo the conclusions of \citet{springel:arepo}. While the ``exact'' correct solution to the SB cluster central entropy problem remains unclear, it almost certainly lies between the results from stationary-grid/AMR codes (which over-mix, predicting too much entropy owing to advection errors and spurious ``gravitational heating'') and traditional SPH codes (which under-mix).

\begin{figure*}
    \plotsidesize{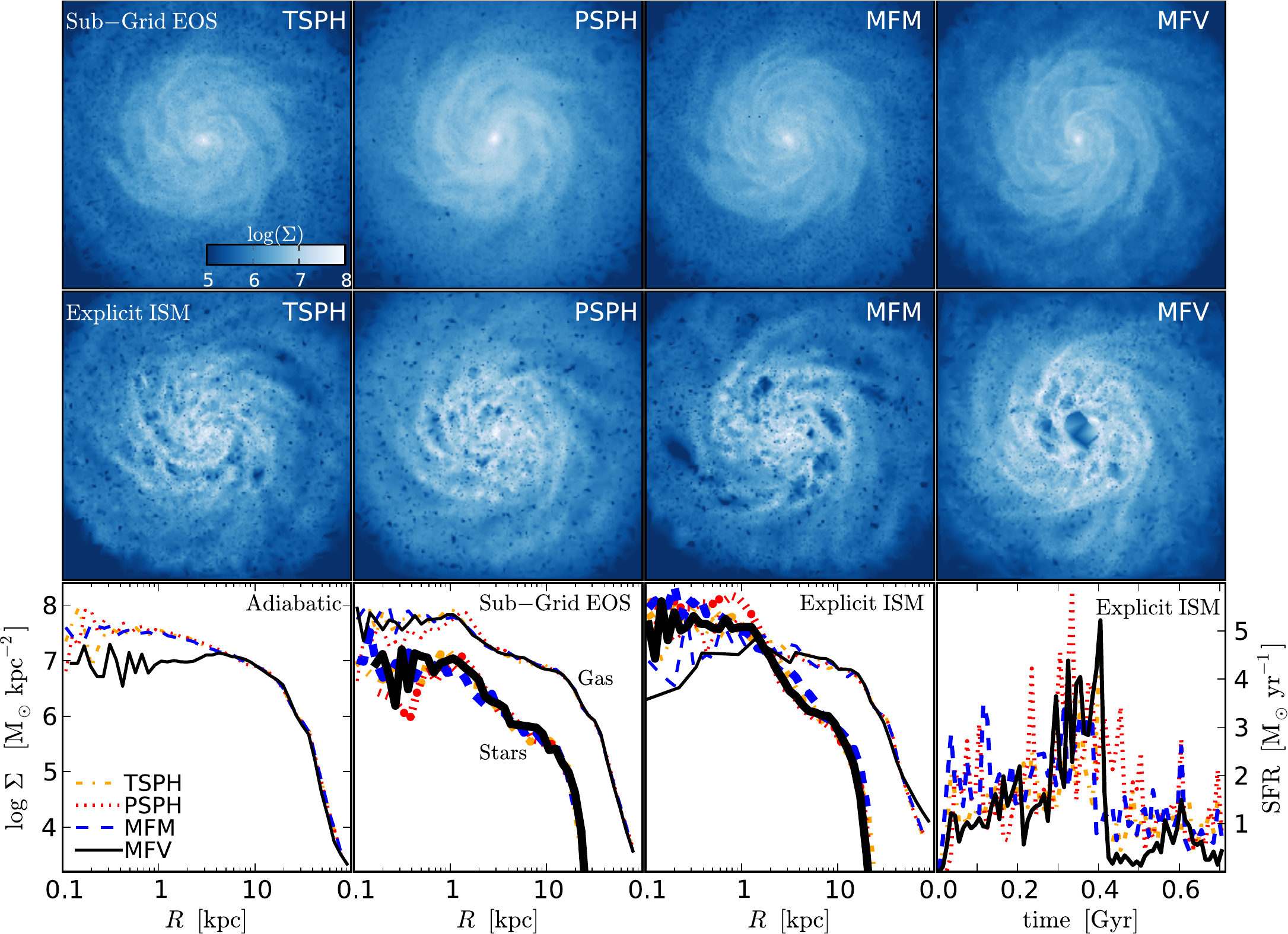}{0.98}
    \vspace{-0.3cm}
    \caption{Isolated Milky Way-like galaxy disk, with stars, gas, and dark matter, evolved for several orbits. The initial conditions and physics are identical in each case, only the hydrodynamic method is varied. {\em Top:} Projected disk gas density at $t=0.7\,$Gyr ($\sim5$ orbits at the effective radius $\sim 5\,$kpc, $\sim150$ orbits at $\sim0.1$\,kpc), in a box $70\,$kpc on a side. The \citet{springel:multiphase} sub-grid model for star formation and the ISM (treating the ISM with an ``effective equation of state'' determined via stellar feedback) is used. {\em Middle:} Same, but using the \citet{hopkins:2013.fire} physics models, which explicitly treat low-temperature cooling, star formation, and stellar feedback via SNe, radiation pressure, photo-heating, and stellar winds, leading to a multi-phase ISM. {\em Bottom:} Mass profiles and SFR in the simulations. We plot the projected (face-on) surface density profiles of gas (thin) and stars formed during the simulation (thick) in runs using only adiabatic hydrodynamics+gravity ({\em left}; no star formation here); using the \citet{springel:multiphase} sub-grid ISM treatment ({\em center-left}); using the \citet{hopkins:2013.fire} explicit treatment of the ISM and star formation ({\em center-right}). Finally, for the explicit ISM case, we plot the SFR vs.\ time for each method ({\em right}). All the methods agree well in morphology, star formation history, and disk angular momentum evolution, with weak second-order differences discussed in the text. The problem is clearly dominated by the input physics rather than numerical methods.\vspacerpostplot
    \label{fig:iso.disk}}
\end{figure*}

\vspace{-0.5cm}
\subsubsection{Isolated Galaxy Disks: Modeling Complex ISM Physics}
\label{sec:galaxy.disks}

We now consider a more practical ``realistic'' problem -- evolving a Milky Way-like galactic disk, with stars, gas, and dark matter. This is not so much a test problem (in that there is no known ``correct'' solution), as it is a means to check whether the methods here are useable on real, complicated problems that involve a wide range of physics including highly non-linear, chaotic processes like stellar feedback and star formation. For this problem, we will invoke a wide range of additional physics beyond just gravity and hydrodynamics; because there is no implementation of these physics in the moving-mesh or stationary-grid codes to which we compare (and, as we show, the choice of physics included dominates the solution), we restrict our comparison to the methods we can run within the same code.

The initial galaxy has a bulge, stellar and gaseous disk, halo, and central black hole. They are initialized in equilibrium following \citet{springel:models} so that in the absence of cooling, star formation, and feedback there are no significant transients. The galaxy has baryonic mass $M_{\rm bar}=7.1\times10^{10}\,\msun$ 
and halo mass $M_{\rm halo}=1.6\times10^{12}\,\msun$ (concentration $c=12$), black hole mass $3\times10^{6}\,\msun$, 
a \citet{hernquist:profile} profile bulge with mass $m_{b}=1.5\times10^{10}\,\msun$, and exponential 
stellar ($m_{d}=4.7\times10^{10}\,\msun$) and gas disks ($m_{g}=0.9\times10^{10}\,\msun$) 
with scale-lengths $h_{d}=3.0$ and $h_{g}=6.0$\,kpc, respectively. The gas disk is initially vertically pressure supported with scale-height $z_{0}=0.3\,$kpc, and the stellar disk scale height and velocity dispersion is such that the Toomre $Q=1$ everywhere. The disk is evolved with vacuum boundary conditions (i.e.\ in isolation, non-cosmologically) for $0.5$\,Gyr (a couple of galactic orbits at $10\,$kpc, but $\sim100$ orbits near our resolution $\sim0.1$\,kpc!). We intentionally focus on a low-resolution example, where differences between methods will be maximized: we use 3.5e4, 5.0e4, 2.0e4, and 1.0e4 particles for the initial gas disk, halo, stellar disk, and bulge. 

We consider three different physics modules, of varying complexity. First, an ``adiabatic'' model, pure hydro+gravity. Here all stars, dark matter, and black holes are collisionless, and the gas obeys a $\gamma=5/3$ equation of state. The test is similar our our Keplerian disk, but for a self-gravitating, thick, three-dimensional gas+stellar disk. All of the methods produce very similar results. The disks develop spiral structure, but do not transfer much angular momentum over this time. At $0.5\,$Gyr, the TSPH, PSPH, and MFM results are nearly identical (and all within $\sim25\%$ of the initial surface density profile at this time, showing the disk is quite stable). The MFV result is somewhat different, with the central density depleted and outermost density enhanced owing to a slow outward diffusion of angular momentum. This has to do with the small angular momentum advection errors associated with the mass fluxes between particles, as in \S~\ref{sec:keplerian}. The central part of the disk at $\sim0.1$\,kpc has executed $\sim100$ orbits by this time, so any preservation of the disk at all is remarkable! We expect, based on \S~\ref{sec:keplerian}, that the MFM method can continue to preserve angular momentum accurately even at such late times; what is more surprising is that the SPH methods show little transfer as well. Recall, in the Keplerian disk problem, the degradation of the disk in SPH was caused by the viscous instability. Here, two effects strongly suppress this. First, the gas is much hotter and more strongly pressure-supported, especially in the center, where it reaches $h/R\sim 1$; so the fractional effect of erroneous viscous forces is much smaller. Second, the disk is relatively gas-poor ($\sim 5-10\%$ gas in the central regions), so the collisionless stellar disk actually dominates the dynamics and the gas disk is stabilized by the mutual interaction with this collisionless component. As a result, the angular momentum errors from pressure forces become negligible, and the only angular momentum errors that build up are those from advection; hence the small effect still visible in the MFV method. This is a serious concern, still, for fixed-grid codes, where advection errors are much larger than in Lagrangian codes. 

The next model we consider uses the \citet{springel:multiphase} sub-grid model for a multi-phase interstellar medium (ISM) and star formation. This has been used in a wide range of previous and current work on cosmological galaxy formation \citep[e.g.][]{springel:lcdm.sfh,robertson:cosmological.disk.formation,
dimatteo:msigma,hopkins:qso.all,hopkins:groups.ell,hopkins:cusps.mergers,narayanan:co.outflows,vogelsberger:2013.illustris.model}. In these models, rather than attempt to resolve the ISM structure or feedback explicitly, the ISM is parameterized by an ``effective equation of state (EOS)'' at high densities ($n\gtrsim0.1\,{\rm cm^{-3}}$), with an adjustable law which turns gas into stars at a fixed efficiency of $\sim1\%$ per gas dynamical time (tuned to be similar to the observed Schmidt-Kennicutt relation). Gas at lower densities follows a standard atomic cooling curve from \citet{katz:treesph}. Here we see similar spiral structure to the pure adiabatic case; the effective EOS is quite ``stiff'' so keeps the gas in the disk smooth. By $0.5$\,Gyr about $\sim20\%$ of the gas has turned into stars, within the disk radius ($\sim10\,$kpc) where the density meets the threshold value above. Here, both the stellar and gas mass profiles agree very well across all methods. There is still a small angular momentum diffusion in the MFV result at large radii, but at small radii we see that the effects are swamped by the effect of slightly enhanced gravitational instability in the disk (it is not quite as ``stiff'' as $\gamma=5/3$ here) leading to gas inflows into the center along the spiral arms which enhances the gas mass within $\sim1\,$kpc. 

The third model we compare is the ``explicit feedback'' model used in the FIRE (Feedback In Realistic Environments) simulations, described in a suite of papers \citep{hopkins:rad.pressure.sf.fb,hopkins:2013.fire,narayanan:2012.mw.x.factor,van-de-voort:2014.rprocess.enrichment,faucher-giguere:2014.fire.neutral.hydrogen.absorption}. Briefly, in these simulations the multi-phase ISM and stellar feedback are treated explicitly: gas can cool to $<100\,$K via fine-structure and molecular cooling \citep{hopkins:dense.gas.tracers}, and star formation occurs in dense regions above a threshold $n>10\,{\rm cm^{-3}}$, which are also molecular (self-shielding), and locally self-gravitating \citep[see][]{hopkins:virial.sf}. The energy, momentum, mass, and metal fluxes from various feedback mechanisms are followed explicitly according to standard stellar evolution models; this includes radiation pressure in the UV and IR \citep[see][]{hopkins:rad.pressure.sf.fb}, supernovae types I \& II, stellar winds (O-star and AGB), and photo-ionization and photo-electric heating \citep{hopkins:fb.ism.prop}. The combination of these physics lead naturally to a self-regulating, multi-phase ISM \citep{hopkins:clumpy.disk.evol,hopkins:2013.accretion.doesnt.drive.turbulence} with strong galactic outflows \citep{hopkins:stellar.fb.winds,hopkins:2013.merger.sb.fb.winds}. We see that the resulting ISM structure in this case shows a more clumpy morphology, as expected, with large GMC complexes and bubbles produced via overlapping SNe explosions. The methods differ in detail, but these differences are consistent with being essentially stochastic -- the interaction of feedback and the ISM is highly chaotic, so we do not expect exact agreement here. The mass profiles are similar in all cases; the central kpc of the galaxy rapidly turns $\sim 50\%$ of its mass into stars, while the outer regions form stars slowly. The gas densities at $>10\,$kpc are elevated by the presence of galactic winds and fountains, which increase the gas mass at large radii considerably. For these models, we also plot the star formation histories. Here we see considerable short-timescale variability, which again relates to the chaotic nature of local star formation and feedback, but the qualitative properties are quite similar: in all cases, there is a mini-burst from a nuclear-bar induced ring which builds up gas at $\sim1$\,kpc and turns into stars at $t\sim0.2-0.3\,$Gyr, after which the system relaxes again. Remarkably, despite the extremely non-linear nature of the physics included, the different numerical methods here produce similar results.

\vspace{-0.5cm}
\section{Performance}
\label{sec:performance}

No methods paper would be complete without some discussion of the speed/computational cost of the method. This is always difficult to quantify, however, since even comparing the identical code with different hydro solvers (as we implement here), the non-linear solutions of the test problems will become different so it is not obvious that we are comparing the ``same'' test anymore (for example, if one method resolves more small-scale structure or higher densities, it will necessarily lead to smaller timesteps, even if it is ``faster'' for identical benchmarks). Nevertheless our suite of simulations gives us some insight. 

First, we compare the MFM and MFV methods to SPH, since these are all run within the same code. Note that while ``traditional'' SPH is computationally very simple, ``modern'' SPH requires higher-order switches which introduce comparable complexity to our method (in complicated pure-hydro tests such as the ``blob'' test, this increases the runtime by $\sim 60\%$ from TSPH to PSPH). At fixed resolution and neighbor number, the hydro loop of SPH is faster because a Riemann solver is not needed. However the performance difference is small: even in a pure hydro problem (ignoring gravity and other code costs), the addition to the hydro adds a fixed multiplier of a factor of a couple. And in fact, because of the timestep requirements which artificial viscosity schemes impose on SPH (and the elimination of various operations needed for the artificial diffusion terms), we are actually able to take larger timesteps in our method. So we actually find that running many of our pure hydro problems with the same particle and neighbor number is slightly ($\sim 10\%$) {\em faster} with the new methods! For example, compare the speeds of our 3D KH problem, normalized to the cpu time to run to the same point with the TSPH method: the runtimes for TSPH, PSPH ($N_{\rm NGB}=32$), PSPH ($N_{\rm NGB}=200$), MFM, and MFV are $1.0$, $1.4$, $2.5$, $0.91$, $1.5$. And in many problems, where gravity is the dominant cost, the differences are small -- e.g.\ in the isolated disk problem, with the Springel \&\ Hernquist equation of state, the respective runtimes for TSPH, PSPH ($N_{\rm NGB}=128$), MFM, and MFV are $1.0$, $1.5$, $1.0$, $1.2$. Moreover, we should really compare performance at fixed {\em accuracy}. This requires at least an order-of-magnitude more neighbors in SPH than in the new method; that in turn means to compare at fixed mass resolution and accuracy means the hydro loop is more expensive by $\sim N_{\rm NGB}^{3/2}$. So it quickly becomes untenable to run even test problems at this accuracy in SPH. 

Comparing our code to {\small AREPO}, in its most-optimized format as of the writing of this paper, shows that both the MFM and MFV methods are somewhat faster on the test problems we have directly compared. The gravity solvers are nearly identical and a Riemann solver is required in both; the typical number of neighbor cells (for a second-order solver) in moving-meshes is usually $\sim13-18$, smaller than even $32$ neighbors, but this trades against the cost of constructing and completeness-testing the mesh, which is substantial (though it is not done every timestep). The bigger difference is in memory cost -- the memory requirements of the MFM and MFV methods are basically identical to SPH (relatively low); however, to avoid reconstructing the Voronoi mesh ``from scratch'' every timestep (which would make the method much slower), moving mesh codes like {\small AREPO} must save the mesh connectivity (or faces) for each particle/mesh generating point. This places some significant limitations on how well the code can be parallelized before communication costs are large. 

Comparing to grid/AMR codes is much more ambiguous, since almost everything ``under the hood'' in these codes is different from the method here and it is not clear how to make a fair speed comparison (after all, different grid codes on the same test problem, with the same method, differ significantly in speed). Purely regular, fixed-grid codes (e.g.\ {\small ATHENA}) are almost certainly faster on problems where the fluid is stationary, if all else (e.g.\ gravity, timestep criterion, choice of Riemann solver) is equal and a second-order method is used, since this minimizes the number of neighbors and means a neighbor ``search'' is unnecessary (the neighbors are always known based on cell position). However, as soon as we run with a higher order stencil, a substantial part of this speed advantage is lost. Moreover, to maintain accuracy, grid codes should limit the timestep based on the speed of the flow over the cell; for super-sonic flows this is far more demanding than the traditional Courant condition. This can reduce the timesteps by factors of $\sim 100-1000$ in some of the problems we consider here, compared to the MFM and MFV methods! Such effects are far larger than the naive algorithmic speed difference. The same is true in AMR codes. Moreover, in AMR the number of neighbors is not so different from our methods, and can sometimes be even larger, so even for a stationary flow the MFM and MFV methods can have a speed advantage. Moreover, it is well-known that AMR methods impose a very large memory cost as they refine; whereas the memory cost of the Lagrangian methods is basically fixed in the initial conditions. 

In short, for a complicated (and probably unfair) comparison problem like a zoom-in simulation (e.g.\ the Santa Barbara cluster), we find the MFM and MFV methods run in comparable (perhaps slightly faster) time than TSPH (comparable to the time for {\small GADGET-3} runs), which is itself substantially faster than ``modern'' SPH and moving mesh methods, which are themselves still faster than the popular AMR methods in e.g.\ {\small RAMSES}, {\small ART}, and {\small ENZO}. The memory costs are similar for SPH, MFM, and MFV methods, and much higher for {\small AREPO} and AMR methods.

\vspace{-0.5cm}
\section{Discussion}
\label{sec:discussion}

We have developed two new, closely related numerical methods for solving the equations of hydrodynamics. The methods are both Lagrangian (move with the fluid flow) and meshless, allowing continuous and automatic adaptive resolution and deformation with the flow, while being simultaneously second-order accurate and manifestly (machine-accurate) conservative of mass, momentum, and energy. We stress that these methods are not a form of SPH. Rather, they are sub-classes of Lagrangian, meshless, finite-volume Godunov-type methods; in a crude sense, like a moving mesh code ``without the mesh.''

We implement these methods in a new code {\small GIZMO}, which couples them to the accurate tree+particle mesh gravity solver, and domain decomposition routines from {\small GADGET-3}, enables adaptive timestepping (while maintaining conservation), and includes cosmological integration, star formation, radiative cooling, and many additional physics (as in {\small GADGET}). 

We have considered an extensive, systematic tests of these methods compared to SPH, moving mesh, and stationary-grid (AMR) methods, and argue they are at least competitive with these methods on all test problems, and appear to capture many of the advantages of both SPH and AMR methods while avoiding many of their disadvantages. More work will be needed, of course, to determine the ultimate utility of these methods, but the results here are promising.

The two new methods here exhibit smaller, but significant, differences between each other. The MFM method exhibits slightly reduced noise, and superior angular momentum conservation, compared to the MFV method; MFM also has the advantage of conserved particle masses, which is very useful for tracing the history of fluid elements and for simulations with complicated self-gravitating interactions (e.g.\ galaxy and star formation), and reduces the ``gravitational heating'' errors in problems like the Santa Barbara cluster. However, this comes at the cost of being slightly more diffusive, and necessarily spreading contact discontinuities over a larger fraction of the kernel width, so that shocks and phase boundaries in e.g.\ the KH or RT instabilities are captured less-sharply.

\vspace{-0.5cm}
\subsection{Comparison to SPH}
\label{sec:discussion:sph}

Both methods we propose avoid many known problems with SPH methods, and as a result give more accurate results in the tests we consider. Even in the ``modern'' SPH,\footnote{``Modern'' SPH defined as those methods using higher-order kernels, pressure-based formulations of the equations of motion, a fully Lagrangian equation of motion, more accurate integral-based gradient approximations, and higher-order dissipation switches for artificial viscosity \&\ conduction.} potentially important issues arise with noise, artificial diffusion, fluid mixing, and sub-sonic flows. While the modern SPH methods have tremendously improved performance in most respects compared to ``traditional'' SPH, there are still fundamental problems related to the zeroth-order errors in the method. Without sacrificing conservation and numerical stability (which leads to disastrously large errors that quickly wipe out any real solutions), these errors can only be ``beaten down'' in SPH by increasing the order of the kernel and number of neighbors. So convergence is very slow. And this entails a loss of resolution (typical mass resolution going as $\sim N_{\rm NGB}^{1/2}$, depending on the choice of kernels). 

Our methods eliminate the need for artificial dissipation terms and so -- despite the use of a Riemann solver -- are substantially {\em less} diffusive than even the high-order modern SPH switches/schemes. They conserve angular momentum more accurately owing to reduced numerical viscosity, allowing gas to be followed in hydrodynamic vortices or gravitational orbits for order-of-magnitude longer timescales. They allow sharper capturing of shocks and discontinuities (to within $<1$ kernel length, instead of $\sim2-3$). They substantially reduce the ``noise'' in the method and so can reliably extend to much smaller Mach numbers. The treatment of fluid instabilities and mixing in the new methods is accurate and robust without requiring any special modifications or artificial diffusion terms. And the new methods eliminate zeroth and first-order errors of SPH, {\em while remaining fully conservative}. This means, most importantly, the methods converge {at fixed neighbor number}. We are therefore able to obtain much higher accuracy with $\sim 32$ neighbors than SPH with $\sim 400$ neighbors, on most problems we consider. And as noted in \S~\ref{sec:performance}, at fixed neighbor and particle number there is little significant performance difference between SPH and our new methods.

SPH may still have some advantages in specific contexts. It naturally handles extremely high Mach number ``cold'' flows such as those in the Zeldovich problem without the need for an explicit switch to reduce noise from a Riemann solver. It is computationally an incredibly simple method. It trivially handles free surfaces with no diffusion into the vacuum, and switching between fluid and particle dynamics is especially simple. And of course, there are many problems where the accuracy of the solution is not limited by convergence or formal numerical integration accuracy, but by {\em physics} missing owing either to their complexity or the resolution required to include them.

\vspace{-0.5cm}
\subsection{Comparison to AMR}
\label{sec:discussion:amr}

Our new methods also avoid many disadvantages of stationary (non-moving) grid methods, for certain classes of problems. In grid methods advection errors are large when the fluid moves with respect to the grid, the errors depend on the bulk velocity (solutions often degrade when the fluid moves), angular momentum is not conserved (unless the grid is designed around a particular geometry), spurious ``grid alignment'' and ``carbuncle'' instabilities can appear, and coupling to N-body gravity solvers is generally ad hoc (introducing new errors and spurious ``grid heating''). 

By moving with the flow, our method minimizes the advection errors that plague grid methods. This leads to sharper and more accurate capturing of contact discontinuities and shocks in moving flows. It also leads to dramatically reduced diffusion in any problems involving non-grid aligned motion. The new methods are Lagrangian and errors are independent of velocity, so they can robustly follow motion of fluid with an arbitrary ``boost''; this is especially important for multi-phase fluids, where, for example, advection errors in grid methods can rapidly diffuse away self-gravitating clouds or structures moving relative to the grid. As we and \citet{springel:arepo} show, this is also important for fluid mixing instabilities: the velocity dependence of errors in grid methods artificially slows down and eventually wipes out the growth of Kelvin-Helmholtz and Rayleigh-Taylor instabilities if the fluid is moving at sufficient bulk velocities (at finite resolution; effectively, the simulation resolution is downgraded). There is also no ``grid alignment'' effect so the carbuncle instability does not appear, disks are not forcibly torqued into alignment with a coordinate axis, and shocks do not preferentially propagate along the grid.

Related to this, our method exhibits excellent angular momentum conservation, and can follow gas in gravitational orbits for hundreds of orbits. In Cartesian grid codes, gas in a rotating disk loses angular momentum and the orbits break down completely in a short time, even with $>10^{7}$ resolution elements in the disk.\footnote{Of course, all of these errors in grid codes (and SPH codes) are resolution-dependent; the methods do formally converge, so they can be reduced by increasing resolution. However, for any practical problem the resolution cannot be infinite so we do care about accuracy at fixed resolution. Moreover, for many problems, the convergence is slow, so formal convergence with some methods may be unattainable. For example, it is well-known that in Cartesian grid codes, the angular momentum converges slowly: even at $\sim 512^{3}$ resolution, a circular gas disk will be strongly torqued to align with one of the coordinate axes, and it will experience strong angular momentum loss, within $\lesssim 3$ orbits \citep[see][for an example in {\small RAMSES}]{hahn:2010.disk.gal.orientations.ramses}. This is already comparable to the best-ever resolution of galaxy formation simulations of a single galaxy! To evolve a disk to $\sim 30-300$ orbits, based on the expected code scalings, would require something like $\sim 10,000^{3}-100,000^{3}$ ($10^{12}-10^{15}$) resolution elements, far out of reach even for exascale computing. Of course, errors can also be reduced by choosing grids with specially designed geometries for a specific problem, but this cannot be generalized to all cases.}

The resolution in our new methods is automatically and continuously adaptive, so provides enhanced resolution where desired, without needing to introduce an ``ad hoc'' refinement scheme (which may or may not correctly capture the desired behavior). Moreover, it is well-known that low-order errors appear at the (necessarily discontinuous) refinement boundaries in AMR, which break the formal higher-order accuracy of the method; since the adaptivity here is continuous and built into our derivation, these do not appear. 

That said, there of course will be contexts where grid codes are particularly useful. It remains to be seen whether the magneto-hydrodynamic treatment in our new method will be competitive with grid codes (this will be the subject of a paper in preparation); it is not obvious, in particular, if constrained-transport methods can be applied. Grid codes, especially fixed (non-adaptive, non-moving) regular (locally orthogonal) meshes minimize certain forms of numerical noise (``grid noise'') and symmetry-breaking compared to any other methods we consider. In highly sub-sonic turbulence (Mach numbers $\sim 0.001 - 0.01$), for example, or other problems where launching of even weak waves sourced by numerical errors could corrupt the desired behavior, this can be quite important. And such simple grids allow for trivially well-optimized parallelization schemes (in the absence of any long-range forces). AMR methods share some, albeit not all, of these advantages. However, in an AMR scheme, one major additional advantage is that refinement can be based on any quantity, in principle, rather than just following mass/density (the usual choice); this means that, unlike our method (unless a special particle-splitting scheme is adopted), AMR methods can be particularly useful when high resolution is desired in low-density regions of a problem (e.g.\ around the reverse shock inside an explosion). 

\vspace{-0.5cm}
\subsection{Comparison to Moving-Mesh Methods}
\label{sec:discussion:movingmesh}

Comparing our new methods to moving mesh approaches, the differences are much more subtle, and more work will be needed to determine the real advantages and disadvantages of each approach (as with any new numerical method). In every test, the methods appear at least competitive with one another. However there are some differences already evident in our comparisons with {\small AREPO} and {\small FVMHD3D}.\footnote{We caution that at least some of the subtle differences we see are not fundamental to the methods, but the result of secondary choices peculiar to each code. For example, we see that shock positions seem to be slightly offset in {\small AREPO} in some tests (Noh, Sedov, interacting blastwaves) relative to the analytic result. We suspect this owes to either a slightly too-aggressive adaptive timestepping or application of the entropy-energy switch, since we find both of these effects can reproduce this error in our own MFM and MFV calculations. The latter effect has been resolved in more recent applications of {\small AREPO} (V.\ Springel, private communication). In some problems, we see reduced post-shock ringing/noise and wall-heating with our new methods, in other tests {\small AREPO} exhibits smaller ``bumps'' at rarefaction fronts and shocks; however these differences are much more sensitive to the slope-limiting procedure than to the method itself.}

From the Gresho test, we see that the exact volume partition and simple faces in moving meshes reduces the ``partition'' and ``mesh deformation'' noise from irregular particle motion in strong shear flows, and hence allows more accurate, smoother tracing of sub-sonic, pressure-dominated rotation (manifest in e.g.\ subsonic turbulence, with Mach numbers $\sim0.01$).

On the other hand, the symmetry and angular momentum conservation in our new methods -- particularly for gas in gravitational orbits (e.g.\ disks) -- may be somewhat superior to that in moving-mesh approaches. Some of this owes to a tradeoff with exactly the errors above: the implicit ``mesh deformation'' in the MFM and MFV methods arises because we map to spherical kernel functions partitioning the volume. This means angular momentum can be well-defined and conserved. In a moving mesh, any irregular (non-spherical) mesh shape means that the total cell angular momentum cannot be defined at higher than second-order quadrature \&\ integration accuracy \citep[see e.g.][]{duffell:2012.disco.method.protoplanetary.disk}; although we stress that moving-mesh methods still have some advantage over Cartesian-grid codes and (highly-viscous) traditional SPH methods. Similarly, the equations of motion in the meshless methods here are manifestly symmetry-maintaining, whereas in moving-mesh approaches the regularization procedures needed to deal with irregular cell shapes may lead to symmetry-breaking ``mesh-bending'' instabilities \citep[see][]{springel:arepo}.

\vspace{-0.5cm}
\subsection{Areas for Improvement \&\ Future Work}
\label{sec:discussion:future}

This is a first study of new methods, and as such there is certainly considerable room for improvement.

For the sake of consistency (and simplicity), in this paper we did not systematically vary things like our slope-limiting procedure, approximate Riemann solver, kernel definition, and timestepping scheme. We have undertaken a limited exploration of these and found (not surprisingly) that for some problems, some choices give better or worse results (although they do not change our qualitative conclusions). However a more thorough study could determine a more ``optimal'' set of choices, especially for cases where the problem structure is known ahead of time.\footnote{For example, we have chosen a simple, commonly-used kernel from the SPH literature. However, the kernel function here has a different meaning from that in SPH and is freed from some of the restrictions of SPH kernels. So studies based on SPH kernels should be revisited, with a more appropriate literature being that on kernel estimation of least-squares field gradients.}

It is also possible to generalize our method to higher order (as in PPM or WENO schemes), using the appropriate matrix-based least-squares gradient estimator. This is useful both if second derivatives are directly needed (for e.g.\ conduction), and to make the method itself more accurate (albeit at additional CPU cost). \citet{gaburov:2011.meshless.dg.particle.method}, for example, show how to generalize this to third-order PPM-like method. Based on their and our own experiments, this produces a smaller improvement than in grid codes (mainly because our advection errors are {already} much smaller than those in arbitrarily high-order grid codes, which is usually the error that motivates higher-order schemes), but it could be useful for some applications.

It would be particularly useful to explore more accurate, higher-order quadrature rules for the volume partition (evaluating $V_{i}\equiv \int \psi_{i}({\bf x})\,d^{\Ddim}{\bf x} \approx \omega({\bf x}_{i})^{-1}$). As we argued above, in many tests, the non-exact nature of our discretized quadrature rule leads to noise which is avoided in moving-mesh and static grid codes; if this can be eliminated, it would represent a considerable improvement in the method.

There is no reason why this method cannot be extended for magnetohydrodynamics (MHD), radiation-hydrodynamics (RHD) and relativistic hydrodynamics, as in many SPH-based and grid-based codes. \citet{gaburov:2011.meshless.dg.particle.method} show one implementation of MHD in an MFV scheme, which we have implemented as well in our code. A systematic comparison of these new methods, SPH-MHD, and grid-MHD methods will be the subject of subsequent work (in preparation). We have only just begun to experiment with radiation-hydro schemes, but this is exciting for many problems of interest. And Lagrangian codes are naturally especially well-suited for relativistic hydrodynamics (many such SPH schemes already exist, and \citealt{duffell:2011.TESS} have developed a moving-mesh implementation). And of course many additional examples of fluid physics (e.g.\ multi-fluid flows, aerodynamic grain-gas coupling, non-ideal MHD, conduction, complicated equations of state, cooling, chemical or nuclear reaction networks) which do not inherently depend on the hydro scheme can be implemented.\footnote{\label{foot:gizmourl2}As noted in \S~\ref{sec:intro}, a public version of our code is available at \\
\gizmourl\\
 Users are encouraged to modify and extend the capabilities of this code; the development version of the code is available upon request from the author.}

\vspace{-0.5cm}
\acknowledgments 
We thank Volker Springel, Romain Teyssier, Lars Hernquist, Jim Stone, Dusan Keres, Paul Duffell, James Wadsley, Oscar Agertz, Thorsten Naab, Richard Bower, Matthieu Schaller, Nick Gnedin, Andrey Kravtsov, Desika Narayanan, Ji-Hoon Kim, Andrew Wetzel, Eliot Quataert, Paul Torrey, Norm Murray, Claude-Andre Faucher-Giguere, and many others for conversations motivating, clarifying, and guiding the development of this work. We thank the anonymous referee for helpful comments, Bert Vandenbroucke for catching a bug in the initial public code release, Evghenii Gaburov for helpful discussions and the initial studies motivating this paper, and Ryan O'Leary for providing additional {\small ATHENA} simulations. I must express my immense debt and gratitude to the many people in the numerical development community who have made their codes and methods public, particularly Volker Springel, Evghenii Gaburov, Jim Stone, and Romain Teyssier, whose excellent codes were used throughout this project. Partial support for PFH was provided by the Gordon and Betty Moore Foundation through Grant \#776 to the Caltech Moore Center for Theoretical Cosmology and Physics, and the Alfred P.\ Sloan Foundation through a Sloan Research Fellowship \#BR2014-022. The simulations here were run on the Stampede supercomputer at TACC through NSF XSEDE allocation \#TG-AST130039.
\\

\vspace{-0.2cm}
\bibliography{/Users/phopkins/Documents/work/papers/ms}

\begin{appendix}

\clearpage

%\vspace{-0.5cm}
\section{The Riemann Problem and Fluxes}
\label{sec:methods:fluxes}

The details of the computation of fluxes between elements in our method (as needed for Eq.~\ref{eqn:final}) is straightforward, and for the most part follows {\small AREPO} (see \citealt{springel:arepo}, \S~3.3). We briefly review it so we can note some subtle differences. 

We compute the solution to the Riemann problem in the rest-frame of the effective face between the two particles/cells $i$ and $j$ at positions ${\bf x}_{i}$ and ${\bf x}_{j}$. Recall, the Riemann problem in our method is always solved at the location ${\bf x}_{ij}$ along the line connecting the particle centers of mass, which moves with velocity ${\bf v}_{{\rm frame},\,ij}$ defined in Eq.~\ref{eqn:vframe}. 

Beginning from a vector of primitive variables ${\bf W} =  (\rho,\,{\bf v},\,P)$ for particles $i$ and $j$, we first (for convenience) boost to the rest-frame of the face $ij$: 
\begin{align}
{\bf W}^{\prime}_{j,\,i} &= {\bf W}_{j,\,i} - 
\left(
\begin{array}{c}
0 \\
{\bf v}_{{\rm frame},\,ij} \\
0 \\
\end{array}
\right)
\end{align}
We then calculate left and right states by linear reconstruction of the values of ${\bf W}^{\prime}$ at ${\bf x}_{ij}$ from particles $i$ and $j$, following \S~\ref{sec:methods:reconstruction}; this gives us left and right ${\bf W}_{{\rm rec},\,L}^{\prime}$ (the ``$j$ side'') and ${\bf W}_{{\rm rec},\,R}^{\prime}$ (the ``$i$ side''). The states are also predicted forward in time by a half-timestep, to obtain time-centered fluxes: 
\begin{align}
{\bf W}_{L,\,R}^{\prime\prime} &= {\bf W}_{{\rm rec},\,L,\,R}^{\prime} + \frac{\partial W^{\prime}_{L,\,R}}{\partial t}\,\frac{\Delta t}{2}\\
&= W^{\prime}_{j,i} + (\nabla W^{\prime})_{{\rm lim},\,j,\,i}\cdot ({\bf x}_{ij} - {\bf x}_{j,\,i})
+ \frac{\partial W^{\prime}_{j,\,i}}{\partial t}\,\frac{\Delta t}{2}
\end{align}
where $(\nabla W^{\prime})_{{\rm lim},\,j,\,i}$ are the slope-limited gradients, and the partial time derivative is estimated based on the spatial derivatives using the Euler equations for an ideal gas, as 
\begin{align}
\frac{\partial {\bf W}}{\partial t} = -
\left(
\begin{array}{ccc}
{\bf v} & \rho & 0 \\
0 & {\bf v} & 1/\rho \\
0 & \gamma\,P & {\bf v} \\
\end{array}
\right)\,
\nabla {\bf W}
\end{align}
(note that it is trivial to modify the pressure equation for a non-ideal gas using the gradients of $\rho$ and $u$). 

We need to solve a 1D Riemann problem in the ${\bf A}_{ij}$ direction; so we transform to a coordinate system aligned with the $\hat{A}_{ij}$ axis; this can be done with the matrix ${\boldsymbol{\Lambda}}$, which rotates the vector (here, velocity) components, but obviously leaves the scalar components intact: 
\begin{align}
{\bf W}_{L,\,R}^{\prime\prime\prime} = {\boldsymbol \Lambda}\,{\bf W}_{L,\,R}^{\prime\prime} = 
\left(
\begin{array}{ccc}
1 & 0 & 0 \\
0 & \boldsymbol{\Lambda}_{\Ddim} & 0 \\
0 & 0 & 1 \\
\end{array}
\right)\,
{\bf W}_{L,\,R}^{\prime\prime}
\end{align}
where $\boldsymbol{\Lambda}_{\Ddim}$ is an ordinary rotation matrix which takes the new coordinate system to coincide with the $x^{\prime}$ axis, i.e.\ ${\bf A}_{ij}^{\prime} = \boldsymbol{\Lambda}_{\Ddim}\,{\bf A}_{ij} = (|{\bf A}_{ij}|,\,0,\,0)$ in 3D. 

We then solve the one-dimensional Riemann problem (see below), to obtain the fluxes 
\begin{align}
\label{eqn:Riemann.problem}
\tilde{\bf F}_{ij}^{\prime\prime} &\equiv  f_{\rm Riemann}\left( {\bf W}_{L}^{\prime\prime\prime}\,,\,{\bf W}_{R}^{\prime\prime\prime} \right)
\end{align}
which we rotate back into the simulation axes (since we solved the one-dimensional Riemann problem in the frame aligned with $\hat{A}_{ij}$, this automatically projects the fluxes appropriately): 
\begin{align}
\tilde{\bf F}_{ij}^{\prime} &\equiv \hat{A}_{ij}\,\cdot \left({\boldsymbol \Lambda}^{-1}\,\tilde{\bf F}_{ij}^{\prime\prime}\right) = 
\left(
\begin{array}{c}
F_{\rho} \\
{\bf F}_{v} \\
F_{e} \\
\end{array}
\right)
\end{align}
and finally ``de-boost'' back into the simulation coordinate frame following \citet{pakmor:2011.arepo.mhd}
\begin{align}
\tilde{\bf F}_{ij}\cdot \hat{A}_{ij} &\equiv \tilde{\bf F}_{ij}^{\prime} +  
\left(
\begin{array}{c}
0 \\
{\bf v}_{\rm frame}\,F_{\rho} \\
\frac{1}{2}\,|{\bf v}_{\rm frame}|^{2}\,F_{\rho} + {\bf v}_{\rm frame}\cdot {\bf F}_{v} \\ 
\end{array}
\right)
\end{align}
Just as in {\small AREPO}, it is easy to verify that this scheme maintains Galilean invariance and eliminates the velocity (boost)-dependent truncation errors in non-moving meshes, if the particles/cells move with the fluid bulk velocity (discussed below). 

To actually solve the Riemann problem, there are many methods commonly used in the literature \citep[see e.g.][]{toro:1997.reimann.solver.book}. We have performed some limited experiments ourselves, but have not rigorously explored the possible parameter space. 

We have implemented an exact Riemann solver, following \citet{toro:1997.reimann.solver.book}. This uses an iterative procedure to exactly solve the Riemann problem for gas described by a local polytropic index. Because the solution is numerical, we must define some numerical tolerance for the deviation between iterations when convergence is assumed; we set this to $10^{-6}$ in the pressure at the contact state $P_{\ast}$. In general, using values as low as $10^{-12}$ we see no improvements beyond this in tests here. Unfortunately, while this method should in principle always return the exact solution, in practice numerical errors mean that, very rarely, the procedure can numerically diverge or fail to converge in many iterations ($>1000$). It is also very expensive to use this for every inter-particle Riemann problem. 

We therefore have also implemented a standard approximate HLLC Riemann solver \citep[see][]{toro:1999.reimann.solvers.book,miyoshi:2005.hlld.reimann.solver}. This method is not exact, but it is accurate at the order we require (and exactly conservative); moreover such methods are extremely well-tested in the literature. HLLC solvers break the problem into a simple set of waves/fronts, and require some initial ``guess'' for certain wavespeeds; we have experimented with a few choices for this following \citet{roe:1981.reimann.solvers,gaburov:2011.meshless.dg.particle.method,toro:1997.reimann.solver.book}. In general, even when we perform convergence tests, we see no measurable loss of accuracy using the HLLC solution as opposed to the exact solver. And the HLLC solver is much faster, and does not require a polytropic index, so can be trivially generalized to non-ideal equations of state. However, in rare examples, this can fail, because of bad estimates for the wavespeeds. This failure is usually assessed by checking whether the pressure returned is everywhere positive in the approximate solution. 

The Riemann solution method therefore proceeds as follows. We begin with our usual piecewise-linear (second-order) reconstruction of left and right states. We then attempt the HLLC Riemann solver. Within the HLLC solver, we first check Roe-average (usually most accurate) wave-speed estimate; if this is bad (returns $P_{\ast}\le 0$), we check the simpler wave-speed estimate from \citet{toro:1999.reimann.solvers.book} as used in \citet{gaburov:2011.meshless.dg.particle.method} (their Eq.~34-36); if this is bad, we check the Rusanov or TVD Lax-Friedrich primitive-variable wave speed estimate. If no good solutions are found, we use the exact Riemann solver. If (in very rare cases) this fails to converge after $1000$ iterations, we go back to the reconstruction step and use a piecewise-constant (first-order) reconstruction, then repeat the process of searching for solutions. If this fails, we print a warning and exit the code. However, in all our tests, we find we always obtain a valid solution so long as physically allowed values for input states are used.

\vspace{-0.5cm}
\section{On the Slope-Limiting Procedure for Unstructured, Meshless Riemann Problems}
\label{sec:slopelimiters}

Reconstruction of fluid quantities for the Riemann problem is straightforward in smooth flows. However, at discontinuities or higher-order divergences, numerical stability requires some slope or flux-limiting procedure.

A common approach is to introduce a slope limiter of the gradients, which ensures that the linearly reconstructed quantities at faces do not exceed the extrema among the interacting neighbor cells \citep[see e.g.][]{barth.jesperson:1989.upwind.schemes.for.unstructured.meshes}. In performing the face reconstruction of some arbitrary quantity $\phi^{i}$ for particle $i$, we replace the ``true'' (matrix-evaluated) gradient $\nabla \phi^{i}_{\rm true}$ with an effective (slope-limited) gradient $\nabla \phi^{i}_{\rm lim}$: 
\begin{align}
\nabla \phi^{i}_{\rm lim} = \alpha_{i}\,\nabla \phi^{i}_{\rm true} 
\end{align}
where
\begin{align}
\alpha_{i} &\equiv {\rm MIN}{\Bigl[}1,\,\beta_{i}\,{\rm MIN}{\Bigl(}\frac{\phi^{\rm max}_{ij\,{\rm ngb}}-\phi_{i}}{\phi^{\rm max}_{ij,\,{\rm mid}}-\phi_{i}},\ \frac{\phi_{i}-\phi^{{\rm min}}_{ij,\,{\rm ngb}}}{\phi_{i}-\phi^{\rm min}_{ij,\,{\rm mid}}}{\Bigr)}{\Bigr]}
\end{align}
where $\phi^{{\rm max}}_{ij,\,{\rm ngb}}$ and $\phi^{{\rm min}}_{ij,\,{\rm ngb}}$ are the maximum and minimum values of $\phi_{j}$ among all neighbors $j$ of the particle $i$, and $\phi^{\rm max}_{ij,\,{\rm mid}}$, $\phi^{\rm min}_{ij,\,{\rm mid}}$ are the maximum and minimum values (over all pairs $ij$ of the $j$ neighbors of $i$) of $\phi$ re-constructed on the ``$i$ side'' of the interface between particles $i$ and $j$ (i.e.\ $\phi^{\rm max}_{ij,\,{\rm mid}} = {\rm MAX}[\phi_{i} + \nabla \phi^{i}_{\rm true} \cdot ({\bf x}_{{\rm face},\,ij}-{\bf x}_{i})]$). 

As noted by \citet{balsara:2004.second.order.accurate.mhd.schemes}, the constant $\beta$ must have a value $\beta > 0.5$ in order to maintain the second-order accuracy of the scheme (with lower values being more stable, but also more diffusive). {\small AREPO}, for example, adopts a scheme very similar to this with $\beta=1$. Ideally, we would like to use a more ``aggressive'' (larger and more-accurate) value of $\beta$ when the gradients are trustworthy and there is good particle/cell order, and a more ``stable'' (diffusive) value when the gradients are less trustworthy (or there are large fluctuations in quantities within the kernel). Fortunately, as noted in \S~\ref{sec:condition.number}, we have an indicator of this already, in the condition number of the gradient matrix. After considerable experimentation, we find a very good mix of stability and accuracy on all problems in this paper with the choice 
\begin{align}
\beta_{i} &= {\rm MAX}[\beta_{\rm min},\,\beta_{\rm max}\,{\rm MIN}(1,\,N_{\rm cond}^{\rm crit}/N_{\rm cond}^{i})]
%\beta &=
%\begin{cases}
%	{\displaystyle 2\ \ \ \ \ \hfill { (N_{\rm cond} < N_{\rm cond}^{\rm crit})}} \\
%	{\displaystyle {\rm MAX}[1,\,2\,N_{\rm cond}^{\rm crit}/N_{\rm cond}]\ \ \ \ \ \hfill { (N_{\rm cond} \ge N_{\rm cond}^{\rm crit})}} 
%\end{cases}
\end{align}
with $\beta_{\rm min}=1$, $\beta_{\rm max}=2$. We find that $\beta_{\rm min}<1$ does not much improve stability, but does begin to introduce noticeable diffusion of discontinuities, while $\beta_{\rm max}>2$ does not much improve accuracy and leads to problems with stability in very strong interacting shocks (though for most other problems, $\beta_{\rm max}=4$ works fine as well with slightly better accuracy). 

We actually find that we achieve slightly greater numerical stability, and are able to eliminate one additional loop over the particle neighbors, at the cost of very little added diffusion, if we make this slope limiter slightly more conservative by replacing the quantities
$\phi_{i}-\phi^{\rm min}_{ij,\,{\rm mid}}$ and $\phi^{\rm max}_{ij,\,{\rm mid}}-\phi_{i}$ by the value $|\nabla\,\phi^{i}_{\rm true}|\cdot |{\bf x}_{{\rm face},ij}-{\bf x}_{i}|^{\rm max}$ (where $|{\bf x}_{{\rm face},ij}-{\bf x}_{i}|$ is the distance between the particle and face for the pair $ij$). In other words we replace the explicitly calculated two extrema which happen to be reconstructed based on the particle positions, with the maximum/minimum value that {\em could be} reconstructed, independent of the geometric arrangement of the particles within the kernel. This is actually closer to what is intended by this such limiters in grid codes. And $|{\bf x}_{{\rm face},ij}-{\bf x}_{i}|^{\rm max}$ can be directly calculated, but given our other definitions is well-approximated by half the maximum size of the local kernel, $h_{i}/2$. 

We note that this limiter, while useful and sufficient for most problems, is not total variation diminishing (TVD), and cannot strictly guarantee stability even if we use very conservative parameters (e.g. $\beta_{i}=0.5$ always). And indeed in some problems with extremely strong shocks (e.g.\ the Zeldovich pancake) or non-hydrodynamic forces (e.g.\ galaxy evolution), we see large errors occur (albeit in a small number of particles) if we only include the above limiter. To ensure stability more generally, it is necessary to adopt a pairwise limiter between interacting particles. This sort of issue has been seen before, especially for unstructured point distribution \citep[see e.g.][]{mocz:2014.galerkin.arepo}. 

There are many choices for this, as in grid codes. For the sake of flexibility, we implement a general form as follows. For the particle pair $ij$, we begin by reconstructing $\phi_{ij,\,{\rm mid}}$ (the re-constructed value on the ``$i$ side'') as above, using the slope-limited gradients $\nabla \phi^{i}_{\rm lim}$. We then apply a second pair-wise limiter to this, replacing our initial estimate $\phi_{ij,{\rm mid}}^{0}$ with a limited $\phi_{ij,\,{\rm mid}}^{\prime}$ based on the values of $\phi_{i}$ and $\phi_{j}$:
\begin{align}
\nonumber\phi_{ij,\,{\rm mid}}^{\prime} &= 
\begin{cases}
	{\displaystyle \phi_{i}\ \ \ \ \ \hfill { (\phi_{i}=\phi_{j})}} \\
	{\displaystyle {\rm MAX}(\phi_{-},\,{\rm MIN}[\bar{\phi}_{ij}+\delta_{2},\,\phi_{ij,{\rm mid}}^{0}])\ \ \ \ \ \hfill { (\phi_{i}<\phi_{j})}} \\
	{\displaystyle {\rm MIN}(\phi_{+},\,{\rm MAX}[\bar{\phi}_{ij}-\delta_{2},\,\phi_{ij,{\rm mid}}^{0}])\ \ \ \ \ \hfill { (\phi_{i}>\phi_{j})}} 
\end{cases}\\
\nonumber\phi_{-} &=
\begin{cases}
	{\displaystyle {\phi_{\rm min}-\delta_{1}}\ \ \hfill { ({\rm SIGN}(\phi_{\rm min} - \delta_{1}) = {\rm SIGN}(\phi_{\rm min}))}} \\
	{\displaystyle \frac{\phi_{\rm min}}{1 + \delta_{1}/|\phi_{\rm min}|}\ \ \hfill { ({\rm SIGN}(\phi_{\rm min} - \delta_{1}) \ne {\rm SIGN}(\phi_{\rm min}))}} 
\end{cases}\\
\nonumber\phi_{+} &=
\begin{cases}
	{\displaystyle {\phi_{\rm max}+\delta_{1}}\ \ \hfill { ({\rm SIGN}(\phi_{\rm max} + \delta_{1}) = {\rm SIGN}(\phi_{\rm max}))}} \\
	{\displaystyle \frac{\phi_{\rm max}}{1 + \delta_{1}/|\phi_{\rm max}|}\ \ \hfill { ({\rm SIGN}(\phi_{\rm max} + \delta_{1}) \ne {\rm SIGN}(\phi_{\rm max}))}} 
\end{cases}\\
\nonumber\bar{\phi}_{ij} &\equiv \phi_{i}+\frac{|{\bf x}_{ij}-{\bf x}_{i}|}{|{\bf x}_{j}-{\bf x}_{i}|}\,(\phi_{j}-\phi_{i}) \\
\nonumber\phi_{\rm min} &\equiv {\rm MIN}(\phi_{i},\,\phi_{j})\\
\nonumber\phi_{\rm max} &\equiv {\rm MAX}(\phi_{i},\,\phi_{j})\\
\nonumber\delta_{1} &\equiv \psi_{1}\,|\phi_{i}-\phi_{j}|\\
\delta_{2} &\equiv \psi_{2}\,|\phi_{i}-\phi_{j}|
\end{align}
While these expressions are somewhat non-intuitive, they are easy to efficiently evaluate, and ultimately allow considerable freedom of slope-limiters, based on our choice of the free parameters $\psi_{1}$ and $\psi_{2}$. Many popular slope limiters can be expressed as variations of these parameters: for example, the monotonized central \citep{van-leer:1977.slopelimiter}, minmod and superbee \citep{roe:1986.upwind.euler.schemes}, \citet{koren:1993.monotonic.slopelimiters}, and \citet{sweby:1984.slopelimiter} limiters all fall in this class. We have experimented with all of these; as always, there is no uniformly ``correct'' choice, but for the problems here we find a good mix of stability and accuracy adopting $\psi_{1}=1/2$, $\psi_{2}=1/4$. As in our convention for $\beta$, these are defined such that smaller values are more conservative/stable but also more diffusive (with $0 \le \psi_{1} \le 1$ and $0 \le \psi_{2} \le 1/2$ being the physically reasonable ranges).

If we make the analogy to a regular Cartesian mesh code, we can directly compare this to the standard limiters defined as a function $\phi_{\rm lim}(r) = \phi_{ij,\,{\rm mid}}\,\psi(r)$ of $r = (\phi_{i} - \phi_{i-1})/(\phi_{i+1}-\phi_{i})$, where following \citet{cha:2010.godunov.sph} we take $\phi_{i-1}=\phi_{j}$ and $\phi_{i+1}$ is calculated by projecting the gradient calculated at $i$ in the opposite direction from $j$ by the same distance. Our default choice ($\psi_{2}=1/4$) is then, for $r>0$, equivalent to $\psi = 2\,r$ for $r<1/2$ and $\psi=1$ for $r\ge1/2$, which is the slope limiter that recovers the ``correct'' ($i$-centered least-squares) gradient most accurately while still satisfying the TVD condition. We do confirm that $\psi_{2} > 1/4$ leads to unstable behavior, with $\psi_{2} > 1/2$ being sufficiently unstable that most Riemann solvers will diverge. Unlike some grid-based slope-limiters, however, we find we do not require $\psi = 0$ for $r<0$ ($\psi_{1}=0$) to ensure stability, because in {\em this} regime, the previous limiter based on the max/min values in the kernel provides stability so long as $\psi_{1} \le 1/2$. For $\psi_{1}>0$, however, we include the ${\rm SIGN}$ terms above to prevent a sign change of extrapolated quantities in the projection (i.e.\ if both $\phi_{i}$ and $\phi_{j}$ are positive, the reconstructed quantity can never be negative, and vice versa). The particular form chosen (which is not unique, but is quite flexible) simply assumes that the derivative measured at $i$, if it were to lead to an implied sign change, actually describes a power-law declining (instead of linearly declining) function. 

Comparing this to the ``standard'' choice of a single, less-flexible limiter such as the Van Leer, minmod, or superbee limiters, we find it enables a significant improvement in accuracy and reduction in numerical diffusion while maintaining stability in every problem considered here. This suggests it might be generally useful for other non-regularly gridded methods, including moving mesh codes (both {\small AREPO} and {\small TESS} find a pair-wise limiter must be used in addition to the global min/max criterion to ensure stability on more complicated problems, but use more diffusive default choices), and even AMR codes (since the usual way of handling cases where the grid is not perfectly uniform but refined more in one direction is to effectively ``down-sample'' to a lower-level grid, increasing numerical diffusion).

\vspace{-0.5cm}
\section{Dealing with Pathological Particle Configurations}
\label{sec:condition.number}

In general, our matrix-based methods for solving the least-squares particle-centered gradients (see \S~\ref{sec:methods:reconstruction}) are very robust, and can deal with arbitrary configurations of particles within the kernel (for example, the proof that the method exactly recovers linear gradients is trivial and independent of the particle spatial locations within the kernel). 

However, in all quasi-Lagrangian methods, there is some possibility that the mesh or particle distribution becomes severely irregular in a way that requires careful consideration (or else errors may increase, and/or the method may crash). In this case, the proof above makes an implicit assumption -- that the matrix in Eq.~\ref{eqn:gradient.matrix} ${\bf E}_{i}^{\alpha\beta} \equiv \sum_{j}\,({\bf x}_{j}-{\bf x}_{i})^{\alpha}\,({\bf x}_{j}-{\bf x}_{i})^{\beta}\,\psi_{j}({\bf x}_{i})$ is non-singular. Consider, for example, the following pathological case. Since the kernel is compact, there are a finite number $N_{\rm NGB}$ of particles inside it; it is conceivable that all $N_{\rm NGB}$ particles lie exactly along one axis (in a 3D simulation). In this case ${\bf E}_{i}$ will be singular, and the gradients in the perpendicular directions will be undefined. This is physically correct, after all, since in this configuration there is no information on these directions! This is analogous to the case in moving mesh codes, when a cell becomes highly deformed so has a very large axis ratio in one direction (leading to divergences and inaccurate gradients).

Such situations are very rare, and clearly pathological, but they can occur in highly non-linear, large simulations (like cosmological simulations) and we must implement some method to deal with them. More likely, we will have situations which are ``close to'' singular (e.g.\ the particles are all on one axis to within some deviation $|\epsilon| \ll h$), in which case the method is formally accurate (the matrix is invertible and stable), but the numerical ``noise'' can be very large (since the inferred gradients become dominated by small offsets of the particles positions). 

Fortunately, there is a well-studied means to properly define ``pathological'' here, which is given by the condition number $N_{\rm cond}$ of the (weighted) position moments matrix. That matrix is just ${\bf E}_{i}$ (Eq.~\ref{eqn:gradient.matrix}), and the condition number is: 
\begin{align}
N_{{\rm cond},\,i} &\equiv \Ddim^{-1}\,{\Bigl[}{\bigl |}{\bigl |} {\bf E}_{i}^{-1} {\bigr |}{\bigr |}\cdot{\bigl |}{\bigl |} {\bf E}_{i} {\bigr |}{\bigr |} {\Bigr]}^{1/2}\\
{\bigl |}{\bigl |} {\bf E}_{i} {\bigr |}{\bigr |} &\equiv \sum_{\alpha=1}^{\alpha=\Ddim}\,\sum_{\beta=1}^{\beta=\Ddim} |{\bf E}_{i}^{\alpha\beta}|^{2}
\end{align}
where $\nu$ is the number of dimensions. It is easy to verify that for a truly singular matrix ${\bf E}_{i}$, $N_{\rm cond}\rightarrow\infty$; at the opposite extreme, if ${\bf E}_{i}$ were the most ``perfectly invertible'' matrix (the identity matrix), $N_{\rm cond}=1$. 

For any configuration of particles, we can measure $N_{\rm cond}$; the problem then reduces to how to deal with large $N_{\rm cond}\gg1$. There are many possible choices. In moving-mesh codes, the usual approach is to ``regularize'' or ``re-mesh'' (drift the mesh-generating points while advecting the fluid over them, until they have regular aspect ratios; see \citealt{springel:arepo}); the analogue in particle-based codes is to split particles, inserting new particles in the directions which are under-sampled in some regular fashion \citep[see e.g.][]{maron:2012.phurbas.algorithm}. We can do this (see \S~\ref{sec:particle.splitting}). Unfortunately, these are highly diffusive operations which can introduce their own lower-order errors; moreover, most of the time in the realistic cases we study here, the pathology is transient (it is a random coincident alignment of particles, with well-sampled particles waiting ``just outside'' the kernel, rather than something systematic and persistent). So in most cases the problem can be addressed without adding errors and diffusion by simply extending the particle search until particles are found in the under-sampled directions and $N_{\rm cond}$ is reduced. We therefore adopt the following approach: if $N_{\rm cond}$ exceeds some critical $N_{\rm cond}^{\rm crit}\gg1$, then we iteratively expand the kernel (increase $N_{\rm NGB}^{\rm eff}$) in small increments until we reduce $N_{\rm cond}$ below 
\begin{align}
N_{\rm cond} \le N_{\rm cond}^{\rm crit}\,{\rm MAX}\left( 1\ ,\ \alpha_{\rm cn}\left[1 - \left(\frac{N_{\rm NGB}^{\rm eff}}{N_{\rm NGB}^{0}}\right)^{2} \right] \right)
\end{align}
where $\alpha_{\rm cn}\approx 10$ and $N_{\rm cond}^{\rm crit}\approx 100-1000$ are set by our own experiments (we find this does the best job of simultaneously minimizing errors and diffusion while stabilizing the code), $N_{\rm NGB}^{0}$ is the ``default'' number of neighbors, and the second term exists only to prevent $N_{\rm NGB}^{\rm eff}$ from running away if, indeed, it cannot find a reduction in $N_{\rm cond}$ with a reasonable augmentation to the neighbor number. 

In the extremely rare cases where this cannot reduce $N_{\rm cond}$ below some threshold, say $\sim 10\,N_{\rm cond}^{\rm crit}$, we simply have the code issue a warning and proceed by replacing the gradient estimators (in both the standard gradient estimation and definition of the ``effective face'' areas for the Riemann problem) for that particle and timestep with the standard SPH gradient estimators, so 
\begin{align}
(\nabla q)_{i} &\approx (\nabla q)_{i}^{\rm SPH} \equiv \sum_{j}\,\frac{1}{\omega_{j}}\,q_{j}\,\nabla _{i}\,W_{ij}(h_{i})
\end{align}
These gradient estimators have low-order errors; however, they are stable in irregular/pathological particle configurations. For example, for the case above (all particles aligned in one axis), this will simply return a gradient of zero in the perpendicular directions. We find that using this method, instead of particle splitting, in these extreme cases, is sufficient to restore stability and produces still less diffusion than particle splitting. However, we stress that this is {\em extremely} rare, occurring only once (for a small number of timesteps around the central caustic in the Zeldovich problem when the analytic density diverges) in all the tests we run.

\vspace{-0.5cm}
\section{Explicit Thermal Energy Evolution and Energy-Entropy Switches at Extremely High Mach Numbers}
\label{sec:methods:switches}

When a Riemann solver is used in an exactly-conservative method, flows which are strongly kinetic-energy dominated (very cold and super-sonic in the frame in which the Riemann problem is solved) exhibit spurious heating in the adiabatic parts of the flow \citep{ryu:1993.entropy.switch.cosmo.grid.hydro,bryan:1995.cosmo.ppm,steinmetz:1997.two.body.nbody.heating}. This ultimately stems from the Riemann problem's use of and conservation of total energy; if the Mach number is high ($\sim 10^{5}$, as in the Zeldovich problem we simulate below), then very small truncation errors (part in $\sim 10^{10}$) appear in the thermal energy. This problem is discussed at length in \citet{springel:arepo} (\S~3.5); it is ubiquitous in cosmology in the early stages of structure formation (where the velocities from gravity produce extremely high Mach numbers), corrupting simulations unless some fix is applied. 

We follow an approach similar to the \citet{bryan:1995.cosmo.ppm} ``dual energy formalism,'' whereby we explicitly evolve the internal energy, in addition to total energy, and when the motion is sufficiently supersonic the temperature and pressure are set based on the results of this equation. Following \citet{gaburov:2011.meshless.dg.particle.method}, \S~3.4, this amounts to explicitly evolving the internal energy $U$ (or internal energy per unit mass $u = U/m$ as 
\begin{align}
\frac{dU}{dt} = \frac{dE}{dt} - {\bf v}\cdot \frac{d {\bf P}}{d t} + \frac{{\bf v}\cdot {\bf v}}{2}\,\frac{d m}{d t}
\end{align}
where $E = U + {\bf P}\cdot {\bf P}/(2\,m)$ is the ``hydrodynamic total energy,'' $P = m\,{\bf v}$ the momentum, and $m$ the particle mass. Note that when this is done, total energy is no longer conserved to machine accuracy, but to the truncation error of the time-integration scheme. However, internal energy is evolved more accurately (otherwise, any errors in the solution are simply shifted into the internal energy). In fact, for every test problem here, we find this produces at least comparable accuracy to the explicitly energy-conserving formalism; and for flows with gravity, where the internal energy would otherwise be determined by the difference between two large numbers, it gives substantially improved accuracy and numerical stability, and actually {\em better} overall energy conservation (to $\ll 1\%$ accuracy) in many cases. The reason for this counter-intuitive result is that when long-range forces like gravity are present, it is no longer possible to conserve total energy to machine precision in any case, because the long-range interactions cannot be perfectly pair-wise symmetric unless an explicit $N^{2}$ (i.e.\ pair wise) method for gravity (with a single time-step) is adopted; this is impossibly expensive for anything but simulations with a tiny number of particles. 

With this choice, most of the problems described above are solved. However, it is possible in the most extreme situations (like the Zeldovich problem) that the numerical convergence accuracy (part in $\sim 10^{8}$) in the Riemann solver still leads to large errors in the thermal energy equation. We can in this case follow \citet{ryu:1993.entropy.switch.cosmo.grid.hydro} and \citet{springel:arepo}, and explicitly calculate the evolution of the system as if it were purely adiabatic in each timestep (see \citealt{springel:arepo}, \S~3.5), with a switch to decide when this solution is used. Experimenting with this, we find that a Mach number switch is unnecessary and can create more problems than it solves (the same is true in {\small AREPO}; V.\ Springel, private communication). However, an energy-based switch is, in rare situations, useful. In each timestep, we determine the expected thermal energy $E_{\rm therm}$ based on the usual update; we compare this to the gravitational energy associated with motion across the particle size ($\delta E_{\rm grav} = m_{i}\,|{\bf a}_{{\rm grav},\,i}|\,h_{i}$) and maximum kinetic energy $\delta E_{\rm kin}^{\rm max}$ of all neighbor cells in the rest-frame of the current cell $i$. If $E_{\rm therm} < \alpha_{\rm kin}\,(\delta E_{\rm kin}^{\rm max}+E_{\rm therm})$, or $E_{\rm therm} < \alpha_{\rm grav}\,\delta E_{\rm grav}$, we use the entropy-based evolution. Because our method is Lagrangian (which minimizes these sorts of errors to begin with), and because of the energy evolution choice above, we can set $\alpha_{\rm kin}$ and $\alpha_{\rm grav}$ to very conservative (low) values, in the simulations here $\approx 0.001$, which means they are almost never triggered but manage to trap the extremely rare pathological cases encountered in some problems.

All of these are choices, and of course it is straightforward to use the method we propose without such switches (and the total energy evolution). However, we find negligible penalty and considerable advantages in this particular form of the method.

\vspace{-0.5cm}
\section{Particle Splitting and Merging}
\label{sec:particle.splitting}

In some problems, it may be necessary to split or merge particles, especially when mass fluxes between them are allowed (as in the MFV method here). For example, gas particles at the center vs.\ in the outskirts of a galactic disk may eventually (over many orbits) develop large (more than order-of-magnitude) differences in their masses. This is fine (and correct, given the nature of the method) if the gas flows are sufficiently smooth, but if galactic winds from say, SNe explosions suddenly expel mass from the center at high velocities, this will lead to particles with very different masses suddenly interacting. The hydro method is formally robust to this (although if the differences are large enough, truncation error in fluxes from one particle could lead to unphysical quantities in the other). However, if self-gravity is also included, this can produce unacceptably large $N$-body scattering effects. 

To deal with this, we have implemented a simple particle splitting/merging algorithm, although we caution that it is not expected to be the most optimal possible algorithm. If a particle falls below a mass $=\epsilon_{\rm min}\,m_{\rm min}(t_{0})$ (where $m_{\rm min}(t_{0})$ is the minimum mass over all particles in the initial conditions), and is the least massive particle currently among its entire neighbor list, it is merged with the second-least-massive particle among the neighbors. The merger is straightforward: the less-massive particle is deleted and conservation requires that the more massive particle inherits the summed mass, momentum, and energy (and their time rates-of-change). For quantities like the signal velocity and kernel length, the larger of the two is chosen, but these will be re-initialized in the next active timestep. The updated particle is moved to the center-of-mass position of the pair. The merge operation is done only on timesteps where the neighbor/gravity tree is being reconstructed, so that no errors in gravity or neighbor searches are introduced. We adopt the somewhat ad-hoc choice $\epsilon_{\rm min}=0.5$. 

Similarly, if a particle is above a mass $=\epsilon_{\rm max}\,m_{\rm max}(t_{0})$ (where $m_{\rm max}(t_{0})$ is the maximum mass over all particles in the initial conditions), and is the most massive particle among its neighbors, it is split into two particles. Each particle has half the mass and inherits the specific (per-unit-mass) properties of the parent. This is straightforward; the ambiguity in particle splitting comes from the positions of the particles. They cannot be placed at identically the parent location, but must be separated by some small amount. However doing so in a way that does not seed fluctuations in the volumetric quantities is highly non-trivial. Here, we adopt a very simple prescription: the two particles are separated by the minimum of $h_{i}/8$ or $|r_{\rm near}|/3$, where $|r_{\rm near}|$ is the distance within the kernel to the closest neighbor particle. They are each moved this distance, in opposite directions along an axis perpendicular to the particle number density gradient (to minimize the perturbation to volumetric quantities). 

We note that these operations are both noisy and diffusive, and we recommend against particle merging/splitting unless absolutely necessary. That said, we have run all the test problems in this paper with and without such splitting and find very little difference in almost every case (because very few particles would be eligible). However for at least one problem -- the isolated disk with the full physics of stellar feedback from the FIRE models -- we simply cannot run the problem using the MFV method without it crashing, if we do not invoke particle splitting and merging (the strong galactic winds led to exactly the $N$-body problems described above). The methods for splitting/merging merit serious, detailed examination in future work, as there are almost certainly ways to improve the simple algorithm we invoke here. 

\vspace{-0.5cm}
\section{The SPH Implementation}
\label{sec:sph.methods}

As discussed in the text, we can run our code as an SPH code, if desired. We implement two ``default'' versions of SPH, and use them throughout the text, so we describe their properties here.

\vspace{-0.5cm}
\subsection{``Traditional'' SPH (TSPH)}
\label{sec:sph.methods:tsph}

The ``traditional'' SPH (TSPH) implementation in our code is particularly simple. As noted in the text, nearly everything in the code remains identical whether we run in SPH mode or one of our new modes. Here we outline the method insofar is it requires something distinct from our other methods.

The TSPH implementation falls within the general class of manifestly conservative, Lagrangian-derived SPH schemes outlined in \citet{hopkins:lagrangian.pressure.sph}. Specifically it is a ``density-energy'' scheme (where the internal energy is explicitly evolved). As shown therein, the choice of ``density-energy'' or ``density-entropy'' scheme (as in {\small GADGET-2}) gives essentially identical results {\em when a Lagrangian-derived scheme is used}, since both simultaneously conserve energy and entropy in global timesteps; we have explicitly confirmed this by comparison to a density-entropy formulation in the tests here. Since the choice is a matter of convenience, we find it more naturally aligns with our other methods, and allows a more flexible equation of state, to use the ``density-energy'' form. 

We also determine the kernel (in this case, the ``smoothing'') length in the same manner as our other methods (\S~\ref{sec:methods:smoothing}) based on the particle number density; in 3D, this means $(4\pi/3)\,h_{i}^{3}\,n_{i} = N_{\rm NGB}$ (where $n_{i}=\sum W({\bf x}_{j}-{\bf x}_{i},\,h_{i})$). This corresponds to the choice $\tilde{x} = 1$ in \citet{hopkins:lagrangian.pressure.sph}, which we argue there provides the most stable and accurate results (as opposed to a ``constant mass in kernel'' or ``constant energy in kernel'' weighting). We use a cubic spline kernel with $N_{\rm NGB} = 4,\,16,\,32$ in $1,\,2,\,3$ dimensions; this is the standard in most traditional SPH formulations.

In TSPH the density is estimated by kernel-smoothing as: 
\begin{align}
\rho_{i}^{\rm TSPH} &\equiv \bar{\rho}_{i}= \sum_{j} m_{j}\,W({\bf x}_{i}-{\bf x}_{j},\,h_{i})
\end{align}
the pressure is then determined from the density as $P_{i}^{\rm TSPH}=P(\bar{\rho}_{i},\,u_{i})=(\gamma-1)\,\bar{\rho}_{i}\,u$ (for a polytropic equation of state). 

Recall, we need to replace our flux calculations. The mass flux in SPH is identically zero. With the choices above, the momentum and internal energy fluxes derived from the particle Lagrangian (see \citealt{hopkins:lagrangian.pressure.sph}, Eq.~12-13 therein) are
\begin{align}
\frac{d{\bf P}_{i}}{d t} &= -\sum_{j}\,m_{i}\,m_{j}\,\left[ \frac{P_{i}}{\bar{\rho}_{i}^{2}}\,f_{i,j}\,\nabla_{i}W_{ij}(h_{i})
+ \frac{P_{j}}{\bar{\rho}_{j}^{2}}\,f_{j,i}\,\nabla_{i}W_{ij}(h_{j})
 \right]\\
\frac{d E}{d t} &= {\bf v}_{i}\cdot \frac{d{\bf P}_{i}}{d t} 
- \sum_{j}\,m_{i}\,m_{j}\,\left({\bf v}_{i} - {\bf v}_{j} \right)\cdot\left[ \frac{P_{i}}{\bar{\rho}_{i}^{2}}\,f_{i,j}\,\nabla_{i}W_{ij}(h_{i}) \right]
\\
f_{i,j} &= 1 - \frac{1}{m_{j}}\,\left( 
\frac{h_{i}}{n_{i}\,\Ddim}\,\frac{\partial \bar{\rho}_{i}}{\partial h_{i}}
\right)\,\left[1 + \frac{h_{i}}{n_{i}\,\Ddim}\,\frac{\partial n_{i}}{\partial h_{i}} \right]^{-1}\\
\frac{\partial n_{i}}{\partial h_{i}}
&= -\sum_{j}\,\frac{1}{h_{i}}\,\left(\Ddim\,W_{ij}(h_{i}) + u_{ij}\,\frac{\partial W(u)}{\partial u}{\Bigr|}_{u=u_{ij}} \right) \\ 
\frac{\partial \bar{\rho}_{i}}{\partial h_{i}}
&= -\sum_{j}\,\frac{m_{j}}{h_{i}}\,\left(\Ddim\,W_{ij}(h_{i}) + u_{ij}\,\frac{\partial W(u)}{\partial u}{\Bigr|}_{u=u_{ij}} \right)
\end{align}
where we abbreviate $W({\bf x}_{i}-{\bf x}_{j},\,h_{k}) = W_{ij}(h_{k})$, $\Ddim$ is the number of dimensions, and $u_{ij} \equiv | {\bf x}_{j} - {\bf x}_{i} | / h_{i}$. 

We also require artificial diffusion terms in SPH, to handle shocks (the equations above only hold for adiabatic flows). In TSPH this is just artificial viscosity, using the \citet{gingold.monaghan:1983.artificial.viscosity} prescription with a \citet{balsara:1989.art.visc.switch} switch. This contributes an additional term to the equations of motion if and only if particles $i$ and $j$ are approaching, i.e.\ $({\bf v}_{i}-{\bf v}_{j})\cdot ({\bf x}_{i}-{\bf x}_{j}) < 0$, in which case:
\begin{align}
\frac{d{\bf P}_{i}}{d t} &= \sum_{j}\,\alpha_{ij}\,\mu_{ij}\,(c_{ij}-2\,\mu_{ij})\,m_{i}\,m_{j}\,\frac{\nabla_{i}W_{ij}(h_{i})+\nabla_{i}W_{ij}(h_{j})}{\bar{\rho}_{i}+\bar{\rho}_{j}}
\\
\frac{d E_{i}}{d t} &= \frac{1}{2}\,\left({\bf v}_{i} + {\bf v}_{j}\right)\cdot \frac{d{\bf P}_{i}}{d t} \\
\mu_{ij} &= \frac{h_{ij}\,({\bf v}_{i}-{\bf v}_{j})\cdot({\bf x}_{i}-{\bf x}_{j})}
{|{\bf x}_{i}-{\bf x}_{j}|^{2} + 0.0001\,h_{ij}^{2}}\\
c_{ij} &= \frac{c_{s,\,i}+c_{s,\,j}}{2}\ , \ \ \ \ \ \alpha_{ij} = \frac{\alpha_{i} + \alpha_{j}}{2}\ , \ \ \ \ \ h_{ij} = f_{\rm kern}\,\frac{h_{i}+h_{j}}{2}\\
\alpha_{i} &= \frac{\alpha_{\rm av}\,|(\nabla\cdot{\bf v})_{i}|}{|(\nabla\cdot{\bf v})_{i}| + |(\nabla\times{\bf v})_{i}| + 0.0001\,c_{s,\,i}/(f_{\rm kern}\,h_{i})}
\end{align}
where $\alpha_{\rm av}=1$ is constant everywhere, $f_{\rm kern}$ depends on the kernel shape but $=1/2$ for the cubic spline here, and the velocity gradients in the \citet{balsara:1989.art.visc.switch} switch are determined by our standard (least-squares) gradient procedure.

Note that, by virtue of our desire to make this implementation as consistent as possible with the rest of our code, there are already a number of subtle improvements of this method over SPH implementations like in {\small GADGET-3}. Our standard (least-squares) gradient estimators are used for predict steps and for quantities like the \citet{balsara:1989.art.visc.switch} switch; these are substantially more accurate than the usual SPH gradient estimators (based on the kernel gradient). We use our manifestly-conservative adaptive timestepping scheme, instead of relying solely on integration accuracy. We include the neighbor and particle-approach-based timestep limiter, which prevents spurious particle inter-penetration in strong shocks. The smoothing length is based on particle number density (not mass density), reducing errors when there are particles of different masses in the same kernel. Gravity includes fully-conservative adaptive force softening. And we use a Lagrangian-derived density-energy formulation, which is necessary to prevent additional errors whenever the smoothing lengths vary in SPH; compared to methods which use a non-Lagrangian (hence non-conservative) SPH equation of motion, the choice here at least ensures that entropy and energy are simultaneous conserved in adiabatic flows.

\vspace{-0.5cm}
\subsection{``Modern'' SPH (PSPH)}
\label{sec:sph.methods:psph}

Our ``modern'' SPH method builds on the TSPH method, using higher-order kernels, pressure-based formulations of the equations of motion, more accurate gradients, and higher-order switches for dissipation terms.

The method is a ``pressure-energy'' scheme, following \citet{hopkins:lagrangian.pressure.sph}, again with $\tilde{x}=1$; so we follow internal energy, and determine $h_{i}$ in the exact same manner. As noted above, ``pressure-energy'' and ``pressure-entropy'' schemes are essentially equivalent if Lagrangian-derived. However, in a pressure-entropy scheme, because the particle-entropies enter the pressure in a non-linear fashion, radiative cooling (if enabled) must be followed in a somewhat complicated iterative manner to ensure proper energy conservation is maintained; pressure-energy formulations avoid this. 

To reduce the E0 errors, we follow standard practice and increase the number of neighbors to $N_{\rm NGB}=128$ in 3D (our default, though we vary this in the text). This cannot be done using the cubic spline kernel without suffering the pairing instability, so we go to a higher-order (in this case, quintic spline) kernel, as advocated in \citet{dehnen.aly:2012.sph.kernels}; we revert to the cubic spline when we run in PSPH mode with $N_{\rm NGB}=32$. The quintic spline is given by: 
\begin{align}
W\left( q\, , \, h_{i} \right)
&= \frac{3^{7}}{40\,\pi\,h_{i}^{3}}\, \times \\
\nonumber&
\begin{cases}
	{ (1-q)^{5} - 6\,\left( \frac{2}{3}-q \right)^{5} + 15\,\left( \frac{1}{3}-q \right)^{5}} & { (0 \le q < \frac{1}{3})} \\
	{ (1-q)^{5} - 6\,\left( \frac{2}{3}-q \right)^{5}} & { (\frac{1}{3} \le q < \frac{2}{3})} \\
	{ (1-q)^{5}} & { (\frac{2}{3} \le q < 1)} \\
	{ 0} & { (q \ge 1)}
\end{cases}
\end{align} 
where $q\equiv {|{\bf x}-{\bf x}_{i}|}/{h_{i}}$. Note that we have also experimented with the Wendland kernels in \citet{dehnen.aly:2012.sph.kernels}; for the neighbor number here, both their experiments and ours find essentially identical behavior to the quintic spline kernel.

In PSPH, both the density and pressure are estimated by kernel smoothing:
\begin{align}
\rho_{i}^{\rm PSPH} &\equiv \bar{\rho}_{i}= \sum_{j} m_{j}\,W_{ij}(h_{i}) \\ 
P_{i}^{\rm PSPH} &\equiv \bar{P}_{i} = \sum_{j} (\gamma-1)\,m_{j}\,u_{j}\,W_{ij}(h_{i})
\end{align}
The momentum and energy equations become
\begin{align}
\nonumber \frac{{\rm d}{\bf P}_{i}}{{\rm d}t} &= -\sum_{j=1}^{N}\,(\gamma-1)^{2}m_{i}\,m_{j}\,u_{i}\,u_{j}\,
{\Bigl[}
\frac{f_{ij}}{\bar{P}_{i}}\,\nabla_{i}W_{ij}(h_{i}) + \frac{f_{ji}}{{\bar{P}_{j}}}\,\nabla_{i}W_{ij}(h_{j})
{\Bigr]}  \\ 
\nonumber\frac{{\rm d}E_{i}}{{\rm d}t} &={\bf v}_{i}\cdot \frac{d{\bf P}_{i}}{d t} 
- \sum_{j=1}^{N} (\gamma-1)^{2}\,m_{i}\,m_{j}\,u_{i}\,u_{j}\,\frac{f_{ij}}{\bar{P}_{i}}\,({\bf v}_{i}-{\bf v}_{j})\cdot\nabla_{i}W_{ij}(h_{i}) \\ 
\nonumber f_{ij} &= 1 - {\Bigl(}\frac{h_{i}}{\Ddim(\gamma-1)\,\bar{n}_{i}\,m_{j}\,u_{j}}\,\frac{\partial \bar{P}_{i}}{\partial h_{i}}{\Bigr)}\,
{\Bigl[}{1 + \frac{h_{i}}{\Ddim\,{n}_{i}}\,\frac{\partial {n}_{i}}{\partial h_{i}}}{\Bigr]}^{-1}\\
\nonumber \frac{\partial n_{i}}{\partial h_{i}}
&= -\sum_{j}\,\frac{1}{h_{i}}\,\left(\Ddim\,W_{ij}(h_{i}) + u_{ij}\,\frac{\partial W(u)}{\partial u}{\Bigr|}_{u=u_{ij}} \right) \\ 
\frac{\partial \bar{P}_{i}}{\partial h_{i}}
&= -\sum_{j}\,\frac{(\gamma-1)\,m_{j}\,u_{j}}{h_{i}}\,\left(\Ddim\,W_{ij}(h_{i}) + u_{ij}\,\frac{\partial W(u)}{\partial u}{\Bigr|}_{u=u_{ij}} \right)
\end{align}

We again require artificial diffusion terms. For the artificial viscosity, we use the higher-order switch from \citet{cullen:2010.inviscid.sph}, as updated in \citet{hopkins:2013.fire}. Once again this contributes if and only if particles $i$ and $j$ are approaching (i.e.\ $({\bf v}_{i}-{\bf v}_{j})\cdot ({\bf x}_{i}-{\bf x}_{j}) < 0$):
\begin{align}
\nonumber \frac{d{\bf P}_{i}}{d t} &= \sum_{j}\,\alpha_{ij}\,\mu_{ij}\,(c_{ij}-\beta_{b}\,\mu_{ij})\,m_{i}\,m_{j}\,\frac{\nabla_{i}W_{ij}(h_{i})+\nabla_{i}W_{ij}(h_{j})}{\bar{\rho}_{i}+\bar{\rho}_{j}}
\\
\nonumber \frac{d E_{i}}{d t} &= \frac{1}{2}\,\left({\bf v}_{i} + {\bf v}_{j}\right)\cdot \frac{d{\bf P}_{i}}{d t} \\
\nonumber \mu_{ij} &= \frac{({\bf v}_{i}-{\bf v}_{j})\cdot({\bf x}_{i}-{\bf x}_{j})} {|{\bf x}_{i}-{\bf x}_{j}|}\, ,\ \ 
c_{ij} = \frac{c_{s,\,i}+c_{s,\,j}}{2}\, , \ \ \alpha_{ij} = \frac{\alpha_{i} + \alpha_{j}}{2} \\
%c_{ij} &= \frac{c_{s,\,i}+c_{s,\,j}}{2}\ , \ \ \ \ \ \alpha_{ij} = \frac{\alpha_{i} + \alpha_{j}}{2}\ , \ \ \ \ \ h_{ij} = f_{\rm kern}\,\frac{h_{i}+h_{j}}{2}\\
\nonumber \alpha_{i} &= {\rm MAX}\left(\frac{|\beta_{\xi}\,\xi_{i}^{4}\,(\nabla\cdot {\bf v})_{i}|^{2}\,\alpha_{0,\,i}(t)}{|\beta_{\xi}\,\xi_{i}^{4}\,(\nabla\cdot {\bf v})_{i}|^{2} + {\rm Trace}({\bf S}_{i}\,{\bf S}_{i}^{T})}\, ,\, \alpha_{\rm min}\right) \\ 
 \xi_{i} &\equiv 1 - \frac{1}{\bar{\rho}_{i}}\sum_{j}\,{\rm SIGN}[(\nabla\cdot{\bf v})_{j}]\,m_{j}\,W_{ij}(h_{i}) 
%\alpha_{t} &= \alpha_{\rm max}\,f_{\rm kern}\,h_{i}
\end{align}
where $\alpha_{0,\,i}(t)$ is set for each particle each timestep by evaluating $\alpha_{\rm tmp}$:
\begin{align}
\nonumber \alpha_{\rm tmp} &= 
\begin{cases}
	{\displaystyle 0\ \ \ \ \  \hfill { ((d[\nabla\cdot{\bf v}]/dt)_{i} \ge 0\ , \ \ \ {\rm or}\ \ \ \  (\nabla\cdot{\bf v})_{i} \ge 0)}} \\
	{\displaystyle \, \, }\\
	{\displaystyle \frac{\alpha_{\rm max}\,|(d[\nabla\cdot{\bf v}]/dt)_{i}|}{|(d[\nabla\cdot{\bf v}]/dt)_{i}| + \beta_{c}\,c_{s,\,i}^{2}/(f_{\rm kern}\,h_{i})^{2}}
	\ \ \ \ \ \hfill { ({\rm otherwise})}} 
\end{cases}\\
\nonumber \\
\alpha_{0,\,i}(t+\Delta t) &= 
\begin{cases}
	{\alpha_{\rm tmp}\ \ \ \ \ \hfill { (\alpha_{\rm tmp} \ge \alpha_{0,\,i}(t))}} \\
%	{\alpha_{\rm tmp} + (\alpha_{0,\,i}(t)-\alpha_{\rm tmp})\,\exp{\left(-\beta_{\rm d}\,\Delta t\,|v_{{\rm sig},\,i}|/(f_{\rm kern}\,h_{i}) \right)}
	{\displaystyle \, \, }\\
	{\alpha_{\rm tmp} + (\alpha_{0,\,i}(t)-\alpha_{\rm tmp})\,e^{-\beta_{\rm d}\,\Delta t\,|v_{{\rm sig},\,i}|/(2\,f_{\rm kern}\,h_{i}) }}\\
	{\ \ \ \ \ \hfill { (\alpha_{\rm tmp} < \alpha_{0,\,i}(t))}} \\
\end{cases}
\end{align}
where after considerable experimentation we find the best mix of accuracy and stability with $\alpha_{\rm min}=0.02$, $\alpha_{\rm max}=2$, $\beta_{c}=0.7$, $\beta_{d}=0.05$, $\beta_{\xi}=1$, $\beta_{b}=1$, ${\bf S}$ is the shear tensor (constructed from our standard velocity derivatives as described in \citealt{cullen:2010.inviscid.sph}), $f_{\rm kern}=1/3$ for the quintic spline kernel, and $(d[\nabla\cdot{\bf v}]/dt)_{i}$ is evaluated using the method in \citet{cullen:2010.inviscid.sph}, which is essentially the same as our least-squares gradient estimation here, applied to the acceleration as well as velocity to obtain the time derivative. Note that there are some very small modifications of this scheme from \citet{cullen:2010.inviscid.sph}; these are motivated by our experiments in \citet{hopkins:2013.fire} and the tests in \citet{hu:2014.psph.galaxy.tests}; they allow the viscosity to be reduced more rapidly in high-shear regions when the flows are complicated (leading to improvements in turbulence), better maintain stability in strong shocks by enforcing a finite $\alpha_{\rm min}$ (necessary in some of our tests), and enhance the detection of weak shocks from high-redshift cosmological structure formation. 

We also include an artificial conductivity term, following \citet{price:2008.sph.contact.discontinuities} with the improvements in \citet{read:2012.sph.w.dissipation.switches}. This enters just the energy equation between $i$ and $j$, when $\tilde{v}_{s} > 0$ where $\tilde{v}_{s} \equiv c_{s,\,i}+c_{s,\,j} - 3\,({\bf v}_{i}-{\bf v}_{j})\cdot ({\bf x}_{i}-{\bf x}_{j})/|{\bf x}_{i}-{\bf x}_{j}|$. We then have
\begin{align}
\frac{d E_{i}}{d t} &= \alpha_{\rm C}\,\sum_{j}\,m_{i}\,m_{j}\,\alpha_{ij}\,\tilde{v}_{s}\,(u_{i} - u_{j})\,
\frac{|P_{i}-P_{j}|}{P_{i}+P_{j}}
\frac{\nabla_{i}W_{ij}(h_{i})+\nabla_{i}W_{ij}(h_{j})}{\bar{\rho}_{i}+\bar{\rho}_{j}}
\end{align}
Here $\alpha_{ij}$ is a similar switch to the above for artificial viscosity; in fact we find essentially identical results using the same switch for both (which means the conductivity only is applied, correctly, in crossing flows). And we set the global constant $\alpha_{C}$ to a relatively conservative value $\alpha_{C}\approx 0.25$; together with the limiters in the equation this leads to greatly reduced diffusion compared to some prescriptions for conductivity in the literature (e.g.\ \citealt{shen:2010.metal.diffusion.model.galaxies}).

\vspace{-0.5cm}
\section{Integration, Timestep Criteria, and Adaptive Time-Stepping}
\label{sec:methods:timesteps}

As noted in the text, the integration scheme here closely follows that in {\small AREPO}, itself similar to that in {\small GADGET-3}. We refer to \citet{springel:arepo}, \S~7, for details, but review the scheme briefly here so we can note some differences in our implementation.

Numerically, the time-integration scheme follows Eq.~\ref{eqn:timestep.basic}, which is second-order accurate. For details see \citep{colella:1990.upwind.methods.conservation.laws.multid,stone:2008.athena}. However, for almost all interesting problems, there is a large dynamic range and so using a global timestep imposes a severe resolution penalty. Therefore we use individual timesteps, following the standard principle in $N$-body problems, SPH simulations \citep{katz:treesph,springel:gadget} and AMR codes. Specifically, we follow the elegant method described in \citet{springel:arepo} (\S~7.2). We discretize allowed timestep sizes into a power-of-two hierarchy (i.e.\ the timestep of particle $i$ is the largest power-of-two subdivision smaller than the locally-calculated timestep criterion), so that there is a nested hierarchy of timestep bins (i.e.\ on a given timestep $\Delta t_{i}$, all particles with timesteps $\Delta t_{j} \le \Delta t_{i}$ are synchronized). Conserved quantities exchanged between cells are always updated synchronously: whenever a flux is calculated between two adjacent particles $i$ and $j$, if the timesteps differ, the conserved quantities on {\em both sides} of the face are updated according to the flux calculation on the smaller of the two timesteps (this is akin to ``sub-cycling'' in AMR codes). Whenever a cell completes its timestep, its primitive variables are updated based on the accumulated change in its conserved quantities; cells which are between timesteps (but interacting with ``active'' cells) use their old primitive variables and gradients (calculated from their last active timestep), drifted according to our predict-step to the synchronous time, to compute the relevant quantities for flux estimation. On the completion of any sub-timestep, the timestep size for any particle may be updated; however, the particle must move into a timestep bin which will be active on the next sub-step. In other words, a the timestep can always be reduced after a sub-step is completed, but it can only increase when the steps are appropriately synchronized. Because the scheme strictly deals with pairwise exchanges of conserved fluid quantities, it remains {\em manifestly} conservative of mass, momentum, and energy, even while adaptive/individual timesteps are used. We can always enforce a global timestep if desired; however, like \citet{springel:arepo}, we actually find that this method performs as accurately in practice (at vastly lower computational cost) to the use of a global timestep, if we use an appropriate particle-based timestep criterion.

For hydrodynamics, we employ a local Courant-Fridrisch-Levy (CFL) timestep criterion as in Eq.~\ref{eqn:CFL}. This, together with our individual timestep method above, is sufficient to ensure numerical stability. However, it is still possible that in e.g.\ a high-velocity shock, particles with a very long timestep will suddenly have neighbor particles with a much shorter timestep; we do not want the conserved properties of the long-timestep particles to change ``too much'' before they are active, since this would entail a loss of accuracy. We therefore combine the CFL condition with the criterion from \citet{saitoh.makino:2009.timestep.limiter} as updated in \citet{durier:2012.timestep.limiter,hopkins:2013.fire}. Specifically, if a particle has an interacting neighbor with a timestep more than two timebins ``lower'' (factor $4$); it is ``woken up'' and moved to the shortest active timebin, with its subsequent timestep criteria re-evaluated. We find this is more than sufficient to completely eliminate artificial particle ``inter-penetration'' even in extreme situations (e.g.\ a collision of two initially well-separated cold blobs with relative $|{\bf v}|/c_{s} \sim 10^{6}$). 

Finally, whenever other physics are present, we note that there are other timestep criteria; the minimum over all such criteria is always chosen. For example, with gravity present, we use a kinematic criterion as in \citet{power:2003.nfw.models.convergence}, $\Delta t_{\rm kin} = ({2\,\alpha_{k}\,\epsilon_{\rm grav} / |{\bf a}|})^{1/2}$, where $|{\bf a}|$ is the total acceleration and $\epsilon_{\rm grav}$ is the force softening (typical $\alpha_{k}\lesssim 0.01$), along with standard restrictions based on the particle displacement relative to the local particle separation; other physics like diffusion, conduction, nuclear reaction networks, chemistry, cooling, star formation and stellar evolution, black hole accretion, and radiative transfer all add their own restrictions.

\vspace{-0.5cm}
\section{Details of the Gravity \&\ Cosmology Algorithms}
\label{sec:methods:gravity}

\subsection{Coupling Hydrodynamics to Gravity}
\label{sec:methods:gravity:hydrocoupling}

When gravity is present, it modifies the Euler equations with the addition of source terms for momentum and energy. As reviewed by \citet{springel:arepo} (\S~5 therein), it has historically been challenging to couple these terms accurately to finite-volume grid codes. The ``standard'' approach from \citet{muller:1995.grid.code.gravity.problems,truelove:1998.gmc.frag} in most current finite-volume codes leads to first-order energy conservation errors (i.e.\ these schemes are no longer actually second-order), which do not converge in time (i.e.\ cannot be controlled with finer timesteps); these can corrupt the true solution on problems like the Evrard test and exacerbate the spurious noise and gravitational heating in problems like the Santa Barbara cluster. A method which explicitly conserves total energy can be constructed (if we use single global timesteps and explicitly calculate the gravitational potentials, which is very expensive), but this leads to catastrophic errors in the thermal energy evolution in gravity-dominated flows which totally corrupt or crash many of the test problems here. A more accurate coupling, which is spatially and temporally second-order accurate in the integration of gravity and conservation of energy (conservation is exact for a linear gravitational force law), is given by retaining the definition of $E_{i}$ as the ``hydrodynamic total energy,'' $E_{i} = U_{i} + {\bf P}_{i}\cdot {\bf P}_{i}/(2\,m_{i})$, and adding the appropriate gravitational work corrections as 
\begin{align}
\label{eqn:grav1} {\bf P}_{i}^{(n+1)} &= {\bf P}_{i}^{(n)} + \Delta {\bf P}_{\rm hydro} \\ 
\nonumber & - \frac{\Delta t}{2}
\left[ m_{i}^{(n)}\,\nabla_{i}\,\Phi^{(n)} + m_{i}^{(n+1)}\,\nabla_{i}\,\Phi^{(n+1)} \right] \\ 
\label{eqn:grav2} E_{i}^{(n+1)} &= E_{i}^{(n)} + \Delta E_{\rm hydro} \\ 
\nonumber & - \frac{\Delta t}{2}
{\Bigl[} m_{i}^{(n)}\,{\bf v}_{i}^{(n)}\cdot\nabla_{i}\,\Phi^{(n)} + m_{i}^{(n+1)}\,{\bf v}_{i}^{(n+1)}\cdot\nabla_{i}\,\Phi^{(n+1)} {\Bigr]}  \\
\nonumber &- \frac{\Delta t}{2}\nabla_{i}\,\Phi^{(n)} \cdot \sum_{j}({\bf x}_{i}-{\bf x}_{j})^{(n)}\,\frac{d m_{ij}}{d t}^{(n)}\\
\nonumber &- \frac{\Delta t}{2}\nabla_{i}\,\Phi^{(n+1)} \cdot \sum_{j}({\bf x}_{i}-{\bf x}_{j})^{(n+1)}\,\frac{d m_{ij}}{d t}^{(n+1)}
\end{align}
where $\Delta {\bf P}_{\rm hydro}$ is the hydrodynamical momentum flux ($\Delta E_{\rm hydro}$ the hydrodynamical energy flux), $-\nabla_{i}\,\Phi$ is the standard gravitational acceleration calculated from our force solver at the beginning $(n)$ and end $(n+1)$ of the timestep, and $dm_{ij}/dt$ is the contribution to the mass flux calculated between particles $j$ and $i$ at the same timestep (this is the extra gravitational work owing to mass fluxes, which vanishes when they are not present). 

Because of the symmetry of the kernel function, it is straightforward to show that the coupling of gravity to the particle at its center (i.e.\ calculating $\nabla_{i}\,\Phi$ at the particle coordinates ${\bf x}_{i}$, as for a point mass), as is done in SPH codes, is accurate to second-order in the gravitational forces (third-order in the potential), as good as the $N$-body solver itself for collisionless particles.

\vspace{-0.5cm}
\subsection{Adaptive, Fully-Conservative Gravitational Softening}
\label{sec:methods:gravity:softening}

To calculate the gravitational forces generated {\em by} gas, (as opposed to coupling to the gas), we need to solve for the potential field. In $N$-body codes, this requires us to decompose the mass density field into the contributions ``from'' each particle/cell. For our new methods this is straightforward: based on our definition of the volume partition, the differential mass at a point ${\bf x}$ which is ``associated with'' a given particle $i$ is just
\begin{align}
\label{eqn:dm.grav}
d m_{i} = d^{\Ddim}{\bf x}\,\rho({\bf x})\,\frac{W({\bf x}-{\bf x}_{i},\,h({\bf x}))}{\omega({\bf x})} 
\end{align}
If we further use our definition of $h^{\Ddim}\,n = h^{\Ddim}\,\omega \propto N_{\rm NGB}$, and note that our normalization of $W$ to integrate to unity requires $W \propto h^{-D}\,\tilde{w}(|{\bf x}-{\bf x}_{i}|/h({\bf x}))$ where $\tilde{w}$ is the dimensionless ``shape function'' set by the kernel choice, then we see $d m_{i} \propto d^{\Ddim}{\bf x}\,\rho({\bf x})\,\tilde{w}(|{\bf x}-{\bf x}_{i}|/h({\bf x}))$. Expanding this to leading order in the gradients of $\rho$ and $h$, we can re-write it $d m_{i} \approx m_{i}\,W({\bf x}-{\bf x}_{i},\,h_{i})\,d^{\Ddim}{\bf x}$. In other words, to leading order, the density distribution ``associated with'' a given particle has the same functional form as the kernel centered at the particle. More exactly, it is straightforward to show that the potential computed by integrating Poisson's equation with the source in Eq.~\ref{eqn:dm.grav} is identical at leading order to the potential we would obtain using just the particle-centered kernel mass distribution (and it is second-order accurate if we average in spherical shells). 

This suggests that we should treat the particles/cells in the $N$-body code as standard $N$-body particles ``softened'' by the kernel function with the same kernel length; as noted in the text (\S~\ref{sec:methods:gravity:brief}) this automatically ensures the resolution of gravity and hydrodynamics are equal and that the two use the same, consistent definition of the volume partition (unlike the case in most grid-based schemes). 

With that assumption, then, the force softening is straightforward. For the cubic spline kernel in 3D, for example, we have 
\begin{align}
W\left( q\, , \, h_{i} \right)
&= \frac{8}{\pi\,h_{i}^{3}}
\begin{cases}
	{ 1 + 6\,q^{2}\,(q-1)} & { (0 \le q < \frac{1}{2})} \\
	{ 2\,(1-q)^{3}} & { (\frac{1}{2} \le q < 1)} \\
	{ 0} & { (q \ge 1)}
\end{cases}\\
\Phi_{i}\left( q\, , \, h_{i} \right) &\equiv G\,m_{i}\,\phi\left(q,\,h_{i}\right)\\
\phi\left( q\, , \, h_{i} \right)
&= -\frac{1}{q\,h_{i}}
\begin{cases}
	{ \frac{14}{5}q-\frac{16}{3}q^{3}+\frac{48}{5}q^{5}-\frac{32}{5}q^{6}} & { (0 \le q < \frac{1}{2})} \\
	{ -\frac{1}{15}+\frac{16}{5}q-\frac{32}{3}q^{3}+16 q^{4}}\\
	{ \ \ \ \ \ \ \ \ \ \ -\frac{48}{5}q^{5} + \frac{32}{15}q^{6}} & { (\frac{1}{2} \le q < 1)} \\
	{ 1} & { (q \ge 1)}
\end{cases}
\end{align} 
where $q\equiv {|{\bf x}-{\bf x}_{i}|}/{h_{i}}$. Note that on scales $>h$, the potential and force are exactly that of a Newtonian point mass. It is common practice to compare softenings to an ``equivalent'' Plummer sphere softening; for this choice, the Plummer equivalent softening is $\sim h/3$. 

Because the kernel lengths change, we must be careful to maintain energy and momentum conservation correctly. Fortunately, \citet{price:2007.lagrangian.adaptive.softening} show how the appropriate terms can be rigorously derived from the particle Lagrangian to maintain manifest conservation with variable softening lengths. If we define the gravitational self-energy of a system of gas cells as 
\begin{align}
E_{\rm grav} = \frac{1}{2}\,\sum_{i,\,j}\,G\,m_{i}\,m_{j}\,\phi(r_{ij},\,h_{j})
\end{align}
and then follow the same derivation as \citet{price:2007.lagrangian.adaptive.softening}, accounting for the fact that we use a slightly different convention to determine the kernel length $h$, we then obtain the acceleration equation for gravity: 
\begin{align}
m_{i}\,\frac{d {\bf v}_{i}}{d t} &= - \nabla_{i} E_{\rm grav} \\
\nonumber &= -\sum_{j}\,\frac{G\,m_{i}\,m_{j}}{2}\,\left(\frac{\partial \phi(r,\,h_{i})}{\partial r}{\Bigr|}_{r_{ij}} + \frac{\partial \phi(r,\,h_{j})}{\partial r}{\Bigr|}_{r_{ij}} \right)\,\frac{{\bf r}_{ij}}{r_{ij}} \\ 
%\nonumber &-\frac{1}{2}\sum_{j}\sum_{k}\,G\,m_{i}\,m_{j}\,\frac{\partial \Phi(r_{jk},\,h)}{\partial h}{\Bigr|}_{h_{j}}\,\nabla_{i}h_{j}\\
\nonumber &\ \ \ \ -\sum_{j}\,\frac{G}{2}\left( \zeta_{i}\,\frac{\partial W(r,\,h_{i})}{\partial r}{\Bigr|}_{r_{ij}}
+\zeta_{j}\,\frac{\partial W(r,\,h_{j})}{\partial r}{\Bigr|}_{r_{ij}} \right)\,\frac{{\bf r}_{ij}}{r_{ij}} \\
\zeta_{a} &\equiv m_{a}\,\frac{h_{a}}{n_{a}\,\Ddim}\,\frac{1}{\Omega_{a}}\,\sum_{b}\,m_{b}\,\frac{\partial{\phi(r_{ab},\,h)}}{\partial h}{\Bigr|}_{h=h_{a}}\\
\Omega_{a} &\equiv 1 + \frac{h_{a}}{n_{a}\,\Ddim}\,\frac{\partial n_{i}}{\partial h_{i}}\\
\nonumber &= 1 - \frac{h_{a}}{n_{a}\,\Ddim}\,\sum_{b}\left(\frac{r_{ab}}{h_{a}}\,\frac{\partial W(r,\,h_{a})}{\partial r}{\Bigr|}_{r_{ab}} + \frac{\Ddim}{h_{a}}\,W(r_{ab},\,h_{a})\right)
\end{align}
where ${\bf r}_{ij} = {\bf x}_{i} - {\bf x}_{j}$ (so $\partial \phi/\partial r = h^{-1}\,\partial \phi/\partial q$). The first part (in $\partial \phi/\partial r$) here is just the usual $\nabla \Phi$ term, assuming $h$ is fixed; the force between each particle pair is symmetrized so forces are always equal and opposite.\footnote{This trivially follows in the point-mass $r > h$ regime, but is non-trivial at small radii if the $h$ differ. As discussed in \citet{price:2007.lagrangian.adaptive.softening}, there are other possible choices for how this can be symmetrized; for example by using a mean $h$ between the particles/cells, or simply using the larger of the two $h$ (as is done in {\small GADGET}). However, these introduce dis-continuous changes in the gravitational softening, which break conservation in the $\nabla_{i}h_{j}$ terms.} The second part (the $\zeta$ terms in $\partial W/\partial r$) accounts for the fact that the $h$ change, so by moving the particles we change $h$, which in turn does additional work by changing the potentials for the particles (note that this term always vanishes outside the kernel radius, as it should). 

We emphasize that these equations for the gravitational forces are {\em manifestly} conservative, based on our definition of the potential and its relationship to the kernel. Moreover, for any given mass distribution which follows the kernel $W$, the equations are also {\em exact}, to {\em all} orders in $h$ and all orders in the gradients of the density field, etc. Of course, these properties assume direct summation, which is carried out in our code for all particles with overlapping kernels or within the same tree node; but the accuracy and conservation properties of the long-range forces (where particle-particle forces are purely Newtonian) depend on the approximate tree-gravity scheme used (see \citealt{springel:gadget} for more details). 

Finally, this suggests that we can and should do the same for other (non-gaseous) volume-filling fluids, like dark matter, when they appear in our simulations. This amounts to making the same assumption as for gas: that the volume is partitioned according to a kernel function. This removes the otherwise ad-hoc assumption that dark matter particles simply represent soft-edged ``spheres''; it simply requires that we determine a dark matter kernel length in an identical manner to how we determine it for gas. This is trivial: we define the kernel function $W$ and $n_{i}^{\rm DM} = \omega_{i}^{\rm DM}$ the same as gas, but based on the neighbor {\em dark matter} particles (not gas particles), and then apply the same constraint equation $h_{{\rm DM},\,i}^{\Ddim}\,n_{{\rm DM},\,i} \propto N_{\rm NGB}^{\rm DM}$. This then produces an identical set of equations for gravity, for these particles, with one important caveat: when gas and/or dark matter particles interact via gravity within their kernel radii, the $\zeta$ terms {only} appear if both particles $ij$ are the same type. This is because these terms stem from how the mutual motion of the particles will change the kernel lengths (based on inter-particle distances for the same types).\footnote{The generalization to multi-fluid simulations is similarly straightforward. However, in some simulations, there are particle types -- for example, stars in cosmological simulations -- for which it is not obvious that an adaptive force softening is always appropriate (as opposed to a fixed force softening), because they do not necessarily represent a volume-filling fluid.}

\vspace{-0.5cm}
\subsection{Cosmological Integration}
\label{sec:methods:gravity:cosmology}

In an expanding (or contracting) Universe, the Euler equations plus gravity, as described above, must be modified to account for the expansion of space; we do so in the same manner as {\small GADGET} \citep[see][for details]{springel:gadget}. If we adopt the useful co-moving coordinates ${\bf x}_{c} \equiv a^{-1}\,{\bf x}$ (our internal position variable; where $a=1/(1+z)$ is the scale factor at redshift $z$), $\rho_{c} = a^{3}\,\rho$, $P_{c}=(\gamma-1)\,\rho_{c}\,u$, and define the peculiar velocity ${\bf v}_{p} = a\,d{\bf x}_{c}/dt$ (distinct from the physical velocity between two points with physical separation ${\bf r}$, ${\bf v}_{\rm phys} = H(a)\,{\bf r} + {\bf v}_{p}$, and canonical momentum ${\bf p}_{c} = m\,a^{2}\,d{\bf x}/d t$, which we use as our internal velocity variable), and hydrodynamic energy $E=U+m\,{\bf v}_{p}\cdot{\bf v}_{p}/2$ (our internal energy variable), together with the Hubble expansion $H(a) = \dot{a}/a$, comoving gravitational potential $\Phi_{c}$, and gradient operator $\nabla_{c}$ acting on co-moving coordinates, then the Euler equations take on a ``normal'' form with simple source terms. Specifically, per \citet{springel:arepo}, the appropriate ``conserved'' variables in hydrodynamic interactions are still mass, momentum, and energy, and the appropriate surface/volume integrals still yield the ``standard'' fluxes of the Euler equations in the non-cosmological frame, evaluated using the appropriate physical fluid quantities and cell effective areas. 

So we do not need to change our Riemann solution method except to make sure the units are correctly converted into physical before the flux computation and back to co-moving after. Quantities like the particle positions and momenta do not need to be explicitly evolved under the influence of the Hubble expansion since the distance units are co-moving (and velocity units are canonical momenta); the cosmological ``integration'' is perfect in this sense. However, we work with the ``hydrodynamic energy,'' as described above, and evolve the internal energy (rather than e.g.\ the particle entropy). This means we do need to include a source term for energy evolution, which is implemented with our usual second-order time integration scheme as gravity (so that the prediction is appropriately included in obtaining time-centered fluxes), giving  
\begin{align}
E_{i}^{(n+1)} = E_{i}^{(n)} - \Delta t\,[H(a^{(n)})\,E_{i}^{(n)} + H(a^{(n+1)})\,E_{i}^{(n+1)}] + ...
\end{align}
where the ``$...$'' represents the usual non-cosmological terms. If we evolve the internal energy directly, the integration can be done exactly, by simply summing the exact adiabatic expansion of the gas under two ``pure cosmological'' half-timesteps with evolution from $a^{(n)}\rightarrow a^{(n+1/2)}$ and $a^{(n+1/2)}\rightarrow a^{(n+1)}$.

\end{appendix}

\end{document}